\documentclass[notitlepage,10pt]{iopart}

\expandafter\let\csname equation*\endcsname\relax
\expandafter\let\csname endequation*\endcsname\relax

\usepackage{cite}
\usepackage{lipsum}
\usepackage{graphicx}
\usepackage{verbatim}
\usepackage{subfigure}
\usepackage{amsmath,amssymb,amsfonts,eufrak,mathrsfs}
\usepackage{color}
\usepackage[bookmarks,
                bookmarksopen = true,
                bookmarksnumbered = true,
                linktocpage,
                colorlinks = true,
                linkcolor = blue,
                urlcolor  = blue,
                citecolor = blue,
                anchorcolor = green,
                hyperindex = true,
                hyperfigures]
                {hyperref}
\allowdisplaybreaks

\graphicspath{{./Figures/}}

\begin{document}

\title[Jet quenching and medium response]{Jet quenching and medium response in high-energy heavy-ion collisions: a review}

\author{Shanshan Cao$^1$ and Xin-Nian Wang $^{2,3}$}

\address{$^1$ Department of Physics and Astronomy, Wayne State University, Detroit, MI 48201, USA}
\address{$^2$ Key Laboratory of Quark and Lepton Physics (MOE) and Institute of Particle Physics, Central China Normal University, Wuhan 430079, China}
\address{$^3$ Nuclear Science Division, Lawrence Berkeley National Laboratory, Berkeley, CA 94720, USA}
\ead{xnwang@lbl.gov}
\ead{sshan.cao@gmail.com}
\vspace{10pt}
\begin{indented}
\item[]December 2019
\end{indented}

\begin{abstract}
Jet quenching has been used successfully as a hard probe to study properties of the quark-gluon plasma (QGP) in high-energy heavy-collisions at both the Relativistic Heavy-Ion Collider (RHIC) and the Large Hadron Collider (LHC). We will review recent progresses in theoretical and phenomenological studies of jet quenching with jet transport models. Special emphasis is given to effects of jet-induced medium response on a wide variety of experimental measurements and their implications on extracting transport properties of the QGP in heavy-ion collisions.
\end{abstract}

%
\vspace{2pc}
\ioptwocol
\thispagestyle{empty}

\tableofcontents
\markboth{}{}


\section{Introduction}
\label{sec:introduction}

It has been established through quantum chromodynamics (QCD) calculations on lattice \cite{Ding:2015ona} that matter with strong interaction governed by QCD under extreme conditions at high temperature and density will go through a transition from hadronic resonant gas to quark-gluon plasma (QGP). The transition is a rapid crossover for matter with nearly zero net baryon density. Many theoretical model studies indicate the transition becomes a first order at high baryon density~\cite{BraunMunzinger:2008tz} which is still yet to be confirmed through first principle calculations with lattice QCD. While experimental search for the first order phase transition and the existence of a critical endpoint (CEP) is been carried out at the beam energy scan (BES) program at the Relativistic Heavy-Ion Collider (RHIC) \cite{Bzdak:2019pkr}, many of the current experimental efforts at RHIC and the Large Hadron Collider (LHC) focus on exploring and extracting properties of the QGP with negligible baryon density formed in high-energy heavy-ion collisions \cite{Gyulassy:2004zy,Muller:2012zq}. These include the extraction of bulk transport properties such as the shear and bulk viscosities of QGP through the study of collective phenomena~\cite{Romatschke:2009im,Braun-Munzinger:2015hba} and the jet transport coefficient through jet quenching~\cite{Majumder:2010qh,Qin:2015srf}. All of these efforts involve systematic comparisons of experimental data and theoretical calculations with ever more realistic simulations of the dynamical evolution of the QGP in high-energy heavy-ion collisions.

Jet quenching is a collection of phenomena in high-energy heavy-ion collisions caused by interactions between energetic jet partons and the QGP medium~\cite{Bjorken:1982tu,Gyulassy:1990ye}. Because of the large transverse momentum transfer in the hard processes, the cross section of the initial jet production can be calculated with perturbative QCD (pQCD) which has been shown to agree with experimental data on jet production in proton-proton (p+p) collisions~\cite{Eskola:1995cj}. These pQCD calculations of the jet production rate can be extended to proton-nucleus (p+A) collisions within the collinear factorized pQCD and agree with experimental data after the nuclear modification of the parton distribution functions is taken into account~\cite{Albacete:2013ei}. Such calculations for nucleus-nucleus (A+A) collisions can be used as baselines for the initial jet production against which medium modification of the jet production due to jet quenching can be obtained and compared to experimental data. These include suppression of single hadron, dihadron and $\gamma$-hadron spectra as well as single jet, dijet and $\gamma$-jet spectra.

Shortly after its initial production, a hard jet parton has to propagate through the dense QGP medium formed in heavy-ion collisions. During the propagation, it will go through multiple interactions with thermal partons in the QGP and experience both energy loss and transverse momentum broadening before hadronization, leading to jet quenching or suppression of the final jet and hadron spectra. For energetic partons, the energy loss is dominated by radiative processes \cite{Gyulassy:1993hr} and is directly proportional to the jet transport coefficient \cite{Baier:1996sk} which is defined as the averaged transverse momentum broadening squared per unit length of propagation,
\begin{equation}
\label{qhat}
\hat q_a=\sum_{b,(cd)} \int{dq_{\bot }^{2} \, \frac{d\sigma_{ab\rightarrow cd}}{dq_{\bot }^2}}\rho_b \, q_{\bot }^{2},
\end{equation}
where $\sigma_{ab\rightarrow cd}$ is the partonic scattering cross section between the jet parton $(a)$ and the medium parton $(b)$ with local density $\rho_b$ (which should contain a degeneracy factor or number of degrees of freedom) and the sum is over the flavor of the thermal parton ($b$) and all possible scattering channels $a+b\rightarrow c+d $ for different flavors of the final partons ($c$, $d$). This jet transport coefficient can be defined for jet parton propagating in a non-thermal medium and can be related to the general gluon distribution density of the QGP medium~\cite{CasalderreySolana:2007sw} and used to characterize a variety of properties of the QGP at finite temperature~\cite{Majumder:2007zh} as well as cold nuclear matter~\cite{Osborne:2002st}. Jet quenching therefore can be used as a hard probe of the QGP properties in high-energy heavy-ion collisions.

\begin{figure}[tbp]
\centering
    \includegraphics[width=0.4\textwidth]{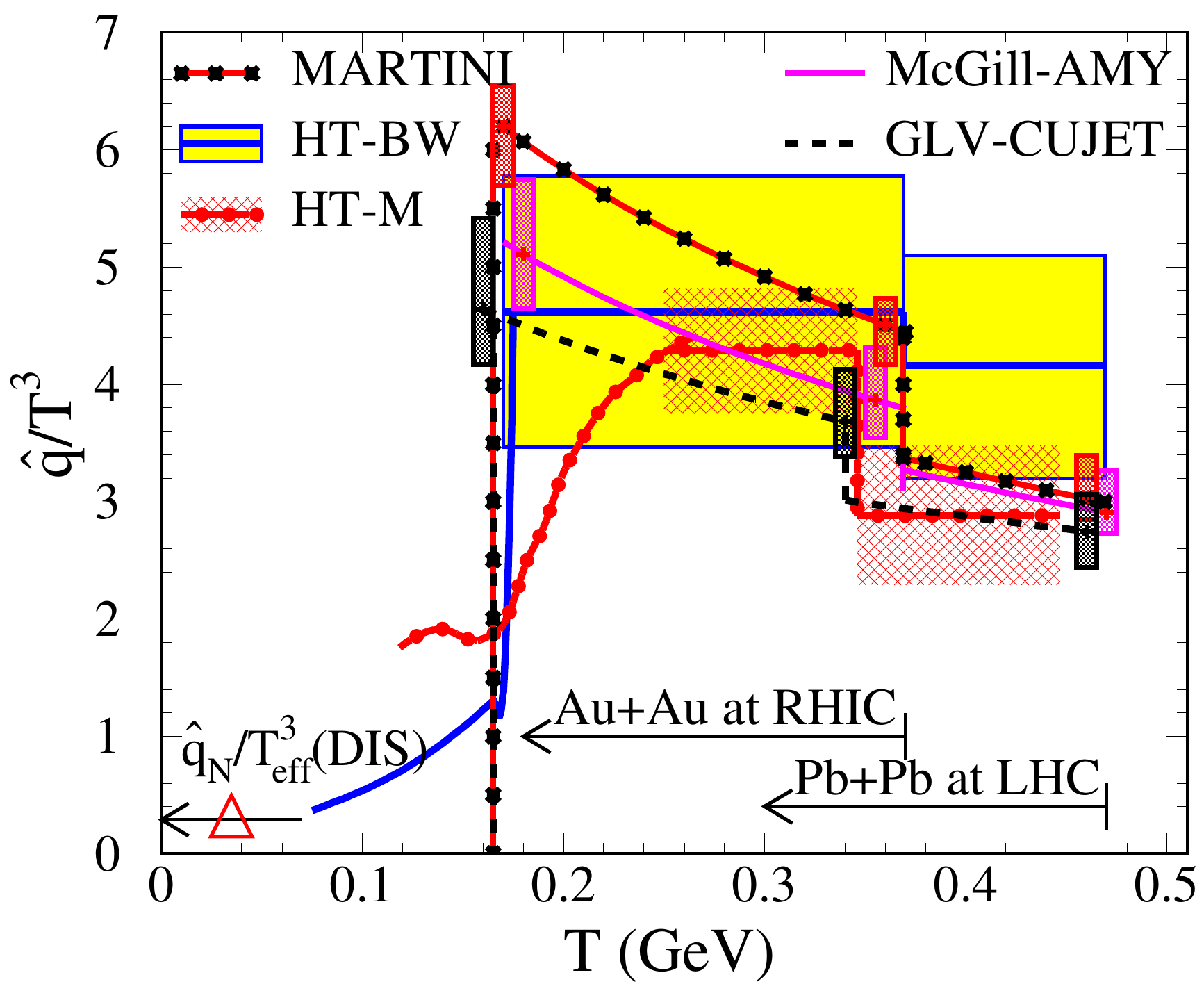}
    \caption{(Color online) The scaled jet transport parameter $\hat q/T^3$ as a function of the initial temperature $T$ for an initial quark jet with energy $E=10$~GeV at the center of the most central A+A collisions at an initial time $\tau_0=0.6$~fm/$c$ extracted by the JET Collaboration from the experimental data on hadron suppression. See Ref.~\cite{Burke:2013yra} for details.} 
    \label{fig:qhat}
\end{figure}

Since the first discovery of the jet quenching phenomena at RHIC~\cite{Gyulassy:2004zy,Wang:2004dn} and the confirmation at the LHC~\cite{Jacobs:2004qv,Muller:2012zq}, the focus of jet quenching studies have shifted to the precision extraction of jet transport coefficients~\cite{Eskola:2004cr,Armesto:2009zi,Chen:2010te} through systematical comparisons of experimental data with theoretical calculations and phenomenological analyses that incorporate up-to-date theories of parton propagation and interaction with the dense medium, the state-of-the-art model for the dynamical evolution of the QGP medium and modern statistical analysis tools. Such a systematical study of the suppression of the single hadron spectra in heavy-ion collisions at both RHIC and LHC has been carried out by the JET Collaboration~\cite{Burke:2013yra}. The extracted jet transport coefficient in the range of the initial temperature achieved at RHIC and LHC, as shown in Fig.~\ref{fig:qhat}, is about 2 orders of magnitude larger than that inside a cold nucleus, indicating the extremely high temperature and density of the QGP formed in the high-energy heavy-ion collisions.

While further phenomenological studies on jet quenching in the final hadron spectra will continue to shed light on the detailed properties of the QGP medium, in particular with the combined experimental data on hadron spectra and azimuthal anisotropy~\cite{Kumar:2017des,Shi:2018vys,Andres:2019eus}, the experimental data on fully reconstructed jets available at RHIC and the LHC will provide unprecedented opportunities to explore the QGP properties with jet quenching. The large rate of jet production in a wide range of transverse momentum at the LHC will allow studies on the energy dependence of jet quenching and jet transport coefficient in correlation with many other aspects of heavy-ion collisions, such as the initial geometry through the final state multiplicity and the bulk hadron azimuthal anisotropy, the energy flow of associated soft hadrons and the flavor tagging. The medium modification of jet substructures such as the jet transverse profile and the jet fragmentation function can further elucidate the QGP properties through parton-medium interaction and parton transport.

Since fully constructed jets are collections of collimated clusters of hadrons within a given jet cone, they consist of not only leading hadrons from the fragmentation of jet partons but also hadrons from the bulk medium within the jet cone that become correlated with jets through jet-medium interactions. Therefore, the medium modification of jets in heavy-ion collisions is not only determined by the energy loss of leading partons but also influenced by the transport of the energy lost by jet partons in the dynamically evolving medium through radiated gluons and recoil partons in the form of jet-induced medium excitation. It is therefore necessary to include the transport of recoil partons in the study of jet quenching with the full jet suppression and medium modification. This will be the focus of this review on the recent developments of jet quenching studies.

The outline of this review is as follows. We will first give a brief review and update on theories of parton propagation and medium-induced energy loss in Sec.~\ref{sec:theory} in a generalized high-twist framework, followed by a review of models of jet evolution in the QGP medium in Sec.~\ref{sec:models} with emphasis on the multi-scale nature of the evolution processes. Then, we will review the description of the jet-induced medium excitation through the transport of recoil partons and the hydrodynamic response to jet-medium interaction in Sec.~\ref{sec:medium_response}. Effects of jet quenching and medium response on hadron and jet spectra and jet substructures will be discussed in Secs.~\ref{sec:hadron_spectra}, \ref{sec:jet_spectra} and~\ref{sec:jet_substructures}. A summary and outlook will be given in Sec.~\ref{sec:summary}.

\section{Parton energy loss in QCD medium}
\label{sec:theory}

During the propagation of an energetic parton inside a QCD medium, it will experience interaction with constituents of the medium involving both elastic and inelastic processes. Within pQCD, one can calculate both elastic and inelastic energy loss of the propagating parton inside the QCD medium when it interacts with the medium partons. 

\subsection{Elastic energy loss}

The elastic energy loss can be calculated from the elastic scattering cross section~\cite{Bjorken:1982tu,Thoma:1990fm},
\begin{equation}
\frac{dE^a_{\rm el}}{dx}=\int\frac{d^3k}{(2\pi)^3}\sum_{b,(cd)} \gamma_bf_b(k) \int{dq_{\bot }^{2} \, \frac{d\sigma_{ab\rightarrow cd}}{dq_{\bot }^2}} \, \nu,
\end{equation}
where $f_b(k)$ is the phase-space distribution (Fermi-Dirac for quarks and Bose-Einstein for gluons) for thermal partons with the degeneracy $\gamma_b$, and $\nu$ is the energy transfer from the jet parton to the thermal medium parton which depends on the propagating parton energy $E$, the thermal parton energy $\omega$ and the transverse momentum transfer $q_\perp$. In the high energy limit $E\gg T$, the $t$-channel gluon and quark scattering cross sections can be approximated by their small angle limits
\begin{equation}
\frac{d\sigma_{ab}}{dq_\perp^2}  = 2\frac{C_2(a)C_2(b)}{N_c^2-1}\frac{2\pi\alpha_\mathrm{s}^2}{(q_\perp^2+\mu^2_{\rm D})^2},
\label{eq-small-el}
\end{equation}
where $C_2$ is the quadratic Casimir factor of parton $a$/$b$ -- $C_2(q)=C_F=(N_c^2-1)/2N_c=4/3$ for quarks and $C_2(g)=C_A=N_c=3$ for gluons.
The color factors $C_{ab}= 2C_2(a)C_2(b)/(N_c^2-1)$ are $C_{gg}=9/4$, $C_{qg}=1$ and $C_{qq}=4/9$ for gluon-gluon, gluon-quark and quark-quark scattering, respectively.  This cross section is collinearly divergent at small angle or transverse momentum which can be regularized by the Debye screening mass $\mu_{\rm D}^2=6\pi\alpha_{\rm s} T^2$ in a medium with 3 flavors of massless quarks. Under the small angle approximation for the parton-parton scattering, $\nu\approx q_\perp^2$, the elastic parton energy loss is
\begin{equation}
\frac{dE^a_{\rm el}}{dx}\approx C_2(a)\frac{3\pi}{2}\alpha_{\rm s}^2 T^2\ln\left(\frac{2.6ET}{4\mu_{\rm D}^2}\right).
\end{equation}

\subsection{Inelastic parton energy loss}

The first calculation of radiative parton energy loss was attempted by Gyulassy and Wang~\cite{Gyulassy:1993hr,Wang:1994fx} within a static potential model of multiple parton interaction for an energetic parton in a thermal medium. However, an important contribution from multiple interaction between the gluon cloud and the medium was not taken into account in this first study which turns out to be the most important in the soft radiation limit. The first calculation that took into account such interaction was by Baier-Dokshitzer-Mueller-Peigne-Schiff (BDMPS)~\cite{Baier:1994bd,Baier:1996sk,Baier:1996kr} in the limit of soft gluon radiation. Later on Zakharov using the path integral formalism \cite{Zakharov:1996fv}, Gyulassy-Levai-Vitev (GLV)~\cite{Gyulassy:1999zd,Gyulassy:2000fs} and Wiedemann~\cite{Wiedemann:2000za} using the opacity expansion method have also calculated the radiative energy loss for a fast parton in a QCD medium. All these calculations assumed the medium as a collection of static scattering centers as in the first study by Gyulassy and Wang~\cite{Gyulassy:1993hr}.  A separate method to calculate parton evolution and energy loss in a thermal QGP medium was carried out by Arnold, Moore and Yaffe (AMY)~\cite{Arnold:2002ja} within the framework of hard thermal loop resummed pQCD at finite temperature. In the high-twist (HT) approach~\cite{Guo:2000nz,Wang:2001ifa,Zhang:2003yn,Schafer:2007xh}, the twist-expansion technique was used in a collinear factorization formalism to calculate the medium-induced gluon spectrum and parton energy loss. Within this approach the information of the medium is embedded in the high-twist parton correlation matrix elements. The relations between these different studies of parton propagation and energy loss have been discussed in detail in Refs.~\cite{Arnold:2008iy,CaronHuot:2010bp,Mehtar-Tani:2019tvy} and numerically compared in Ref.~\cite{Armesto:2011ht}. In the latest study referred to as the SCET\textsubscript{G} formalism~\cite{Ovanesyan:2011xy,Ovanesyan:2011kn}, the soft collinear effective theory (SCET) is supplemented with the Glauber modes of gluon field for parton interaction between a fast parton and static scattering centers. In this subsection, we will briefly review the calculation of induced gluon spectra and parton energy loss within the high-twist framework and its connection to the results in the opacity expansion approach.

\subsubsection{Vacuum bremsstrahlung}

\begin{figure}
\begin{center}
  \includegraphics[scale=0.85]{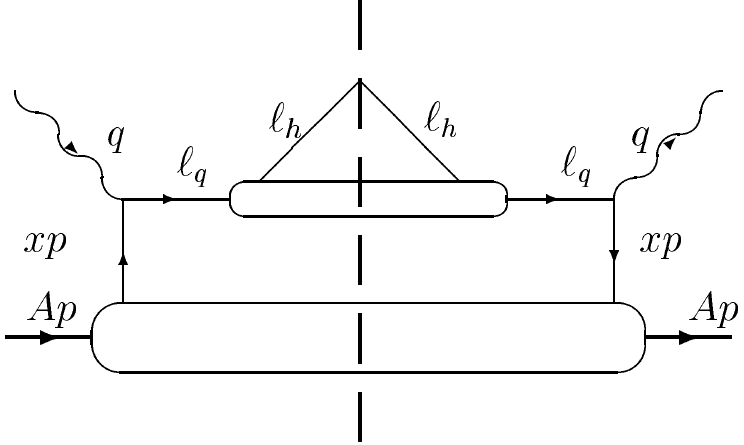}
  \caption{The lowest order and leading twist contribution to the hadronic tensor in the SIDIS process.}
  \label{fig:dis1}
  \end{center}
\end{figure}

Consider the semi-inclusive deeply inelastic scattering (SIDIS) of a lepton off a large nucleus,
\begin{equation}
e(l_1) + A(p) \rightarrow e(l_2) + h(l_h) +\mathcal{X}. \nonumber
\end{equation}
 The cross section of this SIDIS process can be expressed as 
\begin{equation}
d \sigma = \frac{e^4}{2s}\frac{\sum_q e_q^2}{q^4}\int \frac{d^4 l_2}{(2\pi)^4} 2\pi \delta(l_2^2) L_{\mu\nu} W^{\mu\nu},
\end{equation}
where $s =  (l_1 + p)^2$ is the invariant center-of-mass energy squared for the lepton-nucleon scattering, $p$ is the four-momentum per nucleon in the large nucleus with atomic number $A$ and $q$ is the four-momentum of the intermediate virtual photon. The leptonic tensor is 
\begin{equation}
L_{\mu\nu} = \frac{1}{2}{\rm Tr}[\gamma\cdot l_1 \gamma_{\mu} \gamma\cdot l_2  \gamma_{\nu}  ],
\end{equation}
and the semi-inclusive hadronic tensor is,
\begin{equation}
\begin{split}
E_{l_h} \frac{d W^{\mu\nu}}{d^3 l_h} &=\int d^4y e^{-iq\cdot y}  \nonumber \\
&\times \sum_\mathcal{X}  \langle A| J^{\mu}(y)|\mathcal{X}, h \rangle \langle h, \mathcal{X}|J^{\nu}(0)|A\rangle,
\end{split}
\end{equation}
 where $J^{\mu}(0)=\sum_q\bar{\psi}_q(0)\gamma^{\mu}\psi_q(0)$ is the hadronic vector current. 
 The four-momentum of the virtual photon and the initial nucleon are $q=[-Q^2/2q^-, q^-,\vec{0}_{\perp}]$ and $p=[p^+,0,\vec{0}_{\perp}]$, respectively.
The lowest order (LO) contribution to the semi-inclusive hadronic tensor is from the process as shown in Fig.~\ref{fig:dis1} where a quark from the target nucleus is stuck by the virtual photon through a single scattering and fragments into a hadron,
\begin{equation}
\frac{d W^{\mu\nu}_{S(0)}}{d z_h} = \int dx f_q^A(x) H^{\mu\nu}_{(0)}(x) D_{q\rightarrow h}(z_h),
\end{equation}
where the nuclear quark distribution function is  defined as,
 \begin{equation}
 f_q^A(x) = \int \frac{dy^-}{2\pi} e^{-ixp^+y^-} \frac{1}{2} \langle A|\bar{\psi}_q(y^-)\gamma^+ \psi_q(0)|A \rangle ,
 \label{eq:dis0}
  \end{equation}
 and the quark fragmentation function is,
 \begin{eqnarray}
D_{q \rightarrow h} (z_h) = & \dfrac{z_h}{2}\sum_{S}\int \dfrac{dy^+}{2\pi} e^{il_h^-y^+/z_h} \nonumber\\ 
&\hspace{-0.05in}\times {\rm Tr}\left[\dfrac{\gamma^{-}}{2}\langle 0|\psi(y^+)|h,\mathcal{S}\rangle \langle  \mathcal{S},h|\bar{\psi}(0)|0 \rangle\right].
\label{eq:qfrag}
\end{eqnarray}
The hard partonic part is,
\begin{eqnarray}
H^{\mu\nu}_{(0)} (x)& =  \frac{1}{2} {\rm Tr}[\gamma\cdot p \gamma^{\mu} \gamma \cdot (q+xp)\gamma^{\nu}  ] \nonumber \\
&\times \frac{2\pi}{2p^+q^-} \delta(x-x_{\rm B}), \;\;  x_{\rm B} = \frac{Q^2}{2p^+q^-},
\end{eqnarray}
where $x_{\rm B}$ is the Bjorken variable. The momentum fraction of the struck quark is $x=x_{\rm B}$ in this LO process. The fraction of the light-cone momentum carried by the observed hadron with momentum $l_h$ is $z_h = l_h^-/q^-$. 

\begin{figure}
\begin{center}
  \includegraphics[scale=0.75]{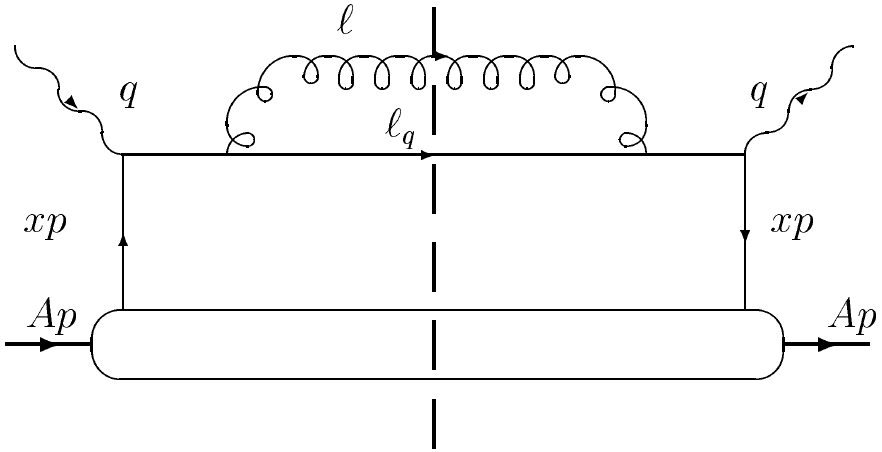}
  \caption{The next-to-leading order and leading twist contribution to the hadronic tensor in the SIDIS process.}
  \label{fig:dis2}
  \end{center}
\end{figure}

At the next-to-leading order (NLO), the struck quark can have vacuum gluon bremsstrahlung after the photon-quark scattering as shown in Fig.~\ref{fig:dis2}. The hard partonic part of this process is
\begin{eqnarray}
&H^{\mu\nu}_{(1)}(x,p,q,z)  =  H^{\mu\nu}_{(0)}(x) \frac{\alpha_\mathrm{s}}{2\pi}  \int  \frac{d l_{\perp}^2}{l_{\perp}^2} P_{q\rightarrow qg}(z), \\
&P_{q\rightarrow qg}(z) = C_F \frac{1+z^2}{1-z},
\end{eqnarray}
where $P_{q\rightarrow qg}(z)$ is the quark splitting function and $z = l_q^-/q^-$ is the quark momentum fraction after the gluon bremsstrahlung.
Considering hadrons from the fragmentation of both the final quark and gluon, the NLO correction to the hadronic tensor is then
\begin{eqnarray}
\frac{d W^{\mu\nu}_{S(1)}}{d z_h} = &\int dx  f_q^A(x)   H^{\mu\nu}_{(0)}(x)  \frac{\alpha_\mathrm{s}}{2\pi}  \int_{z_h}^{1}\frac{dz}{z} \int \frac{d l_{\perp}^2}{l_{\perp}^2}  \\
&\hspace{-0.6in}\times \left[ P_{q\rightarrow qg}(z)  D_{q\rightarrow h}(\frac{z_h}{z}) + P_{q\rightarrow qg}(1-z)  D_{g\rightarrow h}(\frac{z_h}{z})\right], \nonumber
\label{eq:dw1q}
\end{eqnarray}
where the gluon fragmentation function is defined as,
 \begin{eqnarray}
 D_{g \rightarrow h} (z_h)   &=  - \dfrac{z_h^2}{2l_h^-}  \sum_{\mathcal{S}}\int \dfrac{dy^+}{2\pi} e^{il_h^- y^+/z_h} \nonumber \\
  &\times \langle  0|F^{-\alpha}(y^+)|h,\mathcal{S} \rangle \langle  \mathcal{S},h|F^{-}_{\quad \alpha}(0)|0 \rangle.
  \label{eq:gfrag}
 \end{eqnarray}
 
At NLO, one also has to include the virtual corrections to the hadronic tensor,
\begin{eqnarray}
\dfrac{dW^{\mu \nu}_{S(v)}}{dz_h}  &=& -\int dx f_q^A(x)   H^{\mu\nu}_{(0)} (x) \dfrac{\alpha_\mathrm{s}}{2\pi}    \nonumber \\
& \times& \int_0^1 dz  \int \frac{d l_{\perp}^2}{ l_{\perp}^2} P_{q\rightarrow qg}(z) D_{q \rightarrow h} (z_h)   , 
\end{eqnarray}
which can also be obtained from the unitarity requirement on the radiative process.

When summed together, the radiative corrections can be organized into the renormalized fragmentation function $D_{q \rightarrow h} (z_h, \mu^2)$
in the hadronic tensor,
\begin{equation}
\dfrac{dW^{\mu \nu}_{S}}{dz_h} = \int dx f_q^A(x)  H^{\mu\nu}_{(0)}  D_{q \rightarrow h} (z_h, \mu^2),
\end{equation} 
\begin{eqnarray}
&D_{q \rightarrow h} (z_h, \mu^2) =  D_{q \rightarrow h} (z_h) 
+ \frac{\alpha_\mathrm{s}}{2\pi} \int_{z_h}^{1} \frac{dz}{z} \int_0^{\mu^2} \frac{d l_{\perp}^2}{l_{\perp}^2} \\
&\hspace{0.1in}\times  \left[ P_{q\rightarrow qg}^+(z)D_{q\rightarrow h}(\frac{z_h}{z})+P_{q\rightarrow qg}(1-z) D_{g\rightarrow h}(\frac{z_h}{z}) \right], \nonumber\\
&P_{q\rightarrow qg}^+(z)=C_F\left[\frac{1+z^2}{(1-z)_{+}}   + \frac{3}{2}\delta(1-z)\right],
\end{eqnarray}
where the $+$ function is defined as
\begin{equation}
\int_0^1 dz \frac{F(z)}{(1-z)_+}\equiv \int_0^1 dz \frac{F(z)-F(1)}{(1-z)},
\end{equation}
for any function $F(z)$ which is sufficiently smooth at $z=1$. Note that the above renormalized fragmentation function is free of infrared divergence due to the cancellation between the radiative and virtual corrections and it satisfies the DGLAP equation~\cite{Gribov:1972ri,Dokshitzer:1977sg,Altarelli:1977zs}.

\subsubsection{Medium induced gluon spectra}

When the initial jet production occurs inside a medium in DIS off a large nucleus or amid the formation of the dense QCD matter in high-energy heavy-ion collisions, the jet parton will have to interact with other partons as it propagates through the medium. Such final state interaction with medium partons will lead to both elastic energy loss as well as inelastic energy loss through induced gluon radiation or quark-anti-quark pair production. Induced gluon bremsstrahlung turns out to be the most dominant energy loss mechanism. 

For a parton propagating in a medium with large path length, one has to consider many multiple scatterings and the corresponding induced gluon bremsstrahlung. Taking into account of the Landau-Pomeranchuk-Migdal (LPM) interference \cite{Landau:1953gr,Migdal:1956tc} effect, one can calculate the energy loss (per unit length) in the limit of soft gluon radiation which has a unique energy and length dependence. For a medium with finite length, one can consider an expansion in the opacity of the medium as seen by the propagating parton and the leading order contribution comes from processes with one secondary scattering after the initial parton production and the interference between zero and two secondary scattering. This is also equivalent to the twist expansion in the high-twist approach. 

\begin{figure}
\begin{center}
  \includegraphics[scale=0.75]{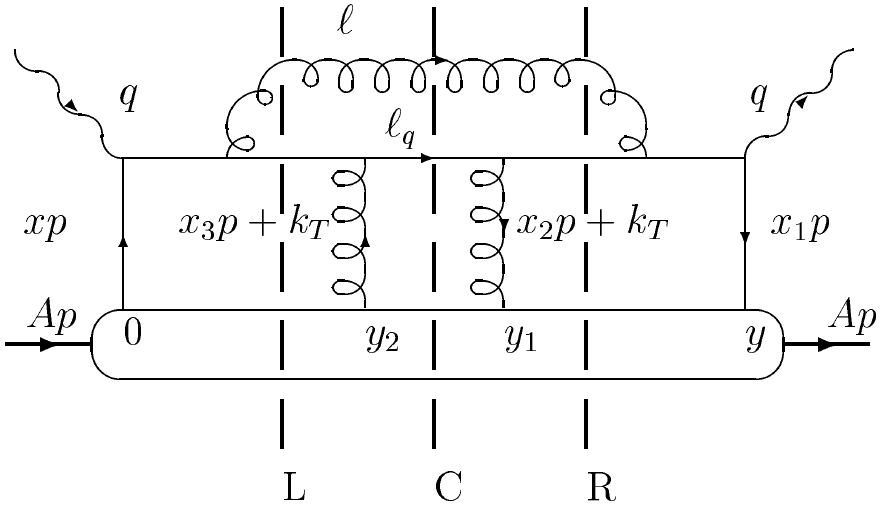}
  \includegraphics[scale=0.75]{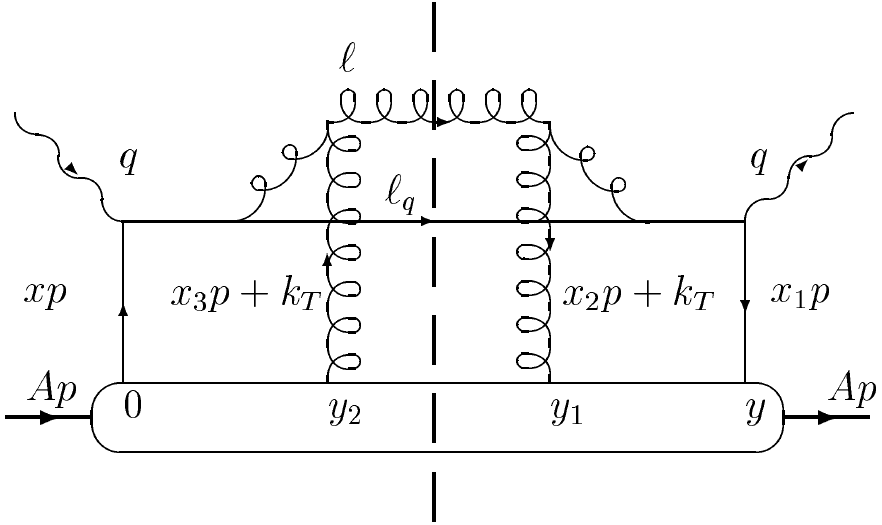}
  \caption{Two example Feynman diagrams for induced gluon radiation induced by double scattering that contribute to the hadronic tensor in the SIDIS process.}
  \label{fig:dis3}
  \end{center}
\end{figure}

Under the opacity or twist expansion, the leading contribution to medium-induced gluon spectra is from processes in which the jet parton undergoes a secondary scattering with the medium after its initial production. In the DIS off a large nucleus, these correspond to gluon radiation induced by a secondary scattering of the quark with another gluon from a different nucleon inside the nucleus after it is knocked out by the hard photon-quark scattering as illustrated by the two example Feynman diagrams in Fig.~\ref{fig:dis3}. There are generally two types of processes represented by these Feynman diagrams. The first one corresponds to final-state gluon radiation induced by the hard photon-quark scattering after which the final quark or gluon have a secondary scattering with another gluon from the nucleus. In the second type of processes, the gluon radiation is induced by the secondary scattering between the final quark and the nucleus after the initial photon-quark interaction. The interferences between these two types of processes give rise to the 
LPM interference effect limiting the gluon radiation whose formation time is longer than the distance between hard photon-quark scattering and the secondary scattering. 

We only illustrate two symmetrical processes with different cuts (left, central and right) in Fig.~\ref{fig:dis3}. One should also consider asymmetrical cut diagrams from interferences of different radiative processes (initial and final radiation from quark and radiation from gluon lines). In recent studies, radiative gluon spectrum from double scattering in DIS off a large nucleus has been calculated \cite{Zhang:2018kkn,Zhang:2018nie,Zhang:2019toi} in which both finite momentum fraction of the gluon and energy-momentum transfer from the medium are taken into account. Under the static-interaction approximation (neglecting energy-momentum transfer from the medium), the final gluon spectrum induced by a secondary scattering with a medium gluon from a nucleon at $(y^-_1,\vec y_\perp)$ after the initial photon-quark scattering at $(y^-,\vec y_\perp)$ can be expressed as, 
\begin{eqnarray}
\dfrac{d N_g }{d l_{\perp}^2 dz}&= \dfrac{\pi \alpha_\mathrm{s}^2}{N_c}\dfrac{1+(1-z)^2}{z}    
\int \frac{d y^-  d^2y_\perp}{A} \rho_A(y^-,\vec{y}_{\perp}) \nonumber\\
&\hspace{-0.3in}\times \int  \dfrac{d ^2k_{\perp}}{(2\pi)^2}  \int d y_1^-
\rho_A(y_1^-,\vec{y}_{\perp})  \left\lbrace  \frac{C_F}{(\vec{l}_{\perp}-z\vec{k}_{\perp})^2} -  \frac{C_F}{{l}_{\perp}^2} \right. \nonumber\\
&\hspace{-0.3in}+C_A \left[  \frac{2}{(\vec{l}_{\perp}-\vec{k}_{\perp})^2} - \frac{\vec{l}_{\perp}\cdot(\vec{l}_{\perp}-\vec{k}_{\perp})}{l_{\perp}^2(\vec{l}_{\perp}-\vec{k}_{\perp})^2}
\right. \nonumber\\
&\hspace{-0.3in} \left. -  \frac{(\vec{l}_{\perp} - \vec{k}_{\perp})\cdot (\vec{l}_{\perp} -z\vec{k}_{\perp})}{(l_{\perp}-k_{\perp})^2(\vec{l}_{\perp}-z\vec{k}_{\perp})^2}\right] (1-\cos[x_{\rm LD}p^+y_1^-])  \nonumber \\
 &\hspace{-0.3in} \left.+ \frac{1}{N_c}\left[   \frac{\vec{l}_{\perp}\cdot (\vec{l}_{\perp} -z\vec{k}_{\perp})}{l_{\perp}^2(\vec{l}_{\perp}-z\vec{k}_{\perp})^2}  - \frac{1}{l_{\perp}^2}  \right] \left(1- \cos[x_{\rm L}p^+y_1^-]\right) \right\rbrace    \nonumber \\
 &\times  \frac{\phi(0,\vec{k}_{\perp})}{{k}_{\perp}^2}.
 \label{eq:gluonspectra}
\end{eqnarray}
The nucleon density inside the nucleus is normalized as $\int dy^-d^2y_\perp \rho_A(y^-,\vec y_\perp)=A$.

The medium gluon in the secondary scattering carries transverse momentum $\vec k_\perp$ and $\phi(0,\vec{k}_{\perp})$ is the transverse-momentum-dependent (TMD) gluon distribution function (with momentum fraction $x\approx 0$) inside the nucleon. It can be related to the jet transport coefficient
\begin{equation}
\hat{q}_R(y) \approx \frac{4\pi^2 \alpha_\mathrm{s} C_2(R)}{N_c^2-1} \rho_A(y) \int \frac{d^2 k_{\perp}}{(2\pi)^2}  \phi(0, \vec{k}_{\perp}),
\label{eq:qhat_nucleon}
\end{equation}
for a propagating parton in the color representation $R$.  Note that the effective quark-nucleon scattering cross section in terms of the TMD gluon distribution is
\begin{equation}
\sigma_{qN}=\dfrac{2\pi^2\alpha_\mathrm{s}}{N_c} \int \dfrac{d ^2k_{\perp}}{(2\pi)^2} \frac{\phi(0,\vec{k}_{\perp})}{{k}_{\perp}^2}.
\end{equation}
The following quantity in the above induced gluon spectra,
\begin{equation}
\dfrac{2\pi^2\alpha_\mathrm{s}}{N_c}\rho_A(y^-,\vec y_\perp) \int \dfrac{d ^2k_{\perp}}{(2\pi)^2} \frac{\phi(0,\vec{k}_{\perp})}{{k}_{\perp}^2},
\end{equation}
or $\sigma_{qN}\rho_A$ can also be interpreted as the inverse of quark's mean-free-path due to quark-gluon scattering inside the nucleus. The integration  $\int d y^-  d^2y_\perp/A$ is the average over the photon-quark scattering position inside the nucleus.

The third and fourth terms in the curly bracket correspond to gluon radiation from the initial gluon (with three-gluon vertex) and quark (both initial and final) during the secondary quark-gluon interaction. There are LPM interference effects for these two kinds of induced gluon radiation that limit the radiation within the formation time
\begin{eqnarray}
\tau_{1f}&= \frac{1}{x_{\rm LD}p^+}, x_{\rm LD}=\frac{(\vec l_\perp -\vec k_\perp)^2}{2p^+q^-z(1-z)}; \\
 \tau_{2f}&=\frac{1}{x_{\rm L}p^+}, x_{\rm L}=\frac{l_\perp^2}{2p^+q^-z(1-z)}.
 \end{eqnarray}
The first two terms in the curly bracket are the remaining contributions from the final and initial state gluon radiation from the quark line during the secondary quark-gluon interaction that are not affected by the LPM interference.

In the limit of soft gluon radiation $z\approx 0$, the induced gluon spectrum becomes much simpler,
\begin{eqnarray}
\dfrac{d N_g}{d l_{\perp}^2 dz}&\approx C_A \dfrac{\alpha_\mathrm{s}}{2\pi}\dfrac{1+(1-z)^2}{z}\dfrac{2\pi^2\alpha_\mathrm{s}}{N_c}    
\int \frac{d y^-  d^2y_\perp}{A}  \nonumber\\
&\hspace{-0.3in}\times \int d y_1^- \int  \dfrac{d ^2k_{\perp}}{(2\pi)^2}  \rho_A(y^-,\vec{y}_{\perp})  \rho_A(y_1^-,\vec{y}_{\perp})  \nonumber \\
&\hspace{-0.3in} \times \frac{2\vec{k}_{\perp}\cdot\vec{l}_{\perp}}{l_\perp^2(\vec{l}_{\perp}-\vec{k}_{\perp})^2} \left( 1-\cos[x_{LD}p^+y_1^-]\right)
 \frac{\phi(0,\vec{k}_{\perp})}{{k}_{\perp}^2},
 \label{eq:GLV}
\end{eqnarray}
which is equivalent to the results from GLV calculations.

\subsubsection{High-twist expansion versus GLV result}

In the original calculation of parton energy loss in the high-twist formalism, one first expands the scattering amplitude in the transverse momentum of the initial gluon, assuming it is much smaller than the transverse momentum of the radiated gluon $k_\perp \ll l_\perp$. One can perform the same procedure of collinear expansion  here for the radiative gluon spectra induced by a secondary  scattering. Keeping the following expansion to the quadratic order in $k_\perp$,
\begin{eqnarray}
&\frac{2\vec{k}_{\perp}\cdot\vec{l}_{\perp}}{l_\perp^2(\vec{l}_{\perp}-\vec{k}_{\perp})^2} \left( 1-\cos[x_{LD}p^+y_1^-]\right)
\approx 4\frac{(\vec k_\perp\cdot\vec l_\perp)^2}{l_\perp^6} \nonumber \\
&\times\left( 1-\cos[x_{L}p^+y_1^-]-x_{L}p^+y_1^-\sin[x_{L}p^+y_1^-]\right) +{\cal O}(k_\perp^4), \nonumber
\end{eqnarray}
one can factor out the TMD gluon distribution from the scattering amplitude. The integration of the TMD gluon distribution over $k_\perp$ leads to the quark transport coefficient $\hat q_F$ as defined in Eq.~(\ref{eq:qhat_nucleon}). The radiative gluon spectrum in the soft gluon approximation in Eq.~(\ref{eq:GLV}) becomes
\begin{eqnarray}
\dfrac{d N_g }{d l_{\perp}^2 dz}&=   \int \frac{d y^-  d^2y_\perp}{A} \rho_A(y^-,\vec{y}_{\perp}) \int d y_1^- \hat q_F(y_1^-,\vec y_\perp) \nonumber\\
&\times \frac{C_A \alpha_\mathrm{s}}{\pi} \dfrac{1+(1-z)^2}{z}\frac{1}{l_\perp^4} \left( 1-\cos[x_{L}p^+y_1^-] \right. \nonumber\\
&\left.- x_{L}p^+y_1^-\sin[x_{L}p^+y_1^-]\right),
   \label{eq:gluonspectra_ht}
\end{eqnarray}
which is the original high-twist result when the last term from the derivative of the phase factor in the scattering amplitude is dropped.

In the above collinear approximation one assumes $k_\perp \ll l_\perp$ for the initial gluon. Such an approximation however missed the contribution to the gluon spectrum from the region $\vec k_\perp \approx \vec l_\perp$ when the formation time $\tau_{1f}= 1/(x_{\rm LD}p^+)$ becomes large.  Such contribution can be recovered approximately when the last term in Eq.~(\ref{eq:gluonspectra_ht}) from the derivative of the phase factor in the scattering amplitude is dropped.

Considering the Gyulassy-Wang (GW) static potential model of the scattering centers inside a medium as in the GLV calculation, the cross section of the quark-medium scattering is
given by Eq.~(\ref{eq-small-el}), 
\begin{equation}
\frac{d\sigma}{dk_\perp^2} = \frac{C_2(R)C_2(T)}{N_c^2-1}\frac{4\pi\alpha_\mathrm{s}^2}{(k_\perp^2+\mu^2_{\rm D})^2},
\end{equation}
where $C_2(R)$ and $C_2(T)$ are the quadratic color Casimir factors of the propagating parton and scattering center, respectively, and $\mu_{\rm D}$ is the Debye screening mass. According to the two definitions of the jet transport coefficient $\hat q_R$ in Eqs.~(\ref{qhat}) and (\ref{eq:qhat_nucleon}),
the TMD gluon distribution of the medium under the GW static potential model is then
\begin{equation}
\frac{\phi(0, \vec{k}_{\perp})}{k_{\perp}^2} =  C_2(T)  \frac{4 \alpha_\mathrm{s}}{(k_{\perp}^2+\mu^2_{\rm D})^2}.
\label{eq:static}
\end{equation}

The radiative gluon spectra without the average over the initial quark production point can be expressed as
\begin{eqnarray}
\dfrac{d N_g}{d l_{\perp}^2 dz}&\approx C_A \dfrac{\alpha_\mathrm{s}}{2\pi}\dfrac{1+(1-z)^2}{z}\dfrac{2\pi^2\alpha_\mathrm{s}}{N_c}    
 \nonumber\\
&\hspace{-0.3in}\times \int d y_1^- \int  \dfrac{d ^2k_{\perp}}{(2\pi)^2}  \rho_A(y_1^-,\vec{y}_{\perp}) 
\frac{2\vec{k}_{\perp}\cdot\vec{l}_{\perp}}{l_\perp^2(\vec{l}_{\perp}-\vec{k}_{\perp})^2}   \nonumber \\
&\hspace{-0.3in} \times\left( 1-\cos[x_{\rm LD}p^+y_1^-]\right)
 C_2(T)  \frac{4 \alpha_\mathrm{s}}{(k_{\perp}^2+\mu^2_{\rm D})^2}.
 \label{eq:GLV2}
\end{eqnarray}
One can make a variable change $\vec l^\prime_\perp=\vec l_\perp - \vec k_\perp$ and then integrate over $\vec k_\perp$,
\begin{eqnarray}
\int d^2k_\perp & \frac{k_\perp l^\prime_\perp\cos\phi+k_\perp^2}{(k_\perp^2+{l^\prime}^2_\perp+2k_\perp l\prime_\perp\cos\phi)(k_\perp^2+\mu^2_{\rm D})^2}
\nonumber \\
&=\pi \int_{{l^\prime}^2_\perp}^\infty \frac{dk_\perp^2}{(k_\perp^2+\mu^2_{\rm D})^2}=\frac{\pi}{{l^\prime}^2_\perp+\mu^2_{\rm D}}.
\end{eqnarray}
The radiative gluon spectra can then be expressed as 
\begin{eqnarray}
\dfrac{d N_g}{d l_{\perp}^2 dz}&\approx C_A \dfrac{\alpha_\mathrm{s}}{2\pi}\dfrac{1+(1-z)^2}{z}\dfrac{2\pi\alpha_\mathrm{s}^2}{N_c}    
 \nonumber\\
&\hspace{-0.3in}\times \int d y_1^- \rho_A(y_1^-,\vec{y}_{\perp}) 
C_2(T)  \frac{1}{l_\perp^2(l_\perp^2+\mu^2_{\rm D})}   \nonumber \\
&\hspace{-0.3in} \times\left( 1-\cos[x_{\rm L}p^+y_1^-]\right),
 \label{eq:GLV3}
\end{eqnarray}
after substituting $l^\prime_\perp\rightarrow l_\perp$. Note that the quark transport coefficient in this static potential model is
\begin{equation}
\hat q_F=\frac{2\pi\alpha_\mathrm{s}^2}{N_c}C_2(T)\rho_A\left[\log(\frac{Q^2}{\mu^2_{\rm D}}+1)-\frac{Q^2}{Q^2+\mu^2_{\rm D}}\right].
\end{equation}
The above induced gluon spectrum is equivalent to the original high-twist result [Eq.~(\ref{eq:gluonspectra_ht})] without the term from the derivative to the phase factor when one rescales the jet transport coefficient by a factor of  $2\log(Q^2/\mu^2_{\rm D})$.

The calculation of induced gluon spectrum and parton radiative energy loss considers only the first order in opacity expansion and is only valid for a few number of parton rescattering in the medium. This is the general assumption for the calculation by Gyulassy-Levai-Vitev (GLV) and Wiedemann ~\cite{Gyulassy:1999zd,Gyulassy:2000fs,Wiedemann:2000za}. In the limit of many soft scatterings, one have to take the approach by Baier-Dokshitzer-Mueller-Peigne-Schiff and Zakharov (BDMPS-Z)~\cite{Zakharov:1996fv,Baier:1996kr,Baier:1996sk} when one considers soft gluon radiation as a result of multiple scatterings. All of these studies assume the medium as a series of static scattering centers as in the Gyulassy-Wang (GW) model \cite{Gyulassy:1993hr}. Alternatively,  Arnold, Moore and Yaffe (AMY)~\cite{Arnold:2001ba,Arnold:2002ja} employed the hard thermal loop improved pQCD at finite temperature to calculate the scattering and gluon radiation rate in a weakly coupled thermal QGP medium. In the recent SCET\textsubscript{G} formalism~\cite{Ovanesyan:2011xy,Ovanesyan:2011kn}, the standard soft collinear effective theory (SCET) is supplemented with the Glauber modes of gluon exchange for interaction between a fast parton and a static scattering medium to study multiple parton scattering and medium-induced gluon splitting. The relations between some of these different studies of parton propagation and energy loss have been discussed in detail in Refs.~\cite{Arnold:2008iy,CaronHuot:2010bp,Mehtar-Tani:2019tvy}. 

One should bear in mind that in most of the past calculations of radiative parton energy loss and gluon spectrum, one assumes the eikonal approximation for the fast parton propagation, in which the energy of the propagating parton $E$ and the radiated gluon $\omega$ are considered larger than the transverse momentum transfer $k_{\perp}$ in the scattering: $E,\omega \gg k_{\perp}$.  The energy of the radiated gluon is also considered larger than its transverse momentum $\omega \gg l_{\perp}$ in the collinear radiation. In addition, the soft gluon approximation $E \gg \omega$ is also assumed in BDMPS-Z and GLV studies. In the models of static scattering centers, interactions between the propagating parton and the medium do not involve energy and longitudinal momentum transfer. The elastic scattering, the radiative process and the corresponding energy loss are calculated separately. To improve these calculations, the GLV calculation has been extended beyond the soft radiation approximation~\cite{Blagojevic:2018nve} and the first order in the opacity expansion~\cite{Sievert:2019cwq}, and now with a dynamical medium through the hard thermal loop resummed gluon propagator~\cite{Djordjevic:2008iz}. The HT approach has also been extended to include the longitudinal momentum diffusion~\cite{Majumder:2009ge,Qin:2014mya}. Further improvements such as effects of color (de)coherence, angular ordering~\cite{MehtarTani:2011tz,Armesto:2011ir,Caucal:2018dla} and overlapping formation time in sequential gluon emissions~\cite{Arnold:2015qya} have also been studied.

\section{Jet evolution models}
\label{sec:models}

\subsection{Vacuum parton showers}
\label{subsec:vac_showers}

High energy partons produced via hard collisions usually start with large virtuality scales (or off-shellness). Then they evolve toward their mass shells through successive splittings, each of which reduces the virtualities of the daughter partons compared to the mother. The parton/hadron fragmentation function $D_a (z, Q^2)$ at a given scale $Q^2$ is typically described by the DGLAP evolution equation~\cite{Gribov:1972ri,Lipatov:1974qm,Dokshitzer:1977sg,Altarelli:1977zs}, where $z$ is the fractional momentum ($p^+$ in the light-cone coordinate if not otherwise specified in this review) of the daughter parton/hadron taken from the initial parton with flavor $a$.

The DGLAP evolution equation can be rewritten in a more convenient fashion using the Sudakov form factor~\cite{Hoche:2014rga},
\begin{align}
\label{eq:sudakov}
\Delta _a(&Q_\mathrm{max}^2, Q^2_a) = \prod_i \Delta_{ai} (Q_\mathrm{max}^2, Q^2_a)\\
=&\prod_i \exp \left[-\int\limits_{Q^2_a}^{Q_\mathrm{max}^2}\frac{d{Q}^2}{{Q}^2}\frac{\alpha_\mathrm{s}({Q}^2)}{2\pi}\int\limits_{z_\mathrm{min}}^{z_\mathrm{max}}dzP_{ai}(z,{Q}^2)\right], \nonumber
\end{align}
that represents the probability of no splitting between scales $Q^2_\mathrm{max}$ and $Q^2_a$, where the former denotes the maximum possible virtuality of the given parton. In Eq.~(\ref{eq:sudakov}), $P_{ai}$ is the parton splitting function in vacuum for a particular channel $i$, and $z_\mathrm{min}$ and $z_\mathrm{max}$ are the lower and upper kinematic limits of the fractional momentum $z$. Note that there is no unique prescription for a ``best" choice of the $Q^2_\mathrm{max}$. It depends on different model assumptions or is treated as a parameter to fit hadron/jet spectra in proton-proton collisions. In addition, the values of $z_\mathrm{min}$ and $z_\mathrm{max}$ depend on the definition of $z$ (momentum fraction or energy fraction), as well as the coordinate one uses. More details on these kinematic cuts will be discussed later when different models are presented. 

With this Sudakov form factor, one may construct an event generator to simulate parton showers. For an $a\rightarrow bc$ splitting process, if a random number $r\in (0,1)$ is smaller than $\Delta_a(Q_\mathrm{max}^2, Q_\mathrm{min}^2)$, parton $a$ is considered stable (no splitting happens) with its virtuality set as $Q_a^2=Q^2_\mathrm{min}$.  Here $Q_\mathrm{min}^2$ denotes the minimum allowed virtuality scale of a parton, which is usually taken as 1~GeV$^2$ (nucleon mass scale) in vacuum parton showers. On the other hand, if $r>\Delta_a(Q_\mathrm{max}^2, Q_\mathrm{min}^2)$, the equation $r = \Delta_a(Q_\mathrm{max}^2, Q_a^2)$ is solved to determine the virtuality scale $Q_a^2$ at which parton $a$ splits. The specific channel $i$ through which $a$ splits is then determined using the branching ratios:
\begin{equation}
\label{eq:branching}
\mathrm{BR}_{ai}(Q_a^2)=\int\limits_{z_\mathrm{min}}^{z_\mathrm{max}} dz P_{ai} (z, Q_a^2).
\end{equation}
Through a particular channel, the longitudinal momentum fractions $z$ and $(1-z)$ of the two daughter partons are determined with the splitting function $P_{ai} (z, Q_a^2)$. Meanwhile, $z^2Q_a^2$ and $(1-z)^2Q_a^2$ are used as the new maximum possible virtualities $Q^2_\mathrm{max}$ of the two daughters respectively, with which their actual virtualities $Q_b^2$ and $Q_c^2$ can be calculated again using Eq.~(\ref{eq:sudakov}) with their own flavors. Finally, the transverse momentum of the daughters with respect to the mother reads
\begin{equation}
\label{eq:transverse}
k_\perp^2=z(1-z)Q_a^2-(1-z)Q_b^2-zQ_c^2,
\end{equation}
which completes one splitting process 
\begin{align}
\label{eq:splittingProcess}
\biggl[p^+,\frac{Q_a^2}{2p^+},0 \biggl] &\rightarrow\biggl[zp^+,\frac{Q_b^2+k_\perp^2}{2zp^+},\vec{k}_\perp\biggl] \nonumber\\
&+\biggl[(1-z)p^+,\frac{Q_c^2+k_\perp^2}{2(1-z)p^+},-\vec{k}_\perp\biggl].
\end{align}
One may iterate this process until all daughter partons reach $Q^2_\mathrm{min}$, generating a virtuality-ordered parton shower where the virtuality scales of single partons decrease through the successive splittings. For simplicity, we have neglected the rest mass of partons in the above discussion. To take this into account, one may replace $Q^2_a$ with $M^2_a=Q^2_a+m^2_a$ in Eqs.~(\ref{eq:transverse}) and (\ref{eq:splittingProcess}), where $m_a$ represents the rest mass of parton $a$. 

This formalism has been successfully implemented in the event generators such as \textsc{Pythia}~\cite{Sjostrand:2006za} for proton-proton (p+p) collisions. Note that in \textsc{Pythia}, $z$ is defined as the fractional energy $E_b/E_a$ instead of the fractional light-cone momentum. With this definition, the kinematic cuts of $z$ in Eqs.~(\ref{eq:sudakov}) and (\ref{eq:branching}) read
\begin{align}
\label{eq:cutWithE}
z_\mathrm{max/min}&=\frac{1}{2}\Bigg[1+\frac{M_b^2-M_c^2}{M_a^2}\nonumber\\
\pm &\frac{|\vec{p}_a|}{E_a}\frac{\sqrt{(M_a^2-M_b^2-M_c^2)^2-4M_b^2M_c^2}}{M_a^2}\Bigg].
\end{align}
This can be obtained by solving the momentum $p$ of parton $b/c$ in the rest frame of parton $a$ via $M_a=\sqrt{p^2+M_b^2}+\sqrt{p^2+M_c^2}$ and then boosting it collinearly with $\pm |\vec{p}_a|/E_a$. Before $M_b$ and $M_c$ are determined, one may assume they are zero compared to $M_a$ (or $Q_a$) and use 
\begin{equation}
\label{eq:cutWithES}
z_\mathrm{max/min}=\frac{1}{2}\left(1\pm \frac{|\vec{p}_a|}{E_a}\right)
\end{equation}
in Eqs.~(\ref{eq:sudakov}) and (\ref{eq:branching}).

In \textsc{Pythia}, the initial maximum possible virtuality scale of each parton is set as $Q_\mathrm{max}^2 = 4Q_\mathrm{hard}^2$ by default, where $Q_\mathrm{hard}^2$ is the scale of transverse momentum exchange square of the initial hard scattering. One may modify this pre-factor of 4 using the \textsc{parp(67)} parameter inside the \textsc{Pythia 6} code.

To take into account the coherence effect on parton splittings, \textsc{Pythia} implements the angular-ordering cut on the virtuality-ordered parton showers. After parton $a$ splits, its daughters $b$ and $c$ are also allowed to further split independently. However, if the opening angle of the splitting from $b$ is larger than the angle between $b$ and $c$, the soft gluon ($b'$) emitted from $b$ can interfere with $c$. In other words, the soft gluon emitted at a large angle corresponds to a large transverse wavelength, which cannot resolve the separate color charges of $b$ and $c$. Therefore, it is treated as being emitted from $a$ directly rather than from $b$ independently. To reduce such soft emission at large angle from the daughter partons, \textsc{Pythia} requires the splitting angle of $b$ (same for $c$) is smaller than the splitting angle of $a$. Using Eq.~(\ref{eq:transverse}) with $Q_a \rightarrow M_a$ and $M_b\approx M_c\approx 0$, one obtains
\begin{equation}
\label{eq:angleA}
\theta_a\approx \frac{k_\perp}{E_b} + \frac{k_\perp}{E_c} \approx \frac{1}{\sqrt{z_a(1-z_a)}}\frac{M_a}{E_a},
\end{equation}
where $E_b = z_a E_a$ and $E_c = (1-z_a)E_a$ are applied. Therefore, $\theta_b < \theta_a$ yields
\begin{equation}
\label{eq:angularOrderingCut}
\frac{z_b(1-z_b)}{M_b^2} > \frac{1-z_a}{z_a M_a^2}.
\end{equation}

\subsection{Medium-modified parton showers}
\label{subsec:med_showers_virtual}

While this virtuality-ordered (or angular-ordered) formalism has been generally accepted for generating vacuum parton showers in p+p collisions, different approaches have been developed to modify this formalism to include medium effects in heavy-ion collisions. In general, approaches in the literature can be categorized into two groups: (1) directly modifying the splitting function $P_{ai}(z,Q^2)$ in Eqs.~(\ref{eq:sudakov}) and (\ref{eq:branching}) with medium effects; and (2) applying medium modification on the parton kinematics between vacuum-like splittings. The former includes \textsc{Matter}~\cite{Majumder:2013re,Cao:2017qpx}, \textsc{Q-Pythia}~\cite{Armesto:2009fj} and \textsc{Yajem-fmed}~\cite{Renk:2009nz}; and the latter includes \textsc{Jewel}~\cite{Zapp:2011ya,Zapp:2012ak,Zapp:2013vla}, \textsc{Hybrid}~\cite{Casalderrey-Solana:2014bpa,Casalderrey-Solana:2016jvj,Hulcher:2017cpt} and \textsc{Yajem-rad/drag/de}~\cite{Renk:2008pp,Renk:2009nz,Renk:2010mf,Renk:2013pua}.


Within the first group, the medium-modified splitting function is written as 
\begin{equation}
\label{eq:splitPtot}
P_{ai}(z,Q^2)=P_{ai}^\mathrm{vac}(z)+P_{ai}^\mathrm{med}(z,Q^2),
\end{equation}
where the first term on the right-hand side is the standard vacuum splitting function, and the second term is known as the medium-induced splitting function. This method is expected to be valid when the medium-induced part can be treated as a small correction to the vacuum part. In the \textsc{Matter} model~\cite{Majumder:2013re,Cao:2017qpx}, the latter is taken from the high-twist energy loss calculation~\cite{Guo:2000nz,Majumder:2009ge},
\begin{align}
\label{eq:medP}
P_{ai}^\mathrm{med}&(z,Q^2)=\frac{C_A}{C_2(a)}\frac{P_{ai}^\mathrm{vac}(z)}{z(1-z)Q^2}\nonumber\\
\times&\int\limits_0^{\tau_f^+}d\zeta^+ \hat{q}_a\left(\vec{r}+\hat{n}\frac{\zeta^+}{\sqrt{2}}\right)\left[2-2\cos \left(\frac{\zeta^+}{\tau_f^+}\right)\right].
\end{align}
Note that the color factor for the medium-induced gluon emission is always $C_A$ which is different from $C_2(a)$ in $P_{ai}^\mathrm{vac}(z)$ for gluon emission in vacuum. Here, $\hat{q}_a$ is the parton jet transport parameter that denotes its transverse momentum broadening square per unit time/length due to elastic scatterings with the medium. It depends on the medium properties -- density and flow velocity -- at the location $\vec{r}+\hat{n}\zeta^+/\sqrt{2}$, where $\vec{r}$ is the production point of the given parton and $\hat{n}=\vec{p}/|\vec{p}|$ denotes its propagation direction. In addition, $\tau_f^+=2p^+/Q^2$ is the mean formation time of the splitting process. 

Compared to vacuum parton showers, one needs to track not only the virtuality scale of each parton, but also its spacetime information. For heavy-ion collisions, the production vertices of high-energy partons in initial hard scatterings are usually distributed according to the binary collision points from the Glauber model~\cite{Miller:2007ri}. After that, each parton is assumed to stream freely for $\tau^+\approx \tau_f^+$ between its production and splitting vertex. One may either directly use the mean formation time $\tau^+ = \tau_f^+$ for each splitting, or sample the splitting time using a Gaussian distribution with a mean value of $\tau^+_f$~\cite{Majumder:2013re,Cao:2017qpx}:
\begin{equation}
\label{eq:tau}
\rho(\xi^+)=\frac{2}{\tau^+_f\pi}\exp\left[-\left(\frac{\xi^+}{\tau^+_f\sqrt{\pi}}\right)^2\right].
\end{equation}
The latter introduces additional effects of quantum fluctuation on parton energy loss into event generator simulations. 

In the \textsc{Matter} model, the spacetime profile of the QGP medium is taken from the (2+1)-dimensional viscous hydrodynamic model \textsc{Vishnew}~\cite{Song:2007fn,Song:2007ux,Qiu:2011hf}. The entropy density distribution of the QGP fireball is initialized with the Monte-Carlo Glauber model. The starting time of the hydrodynamical evolution is set as $\tau_0=0.6$~fm and the specific shear viscosity is tuned as $\eta/s$=0.08 to describe the soft hadron spectra at RHIC and the LHC. This hydrodynamic simulation then provides the spacetime distributions of the temperature ($T$), entropy density ($s$) and flow velocity ($u$) of the QGP. During the QGP stage -- after $\tau_0=0.6$~fm and before jet partons exit the medium (a critical temperature $T_\mathrm{c}\approx 160$~MeV is used for identifying the QGP boundary) -- the splitting function in Eq.~(\ref{eq:splitPtot}) contains both vacuum and medium-induced parts. Before and after the QGP stage, however,  it has only the vacuum contribution. As a jet parton travels inside the QGP, its transport coefficient $\hat{q}_a$ in the local rest frame of the fluid cell is calculated using $\hat{q}_{a,\mathrm{local}} = \hat{q}_{a0} \cdot s/s_0$, where a minimal assumption of its proportionality to the local density of scattering centers (or entropy density) is adopted. Here, $s_0$ is a reference point (e.g. $s_0 \approx 96$~fm$^{-3}$ at the center of the QGP fireballs produced in central $\sqrt{s_\mathrm{NN}}=200$~GeV Au+Au collisions at RHIC), and $\hat{q}_{a0}$ is the jet transport parameter at this reference point. The path length integral in Eq.~(\ref{eq:medP}) is calculated in the center-of-mass frame of collisions. Therefore, one should take into account the effects of the local fluid velocity of the expanding medium by using the rescaled jet transport coefficient $\hat{q}_a=\hat{q}_{a,\mathrm{local}}\cdot p^\mu u_\mu/p^0$~\cite{Baier:2006pt} in Eq.~(\ref{eq:medP}). 

Compared to \textsc{Pythia}, \textsc{Matter} simulates parton showers starting from each individual parton rather than from the entire proton-proton collision system. Without the knowledge of the hard scattering scale, \textsc{Matter}~\cite{Cao:2017qpx} uses the initial parton energy square $E^2$ as its maximum possible virtuality scale $Q_\mathrm{max}^2$ to initiate the time-like parton showers. In a more recent version of \textsc{Matter} that is embedded into the \textsc{Jetscape} framework~\cite{Putschke:2019yrg}, $Q_\mathrm{max}^2=p_\mathrm{T}^2/4$ is set by default to describe the hadron/jet spectra in proton-proton collisions.

In addition, unlike \textsc{Pythia}, \textsc{Matter} uses the light-cone momentum $p^+$ to define the momentum fraction $z$ in the splitting function. This leads to a slightly different form of kinematic cuts $z_\mathrm{max/min}$ compared to Eqs.~(\ref{eq:cutWithE}) and (\ref{eq:cutWithES}). The general idea remains the same: the lower and upper limits of $z$ are obtained when parton $b$ is collinear with parton $a$, i.e., $k_\perp^2=0$ in Eq.~(\ref{eq:transverse}). This yields
\begin{align}
\label{eq:cutWithP}
z_\mathrm{max/min}&=\frac{1}{2}\Bigg[1+\frac{M_b^2-M_c^2}{M_a^2}\nonumber\\
\pm &\frac{\sqrt{(M_a^2-M_b^2-M_c^2)^2-4M_b^2M_c^2}}{M_a^2}\Bigg].
\end{align}
Since there is no $|\vec{p_a}|/E_a$ compared to Eq.~(\ref{eq:cutWithE}), one cannot further simplify the expression by taking $M_b=M_c=0$ which will lead to $z_\mathrm{max}=1$ and $z_\mathrm{min}=0$ where the splitting function may be divergent. A natural alternative approximation would be setting $M_b^2=M_c^2=Q_\mathrm{min}^2$, which gives
\begin{align}
\label{eq:cutWithPS}
z_\mathrm{max/min}&=\frac{1}{2}\left[1 \pm \sqrt{1-\frac{4Q_\mathrm{min}^2}{M_a^2}}\right]\nonumber\\
&\approx \frac{1}{2}\left[1 \pm \left(1-2Q_\mathrm{min}^2/M_a^2\right)\right],
\end{align}
i.e., $z_\mathrm{max}=1-Q_\mathrm{min}^2/M_a^2$ and $z_\mathrm{min}=Q_\mathrm{min}^2/M_a^2$. As a simplification for both Eq.~(\ref{eq:cutWithES}) and Eq.~(\ref{eq:cutWithPS}), the rest masses $m_b$ and $m_c$ have been neglected. If necessary, $M^2_{b/c} = Q_\mathrm{min}^2+m^2_{b/c}$ should be applied instead, especially for heavy quarks. 

Similar to \textsc{Matter}, \textsc{Q-Pythia}~\cite{Armesto:2009fj} also introduces medium effects on parton showers by modifying the splitting function in the Sudakov form factor. The \textsc{Q-Pythia} model directly modifies the parton shower routine \textsc{pyshow} in the \textsc{Pythia 6} code by introducing the medium-induced splitting function from the BDMPS energy loss formalism~\cite{Baier:1996sk,Baier:1996kr,Zakharov:1996fv}. In \textsc{Q-Pythia}, $Q_\mathrm{max}^2=2E^2$ is used as the initial maximum possible virtuality, where $E$ is the energy of the parton that initiates the showers. At this moment, \textsc{Q-Pythia} uses simplified models of the QGP medium that is mainly characterized by two parameters: the jet transport coefficient $\hat{q}_a$ and the medium length $L$. 

One more example within this first group of medium modification approaches is \textsc{Yajem-fmed}~\cite{Renk:2009nz}. Similar to \textsc{Q-Pythia}, \textsc{Yajem} is also based on modifying the \textsc{Pythia 6} shower routine \textsc{pyshow}. Although most modes in \textsc{Yajem} belong to the second group that will be discussed below, its \textsc{fmed} mode implements a factor $(1+f_\mathrm{med})$ to enhance the singular part of the splitting function upon the presence of a medium. For example,
\begin{align}
\label{eq:yajem-fmed}
P_{q\rightarrow qg}&(z) = C_F \frac{1+z^2}{1-z} \nonumber\\
&\Rightarrow C_F \left[\frac{2(1+f_\mathrm{med})}{1-z}-(1+z)\right].
\end{align}
The factor $f_\mathrm{med}$  is assumed to be proportional to the density of scattering centers inside the medium, or $\epsilon^{3/4}$, where $\epsilon$ is the local energy density of the QGP that is simulated with a (3+1)-D ideal hydrodynamic model~\cite{Nonaka:2006yn}:
\begin{align}
\label{eq:yajemKf}
f_\mathrm{med}&=K_f\int d\zeta \left[\epsilon(\zeta)\right]^{3/4}\nonumber\\
&\times \left[\cosh\rho(\zeta)-\sinh\rho(\zeta)\cos\psi\right],
\end{align}
with $K_f$ as a model parameter. Here, $\rho$ represents the local flow rapidity of the medium and $\psi$ represents the angle between the medium flow and the jet parton momentum. When jet parton travels with the speed of light, the $[\cosh\rho(\zeta)-\sinh\rho(\zeta)\cos\psi]$ term is the same as the $p^\mu u_\mu/p^0$ factor to take into account the medium flow effect on jet transport coefficients, as discussed in the \textsc{Matter} model above. This factor is exact for rescaling $\hat{q}_a$ since the transverse momentum broadening square is boost invariant~\cite{Baier:2006pt}, but may not be true for other coefficients.


The second group of approaches introduce medium modification on parton kinematics between vacuum-like splittings without changing the splitting function itself. The aforementioned \textsc{Yajem} model~\cite{Renk:2008pp,Renk:2009nz,Renk:2010mf,Renk:2013pua} provides several modes that belong to this group. For instance, the \textsc{Yajem-rad} mode enhances the parton virtuality before its next splitting based on its scattering with the medium:
\begin{equation}
\label{eq:yajemQ}
\Delta Q^2_a = \int_{\tau_a^0}^{\tau_a^0+\tau_a^f}d\zeta \hat{q}_a(\zeta),
\end{equation}
where $\tau_a^0$ is the production time of parton $a$ and $\tau_a^f$ is its splitting time $2E_a/Q_a^2$. This transverse momentum broadening induced virtuality increase $\Delta Q^2_a$ will shorten the splitting time $\tau_a^f$. Thus, one may need to adjust ($\tau_a^f$, $\Delta Q_a^2$) such that a self-consistent pair is obtained.

The \textsc{Yajem-drag} mode applies a drag to reduce the energy of parton $a$ before its next splitting:
\begin{equation}
\label{eq:yajemE}
\Delta E_a = \int_{\tau_a^0}^{\tau_a^0+\tau_a^f}d\zeta D_a(\zeta),
\end{equation}
where $D_a(\zeta)$ can be understood as a drag coefficient due to parton-medium interactions. Both $\hat{q}_a$ and $D_a$ are assumed to be proportional to $\epsilon^{3/4}[\cosh\rho(\zeta)-\sinh\rho(\zeta)\cos\psi]$, with constant scaling factors in front as model parameters. 

In Ref.~\cite{Renk:2009nz}, different implementations of medium modification are systematically compared within the \textsc{Yajem} model, where no obvious difference in the final state observables has been found between enhancing the parton virtuality (the \textsc{Yajem-rad} mode) and directly modifying the parton splitting function (the \textsc{Yajem-fmed} mode). On the other hand, applying energy loss on jet partons (the \textsc{Yajem-drag} mode) leads to different fragmentation function and angular distribution of final state charged hadrons coming from a single quark. In a later work~\cite{Renk:2013pua}, the $\Delta Q^2_a$ enhancement and $\Delta E_a$ drag methods are combined into a  \textsc{Yajem-de} mode. 

Drag and transverse momentum broadening on jet partons have also been applied in the \textsc{Hybrid} model~\cite{Casalderrey-Solana:2014bpa,Casalderrey-Solana:2016jvj,Hulcher:2017cpt} in a similar way. In this model, \textsc{Pythia}~8 is first applied to generate a chain of vacuum parton showers. To construct the spacetime structure of this chain inside the QGP, the location of the first pair of partons produced from the initial hard collision is distributed according to the Glauber model, after which each parton propagates for a lifetime of $2E/Q^2_a$ before it splits. During its propagation, a drag is applied to the parton, where the form of the drag is taken from holographic calculations of parton energy loss in a strongly coupled plasma using the gauge/gravity duality~\cite{Chesler:2014jva}:
\begin{equation}
\label{eq:adsDrag}
\frac{1}{E_\mathrm{in}}\frac{dE}{dx}=-\frac{4}{\pi}\frac{x^2}{x^2_\mathrm{stop}}\frac{1}{\sqrt{x^2_\mathrm{stop}-x^2}},
\end{equation}
where $E_\mathrm{in}$ is the initial parton energy, $x_\mathrm{stop}$ is the stopping distance given by~\cite{Chesler:2008uy,Gubser:2008as}
\begin{equation}
\label{eq:xStop}
x_\mathrm{stop}=\frac{1}{2\kappa_\mathrm{sc}}\frac{E_\mathrm{in}^{1/3}}{T^{4/3}},
\end{equation}
with $\kappa_\mathrm{sc}$ as a model parameter. Equations~(\ref{eq:adsDrag}) and~(\ref{eq:xStop}) were derived for energetic quarks propagating through the strongly coupled plasma. In the \textsc{Hybird} model, gluons are assumed to follow the same equation, but with a rescaled parameter as $\kappa_\mathrm{sc}^G = \kappa_\mathrm{sc} (C_A/C_F)^{1/3}$ so that within the string-based picture a gluon has the similar stopping distance to a quark with half of the gluon energy~\cite{Casalderrey-Solana:2014bpa}. The transverse momentum kicks on jet partons from the medium is applied with $dk^2_\perp=\hat{q}_a dt$ (sampled with a Gaussian distribution with $dk^2_\perp$ as its width) for each time step $dt$, where $\hat{q}_a=K_aT^3$ is assumed with $K_a$ as a model parameter. 

In the \textsc{Hybrid} model, the local temperature $T$ in the QGP is provided by hydrodynamic models. Both boost invariant ideal hydrodynamic simulations~\cite{Hirano:2010je} and viscous hydrodynamic simulations~\cite{Shen:2014vra} have been used. To take into account the local velocity of the expanding QGP during each time step (or unit length), a given jet parton is first boosted into the local rest frame of the hydrodynamic fluid cell in which its momentum is updated with both longitudinal drag and transverse momentum broadening. It is then boosted back into the center-of-mass frame of collisions for propagation to the next time step. 

Upon each splitting, the \textsc{Hybrid} model assumes that the two daughter partons share the updated momentum of the mother according to the fraction $z$ pre-assigned in the splitting chain based on the \textsc{Pythia} vacuum showers.

A more elaborated treatment of parton-medium scatterings between vacuum-like splittings is implemented in \textsc{Jewel}~\cite{Zapp:2011ya,Zapp:2012ak,Zapp:2013vla}. To take into account the transverse phase in the LPM interference accumulated through multiple scattering, a radiated gluon is allowed to scatter with the medium during its formation time $\tau_f=2\omega/k^2_\perp$, where $\omega$ is the gluon energy and $k_\perp$ is its transverse momentum with respect to its mother. Scatterings between the jet parton and thermal partons inside the medium are described using $2 \rightarrow 2$ pQCD matrix elements. Detailed implementation of such perturbative scatterings will be discussed in the next subsection.  When an additional scattering happens, the total momentum transfer between the jet parton and the medium increases the virtuality scale of the mother, as well as the accumulated $k_\perp$ of the emitted gluon. Its formation time will also be updated accordingly. If this scattering is still within the updated formation time, it is accepted, otherwise rejected. In the end, when there is no more scattering with the medium, this medium-modified gluon emission will be accepted with the probability $1/N_\mathrm{scat}$, where $N_\mathrm{scat}$ is the number of scatterings within the formation time. Note that this $1/N_\mathrm{scat}$ probability is based on the assumption of the ``totally coherent limit"~\cite{Zapp:2011ya}. How to implement the LPM effect between incoherent and totally coherent limits is still challenging. Apart from modifying existing vacuum-like splittings, these parton-medium interactions can also raise the scales and trigger additional splittings for those partons that otherwise cannot split in vacuum when their scales drop below $Q_\mathrm{min}^2$. 

In \textsc{Jewel}, thermal partons inside the QGP can be scattered out of the medium by jets. They are then referred to as ``recoil" partons. In principle, these recoil partons can continue interacting with the medium in the same way as jet partons. However, these rescatterings are not implemented in \textsc{Jewel} as the default setting yet. To ensure the energy-momentum conservation of the entire jet-medium system, the distributions of the thermal partons before being scattered also need to be recorded and subtracted from the final state parton spectra. So far, most calculations using \textsc{Jewel} apply a simplified hydrodynamic model that only describes the boost-invariant longitudinal expansion~\cite{Bjorken:1982qr} of an ideal QGP. In principle, more sophisticated medium can also be applied as long as they provide the phase space density of scattering centers to \textsc{Jewel}.

\subsection{Parton transport}
\label{subsec:transport}

As the virtuality scale of jet partons approaches the scale of parton-medium interactions, the virtuality-ordered parton shower model will no longer be applicable. Instead, transport models become a better choice to describe jet-medium scatterings in this region of the phase space.

A linear Boltzmann transport (\textsc{Lbt}) model is developed to describe jet parton scatterings through a thermal medium~\cite{Wang:2013cia,Cao:2016gvr,Cao:2017hhk,Chen:2017zte,Luo:2018pto,He:2018xjv}, in which the phase space distribution of a jet parton $a$ with $p_a^\mu = (E_a, \vec{p}_a)$ evolves according to the Boltzmann equation,
\begin{equation}
  \label{eq:boltzmann1}
  p_a\cdot\partial f_a(x_a,p_a)=E_a (\mathcal{C}_a^\mathrm{el}+\mathcal{C}_a^\mathrm{inel}).
\end{equation}
On the right-hand side of the above equation, $\mathcal{C}_a^\mathrm{el}$ and $\mathcal{C}_a^\mathrm{inel}$ are the collision integrals for elastic and inelastic scatterings, respectively.

For the elastic scattering ($ab\leftrightarrow cd$) process, the collision integral reads
\begin{align}
\label{eq:collision0}
\mathcal{C}_a^\mathrm{el} &= \sum_{b,(cd)} \int \prod_{i=b,c,d}\frac{d[p_i]}{2E_a}(2\pi)^4\delta^4(p_a+p_b-p_c-p_d) \nonumber\\
&\times \left(\frac{\gamma_c \gamma_d}{\gamma_a}f_c f_d \left|\mathcal{M}_{cd\rightarrow ab}\right|^2 - \gamma_b f_a f_b \left|\mathcal{M}_{ab\rightarrow cd}\right|^2 \right) \nonumber\\
&\times S_2(\hat{s},\hat{t},\hat{u}), 
\end{align}
with a gain term subtracted by a loss term of $f_a$ in the second line. Here, $\sum_{b,(cd)}$ sums over the flavors of parton $b$, and different scattering channels with final parton flavors $c$ and $d$, $d[p_i]\equiv d^3p_i/[2E_i(2\pi)^3]$, $\gamma_i$ is the spin-color degeneracy (6 for a quark and 16 for a gluon), and $f_i$ is the phase space distribution of each parton with a given spin and color. For thermal partons inside the QGP ($i=b, d$), $f_i = 1/(e^{E_i/T}\pm 1)$ with $T$ being the local temperature of the fluid cell; for a jet parton ($i=a, c$) at $(\vec{x},\vec{p})$, $f_i = (2\pi)^3\delta^3(\vec{p}_i-\vec{p})\delta^3(\vec{x}_i-\vec{x})$ is taken. Note that in the gain term, only the production of $a$ from the jet-medium parton scattering between $c$ and $d$ is considered, while that from the thermal-thermal or jet-jet scattering is neglected. The scattering matrix $|\mathcal{M}_{ab\rightarrow cd}|^2$ (see Ref.~\cite{Eichten:1984eu} for massless partons and Ref.~\cite{Combridge:1978kx} for heavy quarks) has been summed over the spin-color degrees of freedom of the final state ($cd$) and averaged over the initial state ($ab$), and similarly for $|\mathcal{M}_{cd\rightarrow ab}|^2$. A double step function $S_2(\hat{s},\hat{t},\hat{u})=\theta(\hat{s}\ge 2\mu_\mathrm{D}^2)\theta(-\hat{s}+\mu^2_\mathrm{D}\le \hat{t} \le -\mu_\mathrm{D}^2)$ is introduced~\cite{Auvinen:2009qm} to regulate the collinear divergence in the leading-order (LO) elastic scattering matrices implemented in \textsc{Lbt}, where $\hat{s}$, $\hat{t}$ and $\hat{u}$ are Mandelstam variables, and $\mu_\mathrm{D}^2=g^2T^2(N_c+N_f/2)/3$ is the Debye screening mass inside the QGP. An alternative method to regulate the divergence is to replace $\hat{t}$ by ($\hat{t}-\mu_\mathrm{D}^2$) in the denominators of $|\mathcal{M}_{ab\rightarrow cd}|^2$ (same for $\hat{u}$). Quantum statistics of the final states of $ab\leftrightarrow cd$ scatterings are neglected in Eq.~(\ref{eq:collision0}).  

Since the detailed balance requires $\gamma_c\gamma_d \left|\mathcal{M}_{cd\rightarrow ab}\right|^2 = \gamma_a\gamma_b \left|\mathcal{M}_{ab\rightarrow cd}\right|^2$, Eq.~(\ref{eq:collision0}) can be simplified as 
\begin{align}
\label{eq:collision}
\mathcal{C}_a^\mathrm{el} &= \sum_{b,(cd)} \int \prod_{i=b,c,d}d[p_i]\frac{\gamma_b}{2E_a}(f_c f_d - f_a f_b) S_2(\hat{s},\hat{t},\hat{u})\nonumber\\
& \times  (2\pi)^4\delta^4(p_a+p_b-p_c-p_d) \left|\mathcal{M}_{ab\rightarrow cd}\right|^2.
\end{align}
Examining Eq.~(\ref{eq:boltzmann1}) and the loss term in Eq.~(\ref{eq:collision}), one obtains the following elastic scattering rate (number of scatterings per unit time) for parton $a$,
\begin{align}
\label{eq:rate}
\Gamma_a^\mathrm{el}&(\vec{p}_a,T)=\sum_{b,(cd)}\frac{\gamma_b}{2E_a}\int \prod_{i=b,c,d}d[p_i] f_b S_2(\hat{s},\hat{t},\hat{u})\nonumber\\
&\times (2\pi)^4\delta^{(4)}(p_a+p_b-p_c-p_d)|\mathcal{M}_{ab\rightarrow cd}|^2.
\end{align}
With the assumption of zero mass for thermal partons $b$ and $d$, this can be further simplified as 
\begin{align}
 \label{eq:rate2}
 \Gamma_a^\mathrm{el}(\vec{p}_a,&T) = \sum_{b,(cd)} \frac{\gamma_b}{16E_a(2\pi)^4}\int dE_b d\theta_b d\theta_d d\phi_{d}\nonumber\\
&\times f_b(E_b,T) S_2(\hat{s},\hat{t},\hat{u})|\mathcal{M}_{ab\rightarrow cd}|^2\nonumber\\
&\times \frac{E_b E_d \sin \theta_b \sin \theta_d}{E_a-|\vec{p}_a| \cos\theta_{d}+E_b-E_b\cos\theta_{bd}},
\end{align}
where
\begin{align}
 \label{eq:E4}
 \cos\theta_{bd}&=\sin\theta_b\sin\theta_d\cos\phi_{d}+\cos\theta_b\cos\theta_d,\\[10pt]
 E_d=&\frac{E_aE_b-p_aE_b\cos\theta_{b}}{E_a-p_a\cos\theta_{d}+E_b-E_b\cos\theta_{bd}}.
\end{align}
Here, we let parton $a$ move in the $+z$ direction, and $b$ in the $x-z$ plane with $\theta_b$ as its polar angle. The outgoing momentum of $d$ has $\theta_d$ and $\phi_d$ as its polar and azimuthal angles, and $\theta_{bd}$ is the angle between $b$ and $d$. Within a time interval $\Delta t$, the average number of elastic scatterings is then $ \Gamma_a^\mathrm{el}\Delta t$. One may allow multiple scatterings between the jet parton and the medium by assuming the number of independent scatterings $n$ to obey the Poisson distribution,
\begin{equation}
\label{eq:poission}
P(n)=\frac{(\Gamma_a^\mathrm{el}\Delta t)^n}{n!}e^{-(\Gamma_a^\mathrm{el}\Delta t)}.
\end{equation}
Thus, the probability of scattering is $P_a^\mathrm{el}=1-\exp(-\Gamma_a^\mathrm{el}\Delta t)$, or just $\Gamma_a^\mathrm{el}\Delta t$ if it is much smaller than 1.

One may extend Eq.~(\ref{eq:rate2}) to the average of a general quantity $X$ per unit time:
\begin{align}
\label{eq:defX}
 \langle\langle X&(\vec{p}_a,T) \rangle\rangle = \sum_{b,(cd)} \frac{\gamma_b}{16E_a(2\pi)^4}\int dE_b d\theta_b d\theta_d d\phi_{d}\nonumber\\
&\times X(\vec{p}_a,T) f_b(E_b,T) S_2(\hat{s},\hat{t},\hat{u})|\mathcal{M}_{ab\rightarrow cd}|^2\nonumber\\
&\times \frac{E_b E_d \sin \theta_b \sin \theta_d}{E_a-|\vec{p}_a| \cos\theta_{d}+E_b-E_b\cos\theta_{bd}}.
\end{align}
Therefore, we have $\Gamma_a^\mathrm{el} = \langle\langle 1 \rangle\rangle$ and
\begin{align}
\hat{q}_a&=\langle\langle \left[\vec{p}_c - (\vec{p}_c \cdot \hat{p}_a)\hat{p}_a\right]^2\rangle\rangle, \label{eq:22qhat}\\
\hat{e}_a&=\langle\langle E_a-E_c\rangle\rangle, \label{eq:22ehat}
\end{align}
where $\hat{q}_a$ and $\hat{e}_a$ denote the transverse momentum broadening square and elastic energy loss of the jet parton $a$, respectively, per unit time due to elastic scattering. For high energy jet partons with the small-angle-scattering approximation, one may obtain~\cite{Wang:1996yf,He:2015pra}:
\begin{align}
\Gamma_a^\mathrm{el} &= C_2(a)\frac{42\zeta(3)}{\pi}\frac{\alpha_\mathrm{s}^2 T^3}{\mu_\mathrm{D}^2}, \label{eq:gammaAnalytic}\\
\hat{q}_a&=C_2(a)\frac{42\zeta(3)}{\pi}\alpha_\mathrm{s}^2 T^3 \ln\left(\frac{C_{\hat{q}} E_a T}{4\mu_\mathrm{D}^2}\right), \label{eq:qhatAnalytic}\\
\hat{e}_a&=C_2(a)\frac{3\pi}{2}\alpha_\mathrm{s}^2 T^2 \ln\left(\frac{C_{\hat{e}} E_a T}{4\mu_\mathrm{D}^2}\right), \label{eq:ehatAnalytic}
\end{align}
where $C_2(a)$ is the quadratic Casimir color factor of parton $a$, $C_{\hat{q}}$ and $C_{\hat{e}}$ are constants depending on kinematic cuts adopted in the calculations. With the implementations discussed above, comparisons between numerical evaluations and these analytical formulae suggest $C_{\hat{q}}\approx 5.7$ and $C_{\hat{e}}\approx 2.6$~\cite{He:2015pra}.

Apart from elastic scattering, the inelastic process, or medium-induced gluon radiation, is included in the \textsc{Lbt} model by relating the inelastic scattering rate to the average number of emitted gluons from parton $a$ per unit time, and is evaluated as \cite{Cao:2013ita,Cao:2015hia,Cao:2016gvr}
\begin{equation}
 \label{eq:gluonnumber}
 \Gamma_a^\mathrm{inel} (E_a,T,t) = \frac{1}{1+\delta_g^a}\int dzdk_\perp^2 \frac{dN_g^a}{dz dk_\perp^2 dt},
\end{equation}
in which the Kronecker delta function $\delta_g^a$ is imposed to avoid double counting in the $g\rightarrow gg$ process, and the medium-induced gluon spectrum in the fluid comoving frame is taken from the high-twist energy loss calculation~\cite{Guo:2000nz,Majumder:2009ge,Zhang:2003wk},
\begin{equation}
\label{eq:gluondistribution}
\frac{dN_g^a}{dz dk_\perp^2 dt}=\frac{2C_A\alpha_\mathrm{s} P^\mathrm{vac}_a(z) k_\perp^4}{\pi C_2(a) (k_\perp^2+x^2 m_a^2)^4}\,\hat{q}_a\, {\sin}^2\left(\frac{t-t_i}{2\tau_f}\right),
\end{equation}
 where $z$ and $k_\perp$ are the fractional energy and transverse momentum of the emitted gluon with respect to its parent parton $a$, $P^\mathrm{vac}_a(z)$ is the vacuum splitting function of $a$ (note again it contains the color factor $C_2(a)$ by our convention here), $\hat{q}_a$ is the parton transport coefficient taken from the elastic scattering process Eq.~(\ref{eq:22qhat}), $t_i$ represents the production time of parton $a$, and $\tau_f={2E_a z(1-z)}/{(k_\perp^2+z^2m_a^2)}$ is the formation time of the emitted gluon with $m_a$ being the mass of parton $a$. In the current \textsc{Lbt} model, zero mass is taken for light flavor quarks and gluon, 1.3~GeV is taken for charm quark mass and 4.2~GeV for beauty quark mass. The lower and upper limits of $z$ are taken as $z_\mathrm{min}=\mu_\mathrm{D}/E_a$ and $z_\mathrm{max}=1-\mu_\mathrm{D}/E_a$ respectively. Note that the medium-induced spectrum Eq.~(\ref{eq:gluondistribution}) here is consistent with the medium-induced splitting function used in the \textsc{Matter} model in Eqs.~(\ref{eq:sudakov}) and (\ref{eq:medP}). 

Multiple gluon emissions within each time interval $\Delta t$ are allowed in the \textsc{Lbt} model. Similar to the elastic scattering process, the number of emitted gluons obeys a Poisson distribution with a mean value of $\Gamma_a^\mathrm{inel}\Delta t$. Thus, the probability of inelastic scattering is $P_a^\mathrm{inel}=1-\exp(-\Gamma_a^\mathrm{inel}\Delta t)$. Note that both multiple elastic scatterings and multiple emitted gluons are assumed incoherent, possible interferences between each other have not been taken into account in \textsc{Lbt} yet.

To combine elastic and inelastic scattering processes, the total scattering probability is divided into two parts: pure elastic scattering without gluon emission $P_a^\mathrm{el}(1-P_a^\mathrm{inel})$ and inelastic scattering $P_a^\mathrm{inel}$. The total probability is then $P_a^\mathrm{tot}=P_a^\mathrm{el}+P_a^\mathrm{inel}-P_a^\mathrm{el} P_a^\mathrm{inel}$. Based on these probabilities, the Monte Carlo method is applied to determine whether a given jet parton $a$ with momentum $\vec{p}_a$ scatters with the thermal medium with local temperature $T$, and whether the scattering is pure elastic or inelastic. With a selected scattering channel, as well as the number of elastic scatterings or emitted gluons given by the Poisson distribution, the energies and momenta of the outgoing partons are sampled using the corresponding differential spectra given by either Eq.~(\ref{eq:rate2}) or Eq.~(\ref{eq:gluondistribution}). In the \textsc{Lbt} model, the emitted gluons are induced by scatterings between jet partons and the thermal medium. Therefore, for an inelastic scattering process, a $2\rightarrow2$ scattering is generated first, after which the four-momenta of the two outgoing partons are adjusted together with the $n$ emitted gluons so that the energy-momentum conservation of the $2 \rightarrow 2 + n$ process is respected.

For realistic heavy-ion collisions, the initial momenta of jet partons are either sampled using spectra from pQCD calculations for initial hard collisions or generated with a pQCD Monte Carlo generator such as \textsc{Pythia} or other programs for  vacuum showers. Their initial positions are either sampled with the Monte-Carlo Glauber models or taken from the \textsc{Ampt}~\cite{Lin:2004en} simulations for the early time evolution. Different hydrodynamic models, (2+1)-D viscous \textsc{Vishnew}~\cite{Song:2007fn,Song:2007ux,Qiu:2011hf} and (3+1)-D viscous \textsc{Clvisc}~\cite{Pang:2014ipa,Pang:2012he} with Monte-Carlo Glauber or \textsc{Ampt} initial conditions, are used to generate the spacetime evolution profiles of the QGP. At the beginning of a given time step, each jet parton is boosted into the rest frame of its local fluid cell, in which its scattering with the medium is simulated using the linear Boltzmann equation. The outgoing partons after scatterings are then boosted back to the global center-of-mass frame of collisions, in which they propagate to the next time step. On the freeze-out hypersurface of the QGP ($T_\mathrm{c}=165$~MeV), jet partons are converted into hadrons using either the \textsc{Pythia} simulation or the recombination model~\cite{Han:2016uhh}.

In the \textsc{Lbt} model, all partons in jet showers are fully tracked, including energetic jet partons and their emitted gluons, as well as ``recoil" partons which are thermal medium partons in the final state of the elastic scattering. All these partons are treated on the same footing and are allowed to re-scatter with the medium. When a recoil parton is generated, it leaves a hole behind in the phase-space inside the medium. These holes are labeled as ``negative" partons in \textsc{Lbt}, denoting the back-reaction of the QGP response to jet propagation. Their energy-momentum will be subtracted from the final-state jet spectra to ensure the energy-momentum conservation of the jet-medium system. A more rigorous treatment of this back-reaction, as well as the subsequent evolution of those soft partons (at or below the thermal scale of the medium) produced in jet-medium scatterings, will be discussed in Sec.~\ref{subsec:concurrent} using the \textsc{CoLbt-Hydro} model.

Another example of jet transport model is \textsc{Martini}~\cite{Schenke:2009gb,Park:2018acg}, which implements the AMY energy loss formalism for the radiative energy loss rates~\cite{Arnold:2002ja,Arnold:2002zm} combined with elastic scattering processes~\cite{Schenke:2009ik}. The medium-induced parton splitting processes are realized by solving a set of coupled rate equations for the time evolution of the energy distribution of quark/gluon jet partons $f_{q/g}(p)$:
\begin{align} 
\label{eq:rateMAR}
	\frac{df_q(p)}{dt} &= \int_k f_q(p+k)\frac{d\Gamma^{q}_{qg}(p+k,k)}{dkdt}\\
	-&f_q(p)\frac{d\Gamma^{q}_{qg}(p,k)}{dkdt} +2f_g(p+k)\frac{d\Gamma^g_{q\bar{q}}(p+k,k)}{dkdt},\nonumber\\[10pt]
	\frac{df_g(p)}{dt} &= \int_k f_q(p+k)\frac{d\Gamma^{q}_{qg}(p+k,p)}{dkdt}\nonumber\\
	+&f_g(p+k)\frac{d\Gamma^{g}_{gg}(p+k,p)}{dkdt}\\
	- &f_g(p)\left[\frac{d\Gamma^g_{q\bar{q}}(p,k)}{dkdt} + \frac{d\Gamma^g_{gg}(p,k)}{dkdt}\theta(2k-p)\right].\nonumber
\end{align}
Here, $d\Gamma^a_{bc}(p,k)/dkdt$ is the transition rate taken from the AMY formalism for parton $a$ (with energy $p$) to split into parton $b$ (with energy $p-k$) and parton $c$ (with energy $k$). The factor of 2 in front of the $g\rightarrow q\bar{q}$ rate takes into account the fact that $q$ and $\bar{q}$ are distinguishable; and the $\theta$ function after the $g \rightarrow gg$ rate is introduced to avoid double counting of its final state. The integration range with $k<0$ corresponds to energy gain of jet partons from the thermal medium; and the range with $k>p$ for the $q \rightarrow qg$ process corresponds to the quark annihilating with an anti-quark with energy $k-p$ from the medium into the gluon.  

The AMY formalism describes the energy loss of hard jets as partons split in a thermal medium. The radiation rates are calculated by means of integral equations~\cite{Arnold:2002ja} with the assumptions that quarks and gluons in the medium are well defined (hard) quasi-particles with momenta much larger than the medium temperature $T$ and thermal masses of the order of $gT$~\cite{Arnold:2002zm}. In the current \textsc{Martini} model, the radiative energy loss mechanism has been improved by implementing the effects of finite formation time~\cite{CaronHuot:2010bp,Park:2018acg} and running coupling~\cite{Young:2012dv}. The formation time of the radiation process is set as $\sim p/p_\mathrm{T}^2$ within which the hard parton and the emitted parton are considered as a coherent state. This effectively reduces the radiation rate at early times after the hard parton is produced. The coupling constant $\alpha_\mathrm{s}(\mu^2)$ runs with the scale of the average momentum transfer square between the jet parton and the medium
\begin{equation}
\label{eq:muMartini}
\mu^2 = {\langle p^2_\perp \rangle} \sim \sqrt{\hat{q}p},
\end{equation}
where $\hat{q}$ is the jet transport parameter for the average momentum transfer square per mean-free path. The daughter partons are strictly collinear with their mother at the splitting vertex; additional transverse momentum broadening is introduced by elastic scattering processes.

For realistic simulations of jet propagation in heavy-ion collisions within the \textsc{Martini} model, the medium background is provided by the (3+1)-D viscous hydrodynamic model \textsc{Music} with IP-Glasma initial condition~\cite{Schenke:2010nt,McDonald:2016vlt}. Jet partons are initialized with \textsc{Pythia} vacuum showers. After their evolution through \textsc{Martini}, they are fed back to \textsc{Pythia} for hadronization.

A recoil parton can be generated within \textsc{Martini} if its momentum -- sum of its original thermal momentum and the momentum transfer from the scattering -- is larger than certain kinematic scale (e.g. $4T$). These recoil partons continue scattering with the medium in the same way as high-energy jet partons. The soft recoil partons below the $4T$ threshold, as well as back-reaction to medium due to the generation of recoil partons, are expected to be deposited into the subsequent QGP evolution in the future work, thus have not been included in the \textsc{Martini} model.

In both \textsc{Lbt} and \textsc{Martini}, while jet parton evolution is described using the linear Boltzmann equation or rate equation, the QGP medium evolution is described using hydrodynamic models. In literature, an alternative approach is applying a full Boltzmann simulation for both jet partons and medium partons, although it is still under debate whether the strongly coupled QGP can be modeled with quasi-particles within the perturbative framework. One example is the \textsc{Ampt} model~\cite{Lin:2004en,Ma:2013pha,Ma:2013bia,Ma:2013gga,Ma:2013uqa,Nie:2014pla,Gao:2016ldo}. In \textsc{Ampt}, the initial spatial and momentum information of high-$p_\mathrm{T}$ jets, mini-jets and soft string excitations are taken from \textsc{Hijing} simulations~\cite{Wang:1991hta,Gyulassy:1994ew}, which are further converted into individual quarks via the string melting model~\cite{Lin:2004en}. These quarks, including both jet partons and medium partons, then scatter with each other through elastic collisions whose interaction strength is controlled by a set of partonic cross sections $\sigma$ that are treated as model parameters. Note that gluon components and inelastic scatterings have not been included in the current \textsc{Ampt} model. At the end of partonic collisions, a simple coalescence model is applied to convert two/three nearest quarks in space into mesons/baryons. These hadrons can further scatter with each other within the \textsc{Art} model~\cite{Li:1995pra}. 

Another example of full Boltzmann transport is the \textsc{Bamps} model~\cite{Xu:2004mz,Xu:2007aa,Senzel:2013dta,Uphoff:2014cba,Senzel:2016qau}, in which both elastic and inelastic scatterings between partons during the QGP phase are simulated with the leading-order perturbative QCD cross sections. For inelastic scatterings, the Gunion-Bertsch approximation is adopted~\cite{Gunion:1981qs,Fochler:2013epa}, and the LPM effect is modeled with a theta function $\theta(\lambda-\tau X_\mathrm{LPM})$ that requires the mean-free path of the parent parton $\lambda$ is larger than the formation time of the emitted gluon $\tau$ scaled with an $X_\mathrm{LPM}$ parameter between 0 and 1. In \textsc{Bamps}, initial partons, including both jet partons and medium partons, are generated by \textsc{Pythia}. At the partonic freeze-out energy density ($\epsilon = 0.6$~GeV/fm$^3$), jet partons are converted into hadrons using the Albino-Kniehl-Kramer (AKK) fragmentation functions~\cite{Albino:2008fy}.

\subsection{Multi-scale jet evolution}
\label{subsec:multi-scale}

Interactions between jets and the QGP differ at various scales. Thus, it is incomplete to apply a single formalism through the entire spacetime history of jet evolution. A first attempt to combine different and complementary theoretical approaches into a multi-scale jet evolution model was implemented in the \textsc{Jetscape} framework~\cite{Cao:2017zih,Putschke:2019yrg,Park:2019sdn,Tachibana:2018yae}, in which medium-modified splittings of jet partons are described using the \textsc{Matter} model at high virtualities as discussed in Sec.~\ref{subsec:med_showers_virtual}, while their subsequent transport via jet-medium interaction at low virtualities are described using either the \textsc{Lbt} model or the \textsc{Martini} model as discussed in Sec.~\ref{subsec:transport}. 

One crucial quantity in this combined approach is the separation scale $Q_0^2$ between the DGLAP-type parton showers and the in-medium transport. In Ref.~\cite{Cao:2017zih}, two different schemes, fixed $Q_0^2$ and dynamical $Q_0^2$, are investigated within a static medium. For the former,  \textsc{Matter} is used to simulate parton splittings when the virtualities are above a fixed value of $Q_0^2$ ($1$, $4$ or $9$~GeV$^2$) while either \textsc{Lbt} or \textsc{Martini} is used to simulate the parton scatterings with the medium when the virtualities are below that separation scale. For the latter, $Q_0^2=\hat{q}\tau_f$ is defined for each parton, in the sense that the virtuality-ordered parton showers should switch to transport when the parton virtuality scale is comparable to the virtuality gain (or transverse momentum broadening square) from scatterings with the medium. With $\tau_f=2E/Q_0^2$, one can obtain $Q_0^2=\sqrt{2E\hat{q}}$, in which $\hat{q}$ can be taken from Eq.~(\ref{eq:qhatAnalytic}) within the picture of perturbative scatterings of jet partons inside a thermal medium. Note that $Q_0^2$ is only applied when it is larger than $Q_\mathrm{min}^2$ (taken as 1~GeV$^2$ in most models) above which the DGLAP evolution is reliable. If not, $Q_\mathrm{min}^2$ is used as the lower boundary for the \textsc{Matter} model. In addition, to calculate the nuclear modification of jets in realistic heavy-ion collisions, if the virtuality scale of a given parton is still larger than $Q_\mathrm{min}^2$ when it exits the QGP, this parton should then continue splitting within \textsc{Matter} with only the vacuum splitting function until all its daughter partons reach $Q_\mathrm{min}^2$. This guarantees a meaningful comparison to the baseline spectra in p+p collisions, which is obtained with vacuum showers in \textsc{Matter} directly down to $Q_\mathrm{min}^2$ for each parton.

Within the \textsc{Jetscape} framework, it is found that energetic partons spend finite times within both the DGLAP-splitting stage and the transport stage. The switching time is delayed as the parton energy increases or the switching scale $Q_0^2$ decreases. Thus, the \textsc{Matter} model dominates the medium modification of parton spectra at high $p_\mathrm{T}$, while \textsc{Lbt}/\textsc{Martini} dominates at low $p_\mathrm{T}$. A larger value of $Q_0^2$ typically weakens the medium modification from the \textsc{Matter} stage, thus enhances the relative contribution from the subsequent \textsc{Lbt}/\textsc{Martini} stage. Increasing the medium size also extends the in-medium path length of partons after they reach $Q_0^2$, thus raises the contribution from \textsc{Lbt}/\textsc{Martini}. To date, there is no theoretical determination of the exact value of the switching scale $Q_0^2$ between different implementations of jet energy loss theory. This is treated as a parameter in this multi-scale jet evolution model when comparing with experimental data~\cite{Park:2019sdn,Tachibana:2018yae}. A similar multi-scale approach has also been applied to investigating heavy quark energy loss~\cite{Cao:2017crw} where the medium-modified parton fragmentation functions at low virtualities (heavy meson mass scales) are first extracted from an in-medium transport model and then evolved up to high virtualities (heavy meson $p_\mathrm{T}$ scales) using the medium-modified DGLAP equation.

\section{Medium response to jet propagation}
\label{sec:medium_response}

\subsection{Hydrodynamic evolution of bulk medium}
\label{subsec:hydro}

Numerical calculations of jet quenching, and in particular the simulation of jet transport, require the space-time profile of the bulk medium evolution in high-energy nuclear collisions. Along the propagation path, one needs the information on the local temperature (or thermal parton density) and fluid velocity in order to evaluate the scattering rate and gluon radiation spectra. Many hydrodynamic models have been developed for this purpose~\cite{Kolb:2000sd,Huovinen:2001cy,Nonaka:2006yn,Romatschke:2007mq,Song:2007fn,Qiu:2011hf,Petersen:2010cw,Werner:2010aa,Schenke:2010nt,Pang:2014ipa,Pang:2012he}. Here we briefly review the \textsc{CLVisc} hydrodynamic model~\cite{Pang:2012he,Pang:2014ipa} which is used for jet transport simulations within the \textsc{Lbt} model discussed in this review.

The second-order hydrodynamic equations for the evolution of QGP with zero baryon density are given by
\begin{equation}
    \partial_\mu T^{\mu\nu} = 0,
   \label{eqn:hydro} 
\end{equation}
with the energy-momentum tensor
\begin{equation}
    T^{\mu\nu}=\varepsilon u^{\mu}u^{\nu}-(P+\Pi) \Delta^{\mu\nu} + \pi^{\mu\nu},
 \label{eqn:tmunu} 
\end{equation}
 where $\varepsilon$ is the
energy density, $P$ the pressure, $u^{\mu}$ the fluid four-velocity, $\pi^{\mu\nu}$ the shear stress tensor, $\Pi$ the bulk pressure
and $\Delta^{\mu\nu}=g^{\mu\nu}-u^{\mu}u^{\nu}$ the projection operator which is orthogonal to the fluid velocity. 
In the Landau frame, the shear stress tensor is traceless ($\pi_{\mu}^{\mu}=0$) and transverse $(u_{\mu}\pi^{\mu\nu}=0)$. 
In the Milne coordinates, $\tau=\sqrt{t^2 - z^2}$ is the proper time and $\eta_s = (1/2)\ln [(t+z)/(t-z)]$ the space-time rapidity.

The \textsc{CLVisc} hydrodynamic model uses the Kurganov-Tadmor algorithm \cite{KURGANOV2000241} to solve the hydrodynamic equation for the bulk medium and the Cooper-Frye particlization for hadron freeze-out with GPU parallelization using the Open Computing Language (OpenCL). With GPU parallelization and Single Instruction Multiple Data (SIMD) vector operations on modern CPUs, \textsc{CLVisc} achieves the best computing performance so far for event-by-event (3+1)D hydrodynamic simulations on heterogeneous computing devices.

The initial energy-momentum density distributions for the event-by-event \textsc{CLVisc} hydrodynamic simulations are obtained from partons given by the \textsc{Ampt} model~\cite{Lin:2004en} with a Gaussian smearing,
\begin{eqnarray}
  T^{\mu\nu} (\tau_{0},x,y,\eta_{s}) & = K\sum_{i}
  \frac{p^{\mu}_{i}p^{\nu}_{i}}{p^{\tau}_{i}}\frac{1}{\tau_{0}\sqrt{2\pi\sigma_{\eta_{s}}^{2}}}\frac{1}{2\pi\sigma_{r}^{2}} \nonumber\\
     	  &\hspace{-0.4in} \times \exp \left[-\frac{(\vec x_\perp-\vec x_{\perp i})^{2}}{2\sigma_{r}^{2}} - \frac{(\eta_{s}-\eta_{i s})^{2}}{2\sigma_{\eta_{s}}^{2}}\right],
  \label{eq:Pmu}
\end{eqnarray}
where $p^{\tau}_{i}=m_{i\mathrm{T}}\cosh(Y_{i}-\eta_{i s})$, $\vec p^{\perp}_{i}=\vec p_{i \perp}$, $p^{\eta}_{i}=m_{i \mathrm{T}}\sinh(Y_{i}-\eta_{i s})/\tau_{0}$, $m_{i \mathrm{T}}=\sqrt{p_{i \perp}^2+m_i^2}$ and the summation runs over all partons $(i)$ produced from the \textsc{Ampt} model simulations. 
The scale factor $K$ and the initial time $\tau_{0}$ are two parameters that are adjusted to fit the experimental data on the central rapidity density of produced hadrons. A parametrized equation of state (EoS) s95p-v1\cite{Huovinen:2009yb} is used in \textsc{CLVisc}.
 
\begin{figure}
 \centering
 \includegraphics[width=8.0cm]{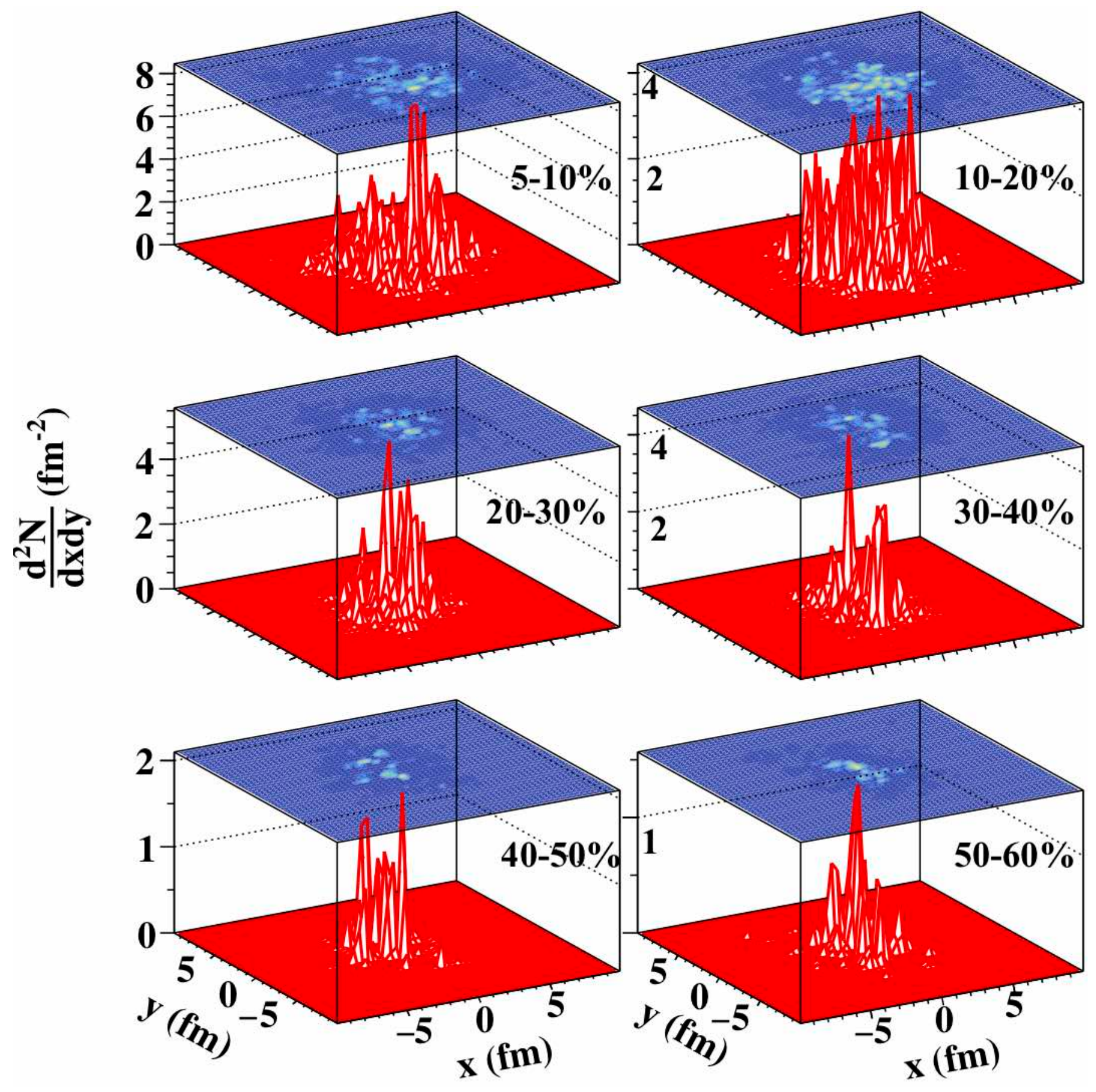}
 \caption{(Color online) The transverse distributions of mini-jets in a typical \textsc{Ampt} event of Pb+Pb collisions with different centralities at $\sqrt{s}=5.02$~TeV. The figure is from Ref.~\cite{He:v2-private}}
 \label{fig:geo}
\end{figure}

The \textsc{Ampt} model employs the \textsc{Hijing} model~\cite{Wang:1991hta,Gyulassy:1994ew} to generate initial bulk partons as well as jets according to the Glauber model of nuclear collisions with the Woods-Saxon nuclear distribution. Bulk parton transport is simulated in \textsc{Ampt} for the whole evolution history in the Cartesian coordinates. Partons at the fixed initial time $\tau_0$ in the Milne coordinates are used to evaluate the initial condition for the \textsc{CLVisc} hydrodynamic evolution according to Eq.~(\ref{eq:Pmu}). The centrality classes of heavy-ion collisions are defined according to the initial parton multiplicity distribution, and the average number of participant nucleons $\langle N_{\rm part}\rangle$ in each centrality class is computed accordingly. 

In the event-by-event simulation of jet transport through the \textsc{CLVisc} hydrodynamic background, a fixed number of \textsc{CLVisc} hydrodynamic events (e.g. 200) is used for each centrality bin of heavy-ion collisions. For  each of these hydrodynamic events, a large number of triggered jets (e.g. 10000) in each bin of transverse momentum and rapidity are simulated, whose initial transverse positions are sampled according to the transverse distribution of mini-jets in the same \textsc{Ampt} event that provides the initial condition for the bulk medium evolution in \textsc{CLVisc}. Shown in Fig.~\ref{fig:geo} are the transverse distributions of mini-jets in a typical \textsc{Ampt} event of Pb+Pb collisions at $\sqrt{s}=5.02$~TeV (per nucleon pair) for different centralities.

\subsection{Medium response in parton transport}
\label{subsec:recoil}

Jet-medium interactions lead not only to medium modification of jets, but also medium response to the energy-momentum deposition from jets. One way to propagate the lost energy and momentum of jets through the medium is via recoil partons. Recoil partons are the final-state partons that are scattered out of their initial phase-space in the thermal background medium by jets. They are fully tracked and allowed to rescatter with the thermal medium in the same way as jet partons do within the \textsc{Lbt} model~\cite{Wang:2013cia,He:2015pra,Luo:2018pto,He:2018xjv}. When these recoil partons are produced, they leave ``holes" in the phase-space of  the thermal medium. These ``holes'' that carry the energy-momentum of the initial-state thermal partons of the scattering are labeled as back-reaction or ``negative" partons, and are propagated inside \textsc{Lbt} as well.

\begin{figure}
\centering
\includegraphics[width=6.5cm]{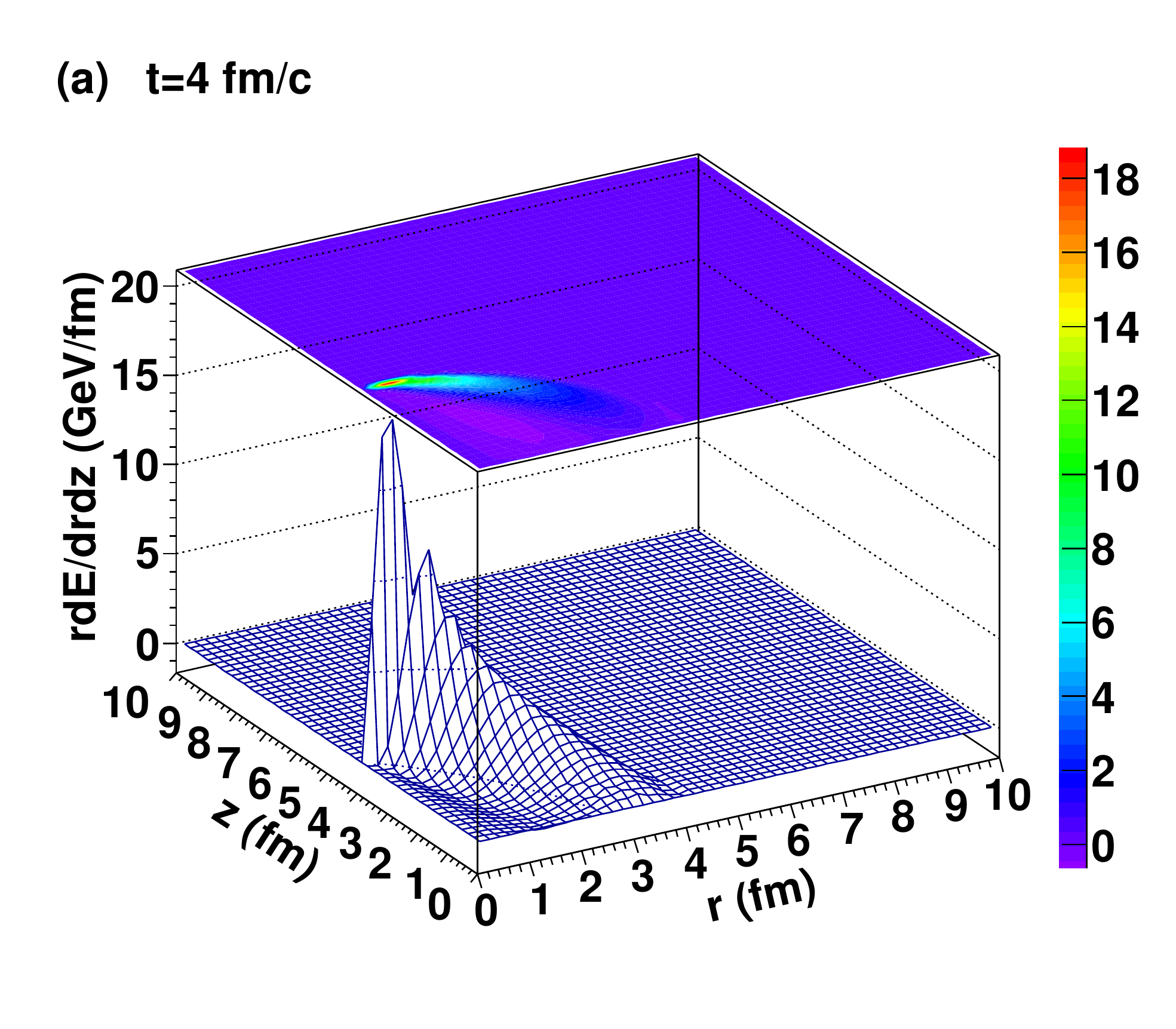}\\
\vspace{-0.4 cm}
\includegraphics[width=6.5cm]{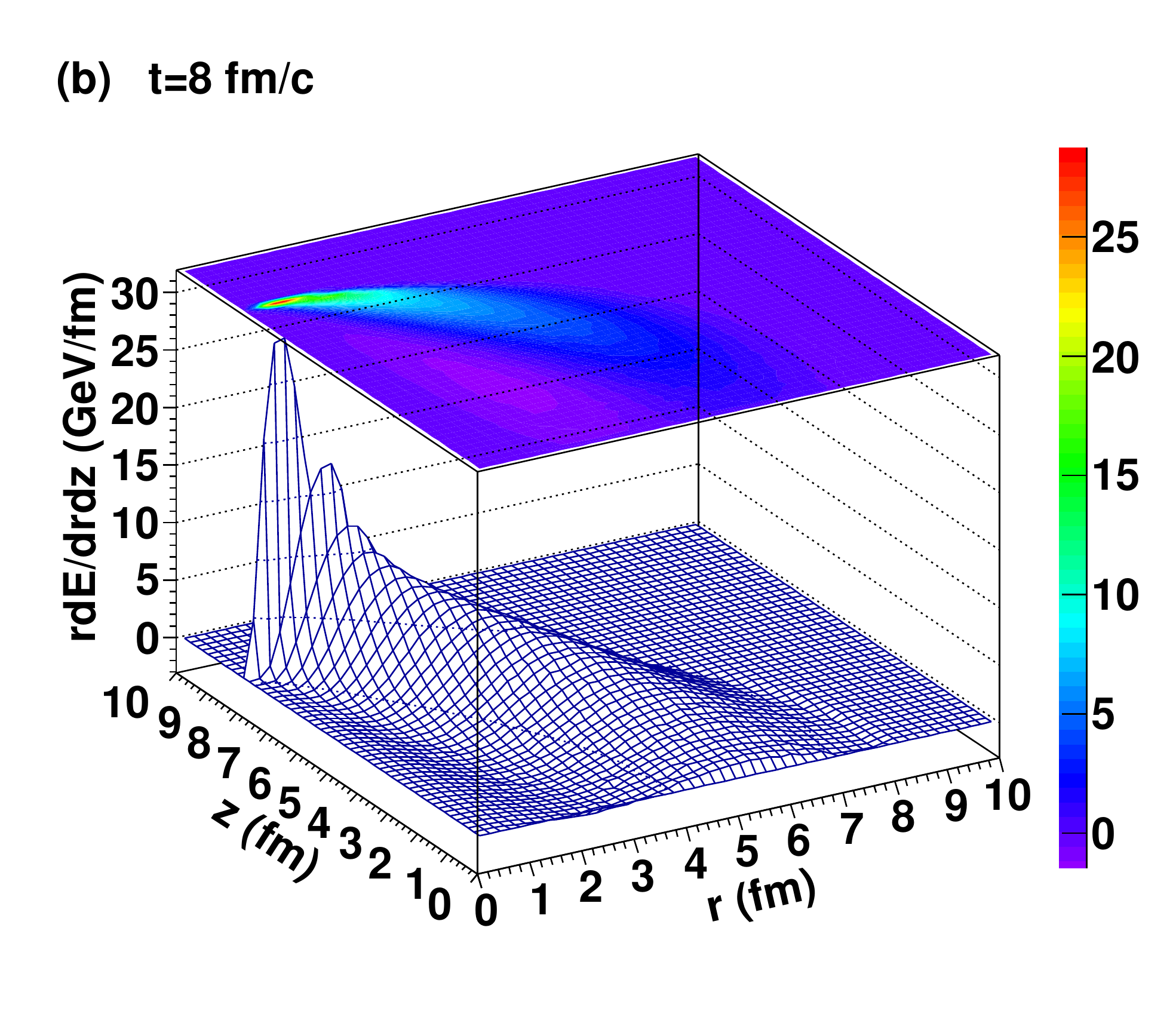}
\caption{(Color online) The energy density distribution of the jet-induced medium response by a gluon with an initial energy $E_0=100$~GeV after (a) 4  and (b) 8 fm of propagation time in a uniform QGP medium at a constant temperature $T=400$~MeV. The gluon propagates along the $+z$-direction from $z=0$ and transverse position $r=0$. The figures are from Ref.~\cite{He:2015pra}.}
\label{fig:LBTcone}
\end{figure}

Shown in Fig.~\ref{fig:LBTcone} are the energy density distributions of the medium response from the LBT simulation~\cite{He:2015pra} of a propagating gluon along the $+z$-direction, starting at $z=0$ and transverse position $r=0$ with an initial energy $E_0=100$~GeV, after 4~fm/$c$ (upper panel) and 8~fm/$c$ (lower panel) of propagation time in a uniform QGP medium at a constant temperature $T=400$~MeV, averaged over many events. One can clearly see the formation and propagation of a Mach-cone-like shock wave induced by parton-medium interactions. The shock wave is rapidly diffused during its propagation because of the dissipation due to the large value of shear viscosity as a result of pQCD parton-parton collisions as implemented in the LBT model.  One can also see the depletion of the energy density behind the propagating parton as a diffusion wake induced by the jet-medium interaction. In realistic calculations, both jet partons and recoil partons are utilized to reconstruct jets, with the energy-momentum of ``negative" partons subtracted. This ensures the energy-momentum conservation of the entire jet-medium system. 

As discussed in the previous section, this treatment of recoil partons in the jet-induced medium excitation has also been applied in \textsc{Martini}~\cite{Schenke:2009gb,Park:2018acg} and \textsc{Jewel}~\cite{Zapp:2011ya,Zapp:2012ak,Zapp:2013vla}. In \textsc{Martini}, only recoil partons above certain kinematic threshold (e.g. 4$T$) are kept in simulations and are allowed to rescatter with the medium. Recoil partons below the threshold, as well as the back-reaction or ``negative" partons, are regarded as part of the medium evolution and have not been included in \textsc{Martini} yet. In \textsc{Jewel}, both recoil and ``negative" partons are included. However, rescatterings of recoil partons with the medium are not implemented yet.

A similar method, though not exactly through recoil particles, has been applied in the \textsc{Hybrid} model~\cite{Casalderrey-Solana:2016jvj} to take into account the jet-induced medium response. The energy-momentum loss from each jet parton $\Delta p^\mu = (\Delta E, \Delta \vec{p})$ is assumed to instantaneously thermalize with the medium background and is directly converted to hadrons using the Cooper-Frye formula. The additional particle production $d\Delta N/d^3p$ due to this energy-momentum deposition is positive along the direction of jet propagation, while can be negative in the opposite direction. Similar to \textsc{Lbt} and \textsc{Jewel}, the latter part is treated as ``negative" particles or back-reaction, representing the diffusion wake behind the jet propagation. To ensure energy-momentum conservation of each jet event, an independent list of hadrons are first sampled using the Cooper-Frye formula until their total energy reaches the lost energy $\Delta E$ of the jet parton. Then the four-momentum of each hadron is re-assigned based on the Cooper-Frye formula again, if this re-assignment improves the the four-momentum conservation according to the Metropolis algorithm, until the total four-momentum of the hadron ensemble is sufficiently close to that lost by the jet parton ($\Delta p^\mu$).

Among these different implementations of recoil and back-reaction, \textsc{Jewel} represents the limit where recoil partons do not rescatter with the medium, while \textsc{Hybrid} represents the opposite limit of sudden thermalization and hadronization of the energy-momentum transfer from jet to medium. In between, \textsc{Lbt} assumes perturbative rescatterings of these recoil partons through the thermal medium before they are converted into hadrons at the QGP boundary. A further improved treatment of the energy-momentum deposition from jet into medium is to evolve this deposition within the hydrodynamic model before hadronization, as will be discussed below.

\subsection{Hydrodynamic response}
\label{subsec:hydroResponse}

One can assume the deposited energy and momentum from jets becomes locally thermalized and evolve hydrodynamically together with the background QGP medium. In this scenario, the medium response to jet propagation is described via solving the hydrodynamic equation with a source term,
\begin{equation}
\label{eq:hydroSource}
\partial_\mu T^{\mu\nu}(x)=J^\nu(x).
\end{equation}
where the source term $J^\nu(x)=[dE/d^4x,d{\vec p}/d^4x]$ represents the space-time profile of the energy-momentum deposition from jets into the medium.
 
Analytical solutions to Eq.~(\ref{eq:hydroSource}) exist under the assumption that the energy-momentum deposition from jets is a small perturbation on top of the medium evolution so that the medium response can be linearized as~\cite{CasalderreySolana:2004qm,Neufeld:2008fi,Neufeld:2008dx}
\begin{equation}
\label{eq:linearHydro}
T^{\mu\nu}\approx T^{\mu\nu}_0+\delta T^{\mu\nu};\;\; \partial_\mu T^{\mu\nu}_0=0, \;\; \partial_\mu \delta T^{\mu\nu}=J^\nu.
\end{equation}
Here $T^{\mu\nu}_0$ is the energy-momentum tensor of the unperturbed medium, and $\delta T^{\mu\nu}$ from the jet-induced medium excitation can be further decomposed as
\begin{equation}
\label{eq:linearHydroDecomp}
\begin{split}
&\delta T^{00}\equiv\delta \epsilon,\;\; \delta T^{0i}\equiv g^i, \\
&\delta T^{ij}=\delta^{ij}c_s^2\delta\epsilon + \frac{3}{4}\Gamma_s (\partial^i g^j + \partial^j g^i + \frac{2}{3} \delta^{ij}  \nabla\cdot{\vec g}),
\end{split}
\end{equation}
where $\delta\epsilon$ is the excess energy density, ${\vec g}$ is the momentum current, $c_s$ denotes the speed of sound, $\Gamma_s\equiv 4\eta/[3(\epsilon_0+p_0)]$ is the sound attenuation length with $\epsilon_0$ and $p_0$ being the unperturbed local energy density and pressure, respectively. Note that in Eq.~(\ref{eq:linearHydroDecomp}), the metric convention $g^{\mu\nu}=\mathrm{diag}\, (1, -1, -1, -1)$ is used as in Ref.~\cite{Song:2007ux}, and $\delta^{ij}=\mathrm{diag}\, (1, 1, 1)$ is the Kronecker delta-function.

With Fourier transformation, one may rewrite the last part of Eq.~(\ref{eq:linearHydro}) in the momentum space as
\begin{equation}
\label{eq:linearHydroPSpace}
\begin{split}
&J^0=-i\omega\delta\epsilon+i{\vec k}\cdot{\vec g}, \\
&{\vec J}=-i\omega{\vec g}+i{\vec k}c_s^2\delta\epsilon+\frac{3}{4}\Gamma_s\left[k^2{\vec g}+\frac{{\vec k}}{3}({\vec k}\cdot{\vec g})\right],
\end{split}
\end{equation}
which yields
\begin{align}
\label{eq:linearHydroSoln1}
\delta\epsilon({\vec k},\omega) &= \frac{(i\omega-\Gamma_s k^2)J^0({\vec k},\omega)+ikJ_L({\vec k},\omega)}{\omega^2-c_s^2 k^2 + i\Gamma_s\omega k^2},\\
\label{eq:linearHydroSoln2}
{\vec g}_L({\vec k},\omega) &= \frac{i c_s^2 {\vec k} J^0({\vec k},\omega)+i\omega \hat{k} J_L({\vec k},\omega)}{\omega^2-c_s^2 k^2 + i\Gamma_s\omega k^2},\\
\label{eq:linearHydroSoln3}
{\vec g}_T({\vec k},\omega) &= \frac{i {\vec J}_T({\vec k},\omega)}{\omega+\frac{3}{4}i\Gamma_s k^2}.
\end{align}
Here both the source term and the perturbed momentum current are divided into transverse and longitudinal components in the momentum space: ${\vec J}=\hat{k}{J}_L+{\vec J}_T$ and ${\vec g}={\vec g}_L+{\vec g}_T$. Therefore, with the knowledge of the source term, one may obtain the variation of the energy-momentum tensor of the medium after Fourier transforming Eqs.~(\ref{eq:linearHydroSoln1})-(\ref{eq:linearHydroSoln3}) back to the coordinate space.

\begin{figure}[tbp]
    \centering
    \includegraphics[width=0.40\textwidth]{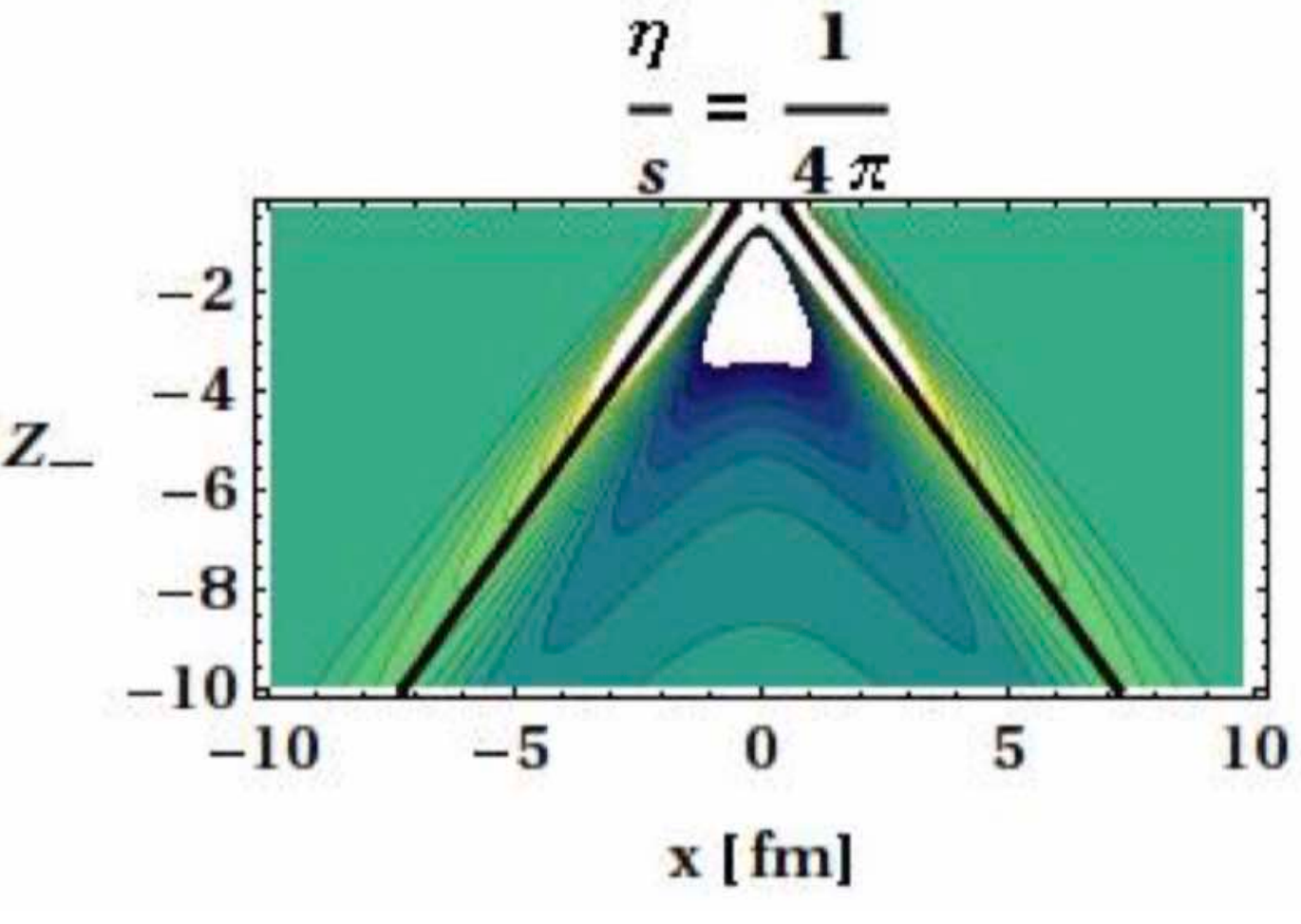}
    \includegraphics[width=0.40\textwidth]{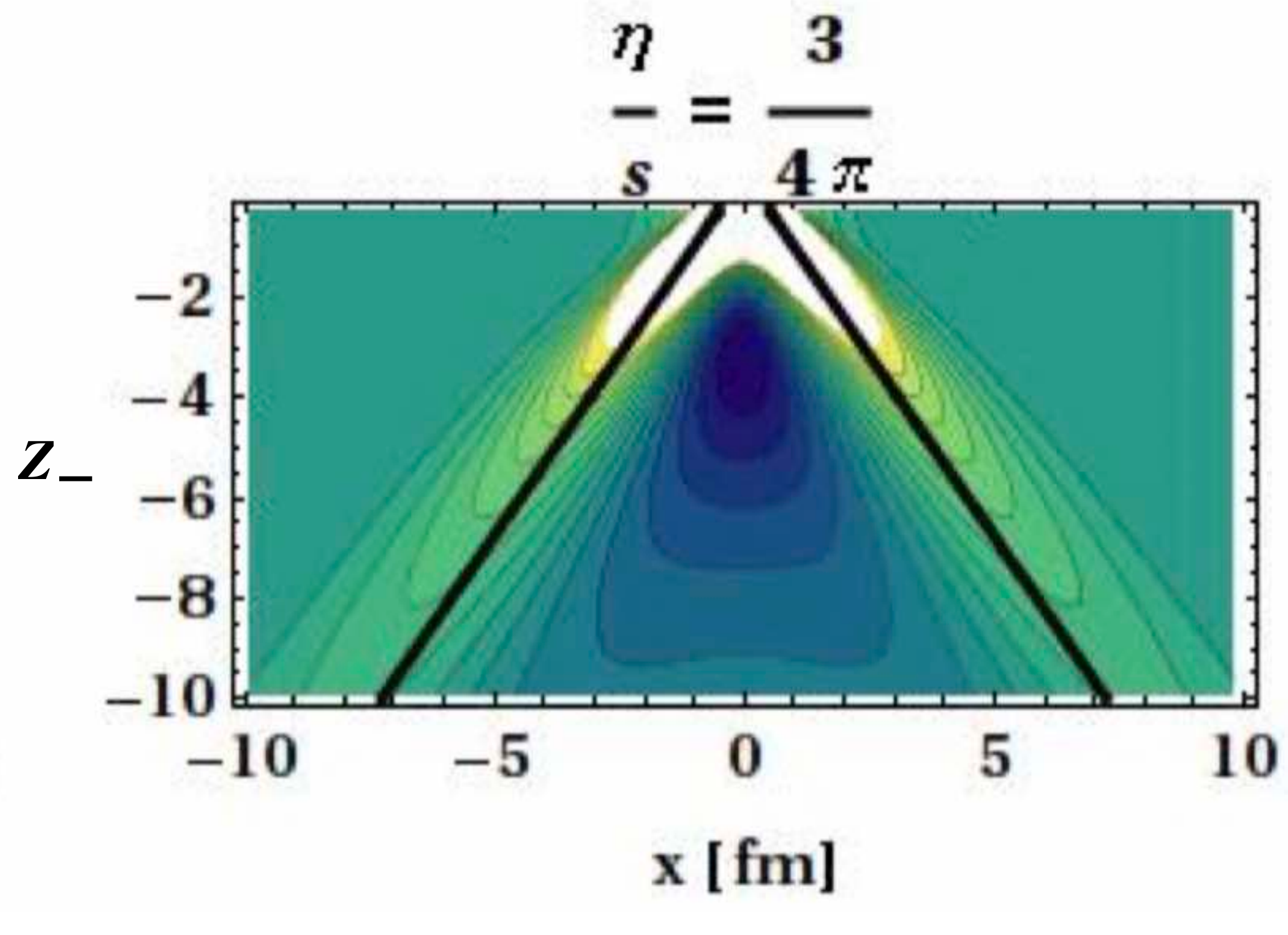}
    \caption{(Color online) The perturbed energy density for different values of the shear-viscosity-to-entropy-density ratio $\eta/s$ when an energetic gluon propagates along the $+z$ direction through a static QGP medium with temperature $T = 350$~MeV and sound velocity $c_s=1/3$. The figures are from Ref.~\cite{Neufeld:2008dx}.}
    \label{fig:linearResponse-Neufeld}
\end{figure}

Many studies have been implemented using this linear response approach. For instance, the Mach cone structure of the perturbed medium induced by jet propagation has been proposed in Refs.~\cite{CasalderreySolana:2004qm,Neufeld:2008fi,Neufeld:2008dx,Yan:2017rku}, as demonstrated in Fig.~\ref{fig:linearResponse-Neufeld}, where the intensity of the Mach cone is found weaker with growing kinematic viscosity. If observed in heavy-ion collisions, these Mach cone structures may provide more direct constraints on the QGP properties, such as the shear viscosity and the speed of sound. However, the strong collective motion of the dynamically evolving QGP with realistic initial geometry may destroy the Mach cone pattern~\cite{Renk:2005si,Bouras:2014rea}, and no sensitive experimental observables have been found so far. Using the linear response theory, different structures of jet-induced excitations inside weakly coupled and strongly coupled QGP are also compared in Ref.~\cite{Ruppert:2005uz}. The relation between hard ($\hat{q}$), soft ($\eta/s$) transport parameters and jet energy loss are also investigated in Ref.~\cite{Ayala:2016pvm}.

\begin{figure}[tbp]
    \centering
    \includegraphics[width=0.40\textwidth]{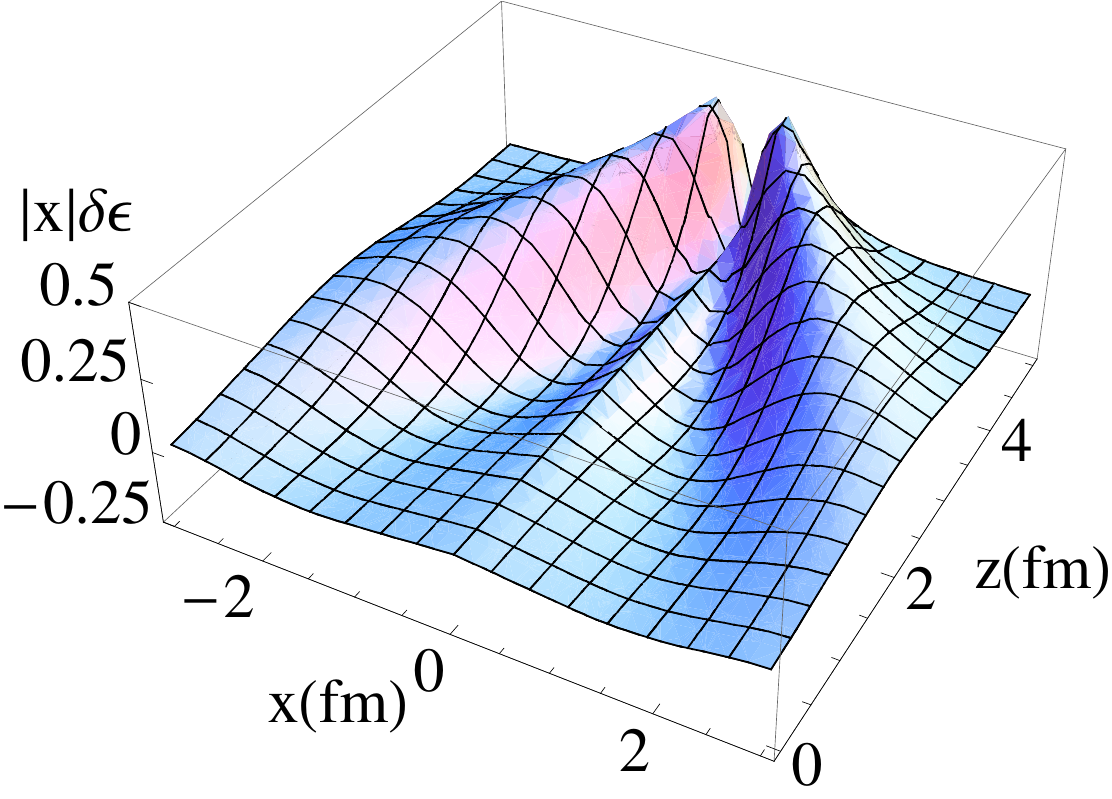}
    \includegraphics[width=0.40\textwidth]{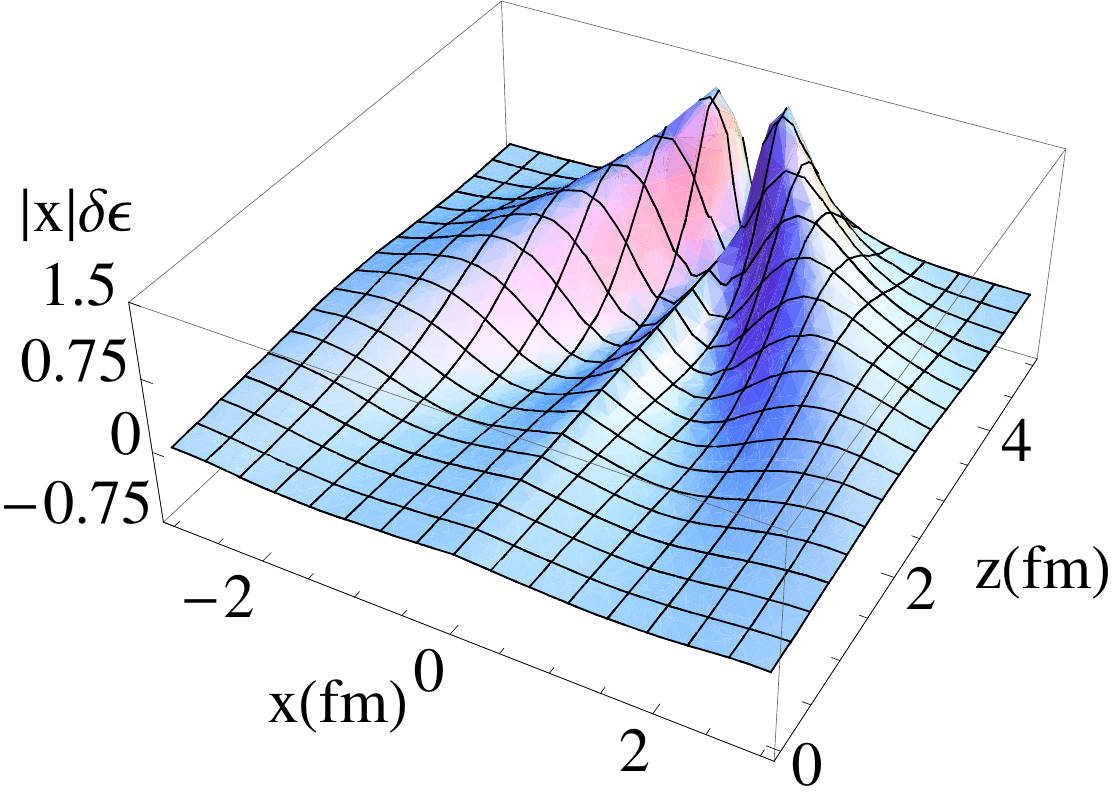}
    \caption{(Color online) The linear hydrodynamical response to energy deposition from a single quark (top) vs. a quark-initiated jet shower (bottom). The figures are from Ref.~\cite{Qin:2009uh}.}
    \label{fig:Qin-shower-source}
\end{figure}

In earlier studies, a simplified model for constructing the source term was usually applied, where the energy-momentum deposition was assumed to come from a single parton. However, jets are collimated showers of partons that may transfer energy into the medium via a combination of elastic and inelastic processes of all constituent partons. Such more realistic modelings of the source term have been proposed in Refs.~\cite{Neufeld:2009ep,Qin:2009uh}. As shown in Fig.~\ref{fig:Qin-shower-source}, a significantly enhanced conical pattern of the hydrodynamical response can be observed when depositing energy with realistic jet parton showers as compared to using energy loss from a single parton. Moreover, quantum interference effects between the primary parton and the radiated gluons within jet showers have been investigated in Ref.~\cite{Neufeld:2011yh} and shown to enhance the energy transfer from jet showers to the QGP and destroy the Mach cone structure of medium response when the gluon emission angle is large.

\begin{figure}[tbp]
    \centering
    \includegraphics[width=0.48\textwidth]{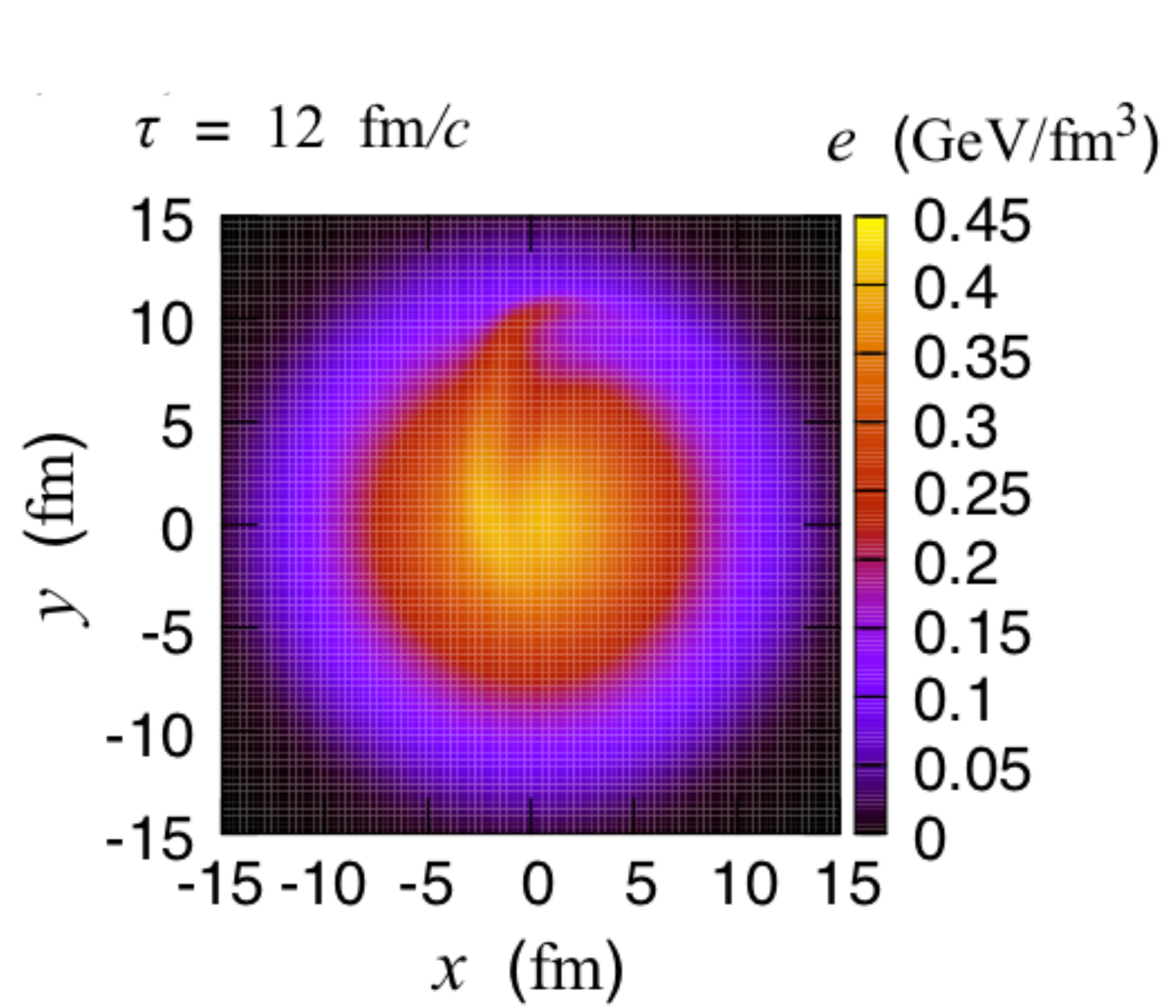}
    \includegraphics[width=0.48\textwidth]{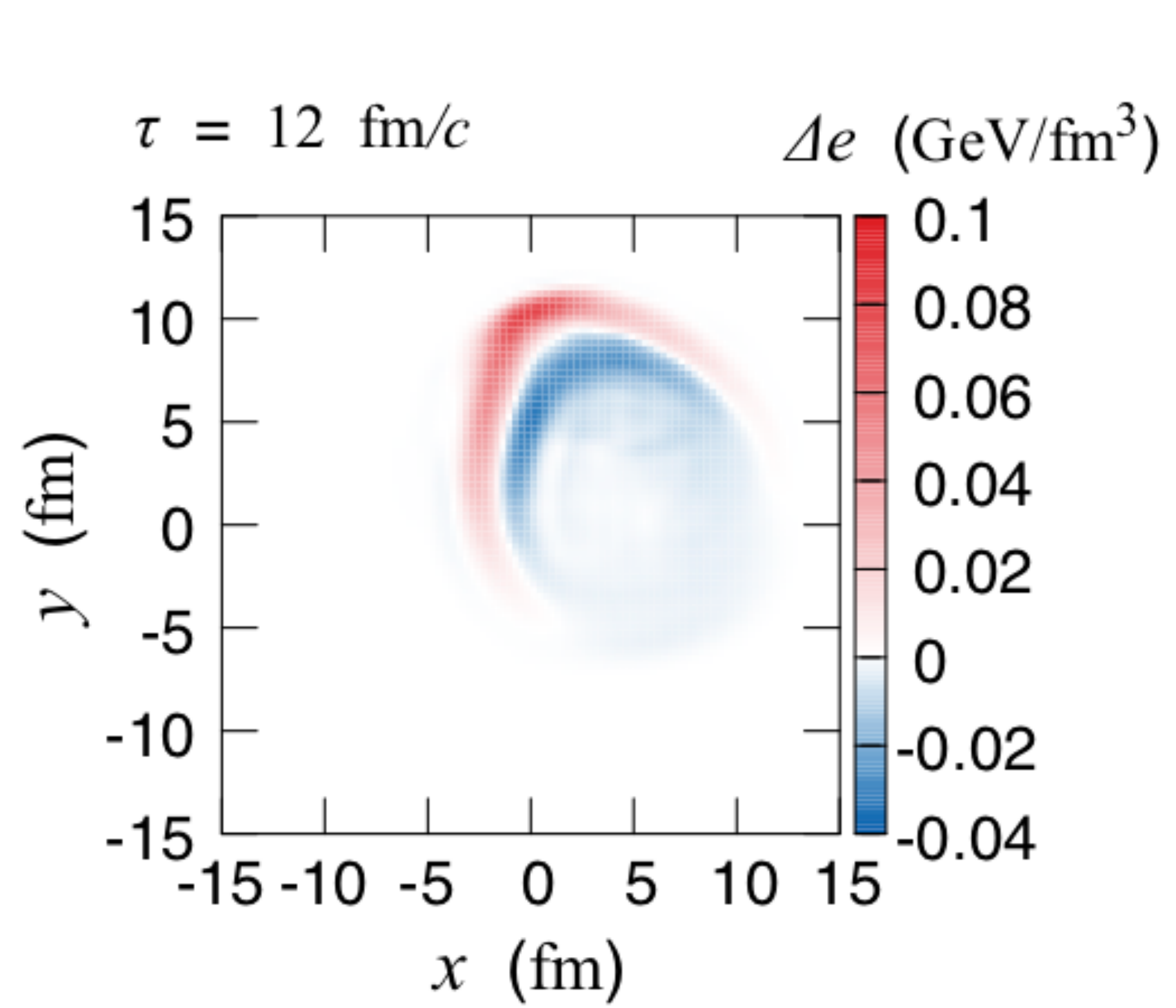}
    \caption{(Color online) The energy density distribution of the QGP in the transverse plane at mid-rapidity in 2.76~TeV central Pb+Pb collisions, with the presence of a single jet initiated at position $(x=0~\textrm{fm},\,y=6.54~\textrm{fm})$ with momentum $(p_\mathrm{T}=150~\textrm{GeV},\, \phi_p=5\pi/8)$, before (top) and after the background subtraction (bottom). The figures are from Ref.~\cite{Tachibana:2017syd}.}
    \label{fig:Tachibana-jime-simulation}
\end{figure}

When the local energy density deposition from jets is comparable to that of the unperturbed medium, linearized hydrodynamic equations [Eq.~(\ref{eq:linearHydro})] are no longer valid. Without these approximations, full solutions to hydrodynamic equations with source terms [Eq.~(\ref{eq:hydroSource})] were provided in Ref.~\cite{Floerchinger:2014yqa} using a (1+1)-D hydrodynamic model, Ref.~\cite{Chaudhuri:2005vc} using a (2+1)-D hydrodynamic model, and Refs.~\cite{Betz:2010qh,Tachibana:2014lja,Tachibana:2020mtb} using full (3+1)-D hydrodynamic models. Within such an improved framework, it is found in Ref.~\cite{Tachibana:2014lja} that while the jet-induced Mach cone is easily distorted in the transverse plane, its pattern remains in the longitudinal direction in the reaction plane due to the expanding $(\tau,\eta)$ coordinates. To obtain the net effects of jet-induced medium excitation, one can subtract the energy density profile from hydrodynamic calculations without the presence of the source term from that with the source, as discussed in Ref.~\cite{Tachibana:2017syd}. Figure~\ref{fig:Tachibana-jime-simulation} shows a snapshot of the QGP evolution that is being disturbed by a single propagating jet before (top) and after (bottom) subtraction of the background medium without energy deposition from jets. One can clearly see a Mach cone induced by the energy and momentum deposition from the jet, as well as a region of energy depletion right behind the wave front, known as the diffusion wake. During the propagation, the wave front is distorted by the radial flow of the medium. Since the jet travels through an off-central path, the Mach cone is deformed asymmetrically in this event. Unfortunately,  these typical conic structures of the jet-induced Mach cone are still hard to identify in current experimental measurements of the final hadron observables as we will discuss in the next few sections.

\subsection{Coupled parton transport and hydrodynamics}
\label{subsec:concurrent}

The parton transport and hydrodynamic description of jet-induced medium response, as presented in Secs.~\ref{subsec:recoil} and \ref{subsec:hydroResponse},  can be considered as two different approaches to modeling how the lost energy-momentum from jets evolves inside the medium in the limit of weakly and strongly coupled system, respectively.  The real scenario may be something in between. Furthermore, neither of them considers how the modified medium in turn affects the subsequent evolution of jets. This could be important when there are multiple jets within one event where one jet travels through the region disturbed by another jet, or when a slowly moving heavy quark may interact with the medium excitation induced by itself. To take these effects into account and bridge the parton transport and hydrodynamic approach, one can develop a coupled approach with concurrent evolution of jets and the QGP medium.

The coupled \textsc{Lbt} and hydrodynamics (\textsc{CoLbt-Hydro}) model~\cite{Chen:2017zte,Chen:2018azc} combines the \textsc{Lbt} model for jet transport and the (3+1)-D viscous hydrodynamic model \textsc{CLVisc}~\cite{Pang:2012he,Pang:2014ipa} for the QGP expansion, and realizes the first concurrent simulation of transport of energetic partons and evolution of the thermal medium. The real-time coupling between  parton transport and hydrodynamic evolution of the medium is through a source term that is updated at each time-step of the  \textsc{Lbt}  and \textsc{CLVisc} simulation.  To achieve this, the Boltzmann equation in \textsc{Lbt} is re-written in the Milne coordinates
as in the  \textsc{CLVisc} hydrodynamic model. At each step of the proper time  $(\tau,\tau+\Delta \tau)$, \textsc{CLVisc} provides the temperature and flow velocity information of the local fluid cell for the simulations of the elastic and inelastic scatterings of hard partons, including both jet shower and recoil partons,  with the background medium.  Among the final-state partons during this step of the proper time, jet and recoil partons below a given energy scale ($p\cdot u < p^0_\mathrm{cut}$), together with all ``negative" partons in the back-reaction, are transferred from the \textsc{Lbt} module to  \textsc{CLVisc}  in a source term constructed as
\begin{equation}
\begin{split}
J^\nu = &\sum_i \frac{\theta(p^0_\mathrm{cut}-p_i \cdot u)dp_i^\nu/d\tau}{\tau (2\pi)^{3/2}\sigma_r^2\sigma_{\eta_s}}\\
&\times \exp\left[-\frac{({\vec x}_\perp-{\vec x}_{\perp i})^2}{2\sigma_r^2}-\frac{(\eta_s-\eta_{si})^2}{2\sigma_{\eta_s}^2}\right].
\end{split}
\end{equation}
Here, an instantaneous thermalization of low-energy partons ($p\cdot u < p^0_\mathrm{cut}$) in the source term for the hydrodynamic evolution of the medium is assumed, and the energy-momentum deposition from each parton is smeared in the coordinate space with Gaussian widths $\sigma_r=0.6$~fm and $\sigma_{\eta_s}=0.6$. This source term enters the \textsc{CLVisc} hydrodynamic evolution [Eq.~(\ref{eq:hydroSource})] for the next step of the proper time. Iteration of this algorithm provides a simultaneous evolution of jets, medium and their interactions. As discussed in Sec.~\ref{subsec:hydro}, in order to break the bottleneck of the computing speed for concurrent simulation of parton transport and hydrodynamic evolution, \textsc{CoLbt-Hydro} parallelizes the hydrodynamic calculations on GPUs, including both the Kurganov-Tadmor algorithm for the space-time evolution of the QGP and the Cooper-Frye particlization, using the Open Computing Language (OpenCL). Benefiting from the large number of computing elements on GPUs and the Single Instruction Multiple Data (SIMD) vector operations on modern CPUs, \textsc{CLVisc} brings the fastest (3+1)D hydrodynamics calculations so far and makes event-by-event \textsc{CoLbt-Hydro} simulations possible.

\begin{figure}[tbp]
    \centering
    \includegraphics[width=0.4\textwidth]{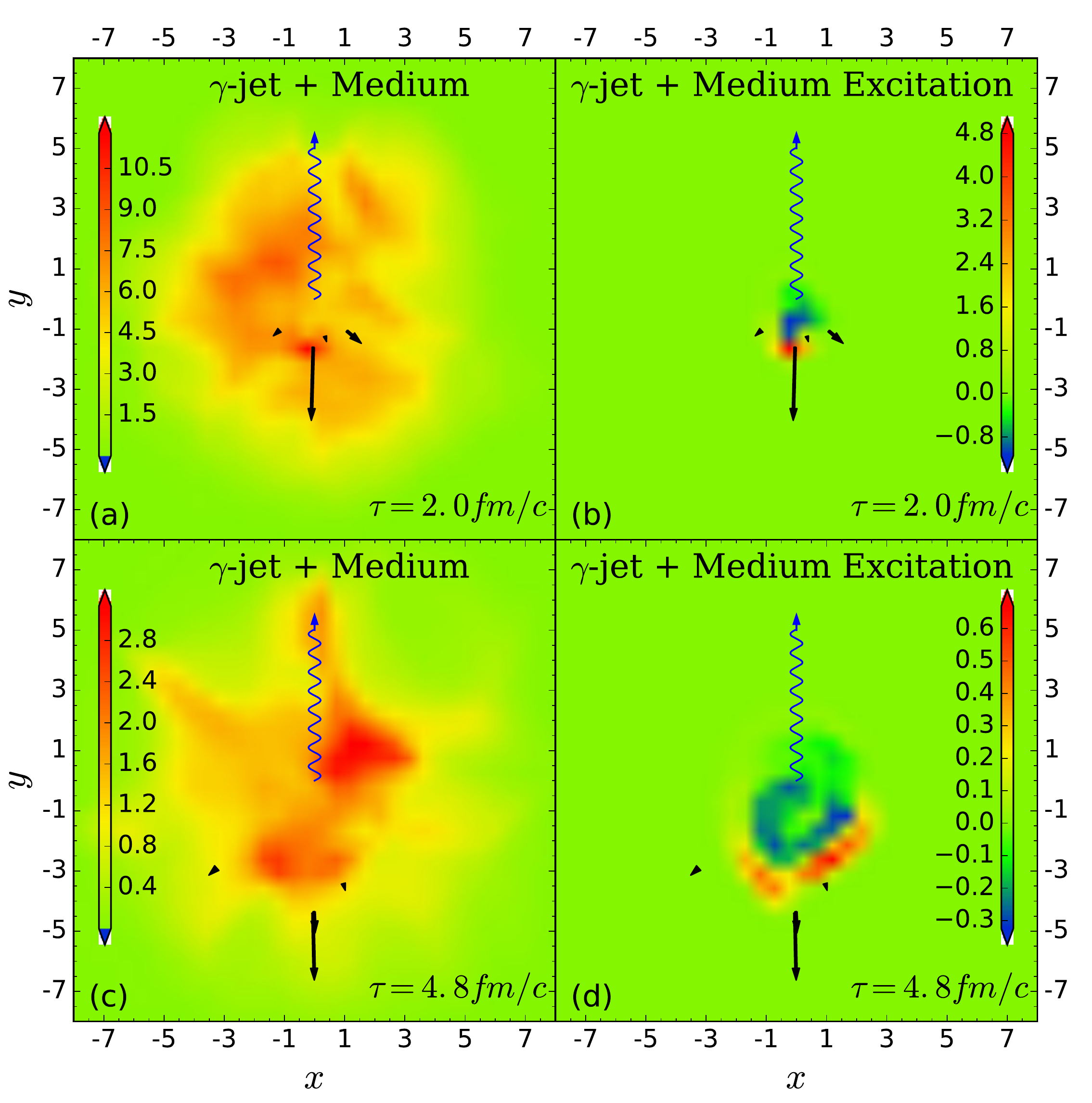}
    \caption{(Color online) The energy density profiles of the QGP and the $\gamma$-jet evolution in the transverse plane at $\eta_s=0$, $\tau = 2.0$ (a, b) and 4.8 fm/c (c, d) in 0-12\% Au+Au collisions at $\sqrt{s}=200$ GeV without (left) and with (right)  background subtraction. Straight and wavy lines represent partons' and $\gamma$'s momenta respectively. The figure is from Ref.~\cite{Chen:2017zte}.}
    \label{fig:Chen-jime-simulation}
\end{figure}

Figure~\ref{fig:Chen-jime-simulation} shows two snapshots, one at $\tau=2.0$ (upper) and the other at $\tau=4.8$~fm (lower), of a $\gamma$-triggered jet event in central Au+Au collisions at $\sqrt{s}=200$~GeV from \textsc{CoLbt-Hydro} simulations. The $\gamma$-jet is produced at the center of the collision and the photon propagates along the $+y$ direction, as indicated by the wavy lines. The left column displays the energy density profiles of the whole collision event, while the right shows the energy density after subtracting the background from the same hydrodynamic event without the presence of the $\gamma$-jet. In Fig.~\ref{fig:Chen-jime-simulation}, one can clearly observe both the medium modification on jet partons, including their splittings and energy loss as shown by the straight lines, as well as the jet-induced modification on the medium in the form of the Mach-cone-like wave fronts of energy deposition, followed by the energy depletion of the diffusion wake behind.

As discussed in Sec.~\ref{subsec:transport}, full Boltzmann transport models have been used to simulate the QGP evolution and propagation of energetic partons through the medium despite the controversy over whether one may use the pQCD-driven transport to describe the strongly coupled QGP matter. Nevertheless, they provide an alternative method that naturally simulates jet and medium evolution concurrently. For instance, by using the \textsc{Ampt} model, Ref.~\cite{Ma:2010dv} investigates how to isolate the effects of jet-induced Mach cone and expanding hot spots on the di-hadron vs. $\gamma$-hadron correlation; Ref.~\cite{Gao:2016ldo} studies how the lost energy from di-jets is redistributed in the lower $p_\mathrm{T}$ hadrons. Similar to Fig.~\ref{fig:linearResponse-Neufeld}, the viscous effects on the Mach cone structures have been studied within the \textsc{Bamps} model~\cite{Bouras:2012mh,Bouras:2014rea}, where effects of different energy loss rates have also been discussed. Moreover, while most studies to date assume instantaneous thermalization of the energy-momentum deposition from jets to the QGP, the detailed thermalization process has been explored within a Boltzmann-based parton cascade model in Ref.~\cite{Iancu:2015uja}.

%
%
%
%

\section{Hadron spectra}
\label{sec:hadron_spectra}

\subsection{Single inclusive hadrons}
\label{subsec:singleHadron}

Nuclear modification of single inclusive hadrons is the most direct measure of the in-medium energy loss of energetic partons. The most frequently used observable is the nuclear modification factor first defined in Ref.~\cite{Wang:1998bha} for jet quenching as,
\begin{equation}
\label{eq:defRAA}
R_\mathrm{AA}(p_\mathrm{T},y,\phi)\equiv\frac{1}{\langle N_\mathrm{coll} \rangle}\frac{\;\;\frac{dN_\mathrm{AA}}{dp_\mathrm{T}dyd\phi}\;\;}{\;\;\frac{dN_\mathrm{pp}}{dp_\mathrm{T}dyd\phi}\;\;},
\end{equation} 
where $\langle N_\mathrm{coll}\rangle$ is the average number of nucleon-nucleon collisions per nucleus-nucleus collision for a given range of centrality, which can be evaluated using the Glauber model. Note that while correctly evaluating $N_\mathrm{coll}$ is important to extract $R_\mathrm{AA}$ from experimental data, it is not necessary in theoretical calculations where the QGP effects are usually implemented on the p+p spectra that have been modified with cold nuclear matter effects. The suppression or nuclear modification factor $R_\mathrm{AA}$  quantifies the variation of hadron spectra in A+A vs. p+p collisions, and has been investigated in nearly all theoretical studies on jets~\cite{Bass:2008rv,Wang:2001ifa,Vitev:2002pf,Salgado:2003gb,Dainese:2004te,Vitev:2004bh,Armesto:2005iq,Wicks:2005gt,Armesto:2009zi,Marquet:2009eq,Chen:2010te,Renk:2010mf,Renk:2011gj,Horowitz:2011gd,Chen:2011vt,Cao:2017hhk,Cao:2017qpx}. 

In general, the hadron spectra produced in high-energy nuclear collisions can be written as
\begin{align}
\label{eq:xsectionFactor}
d\sigma_{pp(AA)\to hX} &= \sum_{abc} \int dx_a \int dx_b \int dz_c f_{a}(x_a) f_{b}(x_b)   \nonumber\\
& \times d\hat{\sigma}_{ab\to c} D^\mathrm{vac(med)}_{h/c}(z_c),
\end{align}
where $\sum_{abc}$ sums over all possible parton flavors, $f_a(x_a)$ and $f_b(x_b)$ are parton distribution functions (PDFs) per nucleon for partons $a$ and $b$ from the two colliding protons (nuclei), $\hat{\sigma}_{ab\to c}$   is the partonic scattering cross section, and $D_{h/c}(z_c)$ is the parton-to-hadron fragmentation function (FF). The PDFs can be taken from CTEQ parameterizations~\cite{Pumplin:2002vw} for p+p collisions, but need to be convoluted with cold nuclear matter effects for A+A collisions, e.g. as implemented in the EPS parameterizations~\cite{Eskola:2009uj} for the nuclear modification of the PDFs. The vacuum (vac) FF is used in Eq.~(\ref{eq:xsectionFactor}) for p+p collisions, while the medium-modified (med) FF should be applied for A+A collisions, as discussed in detail in Sec.~\ref{sec:theory}.

Neglecting hadron production from the hadronization of radiated gluons and recoil partons from the medium response, which contribute mostly to soft hadrons,
the medium-modified FF can be approximated by shifting the momentum of the fragmenting parton~\cite{Wang:1996yh,Salgado:2003gb},
\begin{equation}
\label{eq:medFF}
D_{h/c}^\mathrm{med}(z_c)=\int d\epsilon \frac{P(\epsilon)}{1-\epsilon}D_{h/c}^\mathrm{vac}\left(\frac{z_c}{1-\epsilon}\right),
\end{equation}
where $\epsilon=\Delta E_c / E_c$ is the fractional energy loss of parton $c$ inside the medium, and $P(\epsilon)$ is the probability distribution of the energy loss $\epsilon$. In this approximation, one assumes that a high-energy parton in A+A collisions first loses energy inside the thermal medium, and then fragments into hadrons with its remaining fractional energy $1-\epsilon$ outside the medium (in vacuum).  

\begin{figure}[tbp]
    \centering
    \includegraphics[width=0.23\textwidth]{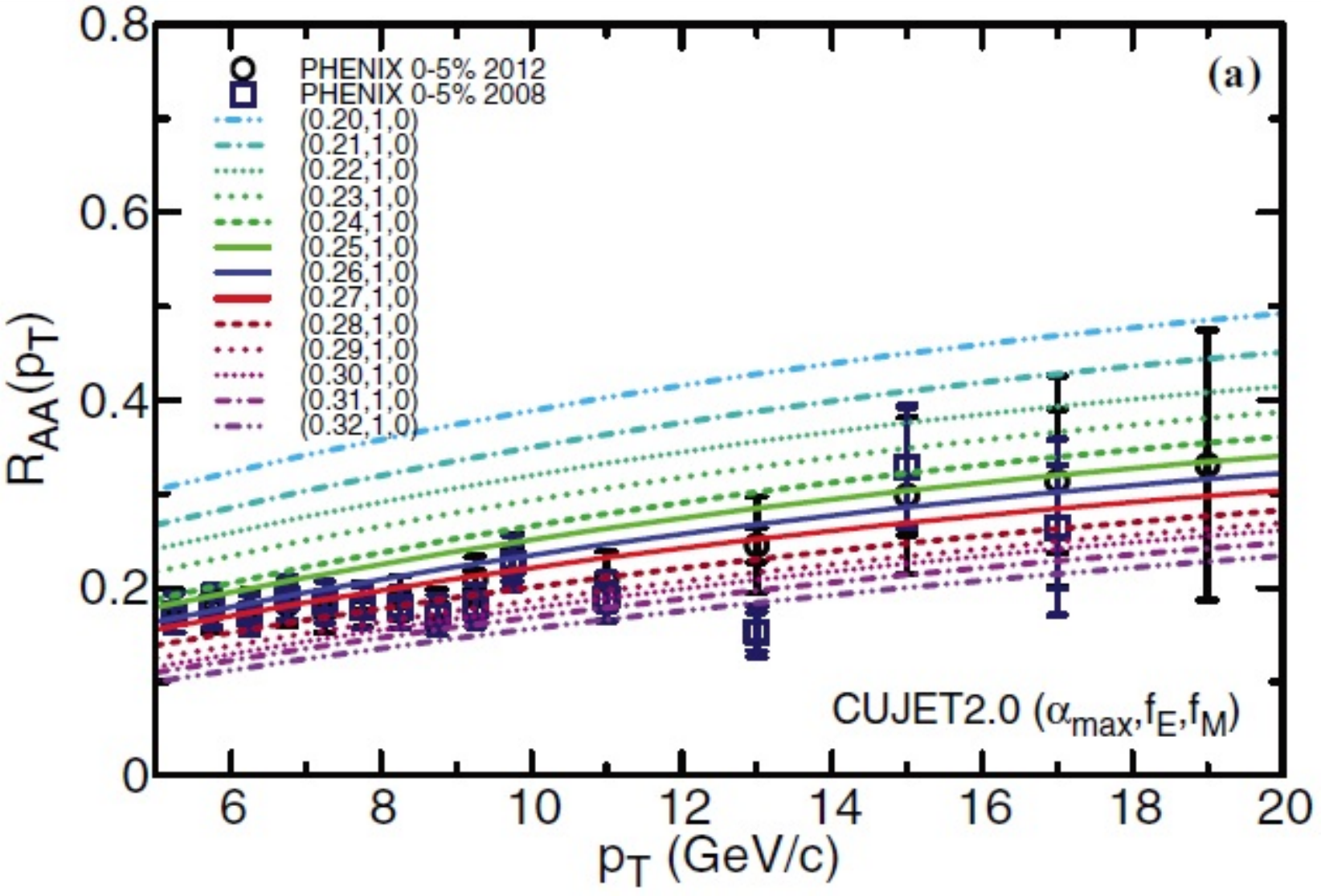}
    \includegraphics[width=0.23\textwidth]{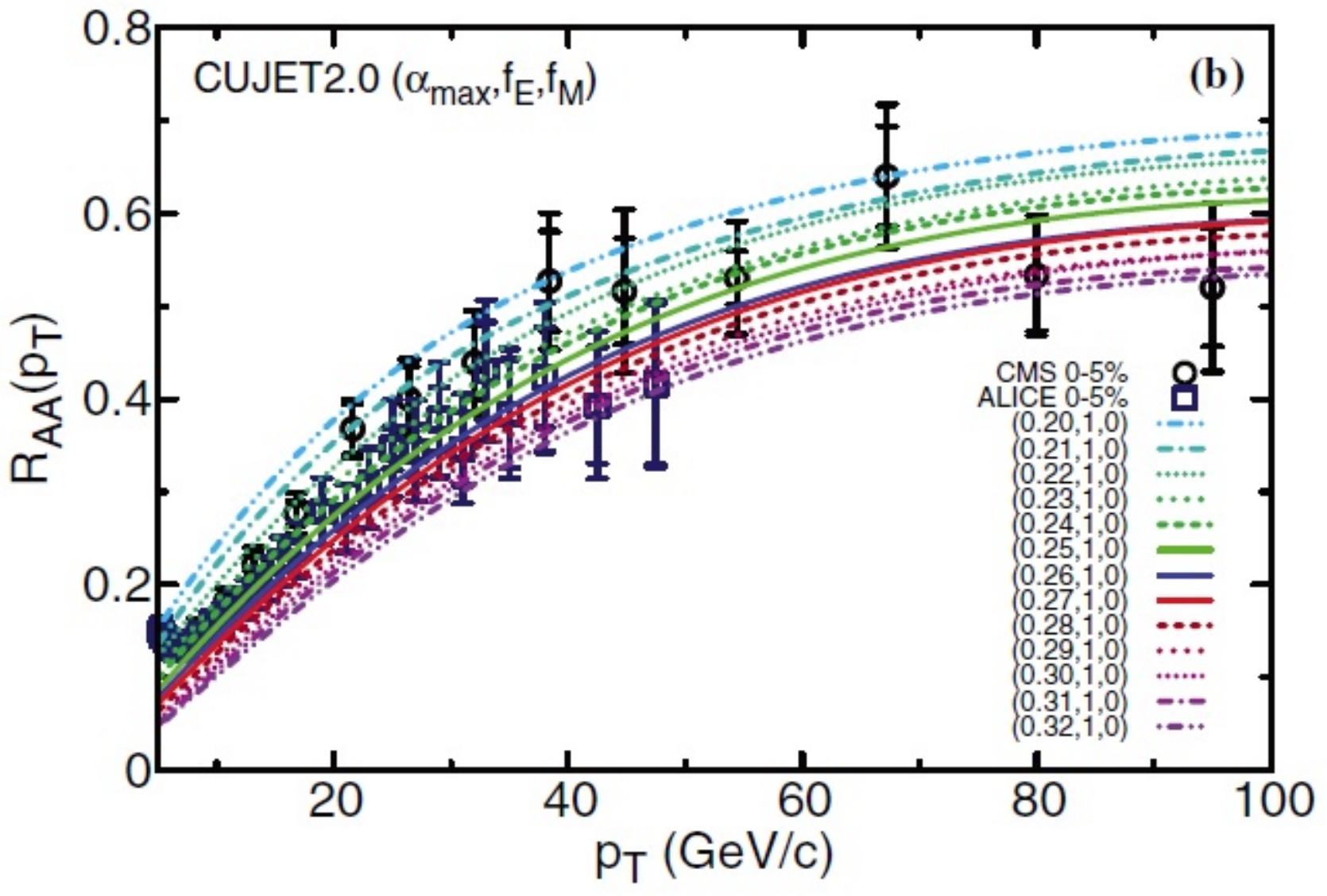}
    \includegraphics[width=0.23\textwidth]{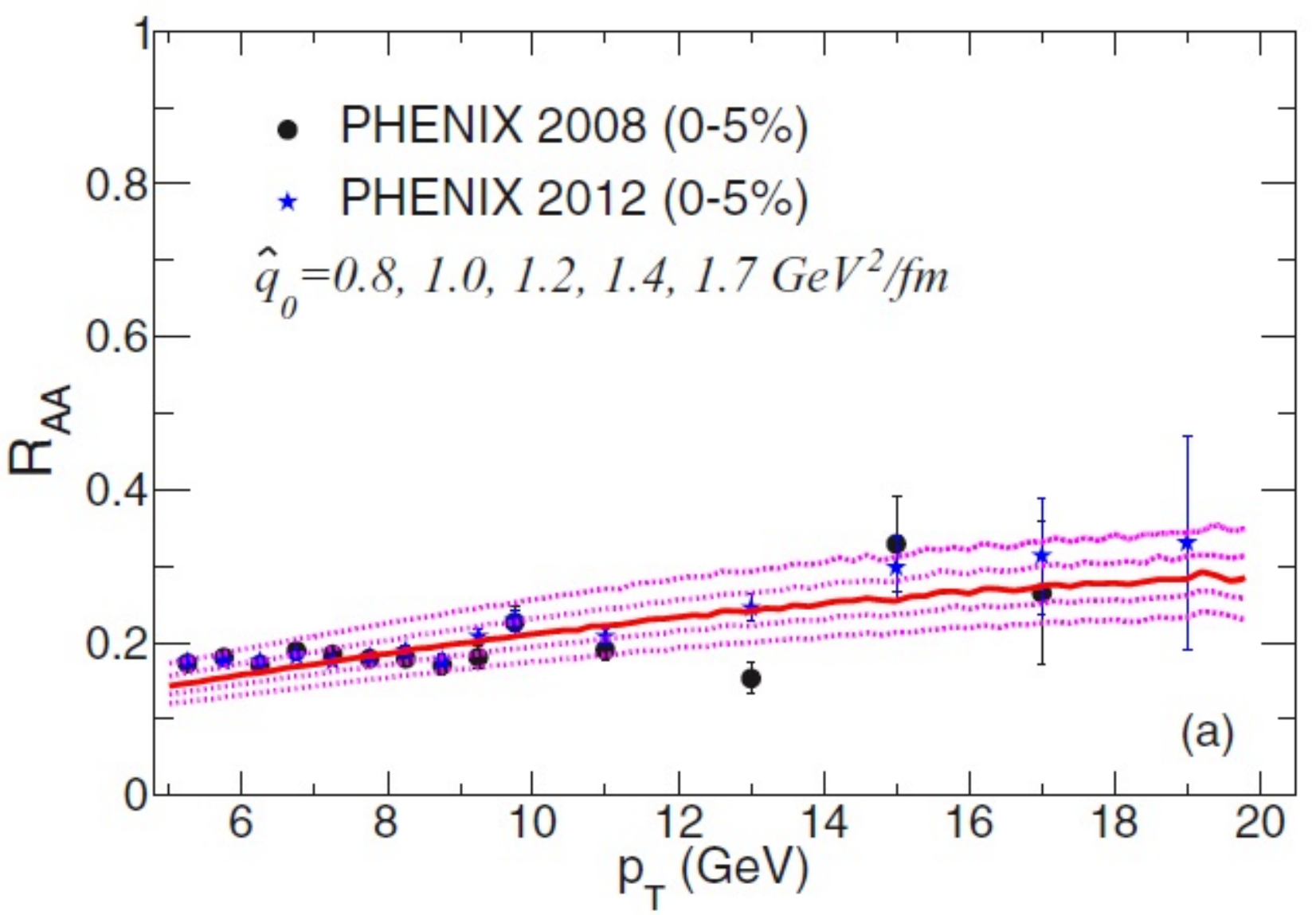}
    \includegraphics[width=0.23\textwidth]{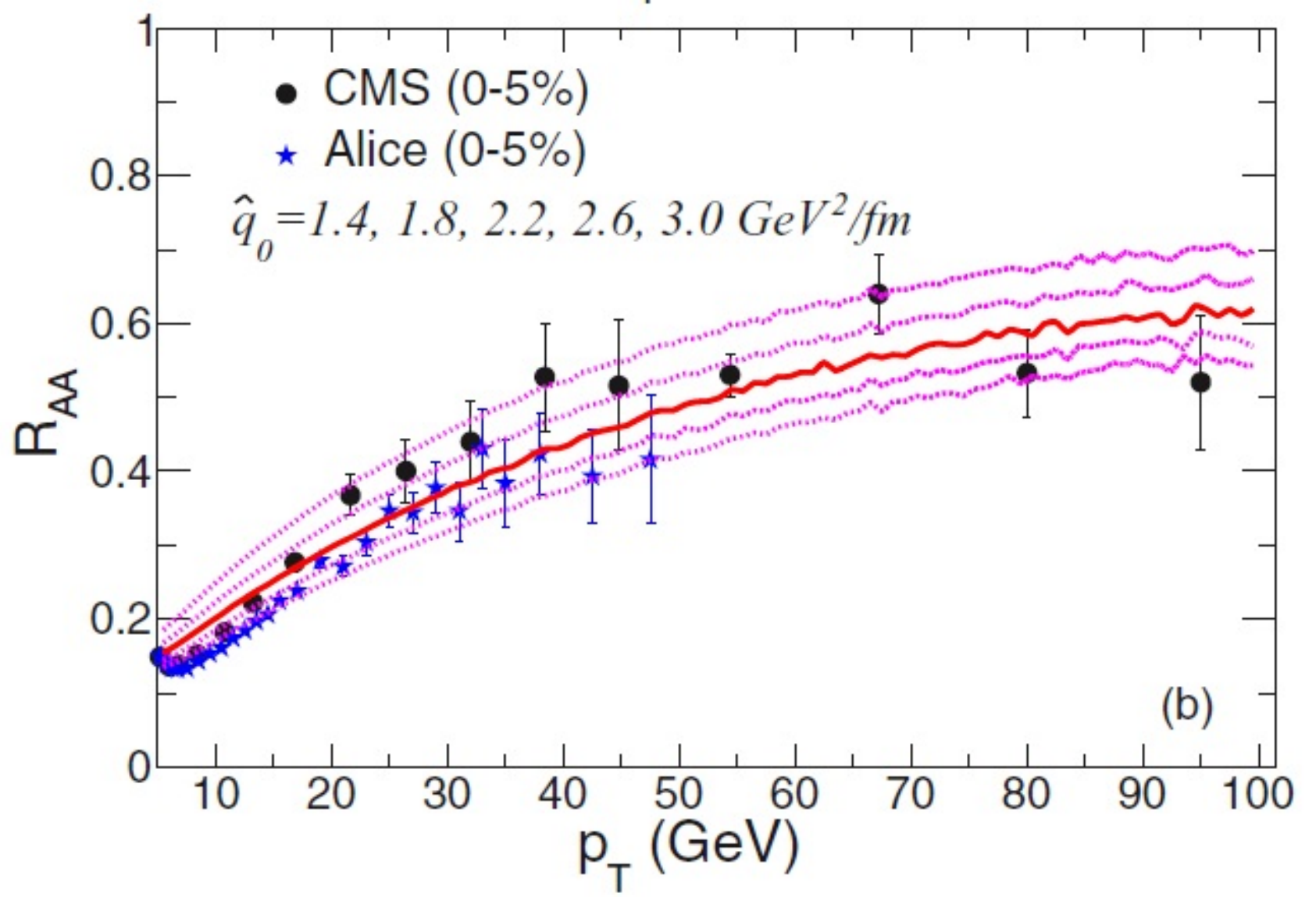}
    \includegraphics[width=0.23\textwidth]{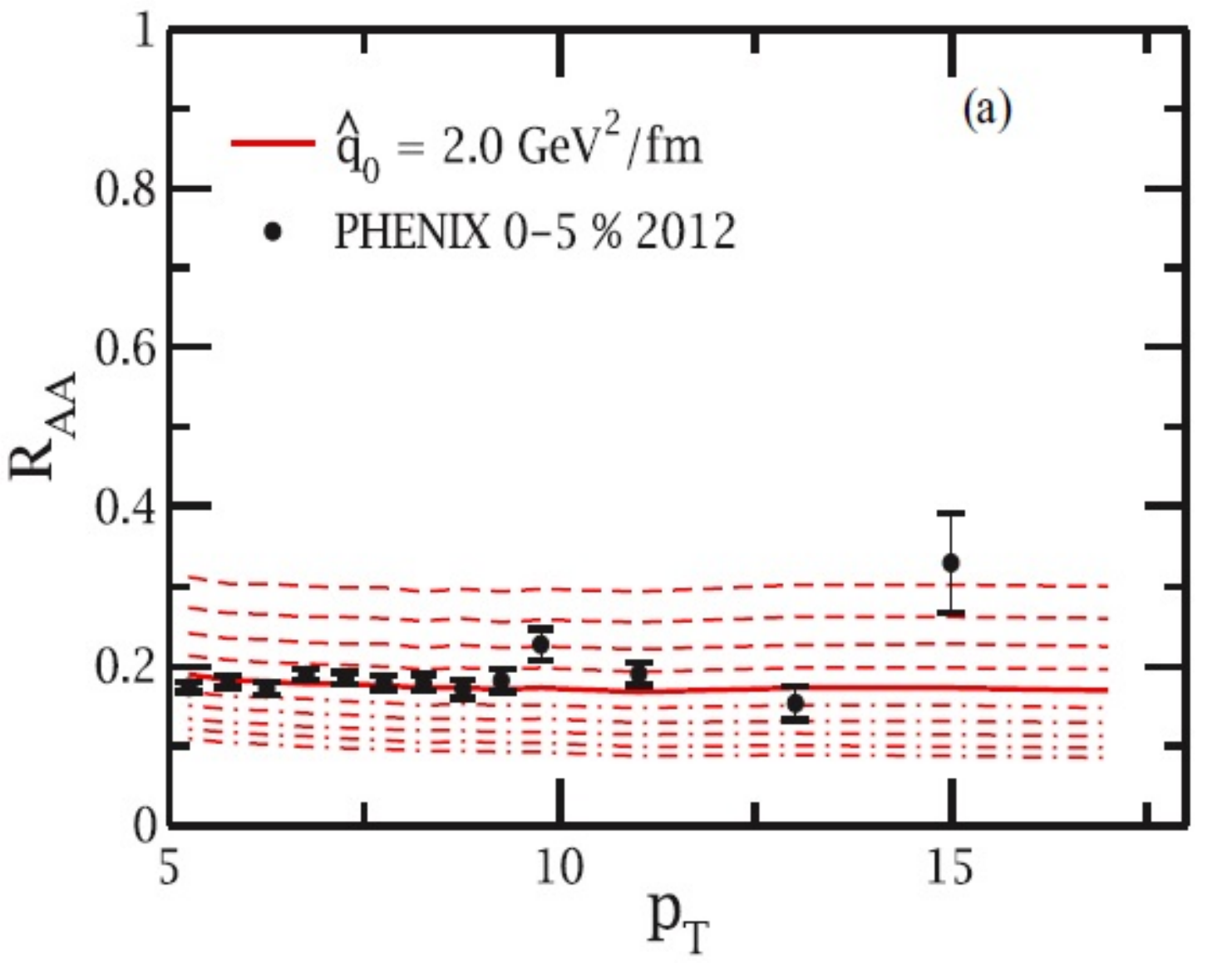}
    \includegraphics[width=0.23\textwidth]{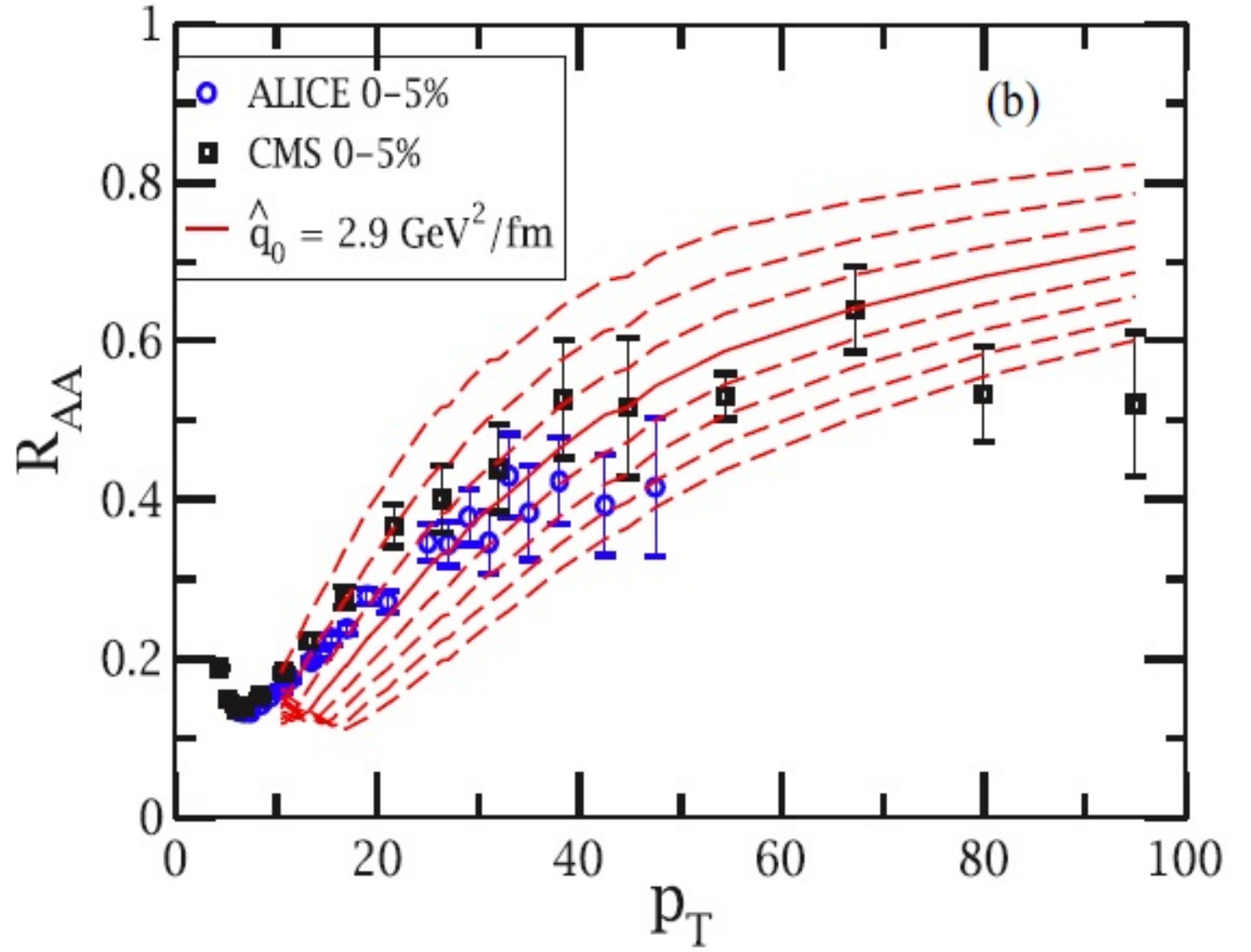}
    \includegraphics[width=0.23\textwidth]{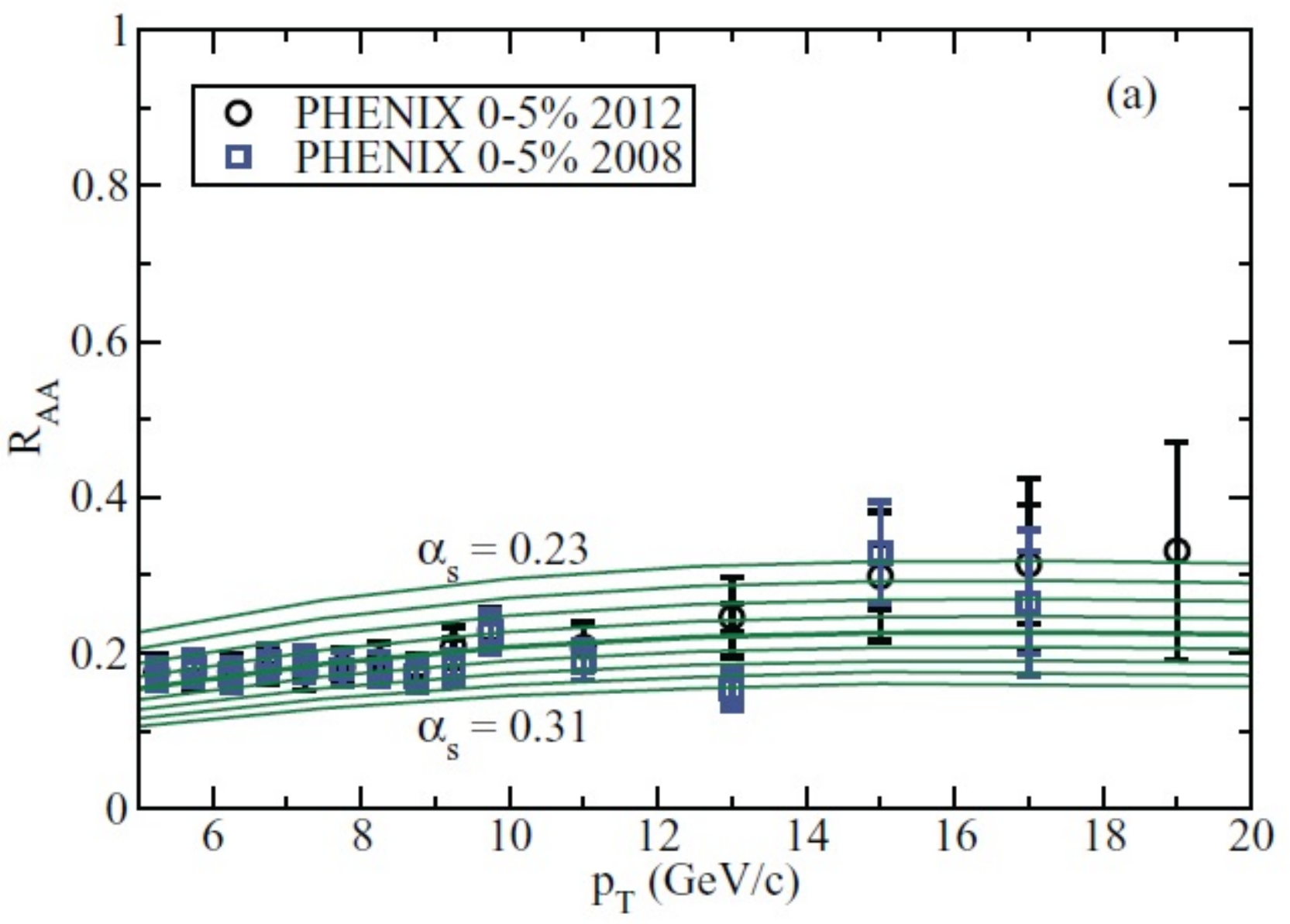}
    \includegraphics[width=0.23\textwidth]{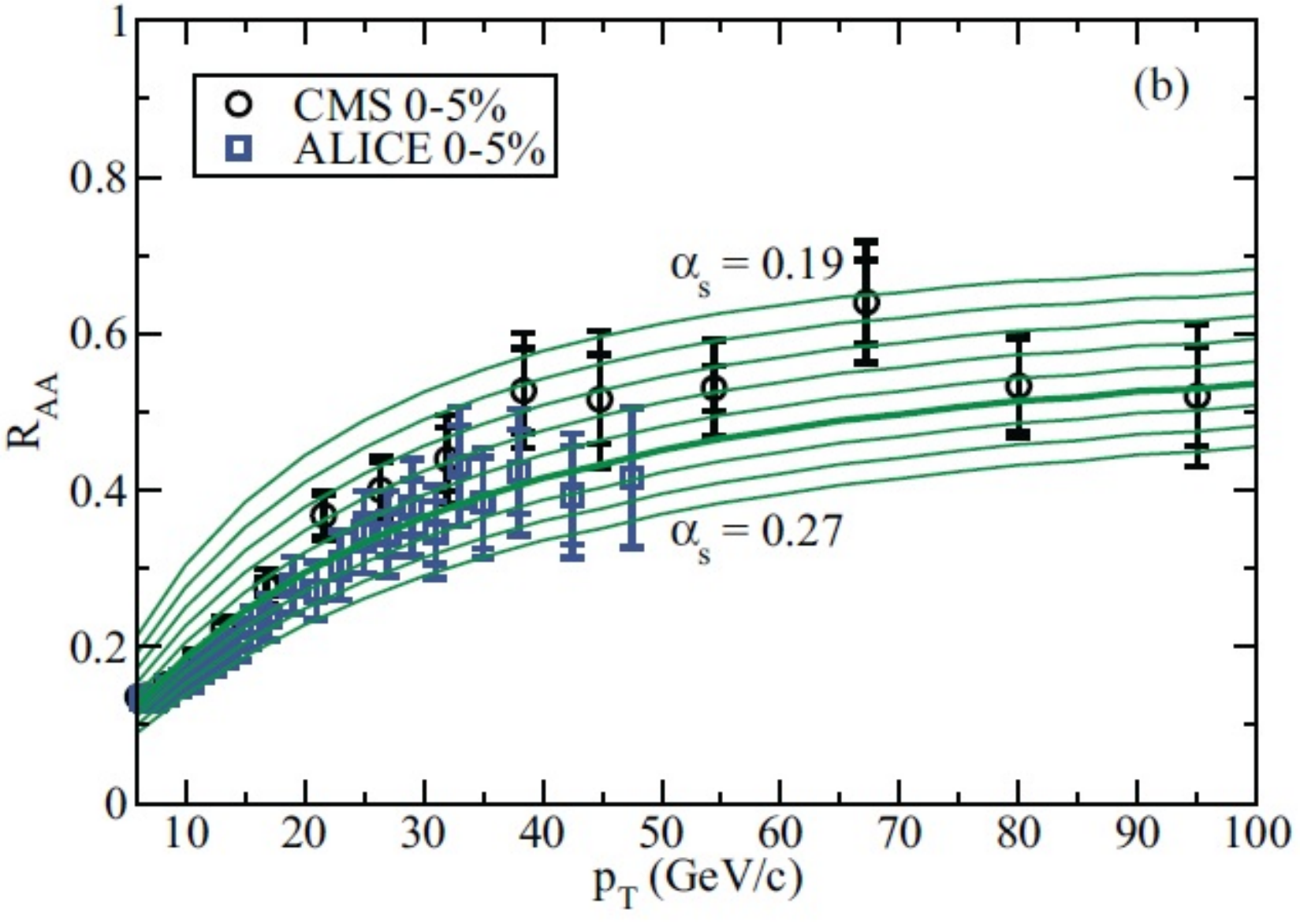}
    \includegraphics[width=0.23\textwidth]{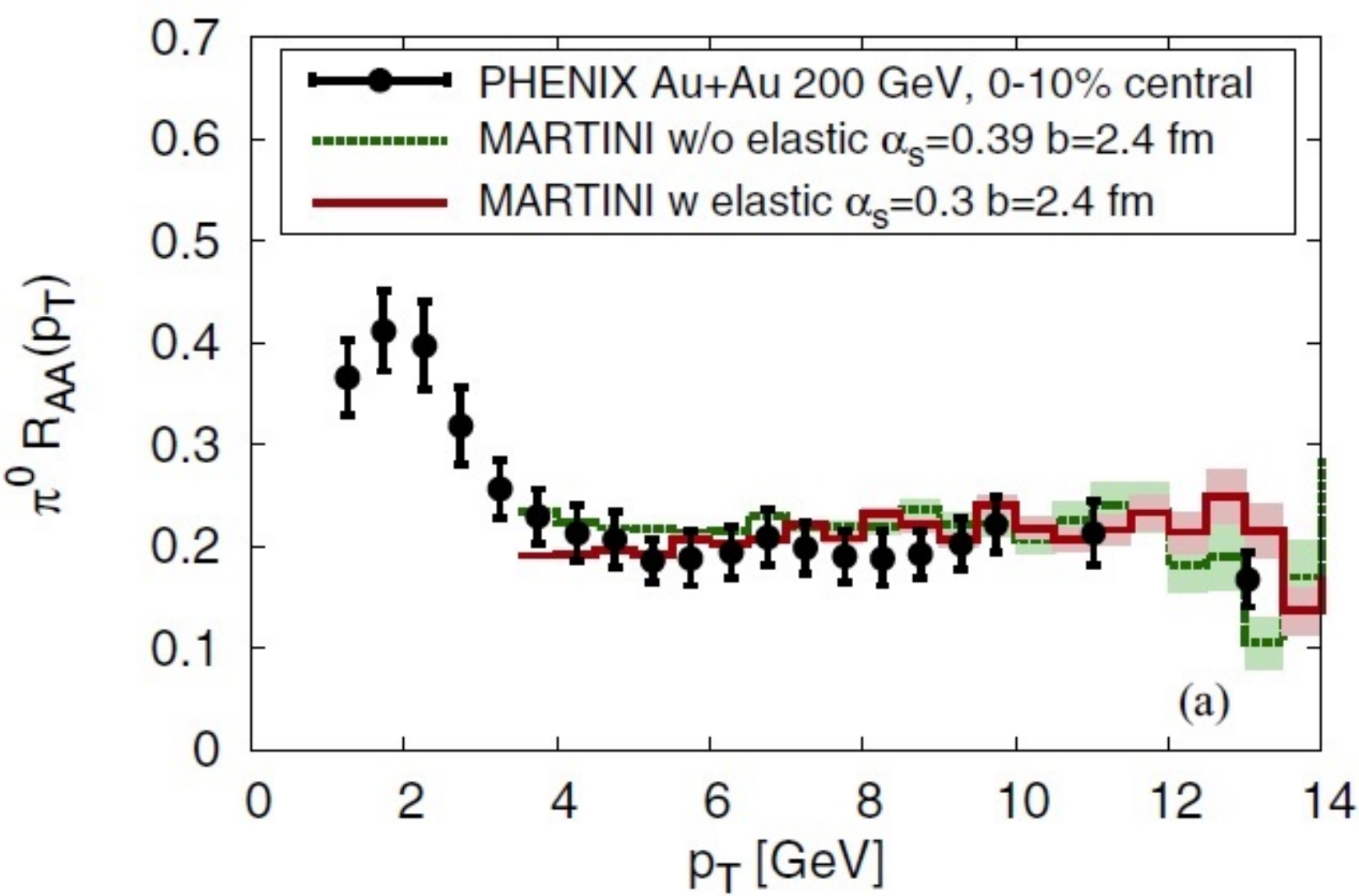}
    \includegraphics[width=0.23\textwidth]{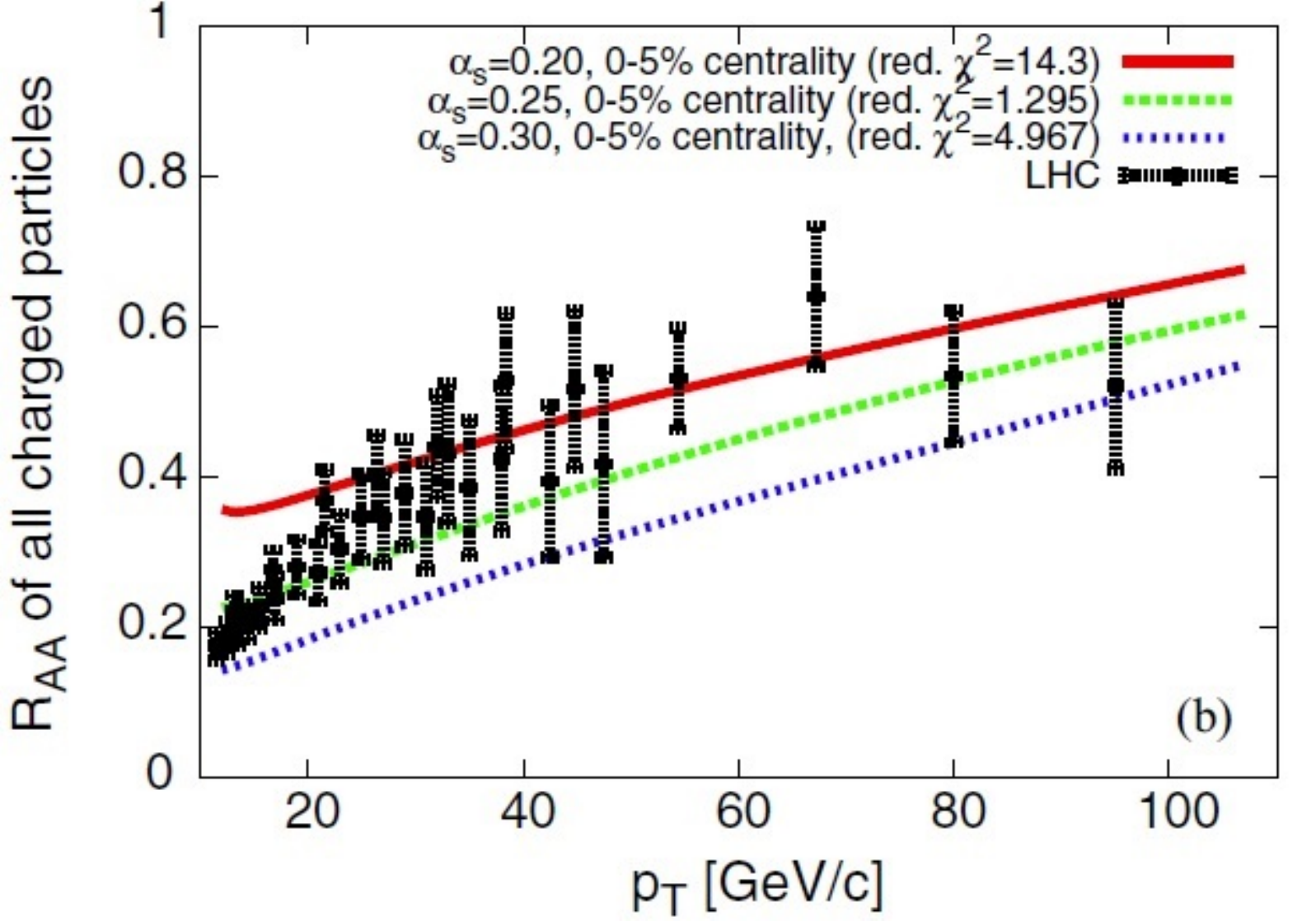}
    \caption{(Color online) Experimental data on the nuclear modification factor $R_\mathrm{AA}$ for single inclusive hadrons at RHIC (left columns) and the LHC (right columns), compared to different model calculations within the JET Collaboration -- (from top to bottom) CUJET-GLV, HT-BW, HT-M, McGill-AMY and MARTINI-AMY. The figures are from Ref.~\cite{Burke:2013yra}.}
    \label{fig:JET-RAA-all}
\end{figure}

In most semi-analytical calculations of the modified hadron spectra, one directly convolutes the production cross section of energetic partons ($\hat{\sigma}_{ab\to c}$) with their medium-modified FFs as shown in Eq.~(\ref{eq:xsectionFactor}).  In this case, one can also include contributions from hadronization of radiated gluons by adding the gluon FF to the right side of Eq.~(\ref{eq:medFF}) with the average energy fraction $\epsilon_g=\epsilon/n$ for each of the final $n$ number of radiated gluons ~\cite{Wang:1996yh}. A more careful approach to the fragmentation of the radiated gluons is to convolute the vacuum gluon FF with the medium-induced splitting function as discussed in Sec.~\ref{sec:theory} within the high-twist approach~\cite{Guo:2000nz,Wang:2001ifa}.

In most Monte-Carlo frameworks, one first generates energetic partons using $\hat{\sigma}_{ab\to c}$ from the initial hard scatterings, then simulates the elastic and inelastic energy loss of these partons through the hot nuclear matter, and in the end converts all partons into hadrons using vacuum FFs outside the medium.  One can include both radiated gluons and medium recoil partons in the final hadronization in these kind of Monte Carlo calculations.

The suppression factor $R_\mathrm{AA}$ of single inclusive hadrons helps to constrain the interaction strength between jet partons and the medium. As shown in Fig.~(\ref{fig:JET-RAA-all}), it decreases with the increase of the jet transport parameter $\hat{q}$ or the strong coupling constant $\alpha_\mathrm{s}$. With a systematical comparison between various model calculations and experimental data at RHIC and the LHC, the JET Collaboration has obtained the constraint on $\hat{q}$ (of quarks) as $\hat{q}/T^3\approx 4.6\pm 1.2$ at RHIC (in the center of central Au+Au collisions at 200~GeV) and $3.7\pm 1.4$ at the LHC (in the center of central Pb+Pb collisions at 2.76~TeV). Details can be found in Fig.~\ref{fig:qhat} and Ref.~\cite{Burke:2013yra}. 

\begin{figure}[tbp]
    \centering
    \includegraphics[width=0.4\textwidth]{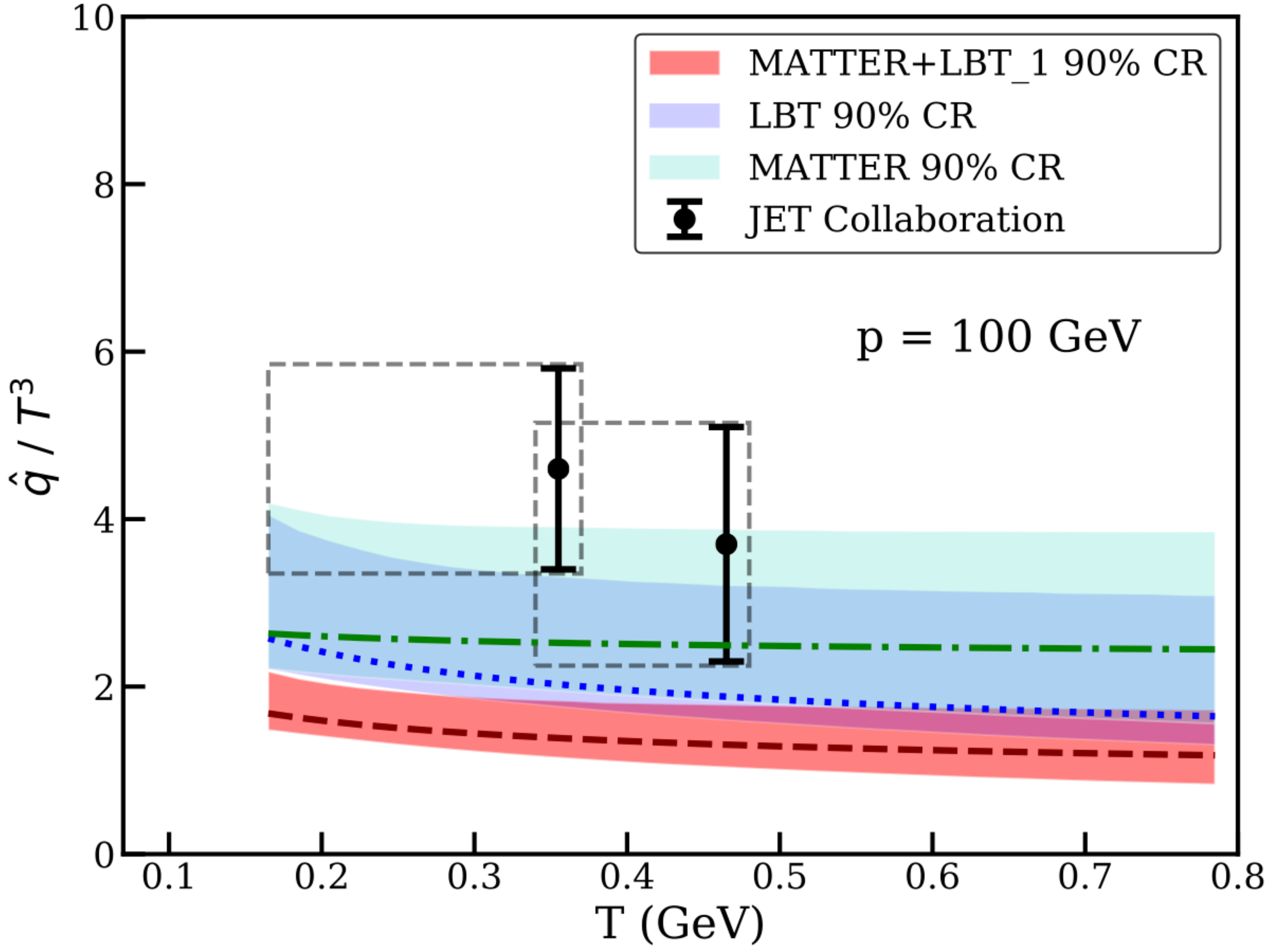}
    \caption{(Color online) The temperature scaled quark transport coefficient $\hat{q}/T^3$ as a function of the medium temperature, extracted by the JETSCAPE Collaboration using \textsc{Matter}, \textsc{Lbt} and \textsc{Matter+Lbt} models as compared to previous JET Collaboration result. The figure is from Ref.~\cite{Soltz:2019aea}.}
    \label{fig:JETSCAPE-qhat}
\end{figure}

This work of the JET Collaboration has recently been further improved by the JETSCAPE Collaboration~\cite{Soltz:2019aea} in several directions: (1) a single parameter (constant $\hat{q}$ or $\alpha_\mathrm{s}$) fit to a single data set has been extended to a simultaneous fit in a multi-dimensional parameter space to multiple data set at RHIC and the LHC in order to obtain $\hat{q}$ as a continuous function of the parton energy and the medium temperature; (2) instead of using a single energy loss formalism through the entire evolution of jets in the medium, a multi-scale approach (\textsc{Matter+Lbt}) as discussed in Sec.~\ref{subsec:multi-scale} has been employed for the first time to describe the nuclear modification of jets; (3) machine learning and Bayesian analysis methods have been introduced to replace the traditional $\chi^2$ fits, which significantly increase the efficiency of calibrating sophisticated model calculations in a wide range of the parameter-space against a vast amount of data. Shown in Fig.~\ref{fig:JETSCAPE-qhat} is the temperature dependence of $\hat{q}/T^3$ obtained within this new framework, with a 4-parameter $(A, B, C, D)$ ansatz,
\begin{align}
\label{eq:JETSCAPE-qhat}
\frac{\hat{q}}{T^3}&=C_2\frac{42\zeta(3)}{\pi}\left(\frac{4\pi}{9}\right)^2\\
\times&\left\{\frac{A\left[\ln\left(\frac{E}{\Lambda}\right)-\ln(B)\right]}{\left[\ln\left(\frac{E}{\Lambda}\right)\right]^2}+\frac{C\left[\ln\left(\frac{E}{T}\right)-\ln(D)\right]}{\left[\ln\left(\frac{ET}{\Lambda^2}\right)\right]^2}\right\}.\nonumber
\end{align}
The second term in the curly bracket of the above ansatz takes the form of Eq.~(\ref{eq:qhatAnalytic}) for $\hat{q}$'s dependence on the jet parton energy ($E$) and the medium temperature ($T$). A running coupling $\alpha_\mathrm{s}(Q^2)$ at the leading order is applied here with the scales set to $Q^2=ET$ and $\Lambda = 0.2$~GeV. Possible dependence of the constant factors  on kinematic cuts  in Eq.~(\ref{eq:qhatAnalytic}) are absorbed in the parameter $D$. 
This form of $\hat{q}$, based on perturbative scatterings between jet partons and medium partons, is usually applied in transport models for low virtuality jet partons. 
When their virtuality is much larger than that of the medium temperature scale, the scaled jet transport parameter $\hat{q}/T^3$ may only depend on the jet energy scale, giving rise to the first term of the ansatz in the curly bracket with a parameter $B$.  Parameters $A$ and $C$ weigh the contributions from these two terms. This 4-parameter ansatz is used in \textsc{Matter} and \textsc{Lbt} model, respectively,  to describe experimental data. In the multi-scale \textsc{Matter+Lbt} model, the separation scale $Q_0$ between \textsc{Matter} and \textsc{Lbt} is introduced as the 5$^\mathrm{th}$ parameter. These setups are applied to calibrate the model calculations against the experimental data on $R_\mathrm{AA}$ for single inclusive hadrons in Au+Au collisions at 200~GeV, Pb+Pb collisions at 2.76  and 5.02~TeV simultaneously. Details can be found in Ref.~\cite{Soltz:2019aea} where different parameterizations of $\hat{q}$ are also discussed and compared.

As shown in Fig.~\ref{fig:JETSCAPE-qhat}, when \textsc{Matter} and \textsc{Lbt} are applied separately, the 90\% credible regions of the jet transport parameter $\hat{q}$ extracted from model-to-data comparisons are still consistent with the previous JET Collaboration result~\cite{Burke:2013yra}. In contrast, combining \textsc{Matter} and \textsc{Lbt} leads to larger jet energy loss inside the QGP than using a single model, and thus yields a smaller value of the extracted $\hat{q}$. In addition, the separation scale $Q_0$ between high-virtuality parton shower and low-virtuality transport model has been found at around $2\sim3$~GeV, which reflects the virtuality scale of the QGP medium produced in heavy-ion collisions. 

The hadron suppression factor $R_\mathrm{AA}(p_\mathrm{T},y)$ quantifies the energy loss of hard partons inside the QGP averaged over the azimuthal angle in the transverse plane. A more differential observable is the elliptic flow coefficient $v_2$, which measures the energy loss anisotropy along different paths through the QGP \cite{Wang:2000fq,Gyulassy:2000gk}. It is defined as the second-order Fourier coefficient of the azimuthal angle distribution of particles in their momentum space
\begin{equation}
\label{eq:def-v2}
v_2(p_\mathrm{T},y)\equiv\frac{\int d\phi \cos(2\phi) \frac{dN}{dp_\mathrm{T} dy d\phi}}{\int d\phi \frac{dN}{dp_\mathrm{T} dy d\phi}}.
\end{equation}
This $v_2$ coefficient can be analyzed for an ensemble of particles as
\begin{equation}
\label{eq:def-v2-simple}
v_2=\left\langle \cos(2\phi) \right\rangle=\left\langle\frac{p_x^2-p_y^2}{p_x^2+p_y^2}\right\rangle,
\end{equation}
where $\langle \ldots \rangle$ denotes the ensemble average. 

\begin{figure}[tbp]
    \centering
    \includegraphics[width=0.4\textwidth]{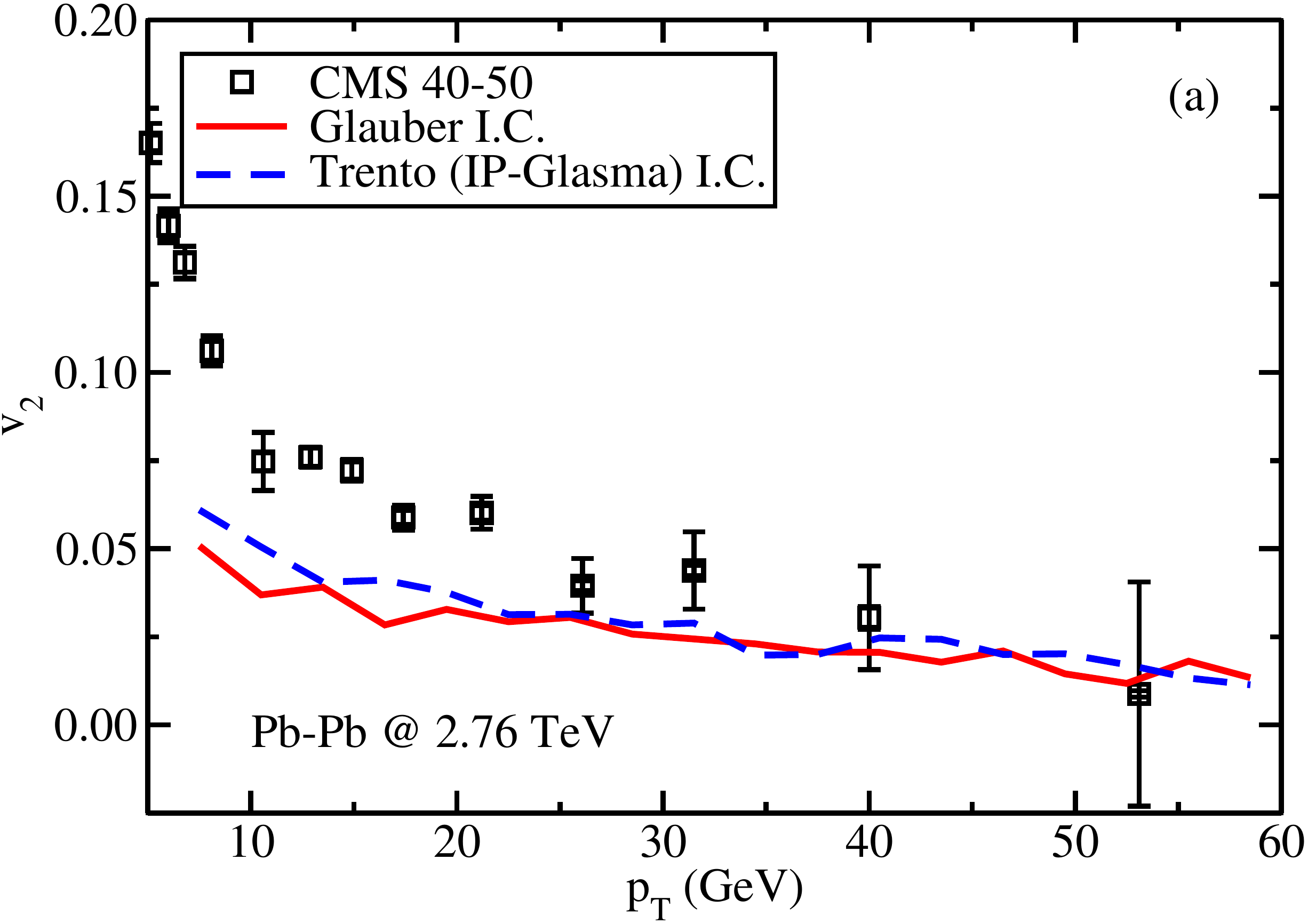}
    \includegraphics[width=0.4\textwidth]{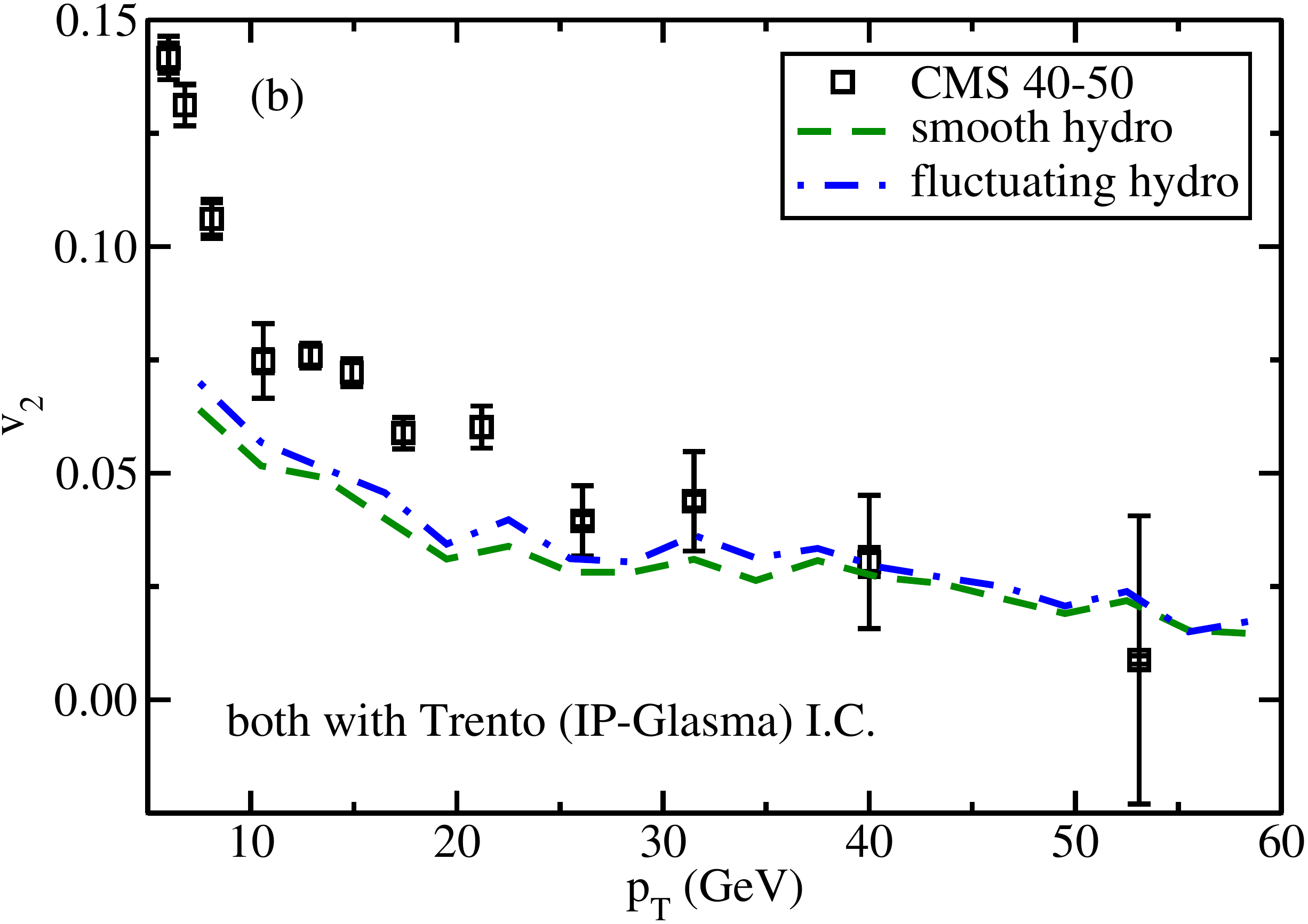}
    \caption{(Color online) Effects of (a) different initial conditions and (b) initial state fluctuations on the hadron $v_2$. The figures are from Ref.~\cite{Cao:2017umt}.}
    \label{fig:Cao-hadron-v2}
\end{figure}

The elliptic flow coefficient is expected to place more stringent constraints on the path length dependence of jet energy loss. However, few models are able to provide satisfactory descriptions of both the hadron suppression factor $R_\mathrm{AA}$ and the elliptic flow coefficient $v_2$ simultaneously as observed in experiments. As shown in Fig.~\ref{fig:Cao-hadron-v2}, while results from model calculations are consistent with experimental data at high $p_\mathrm{T}$, sizable deviation persists as $p_\mathrm{T}$ is around and below 20~GeV. Considerable efforts have been devoted to solving this $v_2$ puzzle. For instance, it has been suggested~\cite{Noronha-Hostler:2016eow} that using a more geometrically anisotropic initial condition of QGP fireballs and including event-by-event fluctuations of the initial profiles can give rise to sufficiently large $v_2$ of energetic hadrons. However, as shown in Fig.~\ref{fig:Cao-hadron-v2} (a) and (b) respectively, both effects turn out to be small when coupling a realistic jet energy loss model (\textsc{Lbt}) to a (3+1)-D viscous hydrodynamic model (\textsc{CLVisc})~\cite{Cao:2017umt}. Other solutions, such as additional enhancement of the jet-medium interaction strength ($\alpha_\mathrm{s}$ or $\hat{q}$) near $T_\mathrm{c}$~\cite{Xu:2014tda,Das:2015ana}, or delaying the starting time of jet-medium interaction~\cite{Andres:2019eus}, have been proposed to increase the $v_2$ of hard probes while keeping the suppression factor $R_\mathrm{AA}$ fixed. Both ideas suggest that with a fixed amount of total energy loss, weighing more jet-medium interaction towards a later evolution time when the QGP collective flow is more anisotropic can effectively enhance the jet $v_2$. However, little agreement has been reached yet on the detailed mechanisms that shift more energy loss to a later time. Most of the current calculations do not take into account medium modification of the hadronization mechanism such as parton recombination which could influence the flavor dependence of the hadron suppression factor and $v_2$ at low and intermediate $p_\mathrm{T}$~\cite{Greco:2003xt,Fries:2003rf,Fries:2003vb}. Such mechanism, however, will have less impact on the full jet spectra which can be described well by many transport models as we will discuss later in this review.

Recent observations of little or no suppression~\cite{Adam:2016ich} but large $v_2$~\cite{Acharya:2017tfn,Sirunyan:2018toe} of hard probes in small colliding (p+Pb) systems have urged us to revisit the parton recombination and initial state effects on jets, because hot nuclear matter effects require a sufficient amount of suppression to accompany a large $v_2$~\cite{Xu:2015iha,Du:2018wsj}. To the contrary, this puzzle can be solved within a model based on the dilute-dense factorization in the Color Glass Condensate (CGC) framework~\cite{Zhang:2019dth}. It starts with a gluon and a quark from the projectile proton: the quark serves as the reference while the energetic gluon fragments into the final state hadron under consideration. The interaction between the incoming partons and the dense gluons in the target nucleus generates correlations between the energetic gluon and the reference quark, leading to the finite $v_2$ of high-energy hadrons. So far, this framework has provided satisfactory descriptions of the $v_2$ of open heavy flavor meson and heavy quarkonium in p+Pb collisions. Further study in this direction may also be essential for solving the puzzle of hard probe $v_2$ in large nucleus-nucleus collisions.

\subsection{Heavy flavor hadrons}
\label{subsec:heavyHadron}

Heavy quarks (charm and beauty quarks) are a special category of hard probe particles. The large mass of heavy quarks suppresses their thermal emission from the QGP, thus most heavy quarks are produced in the primordial hard collisions and then traverse and interact with the QGP with their flavor number conserved. Therefore, they serve as a clean probe of the QGP properties. At low $p_\mathrm{T}$, heavy quarks provide a unique opportunity to study the non-perturbative interaction between hard partons and the thermal medium; at intermediate $p_\mathrm{T}$, heavy quark observables help refine our understanding of the hadronization process from partons to color neutral hadrons; at high $p_\mathrm{T}$, heavy quarks allow us to study the mass and flavor hierarchy of parton energy loss inside the QGP. A more detailed review specializing in heavy quarks, especially their low $p_\mathrm{T}$ dynamics, can be found in Refs.~\cite{Dong:2019byy,Rapp:2018qla,Cao:2018ews,Xu:2018gux}. In this review, we concentrate on the flavor hierarchy of energy loss and heavy quark hadronization which are closely related to high-energy jets.

Due to their different masses and color factors, one would expect the energy losses of beauty, charm, light quarks and gluons have the flavor hierarchy $\Delta E_b < \Delta E_c < \Delta E_q < \Delta E_g$. Therefore, the suppression factor $R_\mathrm{AA}$ of $B$ and $D$ mesons and light flavor hadrons should have the inverted hierarchy $R_\mathrm{AA}^B > R_\mathrm{AA}^D > R_\mathrm{AA}^h$. However, the LHC data~\cite{Khachatryan:2016odn,Sirunyan:2017xss,Sirunyan:2017oug,Sirunyan:2018ktu} reveal comparable $R_\mathrm{AA}$'s for $D$ mesons, $B$ mesons and charged hadrons above $p_\mathrm{T}\sim 8$~GeV. Over the past decade, many theoretical efforts have been devoted to investigating this flavor hierarchy of hadron $R_\mathrm{AA}$~\cite{Qin:2009gw,Buzzatti:2011vt,Djordjevic:2013pba,Cao:2017hhk,Xing:2019xae}.

\begin{figure}[tbp]
    \centering
    \includegraphics[width=0.385\textwidth]{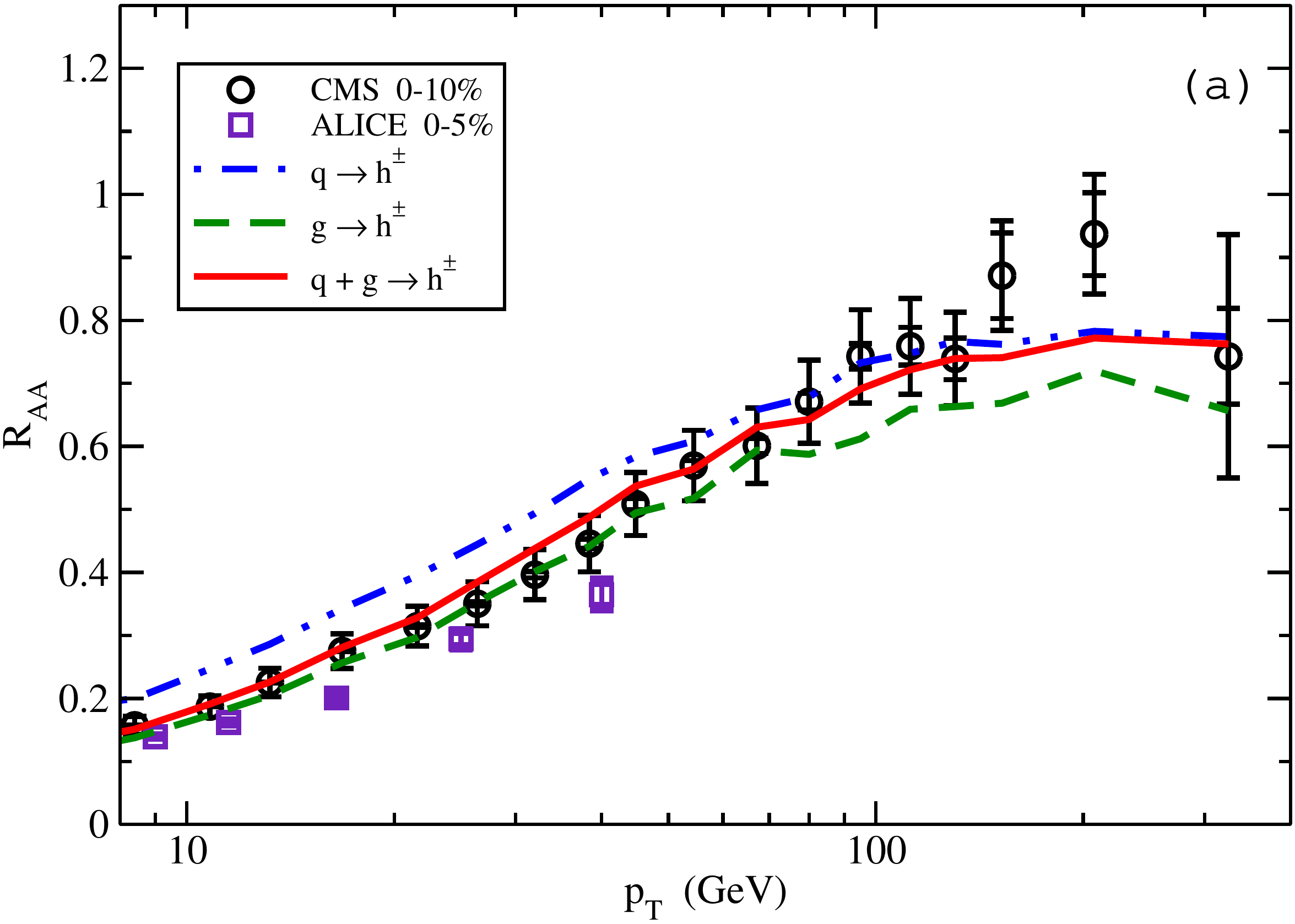}
    \includegraphics[width=0.4\textwidth]{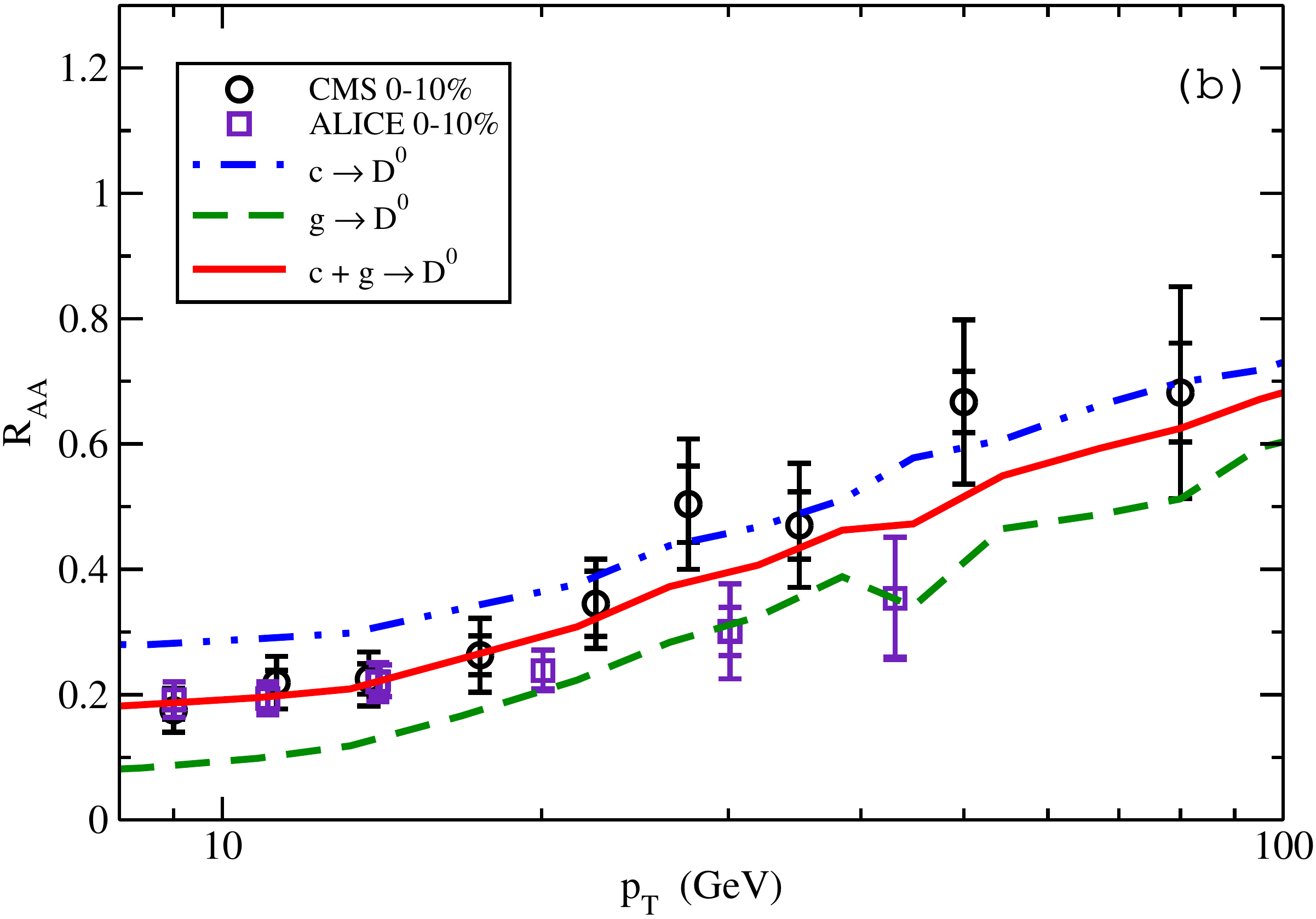}
    \caption{(Color online) The nuclear modification factors of (a) charged hadrons and (b) $D$ mesons in central 5.02~TeV Pb+Pb collisions, compared with contributions from quark and gluon fragmentations. The figures are from Ref.~\cite{Xing:2019xae}.}
    \label{fig:Xing-RAA-separate}
\end{figure}

A full understanding of heavy and light flavor parton energy loss requires a Monte-Carlo framework for realistic jet-medium interactions that treats different species of partons on the same footing. This has been realized in the \textsc{Lbt} model~\cite{Cao:2016gvr,Cao:2017hhk} in which elastic and inelastic energy loss of heavy and light flavor partons are simultaneously described using the Boltzmann transport through a hydrodynamic medium. Within this framework, a recent study~\cite{Xing:2019xae} further shows that the gluon splitting process in the next-to-leading-order (NLO) contribution to parton production is crucial for a simultaneous description of the $R_\mathrm{AA}$'s of different hadron species, which is usually ignored in heavy quark studies. While gluon fragmentation dominates the (light flavor) charged hadron production up to $p_\mathrm{T} \sim 50$~GeV, quark fragmentation starts to dominate beyond that. To the contrary, gluon fragmentation contributes to over 40\% $D$ meson yield up to 100~GeV. In Fig.~\ref{fig:Xing-RAA-separate}, the contributions from quark and gluon fragmentations to $R_\mathrm{AA}$ for charged hadrons and $D$ mesons  are compared in detail. One observes that the $R_\mathrm{AA}$'s for gluon initiated hadrons and $D$ mesons are much smaller than those initiated by quarks, and the $R_\mathrm{AA}$ for light quark initiated hadron  is slightly smaller than that for the charm quark initiated $D$ meson. This supports the flavor hierarchy of parton energy loss -- $\Delta E_c < \Delta E_q < \Delta E_g$ -- as expected. On the other hand, we see that the $R_\mathrm{AA}$ for light hadrons originated from gluon fragmentation is larger than that of $D$ mesons from gluons due to different fragmentation functions. After combining contributions from both quarks and gluons, one obtains similar $R_\mathrm{AA}$ for both charged hadrons and $D$ mesons above $p_\mathrm{T} \sim 8$~GeV.

\begin{figure}[tbp]
    \centering
    \includegraphics[width=0.4\textwidth]{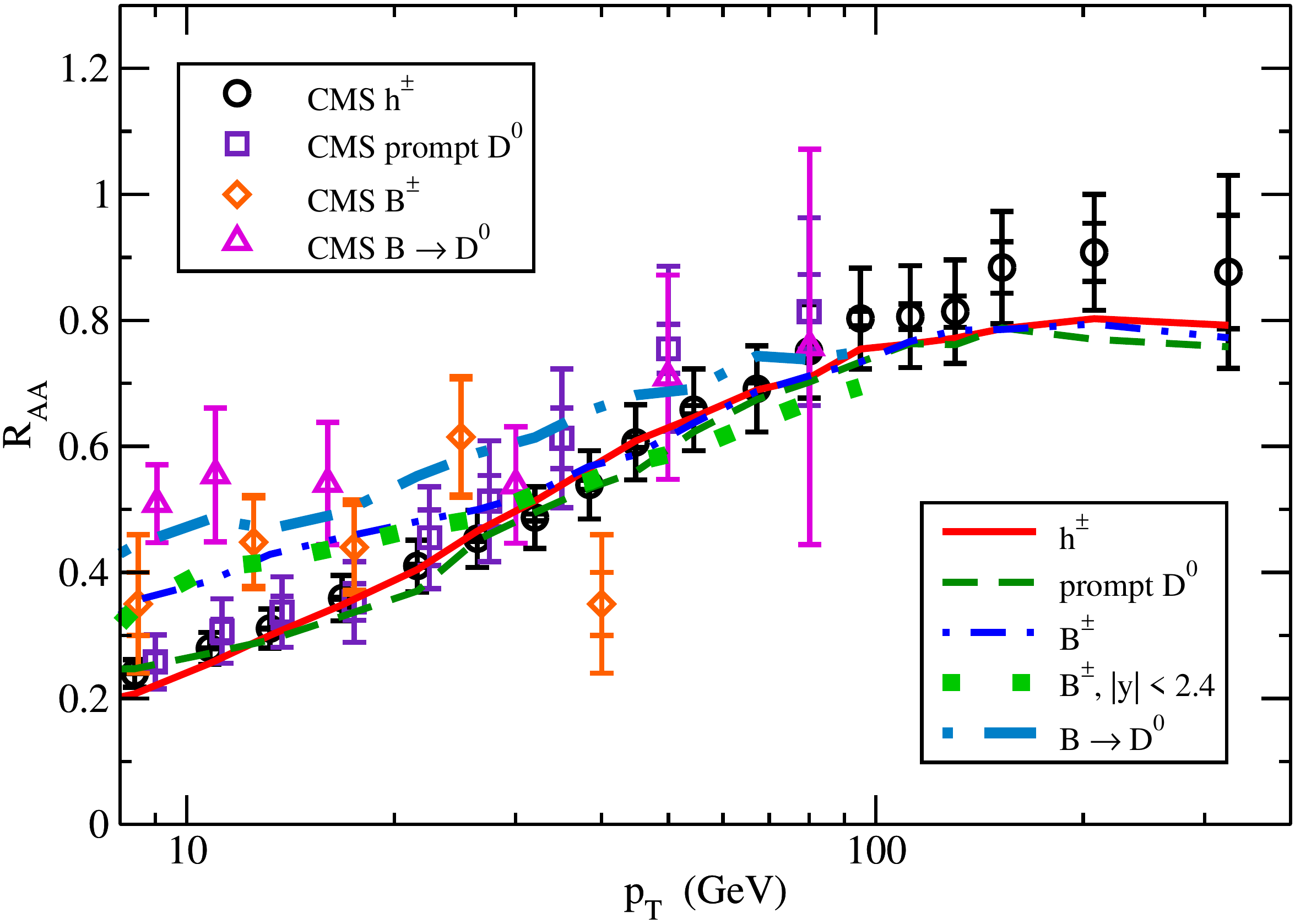}
    \caption{(Color online) The nuclear modification factors of charged hadrons, direct $D$ mesons, $B$ mesons and $B$-decay $D$ mesons in minimum bias 5.02~TeV Pb+Pb collisions. The figure is from Ref.~\cite{Xing:2019xae}.}
    \label{fig:Xing-RAA-all}
\end{figure}

The above findings can be further verified by applying the same calculation to the $B$ meson $R_\mathrm{AA}$. As shown in Fig.~\ref{fig:Xing-RAA-all}, within this perturbative framework that combines the NLO production and fragmentation mechanism with the \textsc{Lbt} simulation of parton energy loss through the QGP, one can naturally obtain a simultaneous description of the $R_\mathrm{AA}$'s of charged hadrons, direct $D$ mesons, $B$ mesons and $D$ mesons from $B$-decay over a wide $p_\mathrm{T}$ region. It also predicts that at intermediate $p_\mathrm{T}$, one should observe a larger $R_\mathrm{AA}$ of $B$ mesons compared to $D$ mesons and charged hadrons. However, this separation disappears above $\sim 40$~GeV. This is expected to be tested by future precision measurement and complete our understanding of the flavor hierarchy of jet quenching inside the QGP.

While the clean perturbative framework is sufficient to describe the nuclear modification of hadrons at high $p_\mathrm{T}$, non-perturbative effects become important in the low $p_\mathrm{T}$ region. The difference between jet spectra at partonic and hadronic levels is non-negligible; it could be as large as the difference between the LO and NLO calculations in some regions of $p_\mathrm{T}$~\cite{Wobisch:1998wt,Abelev:2013fn,Kumar:2019bvr}. 
The hadronization mechanism is a challenging topic because of its non-perturbative nature. Nevertheless, heavy quarks provide us a good opportunity to investigate how partons form hadrons at different momentum scales due to the feasibility of tracking their flavor identity during their evolution. While high $p_\mathrm{T}$ heavy quarks tend to fragment directly into hadrons, it is more probable for low $p_\mathrm{T}$ heavy quarks to combine with thermal partons from the QGP to form hadrons. The latter process is known as coalescence, or recombination, and was first proposed in Ref.~\cite{Oh:2009zj} and found to significantly affect the charmed hadron chemistry (baryon-to-meson ratio) in relativistic heavy-ion collisions. This proposal has been qualitatively confirmed by the recent RHIC and LHC data on the $\Lambda_c/D^0$ ratio~\cite{Adam:2019hpq,Acharya:2018ckj}. Meanwhile, this coalescence model has also been quantitatively improved over the past few years in Refs.~\cite{Plumari:2017ntm,Cho:2019lxb,Cao:2019iqs}.

The coalescence probability from two (three) constituent quarks to a meson (baryon) is given by the wavefunction overlap between the free quark state and the hadronic bound state. If a heavy quark does not coalescence with thermal quarks from the QGP, it fragments. This coalescence formalism has recently been extended from $s$-wave hadronic states to both $s$ and $p$-wave states of charmed hadrons~\cite{Cao:2019iqs}. It is found that adding the $p$-wave contribution significantly increases the total coalescence probability of charm quarks and makes it possible to normalize this probability at zero momentum with a proper in-medium size of charmed hadrons ($r_{D^0}=0.97$~fm) considering that a zero-momentum charm quark is not energetic enough to fragment into hadrons. Additionally, including $p$-wave states naturally incorporates all major charmed hadron states listed in the Particle Data Group (PDG)~\cite{Tanabashi:2018oca} and enhances the $\Lambda_c/D^0$ ratio. A longstanding deficiency of this coalescence formalism is its violation of the energy conservation. This has also been fixed in this work~\cite{Cao:2019iqs} by first coalescing multiple constituent quarks into an off-shell bound state and then decaying it into on-shell hadrons with the 4-momentum of the entire system strictly conserved. 

\begin{figure}[tbp]
    \centering
    \includegraphics[width=0.4\textwidth]{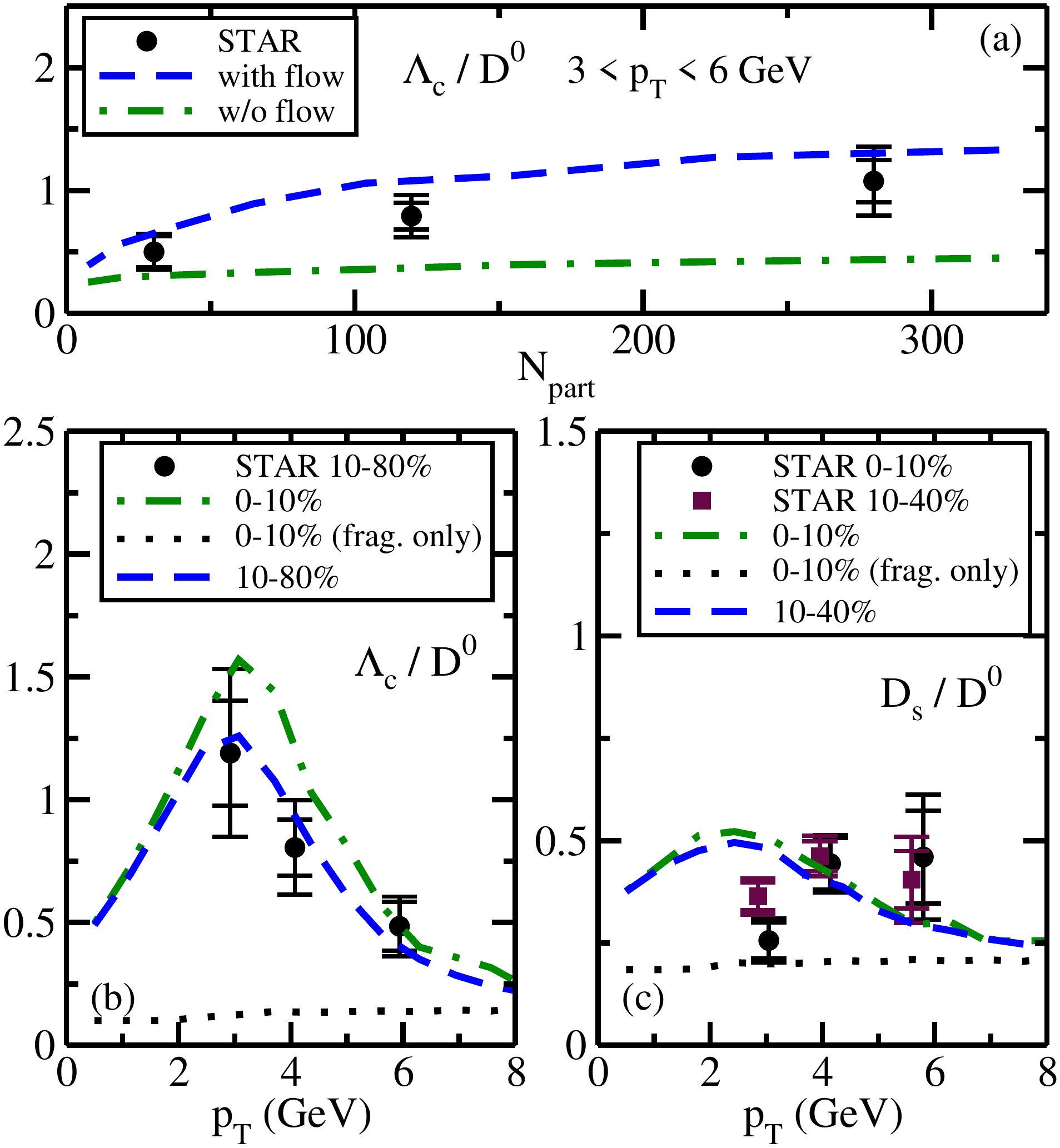}
    \caption{(Color online) The charmed hadron chemistry from a coalescence-fragmentation hadronization model: (a) the $p_\mathrm{T}$ integrated $\Lambda_c/D^0$ ratio, (b) the $p_\mathrm{T}$ differentiated $\Lambda_c/D^0$ ratio, and (c) the $p_\mathrm{T}$ differentiated $D_s/D^0$ ratio. The figure is from Ref.~\cite{Cao:2019iqs}.}
    \label{fig:Cao-hadronization}
\end{figure}

As shown in Fig.~\ref{fig:Cao-hadronization}, after combining this improved hadronization model with a transport-hydrodynamics model that provides the realistic heavy quark distribution after they traverse the QGP, one obtains a satisfactory description of the charmed hadron chemistry as observed in Au+Au collisions at $\sqrt{s}=200$~GeV, including both $p_\mathrm{T}$ integrated and differentiated $\Lambda_c/D^0$ and $D_s/D^0$ ratios. Effects of the QGP flow and fragmentation vs. coalescence mechanism on the charmed hadron chemistry have also been explored in Fig.~\ref{fig:Cao-hadronization}. The $p_\mathrm{T}$ boost from the QGP flow is stronger on heavier hadrons and thus significantly enhances the $\Lambda_c/D^0$ ratio. The coalescence also yields much larger baryon-to-meson ratio than fragmentation. Within this framework, it has been predicted that the in-medium charmed hadron size should be larger than that in vacuum, which may be tested by future hadronic model calculations. There might be other mechanisms affecting the charmed hadron chemistry, such as contributions from possible resonant states beyond the current PDG list~\cite{He:2019vgs} and the sequential coalescence of charmed hadrons at different temperatures~\cite{Zhao:2018jlw}.

\subsection{Dihadron and $\gamma$/$Z^0$-triggered hadrons}
\label{subsec:diHadron}

In addition to single inclusive hadrons, dihadron~\cite{Majumder:2004pt,Zhang:2007ja,Renk:2008xq,Cao:2015cba} and $\gamma$/$Z^0$-triggered hadrons~\cite{Zhang:2009rn,Qin:2009bk,Chen:2017zte} provide additional tools to place more stringent constraints on our understanding of parton energy loss inside the QGP. For instance, one may measure the medium modification of the momentum imbalance between the associated hadron and the triggered hadron or $\gamma$/$Z^0$ in  A+A collisions relative to that in p+p. Such observables are also independent of $\langle N_\mathrm{coll} \rangle$ in Eq.~(\ref{eq:defRAA}) for the $R_\mathrm{AA}$ of single inclusive hadrons and therefore are free of the associated systematic uncertainties. The $\gamma$/$Z^0$-triggered hadrons/jets are in particular considered ``golden channels" for the study of jet quenching since the triggered $\gamma$/$Z^0$ does not lose energy inside the medium, and therefore serves as an ideal reference for the energy loss of the associated jet partons.

\begin{figure}[tbp]
    \centering
    \includegraphics[width=0.4\textwidth]{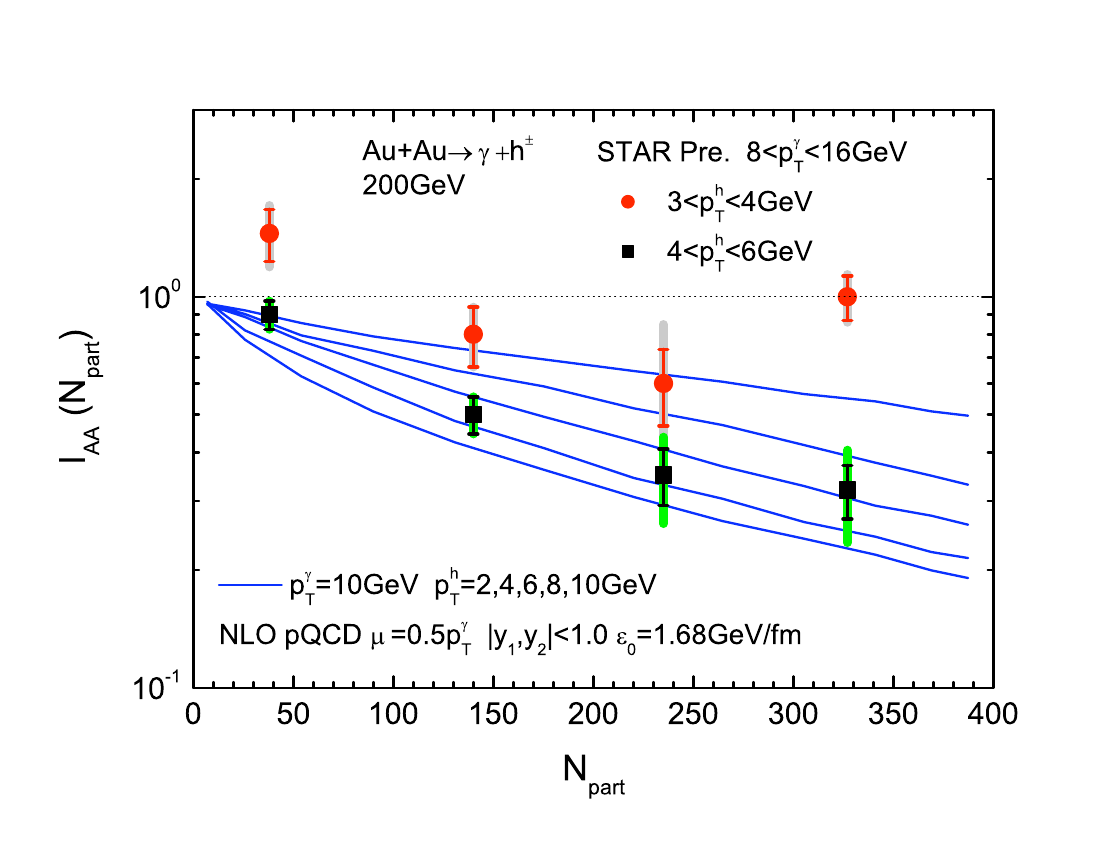}
    \caption{(Color online) The participant number dependence of the $\gamma$-triggered hadron suppression factor $I_\mathrm{AA}$ in Au+Au collisions at RHIC. Curves from top to bottom correspond to the hadron $p_\mathrm{T}$ at 2, 4, 6, 8, 10~GeV respectively. The figure is from Ref.~\cite{Zhang:2009rn}.}
    \label{fig:Zhang-photon-hadron-IAA}
\end{figure}

One quantity being commonly investigated is the triggered nuclear modification factor $I_\mathrm{AA}$ \cite{Adler:2002tq} defined as
\begin{equation}
\label{eq:defIAA}
I_\mathrm{AA}(z_\mathrm{T})\equiv\frac{\frac{1}{N_\mathrm{trig}}\frac{dN^\mathrm{asso}}{dz_\mathrm{T}}|_\mathrm{AA}}{\frac{1}{N_\mathrm{trig}}\frac{dN^\mathrm{asso}}{dz_\mathrm{T}}|_\mathrm{pp}},
\end{equation}
where $z_\mathrm{T}=p^{\rm asso}_\mathrm{T}/p_\mathrm{T}^{\rm trig}$ is the $p_\mathrm{T}$ ratio between the associated hadron and the triggered particle (hadron, $\gamma$ or $Z^0$ boson), measuring the transverse momentum imbalance between them. The numerator and denominator in the above equation are both normalized to yields of triggered particles, and are called hadron or $\gamma$/$Z^0$-triggered fragmentation functions in literature.

One would expect that with an increase of parton energy loss inside the QGP, there will be a suppression of the associated hadrons at high $p_\mathrm{T}$ ($z_\mathrm{T}$), thus larger imbalance between the associated and the triggered particles. In Fig.~\ref{fig:Zhang-photon-hadron-IAA}, we present the $z_\mathrm{T}$-integrated $I_\mathrm{AA}$ of $\gamma$-triggered hadrons as a function of the participant number ($N_\mathrm{part}$) in Au+Au collisions at $\sqrt{s}=200$~GeV. For a fixed $p_\mathrm{T}$ range of the triggered photon, a stronger suppression of the associated hadron is observed at larger $p_\mathrm{T}$~\cite{Zhang:2009rn}. Due to larger energy loss of the associated jet parton in more central collisions, the $I_\mathrm{AA}$ also decreases with the increase of $N_\mathrm{part}$.

\begin{figure}[tbp]
    \centering
    \includegraphics[width=0.4\textwidth]{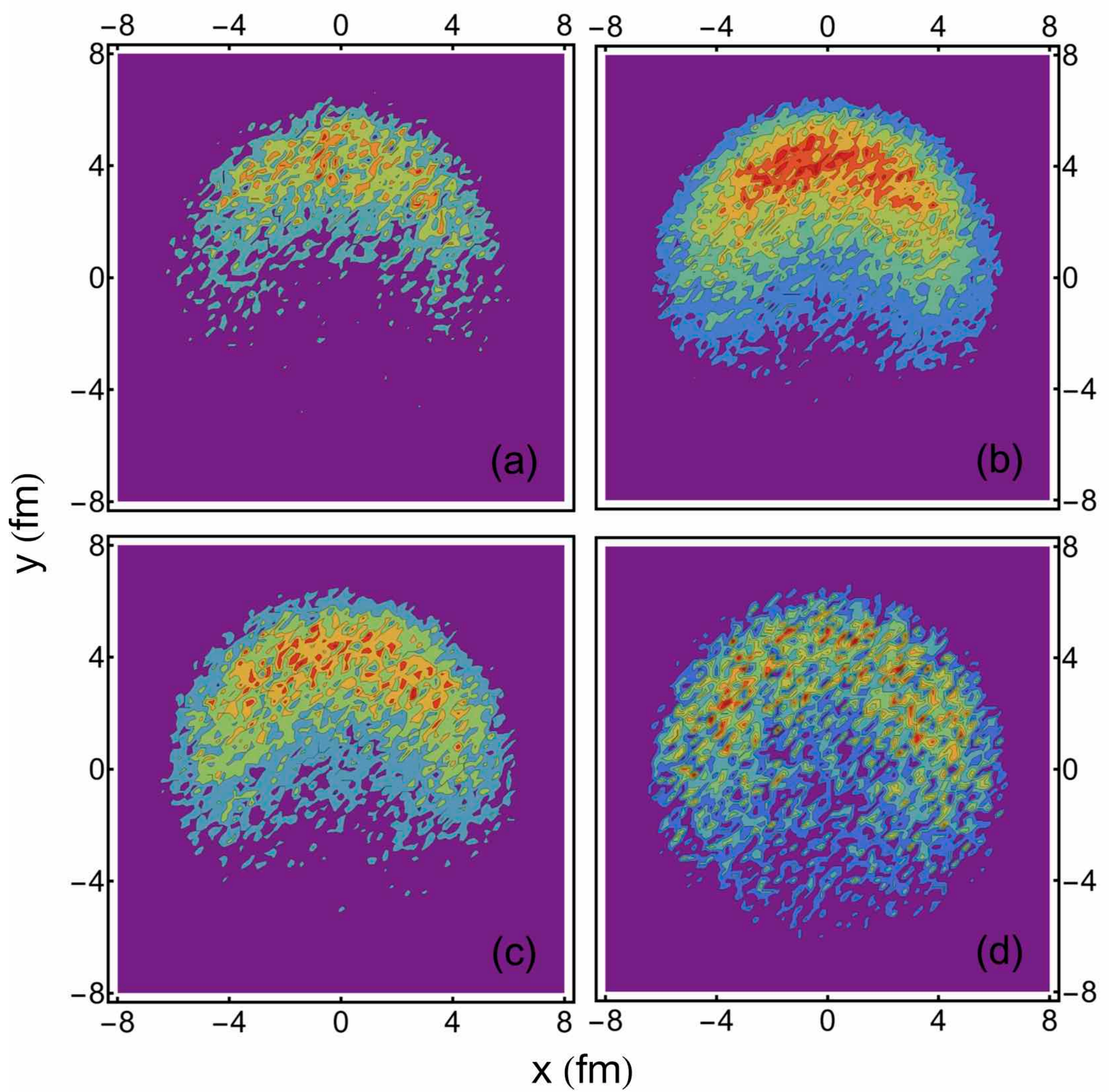}
    \caption{(Color online) The density distribution of the initial $c\bar{c}$ production positions in the transverse ($x$-$y$) plane for different $z_\mathrm{T}$ values between the final state $D\overline{D}$ pairs: (a) $z_\mathrm{T}\in (0.2, 0.4)$, (b) $z_\mathrm{T}\in (0.4, 0.6)$, (c) $z_\mathrm{T}\in (0.6, 0.8)$, and (d) $z_\mathrm{T}\in (0.8, 1.0)$. The triggered $D$ or $\overline{D}$ mesons are taken along the out-of-plane directions ($|\phi_\mathrm{trig} - \pi/2| < \pi/6$) for central Au+Au collisions at 200 GeV ($p_\mathrm{T,trig} > 4$~GeV and $p_\mathrm{T,asso} > 2$~GeV). The figure is from Ref.~\cite{Cao:2015cba}.}
    \label{fig:Cao-initXY-zT}
\end{figure}

It has also been found in Refs.~\cite{Zhang:2007ja,Zhang:2009rn,Qin:2012gp,Cao:2015cba} that the $z_\mathrm{T}$ value can help identify the position from which the jet event is produced in the initial nucleus-nucleus collision. This is demonstrated in Fig.~\ref{fig:Cao-initXY-zT} by the transverse distribution of  the initial hard processes leading to heavy flavor meson pairs as an example, where within the 0-10\% centrality of Au+Au collisions at $\sqrt{s}=200$~GeV, $D$ or $\overline{D}$ mesons with $p_\mathrm{T} > 4$~GeV are triggered, and $p_\mathrm{T} > 2$~GeV is required for their associated anti-particles. One may observe for smaller values of $z_\mathrm{T}$, the initial charm quark pairs are more biased toward the edge of the overlap region of heavy-ion collisions so that the difference in the path lengths and thus energy loss is larger between triggered and associated particles. To the contrary, for larger $z_\mathrm{T}$ values, initial charm quark pairs are more likely to spread smoothly over the whole overlap region. Similar analyses have been done for dihadron~\cite{Zhang:2007ja}, $\gamma$-hadron~\cite{Zhang:2009rn} and $\gamma$-jet~\cite{Qin:2012gp} events. This allows us to use the momentum imbalance of triggered particle/jet pairs to probe different regions of the hot nuclear matter and also obtain better understanding of the path length dependence of parton energy loss inside the QGP. 

Apart from the momentum imbalance, the angular correlation between the associated hadrons and the triggered particle is another interesting observable for quantifying the transverse momentum broadening of jet partons. This may provide a more direct constraint on $\hat{q}$. In Ref.~\cite{Chen:2016vem}, a systematical resummation formalism has been employed for the first time to calculate the dihadron and hadron-jet angular correlation in p+p and peripheral A+A collisions. With a global $\chi^2$ fit to experimental data, the authors obtain the medium-induced broadening of a quark jet around $\langle p_\perp^2\rangle\sim 13$~GeV$^2$, and the jet transport parameter $\hat{q}_0=3.9^{+1.5}_{-1.2}$~GeV$^2$/fm at the top RHIC temperature. In addition, Ref.~\cite{Cao:2015cba} proposes that using the angular correlation between heavy meson pairs can help constrain the detailed energy loss mechanism of heavy quarks, which cannot be uniquely identified with the single inclusive hadron observables. It has been found that with the same $R_\mathrm{AA}$ factor of $D$ mesons, collisional energy loss is much more effective in smearing the angular distribution between the $D\overline{D}$ pairs compared to energy loss from collinear gluon radiation. In this work, different momentum imbalance ($z_\mathrm{T}$) cuts have been applied to separate the energy loss effect from the momentum broadening effect on the angular correlation, so that the best kinematic region has been suggested for future experimental measurements on constraining the heavy quark dynamics in the QGP.

\begin{figure}[tbp]
    \centering
    \includegraphics[width=0.4\textwidth]{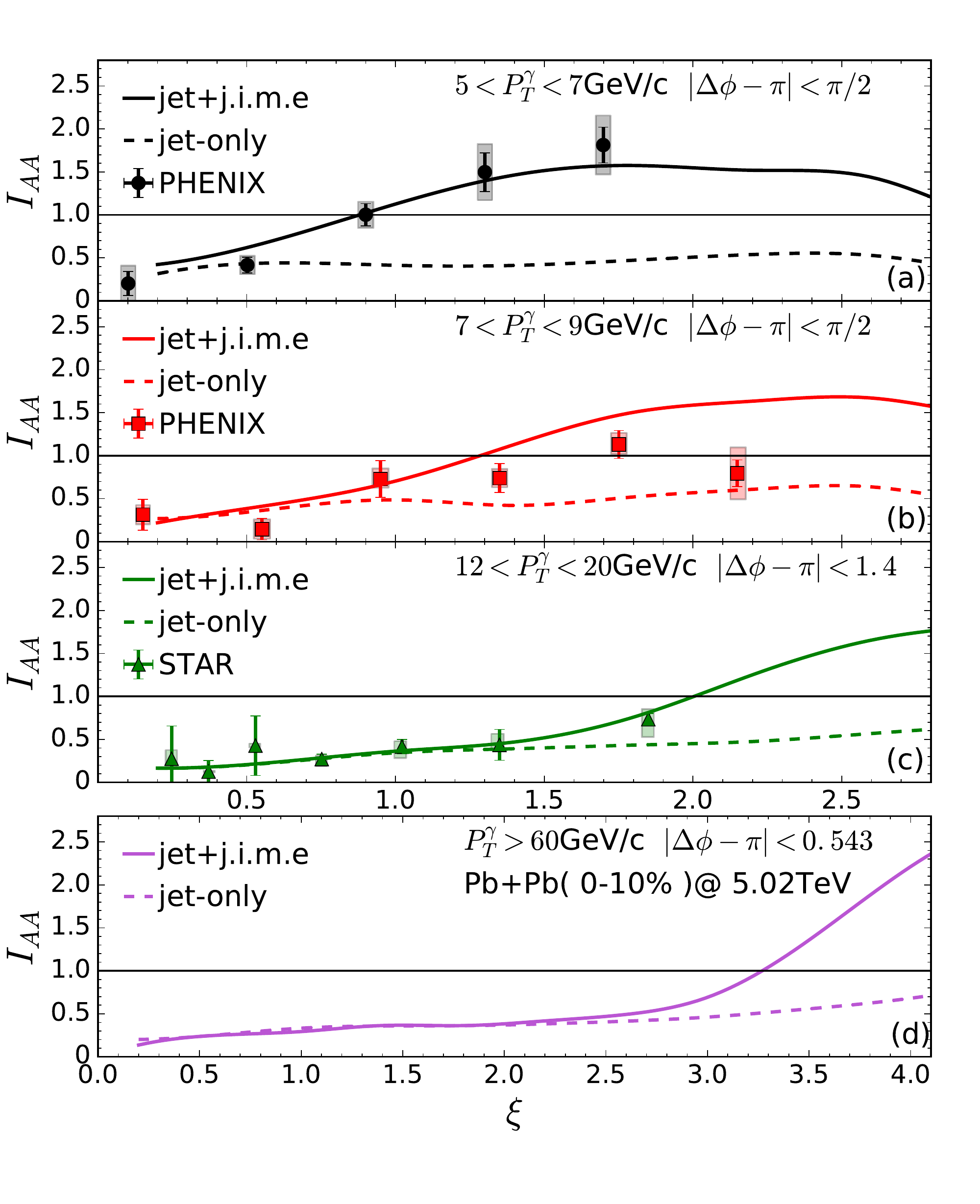}
    \caption{(Color online) Nuclear modification of the $\gamma$-triggered hadron yield in different $p_\mathrm{T}^\gamma$ regions in  Au+Au collisions at 200~GeV and Pb+Pb collisions at 5.02~TeV, with $|\eta_{h,\gamma}|<0.35$. The figure is  from Ref.~\cite{Chen:2017zte}.}
    \label{fig:Chen-gamma-hadron-IAA-xi}
\end{figure}

While most earlier work concentrated on the suppression of the high $z_\mathrm{T}$ hadron yield, a recent study~\cite{Chen:2017zte} has found that the enhancement of the soft hadron production at low $z_\mathrm{T}$ in $\gamma$-triggered hadron events could serve as a smoking-gun signal of the QGP response to jet propagation.  Such soft hadron enhancement is investigated within the \textsc{CoLbt-Hydro} model that realizes an event-by-event concurrent simulation of jet and QGP evolution in relativistic nuclear collisions as discussed in Sec.~\ref{subsec:concurrent}. The nuclear modification of the $\gamma$-triggered hadron yield in different $p_\mathrm{T}$ ranges of the triggered photon from this study is shown in Fig.~\ref{fig:Chen-gamma-hadron-IAA-xi}. To compare to different data set, the 0-40\% centrality bin is used in panels~(a) and (b), 0-12\% is used in panel~(c), and 0-10\% is used in panel~(d).
In order to investigate $I_\mathrm{AA}$ at low $z_\mathrm{T}$, the variable $\xi=\ln(1/z_\mathrm{T})$ is used for the horizontal axis. 

From Fig.~\ref{fig:Chen-gamma-hadron-IAA-xi}, one clearly observes a suppression of $I_\mathrm{AA}$ at small $\xi$ (large $z_\mathrm{T}$) due to parton energy loss before fragmenting into hadrons.  On the other hand, jet-induced medium excitation is clearly shown to lead to an enhancement of soft hadron yield at large $\xi$ (small $z_\mathrm{T}$).  The onset of the soft hadron enhancement ($I_\mathrm{AA} \ge 1$) shifts towards a larger $\xi$ value with the increase of $p_\mathrm{T}^\gamma$ [from Fig.~\ref{fig:Chen-gamma-hadron-IAA-xi} (a) to (d)], corresponding to a fixed hadron transverse momentum $p_\mathrm{T}^h = z_\mathrm{T}p_\mathrm{T}^\gamma\sim 2$~GeV.  This scale reflects the thermal momentum of hadrons from the jet-induced medium response in QGP which is approximately independent of the jet energy.  This is a unique feature of the jet-induced medium response from the \textsc{CoLbt-Hydro} model. The \textsc{CoLbt-Hydro}'s predictions on the high $p_\mathrm{T}^\gamma$ $\gamma$-triggered jets have also been confirmed by the recent LHC data~\cite{Aaboud:2019oac}. Jet-induced medium response in \textsc{CoLbt-Hydro} also explains the enhancement of the jet fragmentation function at small $z_\mathrm{T}$ in $\gamma$-jet events as we will show in the discussion about medium modification of jet substructures in Sec.~\ref{subsec:jetFragmentation}. Similar experimental measurements~\cite{ATLAS:2019gif} together with new theoretical calculations~\cite{Casalderrey-Solana:2015vaa,Zhang:2018urd} on $Z^0$-triggered hadrons/jets also become available.

\begin{figure}[tbp]
    \centering
    \includegraphics[width=0.45\textwidth]{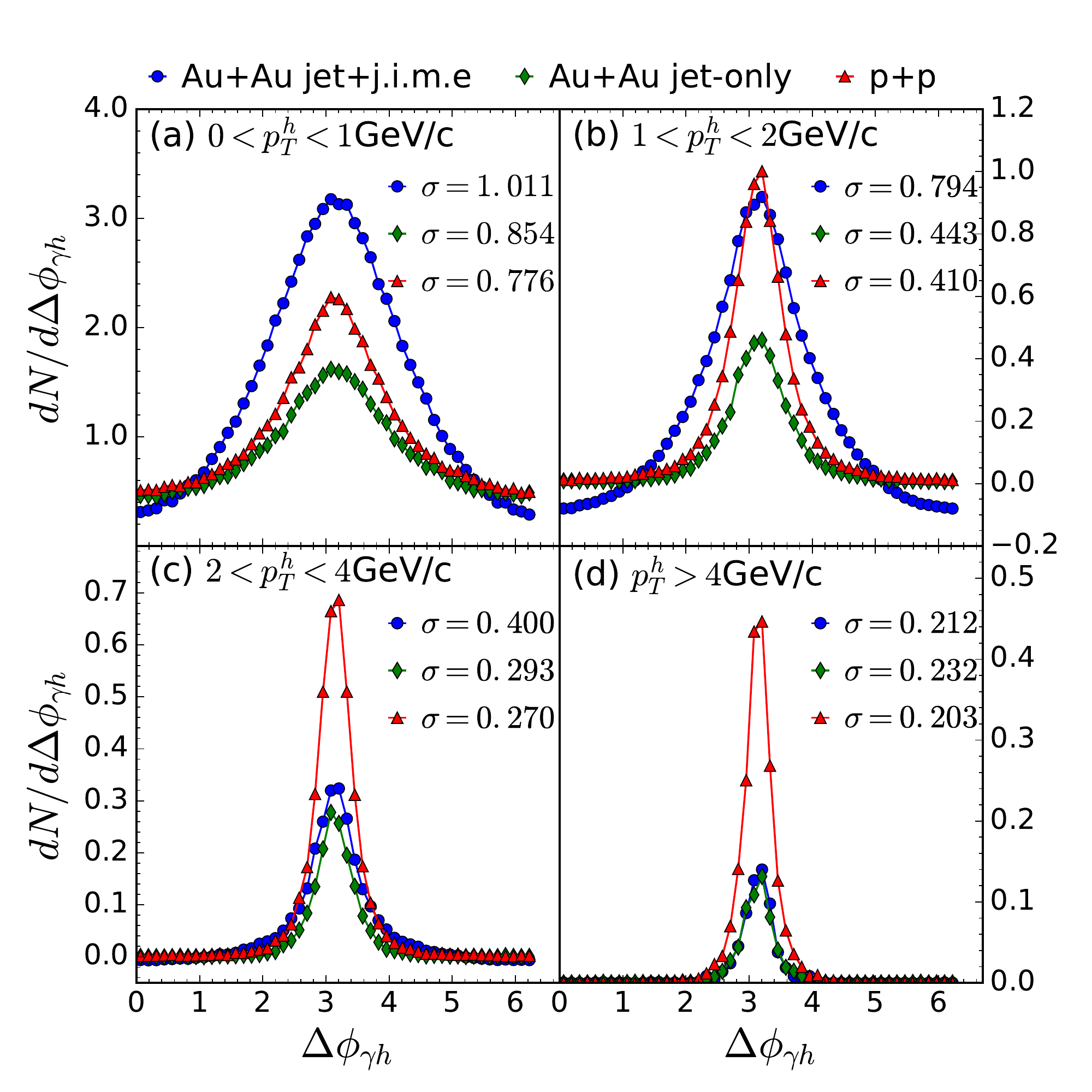}
    \caption{(Color online) The $\gamma$-hadron azimuthal correlation in different $p_\mathrm{T}^h$ region in p+p and 0-12\% Au+Au collisions at 200~GeV with $|\eta_{h,\gamma}|<1.0$ and $12<p_\mathrm{T}^\gamma < 20$~GeV, normalized to per triggered photon yield. The half width $\sigma$ is obtained from a Gaussian fit within $|\Delta \phi_{\gamma h}-\pi| < 1.4$. The figure is from Ref.~\cite{Chen:2017zte}.}
    \label{fig:Chen-gamma-hadron-angular-corr}
\end{figure}

The effect of jet-induced medium response can also be investigated in the $\gamma$-hadron angular correlation as shown in Fig.~\ref{fig:Chen-gamma-hadron-angular-corr}, where results for central Au+Au collisions at 200~GeV with and without contributions from the jet-induced medium excitation
are compared to that in p+p collisions. One can see that contribution from medium response is negligible for high $p_\mathrm{T}^h$. The widths of the angular correlation $\sigma$ in Au+Au and p+p collisions from fitting to a Gaussian function within the $|\Delta \phi_{\gamma h}-\pi| < 1.4$ region are comparable though there is an obvious suppression of the hadron yield at large $p_\mathrm{T}^h$ [Fig.~\ref{fig:Chen-gamma-hadron-angular-corr} (d)]. 
At low $p_\mathrm{T}^h$ [Fig.~\ref{fig:Chen-gamma-hadron-angular-corr} (a)], on the other hand, there is a significant enhancement of the hadron yield and broadening of their angular distribution in Au+Au collisions due to jet-induced medium excitation. The most interesting feature in the angular distribution of soft hadrons is their depletion near $\Delta \phi_{\gamma h}=0$ along the direction of triggered photon due to the diffusion wake left behind the jet. This is consistent with the snapshots of the \textsc{CoLbt-Hydro} simulation in Fig.~\ref{fig:Chen-jime-simulation}. Experimental verification of such depletion will be an unambiguous signal of jet-induced medium response.

\section{Jet spectra}
\label{sec:jet_spectra}

In the study of the suppression of jet spectra with a given jet-cone size $R$ in heavy-ion collisions, one should consider not only the jet energy loss due to transport of partons to the outside of the jet-cone through elastic scattering and induced gluon radiation, but also the effect of jet-induced medium response that can also contribute to the total energy inside the jet-cone as constructed by a jet-finding algorithm. This contribution from jet-induced medium response will affect the transverse momentum and jet-cone size dependence of jet energy loss and thus jet spectrum suppression. In this section, we will review the suppression of single and $\gamma$/$Z$-triggered jets in heavy-ion collisions 
and effects of jet-induced medium response.

\subsection{Single inclusive jets}
\label{subsec:singleJet}

To calculate the suppression of single inclusive jet spectra in high-energy heavy-ion collisions, we can first use \textsc{Pythia~8}~\cite{Sjostrand:2006za} or other Monte Carlo programs to generate the initial jet shower parton distributions from elementary nucleon-nucleon collisions and then use transport models such as \textsc{Lbt} to simulate the transport of these jet shower partons through the bulk medium that evolves according to a hydrodynamic model. The \textsc{Fastjet} program~\cite{Cacciari:2011ma}, modified to include the subtraction of  ``negative" partons from the total energy inside a jet-cone, is utilized with the anti-$k_\mathrm{T}$ algorithm to reconstruct jets and calculate the final single inclusive jet spectra. 

In practice as we discuss in this section, we use \textsc{Pythia~8} to generate the initial jet shower partons (with both initial and final state radiation) for a given number of events within the interval of the transverse momentum transfer $p_{\mathrm{T}c} \in (p_{\mathrm{T} c}-d p_{\mathrm{T} c}/2, p_{\mathrm{T} c}+dp_{\mathrm{T} c}/2)$ and the cross section $d\sigma_{\rm LO}^{{\rm pp}(c)}/dp_{\mathrm{T} c}$ in the leading-order (LO) perturbative QCD (pQCD) in p+p collisions. Using \textsc{Fastjet} with a given jet-cone radius $R$, one can get an event-averaged single inclusive jet spectrum $dN^{\rm jet}_{c}(p_\mathrm{T},p_{\mathrm{T}c})/dydp_\mathrm{T}$, here $p_\mathrm{T}$ and $y$ are the transverse momentum and rapidity of the final jet, respectively. The final single inclusive jet cross section in p+p collisions is given by
\begin{equation}
    \frac{d^2\sigma^{\rm jet}_{\rm pp}}{dp_\mathrm{T}dy} = \sum_c\int dp_{\mathrm{T}c}  \frac{d\sigma_{\rm LO}^{{\rm pp}(c)} }{dp_{\mathrm{T}c}}
     \frac{d^2N^{\rm jet}_{c}(p_\mathrm{T}, p_{\mathrm{T}c})} {dp_\mathrm{T} dy},
    \label{eq-jetcrs}
\end{equation}
where the LO pQCD cross section for the production of initial hard parton $c$ in p+p collisions is given by
\begin{eqnarray}
\frac{d \sigma^{{\rm pp}(c)}_{\rm LO}}{dp_{\mathrm{T}c}} & = & 2 p_{\mathrm{T}c}\sum_{a,b,d}  \int dy_c dy_d  x_a f_{a/p} (x_a, \mu^2) 
 \nonumber\\ 
 & &  \times x_b f_{b/p} (x_b, \mu^2) \frac{d\hat\sigma_{ab\to cd}}{dt},
\label{eq:cs.pp}
\end{eqnarray}
where $y_c$ and $y_d$ are rapidities of the final hard partons in the $a+b\rightarrow c+d$ processes, $x_a=x_{\mathrm{T}c}(e^{y_c}+e^{y_d})$ and $x_b=x_{\mathrm{T}c}(e^{-y_c}+e^{-y_d})$ are the momentum fractions carried by the initial partons from the two colliding protons with $x_{\mathrm{T}c}=2p_{\mathrm{T}c}/\sqrt{s}$, $f_{a/p}(x,\mu^2)$ is the parton distribution inside a proton at the scale $\mu^2=p_{\mathrm{T}c}^2$ and 
$d\hat\sigma_{ab\to cd}/dt$ is the parton level LO cross section which depends on the Mandelstam variables 
$\hat s=x_ax_bs$, $\hat t=-p_{\mathrm{T}c}^2(1+e^{y_d-y_c})$ and $\hat u=-p_{\mathrm{T}c}^2(1+e^{y_c-y_d})$. Because of the initial and final state radiations, there can be more than two jets in the final state and the transverse momentum $p_\mathrm{T}$ of the final leading jet is normally different from the trigger $p_{\mathrm{T}c}$.  

\begin{figure}[tbp]
    \centering
    \includegraphics[width=7.5cm]{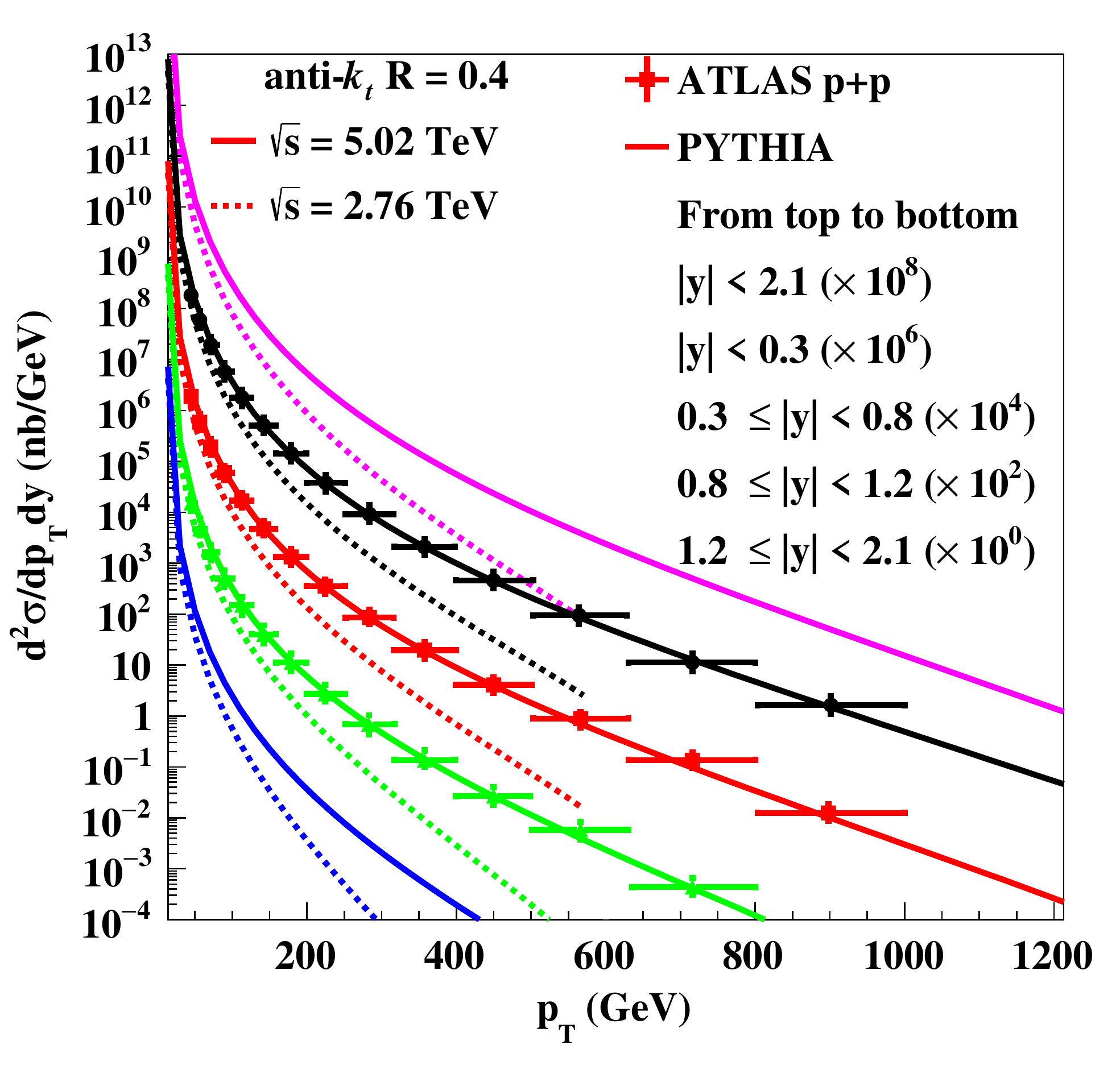}
    \caption{(Color online) The inclusive jet cross section from \textsc{Pythia~8} as a function of the jet transverse momentum $p_\mathrm{T}$ in different rapidity bins in p+p collisions at $\sqrt{s} = 5.02$ TeV (solid)  and 2.76~TeV (dashed), using the anti-$k_\mathrm{T}$ algorithm with the jet cone radius $R = 0.4$ as compared to the ATLAS experimental data~\cite{Aaboud:2018twu}. Results for different rapidities are scaled by successive powers of 100. The figure is from Ref.~\cite{He:2018xjv}.}
    \label{jetCS}
\end{figure}

Shown in Fig.~\ref{jetCS} are the single inclusive jet cross sections as a function of the final jet transverse momentum $p_\mathrm{T}$ in different rapidity bins of p+p collisions at $\sqrt{s}=2.76$ and 5.02~TeV from \textsc{Pythia~8} as compared to the ATLAS experimental data~\cite{Aad:2014bxa,Aaboud:2018twu}. We see \textsc{Pythia~8} can describe the experimental data well. The shape of the single inclusive jet spectra at $\sqrt{s}=5.02$~TeV are much flatter than that at 2.76~TeV as determined mainly by the parton distribution functions.

To calculate the jet spectra in heavy-ion collisions, one first needs to consider the nuclear modification of the initial parton distributions~\cite{Eskola:2009uj,Ru:2016wfx}. One then lets the initial jet shower partons to propagate through the QGP medium within the \textsc{Lbt} model. Using \textsc{Fastjet} for jet reconstruction, one gets an event-averaged final single inclusive jet distribution $d\widetilde{N}^{\rm jet}_{(c)}(p_\mathrm{T}, p_{\mathrm{T}c},\phi_c,{\vec r},{\vec b})/dydp_\mathrm{T}$ for a given initial production point $\vec r$, azimuthal angle $\phi_c$ of the initially produced hard parton $c$ and impact parameter $\vec b$ of the nucleus-nucleus collisions. The cross section for the single inclusive jet production in A+A collision is then given by,
\begin{eqnarray}
\frac{d \sigma^{\rm jet}_{\rm AA}}{dp_\mathrm{T}dy} & = &\sum_{a,b,c,d}  \int d^2{r} d^2{b}  t_A(r) t_A(|{\vec b}-{\vec r}|) \frac{d\phi_c}{\pi}  dy_c dy_d \nonumber\\
&& \times  \int dp_{\mathrm{T}c} p_{\mathrm{T}c}  x_a f_{a/A} (x_a, \mu^2) x_b f_{b/B} (x_b, \mu^2)
\nonumber \\
&& \times \frac{d\hat\sigma_{ab\to cd}}{dt} \frac{d\widetilde{N}^{\rm jet}_{(c)}(p_\mathrm{T},p_{\mathrm{T}c},\phi_c,{\vec r},{\vec b},\phi_c)}{dydp_\mathrm{T}},
\label{eq:cs.aa}
\end{eqnarray}
where $t_{A}(r)$ is the nuclear thickness function with normalization $\int d^2{ r} t_A(r)=A$ and $f_{a/A}(x,\mu^2)$ is the nuclear modified parton distribution function \cite{Eskola:2009uj,Ru:2016wfx} per nucleon. The range of the impact parameter $b$ is determined by the centrality of the nucleus-nucleus collisions according to experimental measurements.

The suppression factor due to interactions between shower and medium partons in heavy-ion collisions is given by the ratio of the jet cross sections for A+A and p+p collisions normalized by the averaged number of binary nucleon-nucleon collisions,
\begin{equation}
R_{\rm AA}=\frac{1}{\int d^2rd^2b  t_A(r) t_A(|{\vec b}-{\vec r}|)} \frac{d\sigma^{\rm jet}_{\rm AA}}{d\sigma^{\rm jet}_{\rm pp}}.
\label{eq:raa}
\end{equation}

In the jet reconstruction using \textsc{Fastjet} one should also subtract the underlying event (UE) background. In the \textsc{Lbt} study presented here, a scheme inspired by the method in the experimental studies~\cite{Aad:2012vca} is used. In \textsc{Lbt} simulations, only jet shower partons, radiated gluons and recoil medium partons including ``negative" partons are used for jet reconstruction. The UE background is very small as compared to the full hydrodynamic UE background. The contribution of UE to the jet energy before the subtraction in \textsc{Lbt} simulations is only a few percent in central Pb+Pb collisions and much smaller in p+p collisions.

\begin{figure}[tbp]
    \centering
    \includegraphics[width=7.5cm]{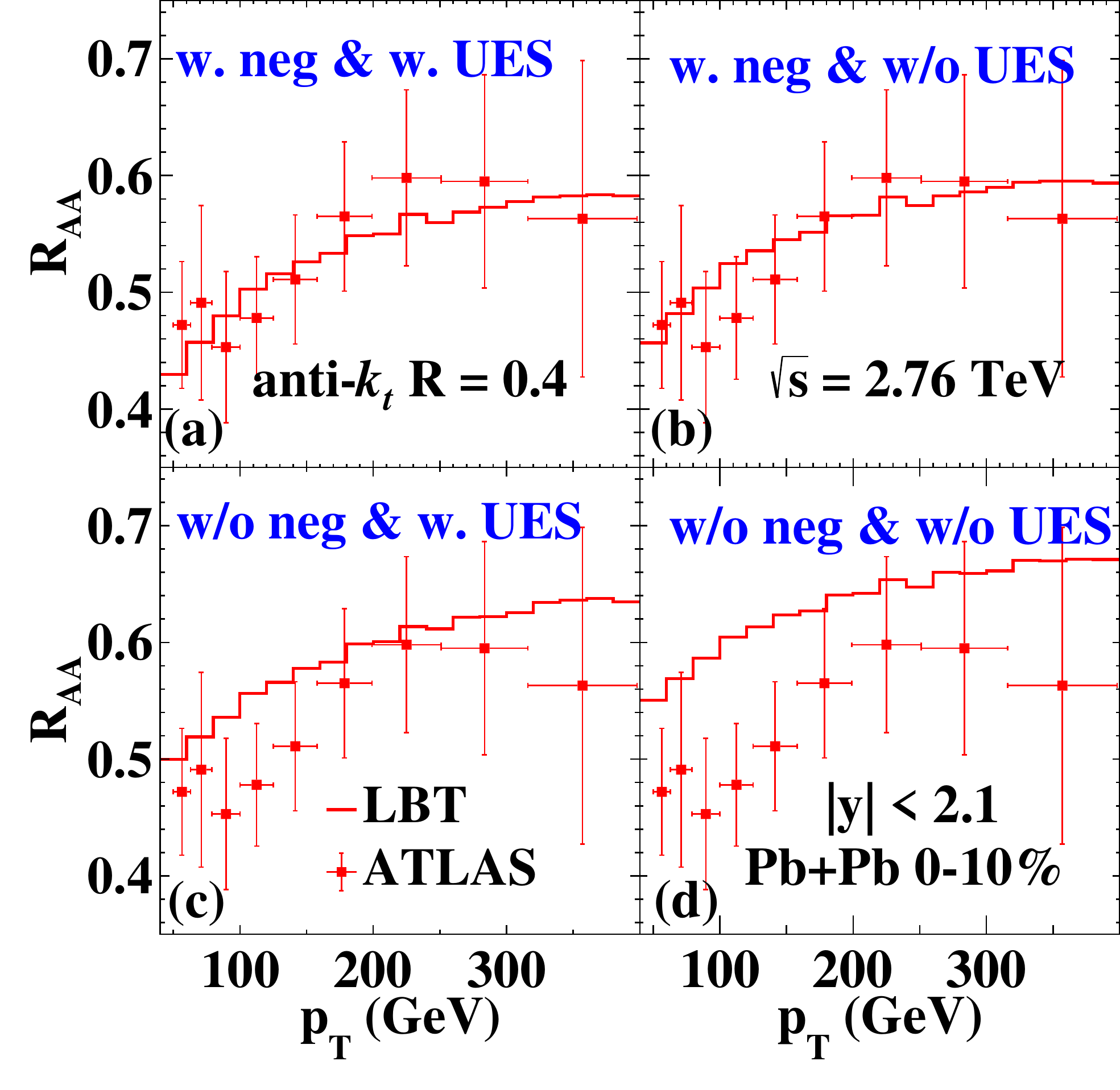}
    \caption{(Color online) The suppression factor $R_{\rm AA}$ of single inclusive jet spectra in the central rapidity $|y|<2.1$ region of 0-10\% central Pb+Pb collisions at $\sqrt{s}=2.76$~TeV from \textsc{Lbt} simulations with fixed $\alpha_{\rm s}=0.15$ as compared to the ATLAS data at the LHC \cite{Aad:2014bxa}. The jet reconstruction with $R=0.4$ and anti-$k_\mathrm{T}$ algorithm includes four different options on ``negative" partons and UES: (a) with both ``negative" partons and UES, (b) with ``negative" partons but without UES, (c) with UES but without ``negative" partons, and (d) without ``negative" partons and UES. The figure is from Ref.~\cite{He:2018xjv}.} 
    \label{RAA_4opts}
\end{figure}

Shown in Fig.~\ref{RAA_4opts} are the suppression factors $R_{\rm AA}(p_\mathrm{T})$ in the central rapidity $|y|<2.1$ region of 0-10\% central Pb+Pb collisions at $\sqrt{s}=2.76$~TeV from \textsc{Lbt} simulations with different options on  ``negative" partons and UE subtraction (UES)  as compared to the ATLAS data.  The fixed value of $\alpha_{\rm s}=0.15$ is used which minimizes the $\chi^2$/d.o.f. from fitting to the ATLAS data when ``negative" partons and UES are both considered [Fig.~\ref{RAA_4opts} (a)]. The fixed value of $\alpha_{\rm s}$ is only an effective strong coupling constant in the \textsc{Lbt} model in which the perturbative Debye screening mass is used to regularize the collinear divergence in elastic scattering and radiative gluon spectrum. Other non-perturbative physics such as chromo-magnetic monopoles can effectively increase the screening mass~\cite{Liao:2008jg,Liao:2008dk,Xu:2015bbz,Xu:2014tda}. Furthermore, the thermal mass of medium partons can also reduce the effective thermal parton density significantly in the interaction rate. These can all increase the value of the effective strong coupling constant in \textsc{Lbt} in order to fit experimental data.

As we discussed before, the inclusion of recoil partons contributes to the total energy inside the jet-cone and therefore significantly reduces the final jet energy loss. The ``negative" partons from the diffusion wake of the jet-induced medium response, however, will reduce the energy inside the jet-cone. One can consider this as jet-induced modification of the background. It will increase the net jet energy loss.  The UES also similarly increases the net jet energy loss. Therefore, they both lead to smaller values of the suppression factor as seen in Fig.~\ref{RAA_4opts}, though the effect of ``negative" partons is relatively larger. Without ``negative" partons, the effect of UES is also larger than with ``negative" partons. 

\begin{figure}[tbp]
    \centering
    \includegraphics[width=7.5cm]{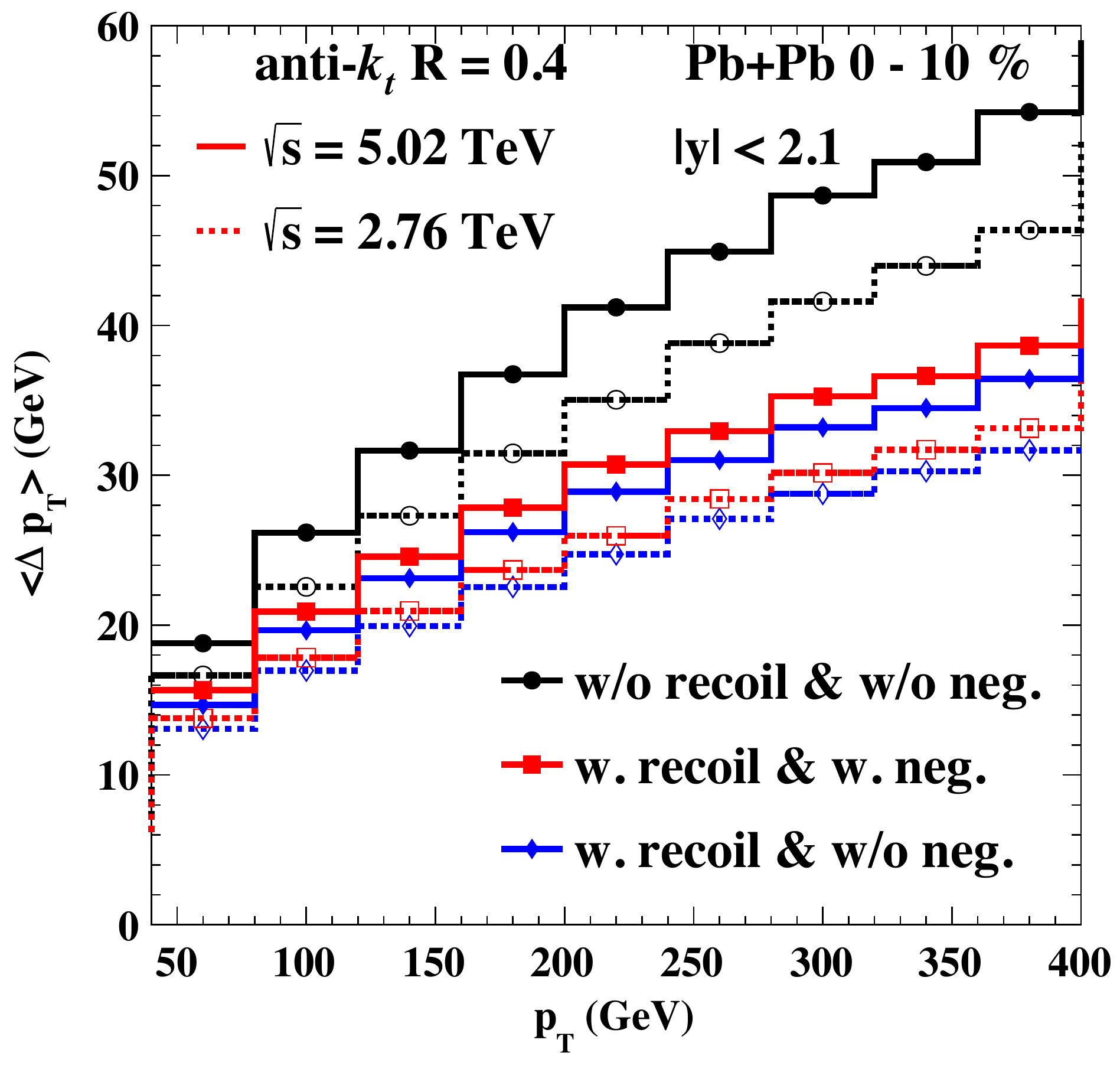}
    \caption{(Color online) The average jet transverse momentum loss as a function of its initial $p_\mathrm{T}$, compared between with and without including recoiled/``negative" parton contributions. The figure is from Ref.~\cite{He:2018xjv}.}
    \label{fig:plot-pTloss_recoil}
\end{figure}

To illustrate the effect of jet-induced medium response on the suppression of single inclusive jet spectra, one can examine the colliding energy and transverse momentum dependence of the jet energy loss as shown in Fig.~\ref{fig:plot-pTloss_recoil} for central Pb+Pb collisions at both $\sqrt{s}=2.76$~TeV and 5.02~TeV. Apparently, the average transverse momentum loss of jets depends on whether the jet-induced medium excitation is taken into account.  At both colliding energies, one observes that including recoil partons in jet reconstruction significantly enhances the final jet $p_\mathrm{T}$ as compared to that without, thus reduces their $p_\mathrm{T}$ loss. On the other hand, the subtraction of ``negative" partons reduces the jet $p_\mathrm{T}$, thus enhances their $p_\mathrm{T}$ loss. The time (or path length) dependence of these medium response effects on jet energy loss is also interesting~\cite{He:2015pra}. One observes that the contribution from ``negative" partons is negligible at the early stage of parton propagation. However, as the number of partons within the jet shower grows with time, so does the number of jet-medium scatterings and the number of ``negative" partons. At later times, the ``negative" partons significantly deplete the thermal medium behind the propagating jet  and effectively modify the background underlying the jet. Only with the subtraction of the ``negative" parton energy, one is able to obtain a linear increase of elastic energy loss with the path length as expected. Therefore, a proper treatment of these medium response effects is crucial for correctly describing jet observables.

\begin{figure}[tbp]
    \centering
    \includegraphics[width=7.5cm]{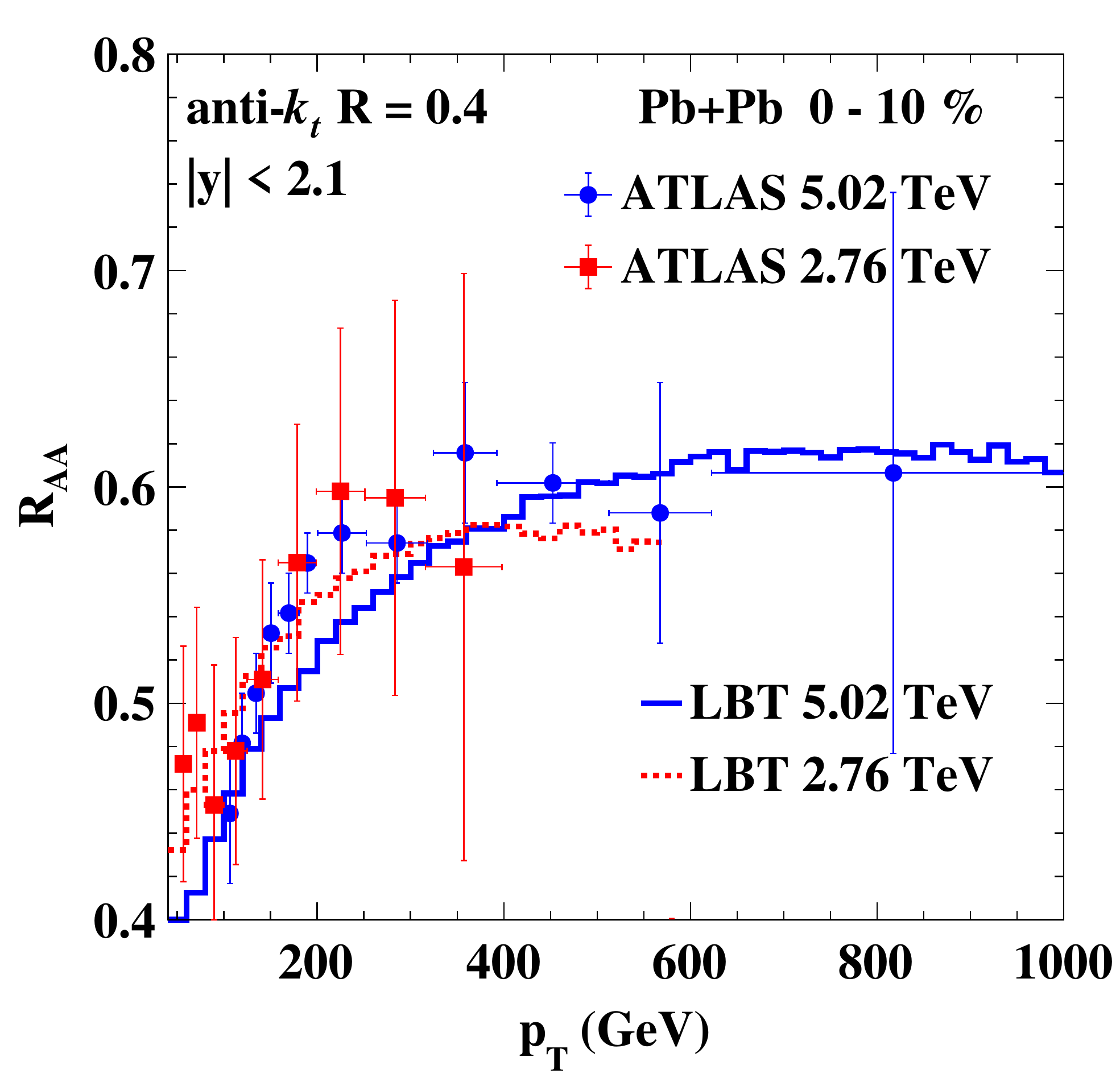}
    \caption{(Color online) The \textsc{Lbt} results on $R_{\rm AA}(p_\mathrm{T})$ in the central rapidity $|y|<2.1$ for single inclusive jet spectra in 0-10\% central Pb+Pb collisions at $\sqrt{s} = 2.76$ (red dashed line) and 5.02~TeV (blue solid line) as compared to the ATLAS data~\cite{Aad:2014bxa,Aaboud:2018twu}. The figure is from Ref.~\cite{He:2018xjv}.}
    \label{RAA_twoEnergy}
\end{figure}

The relatively weak transverse momentum dependence of the jet energy loss in Fig.~\ref{fig:plot-pTloss_recoil} is also shown \cite{He:2018xjv} to be influenced by the combination of many interplaying effects such as radial flow and medium modified flavor composition in addition to jet-induced medium response.  Since the initial parton density increases with the colliding energy as reflected in the 20\% increase of the measured hadron multiplicity in the central rapidity~\cite{Abbas:2013bpa,Adam:2016ddh}, the net jet energy loss at $\sqrt{s}=5.02$~TeV is indeed about 15\% larger than at $\sqrt{s}=2.76$~TeV in the $p_\mathrm{T}=50-400$~GeV range when the medium response is taken into account as shown in Fig.~\ref{fig:plot-pTloss_recoil}. Assuming the effective strong coupling constant in \textsc{Lbt} is independent of the local temperature, the predicted suppression factor for single inclusive jet spectra in Pb+Pb collisions at $\sqrt{s}=5.02$~TeV, shown in Fig.~\ref{RAA_twoEnergy} together with ATLAS the data~\cite{Aad:2014bxa,Aaboud:2018twu}, is almost the same as that at 2.76~TeV. The transverse momentum dependence of the jet suppression factor in this range of $p_\mathrm{T}$ is also very weak, which is very different from the suppression factor of single inclusive charged hadrons~\cite{Aamodt:2010jd,CMS:2012aa,Khachatryan:2016odn,Acharya:2018qsh}. Both of these two features are the consequences of the energy and transverse momentum dependence of the jet energy loss and the initial jet spectra. The increased jet energy loss at higher colliding energy is offset by the flatter initial jet spectra (see Fig.~\ref{jetCS}) to give rise to almost the same $R_\mathrm{AA}(p_\mathrm{T})$.

\begin{figure}[tbp]
    \centering
    \includegraphics[width=7.5cm]{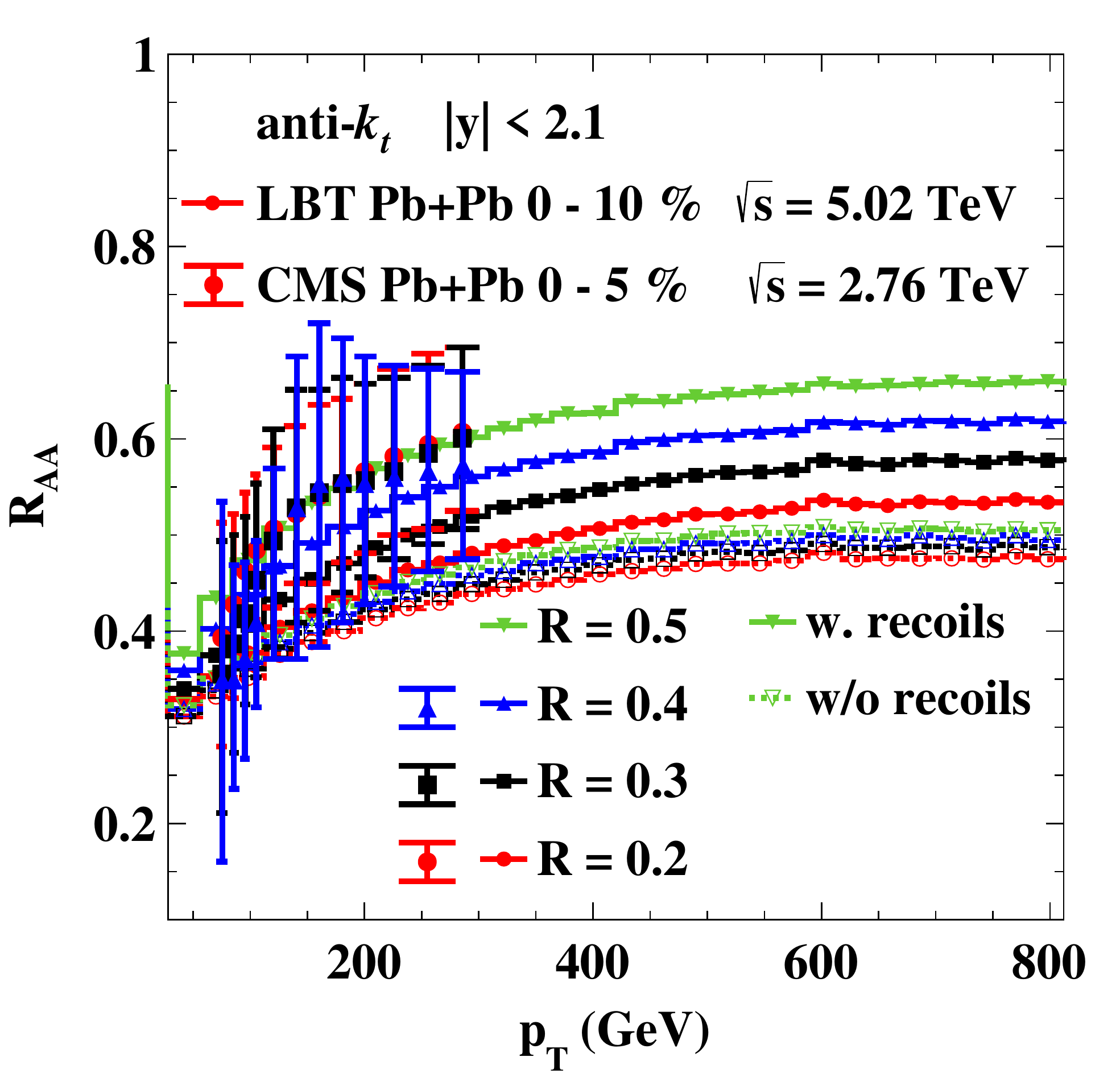}
    \caption{(Color online) The jet suppression factor $R_{\rm AA}$ as a function of $p_\mathrm{T}$ in 0-10\% Pb+Pb collisions at $\sqrt{s}=5.02$~TeV from \textsc{Lbt} with (solid) and without medium response (dashed) for different jet-cone sizes, $R$=0.5, 0.4, 0.3 and 0.2 as compared to the CMS data~\cite{Khachatryan:2016jfl} in 0-5\% Pb+Pb collisions at $\sqrt{s}=2.76$~TeV. The figure is from Ref.~\cite{He:2018xjv}.}
    \label{RAA-cone}
\end{figure}

Since the net jet energy loss decreases with the jet-cone size as a bigger cone size includes more medium recoil partons and radiated gluons. Inclusion of medium response should lead to a unique cone size dependence of the jet suppression. Shown in Fig.~\ref{RAA-cone} are the jet suppression factors from \textsc{Lbt} with (solid) and without medium recoil (dashed) in the central rapidity region of 0-10\% Pb+Pb collisions at $\sqrt{s}=5.02$~TeV for different jet-cone sizes, $R$=0.5, 0.4, 0.3 and 0.2. As expected, the suppression factor increases with the jet-cone size as the net jet energy loss gets smaller for bigger jet-cone size.  Without medium response, the suppression factors are both significantly smaller due to increased energy loss and much less sensitive to the jet-cone size.  Similar behavior was also predicted in Refs.~\cite{Vitev:2008rz,Vitev:2009rd,Zapp:2012ak,Kang:2017frl} but with different $p_\mathrm{T}$-dependence because of lacking influence from the medium response and the radial expansion of the QGP. The systematic uncertainties of the CMS data~\cite{Khachatryan:2016jfl} in Fig.~\ref{RAA-cone} for Pb+Pb collisions at $\sqrt{s}=2.76$~TeV are too big to indicate any jet-cone size dependence.

\begin{figure*}[tbp]
    \centering
    \includegraphics[width=12.0cm]{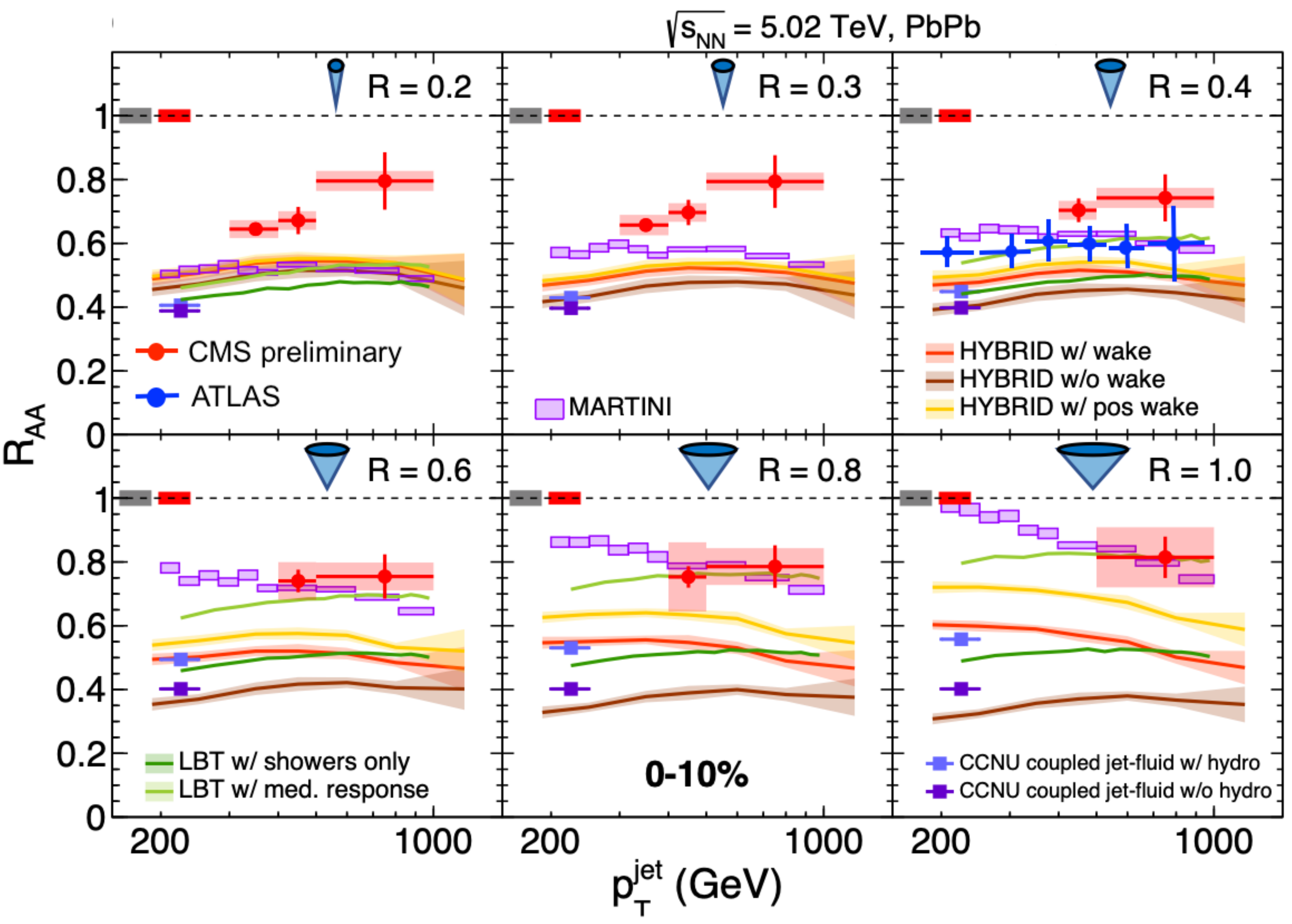}
    \caption{(Color online) The preliminary CMS data on the jet suppression factor $R_{\rm AA}(p_\mathrm{T})$ (red solid circles) in 0-10\% central Pb+Pb collisions at $\sqrt{s} = 5.02$~TeV for different jet cone sizes as compared to the ATLAS data~\cite{Aaboud:2018twu} (blue solid circles) and results from different transport model calculations. The figure is from Ref.~\cite{CMS:2019btm}.}
    \label{RAA-cone-new}
\end{figure*}

The CMS preliminary data on the jet-cone size dependence of the jet suppression factor with high precision in 0-10\% Pb+Pb collisions at $\sqrt{s}=5.02$~TeV become available recently~\cite{CMS:2019btm}. They are compared with many transport model simulations in Fig.~\ref{RAA-cone-new}. While most of the transport models fail to describe the preliminary CMS data, \textsc{Lbt} (with medium response) and \textsc{Martini} results agree with the data well for large jet-cone size ($R$=0.6, 0.8, 1.0). However, \textsc{Lbt} and \textsc{Martini} results on $R_\mathrm{AA}$ continue to decrease slightly with smaller jet cone-size ($R$=0.4,0.3,0.,2) while the CMS data remain the same and even increase slightly. Other theory calculations without medium response appear to agree with the preliminary data even for small jet-cone size~\cite{CMS:2019btm}. It is important to note that the ATLAS data for $R=0.4$ agree with \textsc{Lbt} and \textsc{Martini} results and are systematically smaller than the CMS preliminary data. Therefore, in order to verify the effect of medium response on the cone-size dependence of jet suppression, it is necessary for CMS and ATLAS to reconcile the discrepancy between their measurements by extending the CMS's coverage to small $p_\mathrm{T}$ and ATLAS's analyses to different (both small and large) jet-cone sizes.

\subsection{$\gamma$/$Z^0$-jet correlation}
\label{subsec:dijet}

Similar to $\gamma$-hadron correlation, $\gamma/Z^0$-jet correlations are excellent probes to study jet quenching and jet-induced medium response because the energy of the triggered $\gamma$/$Z^0$ boson provides an approximate proxy of the initial jet energy before its propagation and transport through the QGP medium. Jet yields per trigger are also free of the uncertainties related to the estimate of the number of binary collisions in the study of quenching of single inclusive hadrons and jets. One can also measure jet transport coefficient directly through di-hadron, $\gamma$-hadron, or $\gamma/Z^0$-jet correlation in the azimuthal angle~\cite{Appel:1985dq,Blaizot:1986ma}. Even though the Sudakov form factor from initial state radiation dominates the azimuthal angle correlation between $\gamma/Z^0$ and jets with large transverse momentum~\cite{Mueller:2016gko,Chen:2016vem}, the large angle correlation could be influenced by large angle scattering between jet shower and medium partons~\cite{DEramo:2012uzl}, in particular when the transverse momentum scale is not too large.

There have been many theoretical studies on jet quenching with $\gamma/Z^0$-jets in heavy-ion collisions~\cite{Li:2010ts,Dai:2012am,Wang:2013cia,Kang:2017xnc}. We will review here recent work on $\gamma/Z^0$-jets using the \textsc{Lbt} model with special emphasis on the effect of medium response, multiple jet production and suppression \cite{Luo:2018pto,Zhang:2018urd}. Similar to the simulation of single inclusive jets,  \textsc{Pythia~8}~\cite{Sjostrand:2006za} is used to generate the initial jet shower partons in $\gamma$-jet events in p+p collisions with a minimum transverse momentum transfer that is half of the transverse momentum of the triggered photons. Events with bremsstrahlung photons from QCD processes are also included. The distribution of the initial $\gamma$-jet production in the transverse plane is sampled according to the distribution of hard processes in the initial condition for the underlying hydrodynamic evolution of the bulk matter. Parton transport is simulated within the \textsc{Lbt} model until the hadronic phase of the bulk matter. The \textsc{Fastjet}~\cite{Cacciari:2011ma} package is again used to reconstruct jets from the final partons.

\begin{figure}[tbp]
\centering
\includegraphics[width=8.0cm]{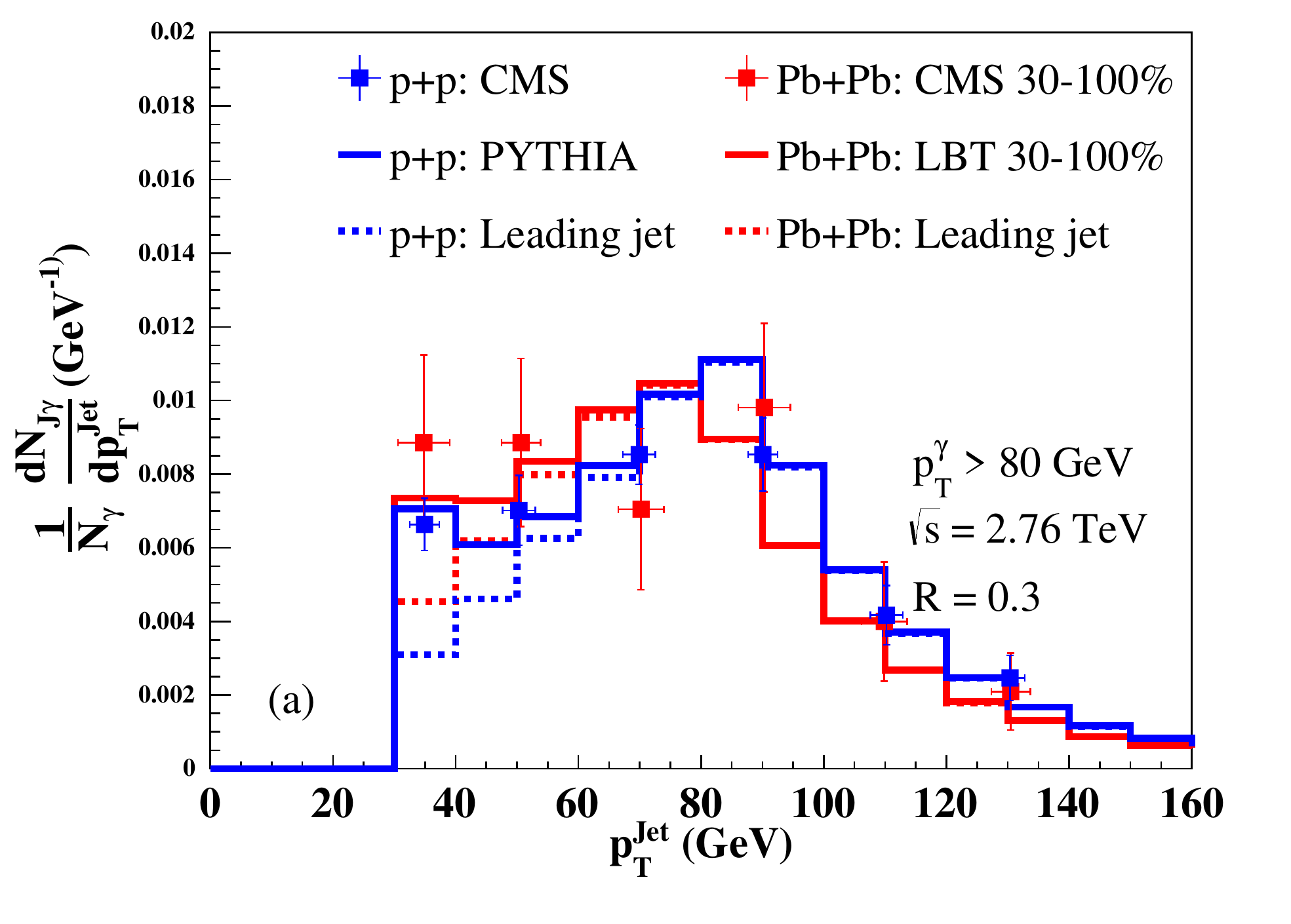}\\
\vspace{-0.35in}
\includegraphics[width=8.0cm]{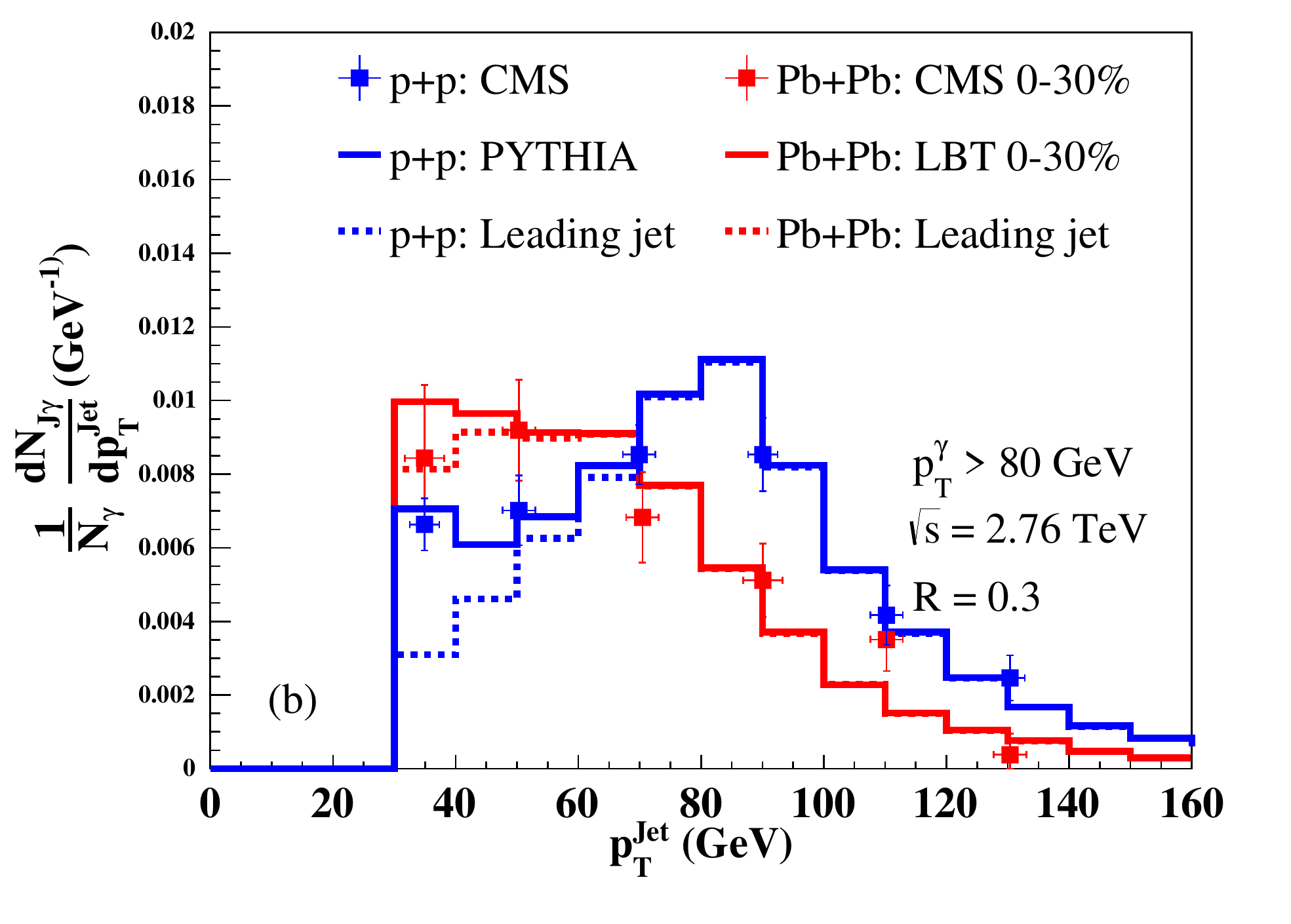} 
\caption{(Color online) The transverse momentum distribution of $\gamma$-jet in (a) peripheral (30-100\%) and (b) central (0-30\%) Pb+Pb (red) and p+p collisions (blue) at $\sqrt{s}=2.76$~TeV from \textsc{Lbt} simulations as compared to the CMS experimental data~\cite{Chatrchyan:2012gt}. Dashed lines are the transverse momentum distributions for leading jets only. The figures are from Ref.~\cite{Luo:2018pto}.}
\label{ptjet}
\end{figure}

Shown in Fig.~\ref{ptjet} are the distributions of the associated jets from \textsc{Lbt} simulations as a function of $p_\mathrm{T}^{\rm jet}$ for fixed $p_\mathrm{T}^\gamma \ge 80$ GeV in both p+p and Pb+Pb collisions at $\sqrt{s}=2.76$~TeV which compare fairly well with the experimental data from CMS~\cite{Chatrchyan:2012gt}. The same kinematic cuts $|\eta_\gamma|<1.44$, $|\eta_{\rm jet}|<1.6$ and $|\phi_\gamma-\phi_{\rm jet}|>(7/8)\pi$ are imposed as in the CMS experiments. A lower threshold of the transverse momentum of reconstructed jets is set at $p_\mathrm{T}^{\rm jet}>30$~GeV. To compare to the CMS data, the \textsc{Lbt} results in both Pb+Pb and p+p collisions are convoluted with a Gaussian smearing with the same jet energy resolution in each centrality class of Pb+Pb collisions as in the CMS data. A complete subtraction of the uncorrelated underlying event background is assumed for the \textsc{Lbt} results. 

\begin{figure}[tbp]
\centering
\includegraphics[width=8.0cm]{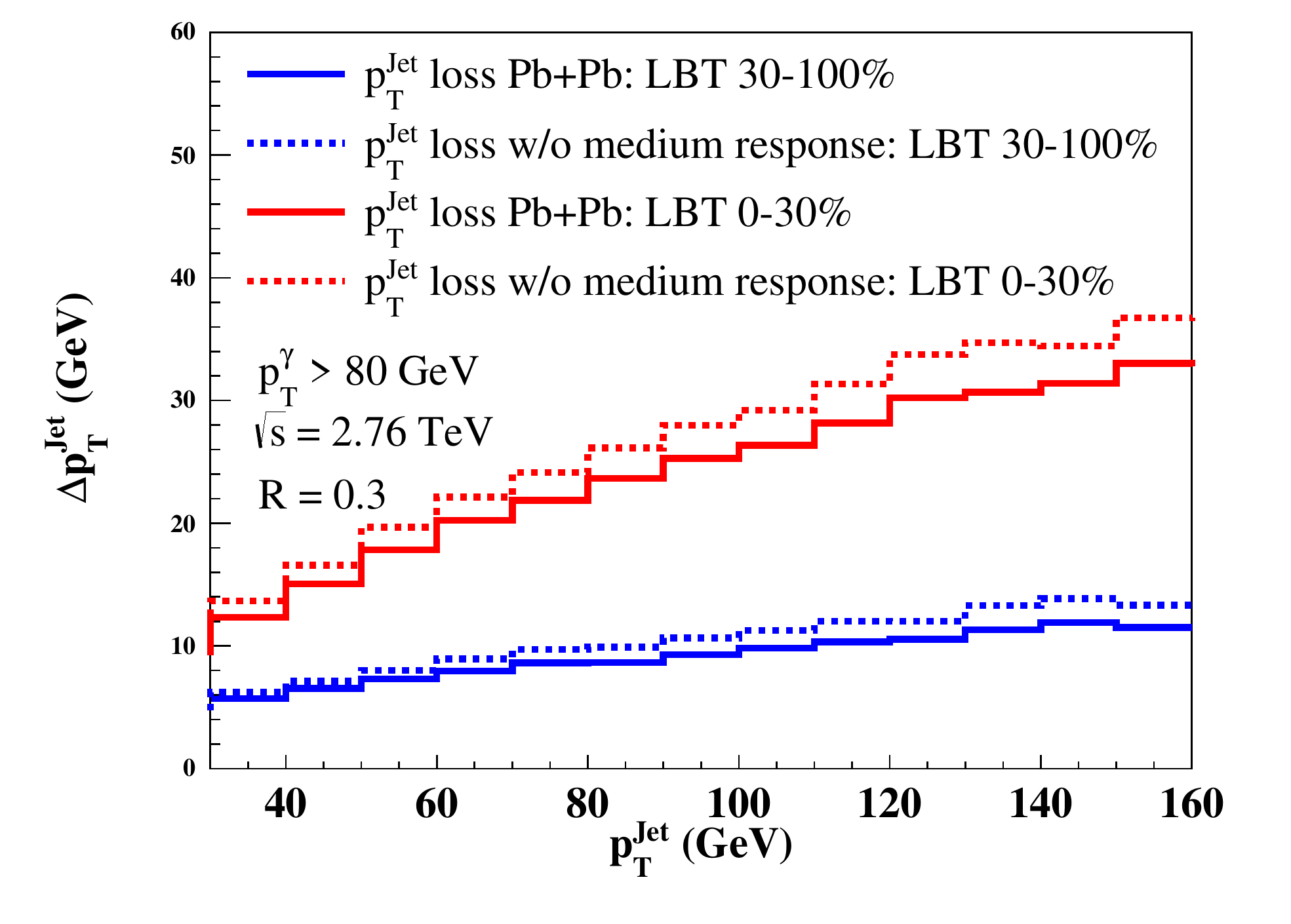} \\
\caption{(Color online) The average transverse momentum loss of the leading $\gamma$-jet in Pb+Pb collisions at $\sqrt{s}=2.76$~TeV calculated within \textsc{Lbt} as a function of the initial jet transverse momentum with (solid) and without (dashed) contributions from recoil and ``negative" partons. The figure is from Ref.~\cite{Luo:2018pto}.}
\label{ptloss}
\end{figure}

In the $\gamma$-jet events, the transverse momentum of the triggered photon is not completely balanced by the jet because of the initial-state gluon bremsstrahlung. However, the peak position of the jet $p_\mathrm{T}^{\rm jet}$ distribution in p+p collisions reflects the average initial $p_\mathrm{T}^{\rm jet}$ value of the $\gamma$-jet.  From Fig.~\ref{ptjet} one can clearly see the shift of the peak position to a smaller $p_\mathrm{T}^{\rm jet}$ value due to jet energy loss in Pb+Pb collisions. To illustrate this shift in detail, we show in Fig.~\ref{ptloss} the average transverse momentum loss of the leading jet in $\gamma$-jet events in two centrality classes of Pb+Pb collisions as a function of the initial transverse momentum of the leading jet in p+p collisions. As one can see, including recoil and ``negative" partons from medium response in jet reconstruction reduces the net jet energy loss. The jet transverse momentum loss increases with the initial jet transverse momentum and the dependence is slightly weaker than a linear increase due to a combined effect of the jet energy loss for a given jet flavor (quark or gluon) and the transverse momentum dependence of the initial jet flavor composition. The initial quark fraction increases with transverse momentum in $\gamma$-jets and the energy loss of a gluon jet is found to be about 1.5 times bigger than that of a quark for a jet-cone size $R=0.3$ in this range of $p_\mathrm{T}^{\rm jet}$~\cite{He:2018xjv}. By shifting the transverse momentum distribution of $\gamma$-jet in p+p collisions, one can approximately describe the modification of the $\gamma$-jet spectra in Pb+Pb collisions~\cite{Luo:2018pto}.

\begin{figure}[tbp]
\centering
\includegraphics[width=6.9cm]{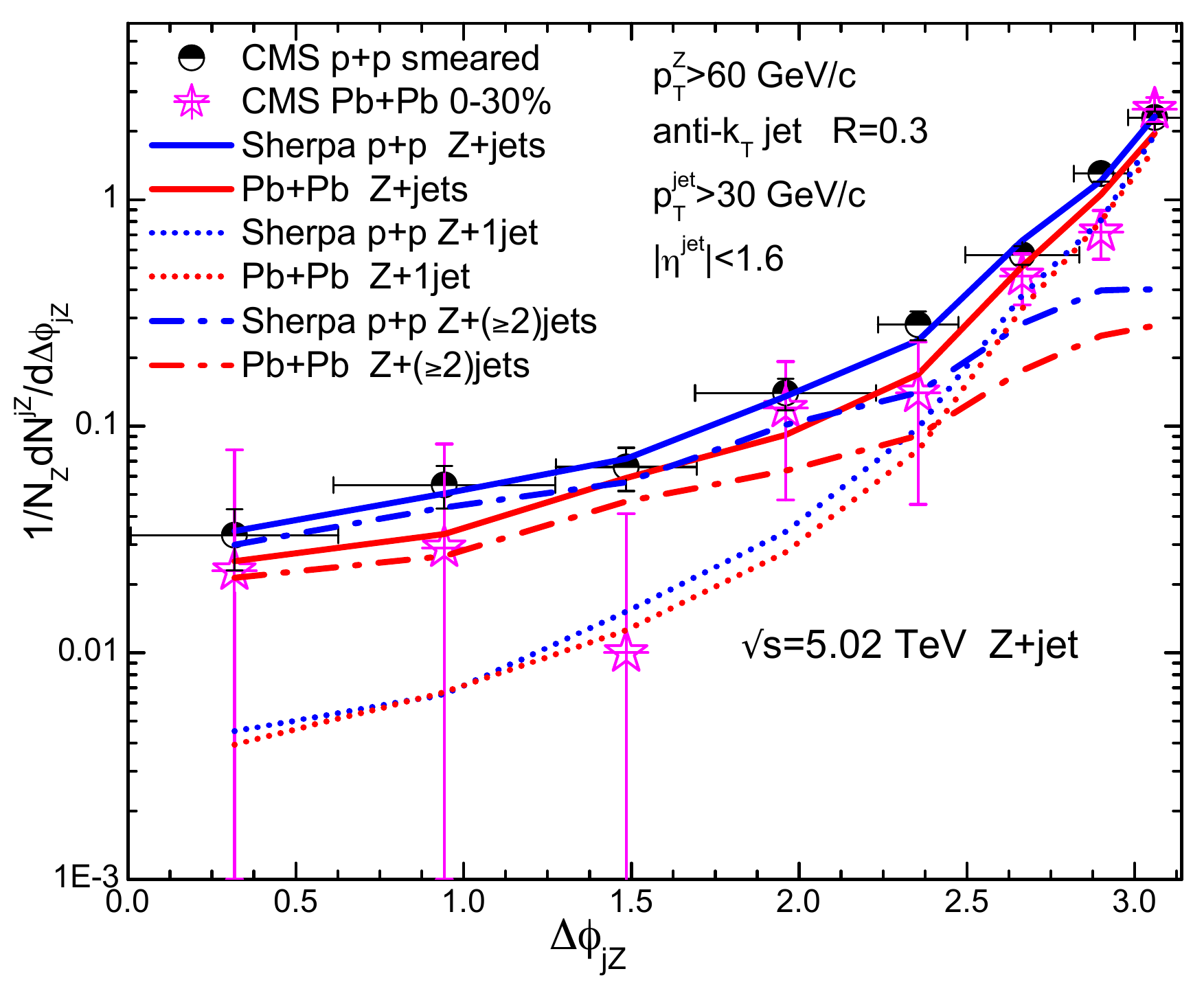}
\caption{(Color Online) The $Z^0$-jet correlation in the azimuthal angle $\Delta \phi_{\rm jZ}$ from \textsc{Lbt} simulations in p+p (blue) and Pb+Pb collisions (red) at $\sqrt{s} = 5.02$~TeV as compared to the CMS data~\cite{Sirunyan:2017jic}. The dotted (dash-dotted) lines show the contributions from $Z^0+1$ jets ($Z^0+(\ge 2)$ jets). The figure is from Ref.~\cite{Zhang:2018urd}.}
\label{zjet_deltaphi}
\end{figure}

In Fig.~\ref{ptjet}, the \textsc{Lbt} results for the associated leading jet are shown (dashed lines) to deviate from the inclusive associated jet yields at small values of $p_\mathrm{T}^{\rm jet}$. The difference at low $p_\mathrm{T}^{\rm jet}<p_\mathrm{T}^\gamma$ is mainly caused by secondary jets associated with the triggered photon. Energy loss and suppression of the sub-leading jets lead to medium modification of the $\gamma$-jet correlation at lower $p_\mathrm{T}^{\rm jet}$ in addition to the modification caused by energy loss of the leading jets in $\gamma$-jet events.

Multiple jets are produced from the large angle radiative processes in the initial hard processes. Their contributions can become significant in the region of large momentum imbalance $p_\mathrm{T}^{\rm jet}<p_\mathrm{T}^\gamma$ and even become dominant at large azimuthal angle difference $|\phi^\gamma-\phi^{\rm jet}-\pi|$. One can see this from the $Z^0$-jet correlation in the azimuthal angle $\Delta \phi_{\rm jZ}$ in p+p and Pb+Pb collisions at $\sqrt{s} = 5.02$~TeV as compared to the CMS data~\cite{Sirunyan:2017jic} in Fig.~\ref{zjet_deltaphi}. Here, the initial $Z^0$-jet showers in p+p collisions are simulated with the \textsc{Sherpa} Monte Carlo program~\cite{Gleisberg:2008ta} that combines the NLO pQCD with resummation of a matched parton shower. This model has a much better agreement with the experimental data on the large angle $Z^0$-jet correlation~\cite{Zhang:2018urd} in p+p collisions.  As one can see, contributions from $Z^0$+($\ge 2$) jets from the NLO processes are much broader than that of $Z^0$+1 jets and dominate in the large angle $|\Delta \phi_{\rm jZ}-\pi| $ region.  The $Z^0$+1 jets contribute mostly in the small angle $|\Delta \phi_{\rm jZ}-\pi|$ region where soft/collinear radiation from parton shower dominates. Jet quenching has negligible effects on the azimuthal correlation contributed by $Z^0$+1 jets because of the trigger bias. However, it suppresses the contribution from $Z^0$+($\ge 2$) jets and therefore leads to the suppression of the $Z^0$-jet correlation at large angle $|\Delta \phi_{\rm jZ}-\pi|$.  Similar effects are also seen in the $\gamma$-jet correlation in the azimuthal angle~\cite{Luo:2018pto}.


%
%
%
%

\section{Jet substructures}
\label{sec:jet_substructures}

\subsection{Jet fragmentation function}
\label{subsec:jetFragmentation}

\begin{figure*}[tbp]
    \centering
    \includegraphics[width=0.9\textwidth]{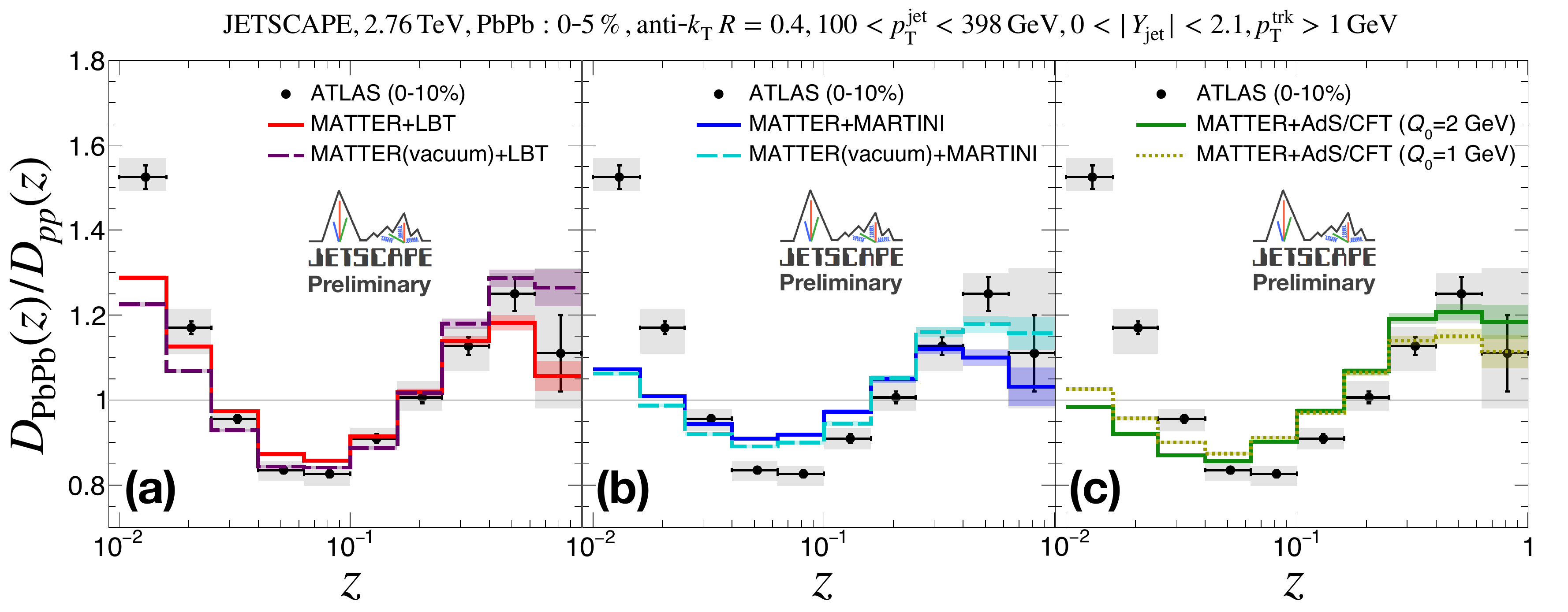}
    \caption{(Color online) Nuclear modification of the jet fragmentation function from (a) \textsc{Matter+Lbt}, (b) \textsc{Matter+Martini} and (c) \textsc{Matter}+AdS/CFT calculations, compared to experimental data. The solid and dashed lines in (a) and (b) are with and without medium effects in \textsc{Matter} respectively, both with separation scale $Q_0=2$~GeV. Different separation scales in \textsc{Matter}+AdS/CFT are compared in (c). The figure is from Ref.~\cite{Tachibana:2018yae}.}
    \label{fig:JETSCAPE-jetFF}
\end{figure*}

Recent developments on experimental and computational techniques allow us to measure/calculate not only the total energy loss of full jets, but also how energy-momentum is distributed within jets. The latter is known as the jet substructure (or inner structure), which helps place more stringent constraints on our knowledge of parton-medium interactions, especially how jet-induced medium excitation modifies the soft hadron distribution within jets.

The first observable of jet substructure is the jet fragmentation function~\cite{Chatrchyan:2012gw,Chatrchyan:2014ava,Aaboud:2017bzv,Aaboud:2018hpb}:
\begin{equation}
\label{eq:defJetFF}
D(z)\equiv\frac{1}{N_\mathrm{jet}}\frac{dN_h}{dz},
\end{equation}
which quantifies the hadron number ($N_h$) distribution as a function of their longitudinal momentum fraction ($z$) with respect to the jet  ($z={\vec p}_\mathrm{T}^{\;h}\cdot {\vec p}_\mathrm{T}^\mathrm{\; jet}/|{\vec p}_\mathrm{T}^\mathrm{\; jet}|^2$), normalized to one jet.  Note that the fragmentation function defined this way should not be confused with that for hadron distributions from the fragmentation of a single parton as defined in Eqs.~(\ref{eq:qfrag}) and (\ref{eq:gfrag}) and used to calculate single inclusive hadron spectra from hard processes [see e.g., Eqs.~(\ref{eq:dis0}) and (\ref{eq:xsectionFactor})].

Shown in Fig.~\ref{fig:JETSCAPE-jetFF} is a comparison between model calculations and experimental data on the nuclear modification factor of the jet fragmentation function within the \textsc{Jetscape} framework~\cite{Tachibana:2018yae}. Here, charged particles (tracks) with $p_\mathrm{T}^\mathrm{trk}$ are used to calculate the fragmentation function of jets with $R=0.4$, $100<p_\mathrm{T}^\mathrm{jet}<398$~GeV and $|y^\mathrm{jet}|<2.1$ in central Pb+Pb collisions at 2.76~TeV. A general feature of this nuclear modification factor is its enhancement at both small and large $z$ values with a suppression in the intermediate region of $z$. 

The enhancement near $z\rightarrow 1$ results from the different energy loss mechanisms between full jets and leading partons inside the jets after  their transport through the QGP. For a full jet with a given cone size $R$, its energy loss mainly arises from losing soft components either through transport outside the jet-cone or ``absorption" by the medium of soft partons. The absorption occurs when soft hadrons are excluded from jets either through background subtraction or $p_\mathrm{T}$ cuts on particle tracks in the jet reconstruction. Consequently, even after losing an amount of energy, a full jet in A+A collisions can contain more leading partons at large $z$ than a jet in p+p collisions with the same $p_\mathrm{T}^{\rm jet}$.   Of course, this enhancement at large $z$ depends on both the jet-cone size and the lower $p_\mathrm{T}^\mathrm{trk}$ cut for soft particles that are used for jet construction. This should be further investigated with more detailed calculations and measurements.  As we will show later, such an enhancement at large $z$ disappears and is replaced by a suppression instead if one defines the momentum fraction by the transverse momentum of a triggered particle (e.g. $\gamma/Z^0$) which is approximately the initial jet energy before jet propagation through the QGP medium. Meanwhile, the medium-induced parton splitting and jet-induced medium response generate a large number of low $p_\mathrm{T}$ hadrons, leading to the low $z$ enhancement of the jet fragmentation function. Due to the energy conservation, enhancements at both large and small $z$ must be compensated by a depletion in the intermediate region of $z$.

Different model implementations of jet transport in QCD are compared in Fig.~\ref{fig:JETSCAPE-jetFF}. 
In Figs.~\ref{fig:JETSCAPE-jetFF}~(a) and (b), results with and without medium modification of jets at high virtualities are compared, where the separation scale between the high-virtuality \textsc{Matter} shower and the low-virtuality \textsc{Lbt}/\textsc{Martini} transport is set as $Q_0=2$~GeV.
The medium-modified splittings in \textsc{Matter} (solid lines) leads to both suppression of leading hadrons at large $z$ and enhancement of soft hadrons at small $z$ relative to the results with only the vacuum splittings in \textsc{Matter}  (dashed line). The medium modification in the \textsc{Matter} phase has a larger impact on the hard cores (leading partons) of jets than on the soft coronas. Including this modification is more effective in increasing the energy loss of leading partons than the full jets. It therefore suppresses the enhancement of the jet fragmentation function at large $z$. 
For the same reason, reducing $Q_0$ from 2 to 1~GeV in Fig.~\ref{fig:JETSCAPE-jetFF}~(c) increases the \textsc{Matter} contribution in this multi-scale jet transport, hence also suppresses the large $z$ enhancement. 

Comparing results from the three different combinations of models within the JETSCAPE framework in Figs.~\ref{fig:JETSCAPE-jetFF}~(a), (b) and (c), 
one can see  a much stronger low-$z$ enhancement of the jet fragmentation function from  \textsc{Matter}+\textsc{Lbt} simulations in Fig.~\ref{fig:JETSCAPE-jetFF}~(a) because all recoil partons and their transport are fully tracked within the \textsc{Lbt} model to account for jet-induced medium excitation. These are not included in the results from \textsc{Matter}+\textsc{Marini} and \textsc{Matter}+AdS/CFT in Figs.~\ref{fig:JETSCAPE-jetFF}~(b) and (c).
Similar effects from medium response on soft hadron production have also been found in Ref.~\cite{Casalderrey-Solana:2016jvj} within the \textsc{Hybrid} model and Ref.~\cite{KunnawalkamElayavalli:2017hxo} within the \textsc{Jewel} model.

\begin{figure}[t]
\centering
\includegraphics[scale=0.36]{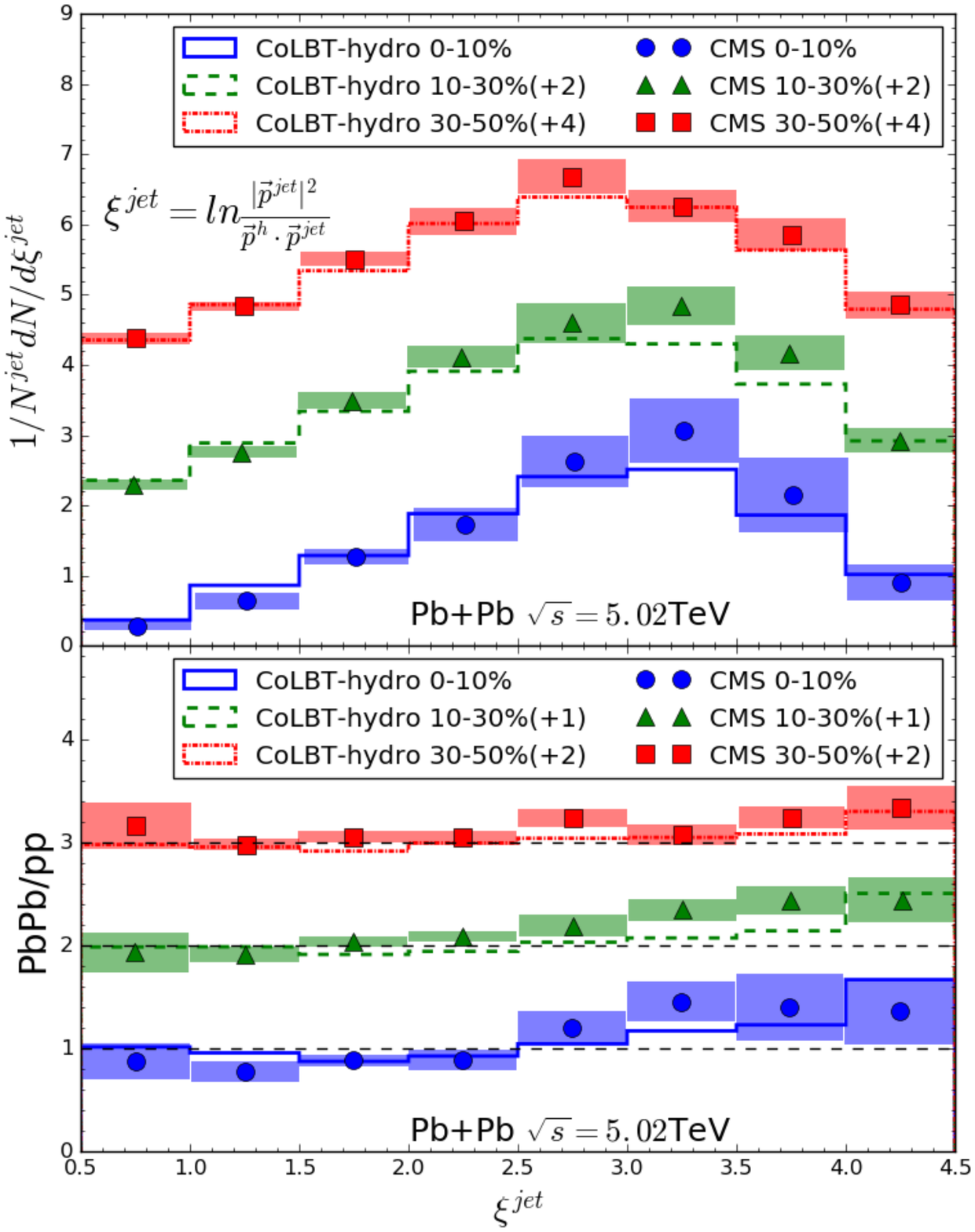} 
\caption{(Color online) The $\gamma$-jet fragmentation function as a function of $\xi^{\rm jet}$ in Pb+Pb collisions at $\sqrt{s}$=5.02 TeV for different centrality classes (upper panel) and the corresponding ratio of the Pb+Pb to p+p results (lower panel) for $p_\mathrm{T}^\gamma>60$~GeV, $p_\mathrm{T}^{\rm jet}>30$~GeV, $|\phi^{\rm jet}-\phi^\gamma|<7\pi/8$ and jet cone-size $R=0.3$. The figure is from Ref.~\cite{Chen:2018azc}.}
\label{fig-ratio-jet}
\end{figure}

\begin{figure}[t]
\centering
\includegraphics[scale=0.36]{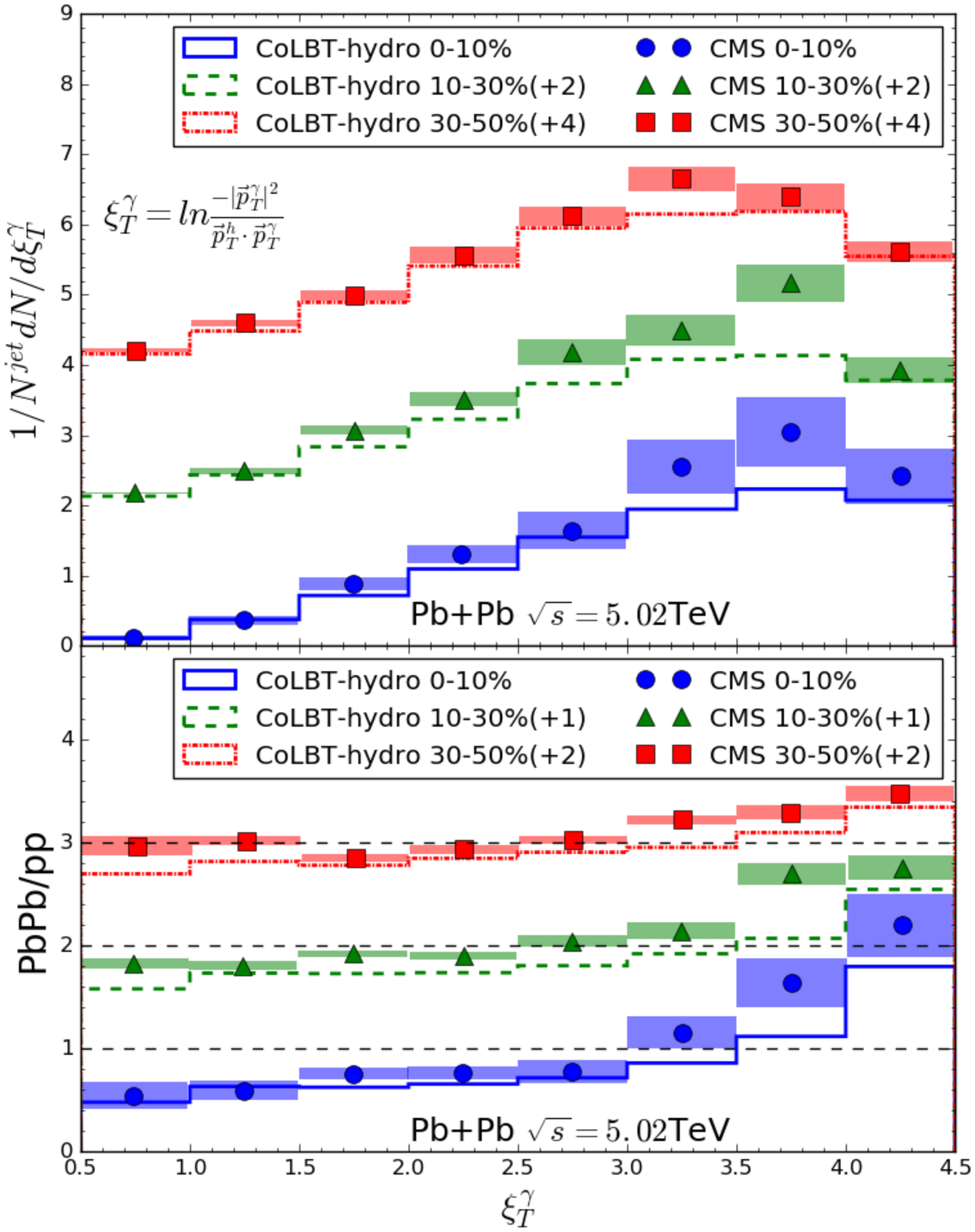} 
\caption{(Color online) The $\gamma$-jet fragmentation function as a function of $\xi^{\gamma}$ in Pb+Pb collisions at $\sqrt{s}$=5.02 TeV for different centrality classes (upper panel) and the corresponding ratio of the Pb+Pb to p+p results (lower panel) for $p_\mathrm{T}^\gamma>60$~GeV, $p_\mathrm{T}^{\rm jet}>30$~GeV, $|\phi^{\rm jet}-\phi^\gamma|<7\pi/8$ and jet cone-size $R=0.3$. The figure is from Ref.~\cite{Chen:2018azc}.} 
\label{fig-ratio-gamma}
\end{figure}

The jet fragmentation function has also been studied within many other theoretical approaches. For instance, a simplified energy loss model~\cite{Spousta:2015fca} that shifts the jet $p_\mathrm{T}$ in A+A collisions can also grasp certain features of the in-medium jet fragmentation function at high $z$, though it fails at low $z$. Using the \textsc{Yajem} model~\cite{Perez-Ramos:2014mna}, one finds that the data on the jet fragmentation function prefer the \textsc{Yajem-rad} module to the \textsc{Yajem-fmed} module of jet energy loss -- the former assumes virtuality enhancement of parton splitting inside the QGP while the latter assumes a simple parametrization of the medium-modified splitting function. Within the \textsc{Pyquen} model~\cite{Lokhtin:2014vda}, the experimental data are found to favor wide angle over small angle radiation in parton showers. The jet fragmentation function is also suggested as a valuable tool to extract the color (de)coherence effects in jet-QGP interactions in Ref.~\cite{Mehtar-Tani:2014yea}.

Apart from single inclusive jets, the fragmentation function has also been explored for $\gamma$-triggered jets~\cite{Aaboud:2019oac} in which one can test sophisticated modelings of both parton energy loss and jet-induced medium excitation for fully understanding the $z$-dependence of the nuclear modification factor.  Shown in Fig.~\ref{fig-ratio-jet} are the hadron yields of $\gamma$-jets as a function of $\xi^{\rm jet}=\ln(|{\vec p}_\mathrm{T}^{\rm\; jet}|^2/{\vec p}_\mathrm{T}^{\rm\; jet}\cdot{\vec p}_\mathrm{T}^{\;h})$ and their nuclear modification factors in Pb+Pb collisions with different centralities at $\sqrt{s}=5.02$~TeV from \textsc{CoLbt-Hydro} simulations~\cite{Chen:2018azc} as compared to the CMS data~\cite{Sirunyan:2018qec}. One can see there is a significant enhancement of the hadron yields at large $\xi^{\rm jet}$ (or small momentum fraction) due to the contribution from medium response. However, there is little change of hadron yields at small $\xi^{\rm jet}$ (or large momentum fraction). This is because of the trigger bias in the selection of jets with fixed values of $p_\mathrm{T}^{\rm jet}$ as we have just discussed before. One can also calculate the hadron yields per jet as a function of $\xi^\gamma=\ln(|{\vec p}_\mathrm{T}^{\;\gamma}|^2/{\vec p}_\mathrm{T}^{\;\gamma}\cdot{\vec p}_\mathrm{T}^{\;h})$ without fixing the transverse momentum of jets as shown in Fig.~\ref{fig-ratio-gamma}. In this case, there is a suppression of the hard (leading) hadrons at small $\xi^\gamma$ (or large momentum fraction) due to jet quenching and jet energy loss as well as an enhancement of soft hadrons at large $\xi^\gamma$ (or small momentum fraction) due to medium response. The enhancement is much bigger at large $\xi^\gamma$ than at large $\xi^{\rm jet}$ due to the fluctuation of the initial jet energy when $p_\mathrm{T}^\gamma$ is fixed.

Note that it has been suggested in Ref.~\cite{Zhang:2015trf} that our current poor knowledge on the hadronization process limits the precision of our description of the hadron number within jets, thus also the fragmentation function. As a result, sub-jet structures are proposed, which measure the distribution of small-cone sub-jets, instead of hadrons, within large-cone jets. These new observables are expected to provide a larger discriminating power between different theoretical models and lead to the discovery of new features in medium modification of jet structures.

\subsection{Jet shape}
\label{subsec:jetShape}

While the jet fragmentation function measures the longitudinal momentum distribution within a full jet, a complimentary observable is the jet shape~\cite{Chatrchyan:2013kwa,Khachatryan:2016tfj,Sirunyan:2018ncy}. It measures the momentum distribution transverse to the jet axis and should be sensitive to jet-induced medium response. It is also known as the jet energy density profile defined as
\begin{equation}
\label{eq:defJetShape}
\rho(r)\equiv\frac{1}{\Delta r}\frac{1}{N_\mathrm{jet}}\sum_\mathrm{jet}\frac{p_\mathrm{T}^\mathrm{jet}(r-\Delta r/2,r+\Delta r/2)}{p_\mathrm{T}^\mathrm{jet}(0,R)},
\end{equation}
where $r=\sqrt{(\eta-\eta_\mathrm{jet})^2+(\phi-\phi_\mathrm{jet})^2}$ is the radius to the center of the jet located at $(\eta_\mathrm{jet},\phi_\mathrm{jet})$, and 
\begin{equation}
\label{eq:defJetShapePT}
p_\mathrm{T}^\mathrm{jet}(r_1,r_2)=\sum_{\mathrm{trk}\in(r_1, r_2)} p_\mathrm{T}^\mathrm{trk}
\end{equation}
represents the summed energy of particle tracks within the circular annulus between $(r_1,r_2)$. The jet profile is normalized by the total energy within $(0,R)$ of each jet, with $R$ being the cone size utilized to reconstruct the jet. The above equation is also normalized to per jet event. Note that the study of the jet shape can be extended to the $r>R$ region.

\begin{figure}[tbp]
    \centering
    \includegraphics[width=0.43\textwidth]{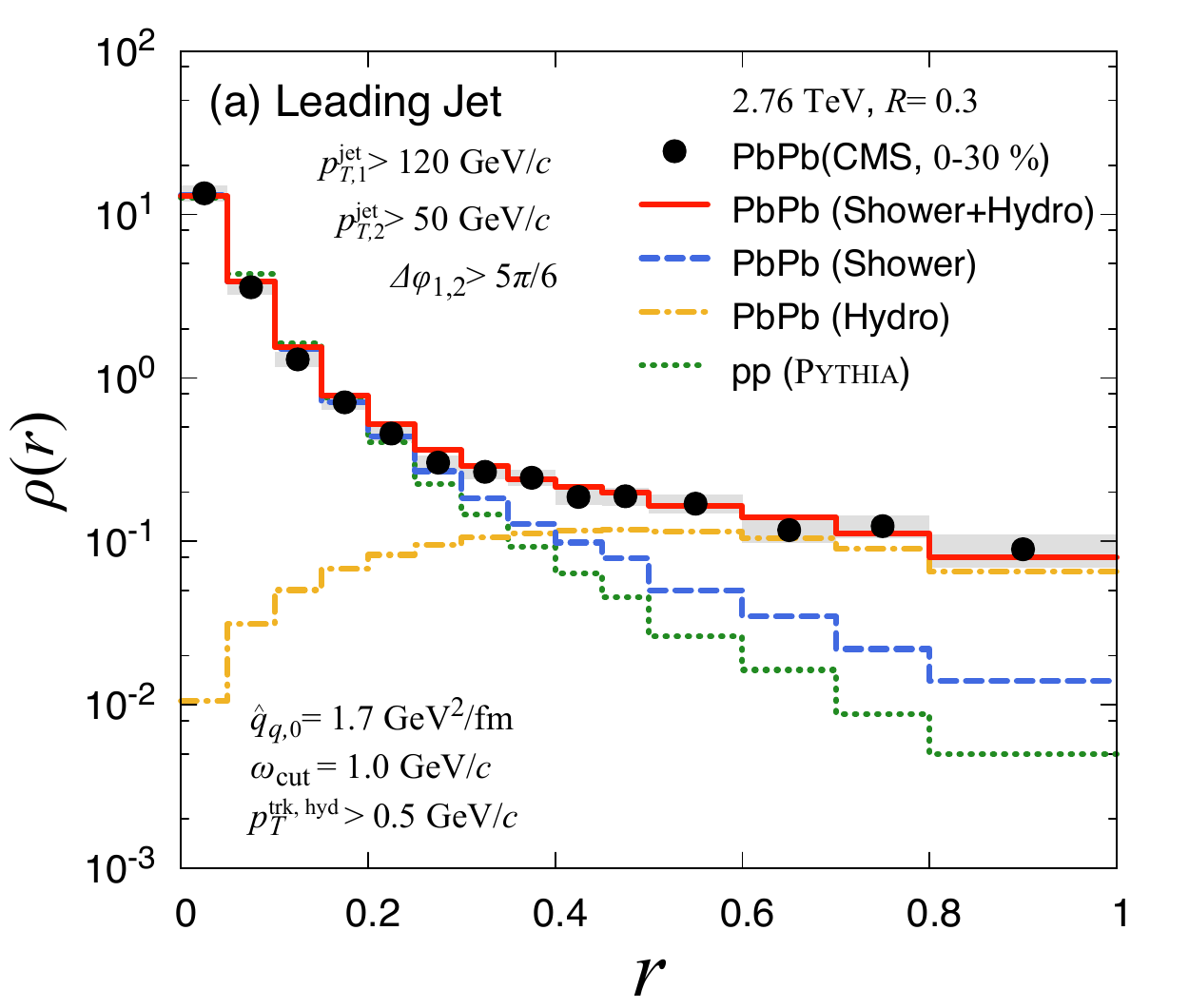}\\
    \vspace{-0.3cm}
    \includegraphics[width=0.43\textwidth]{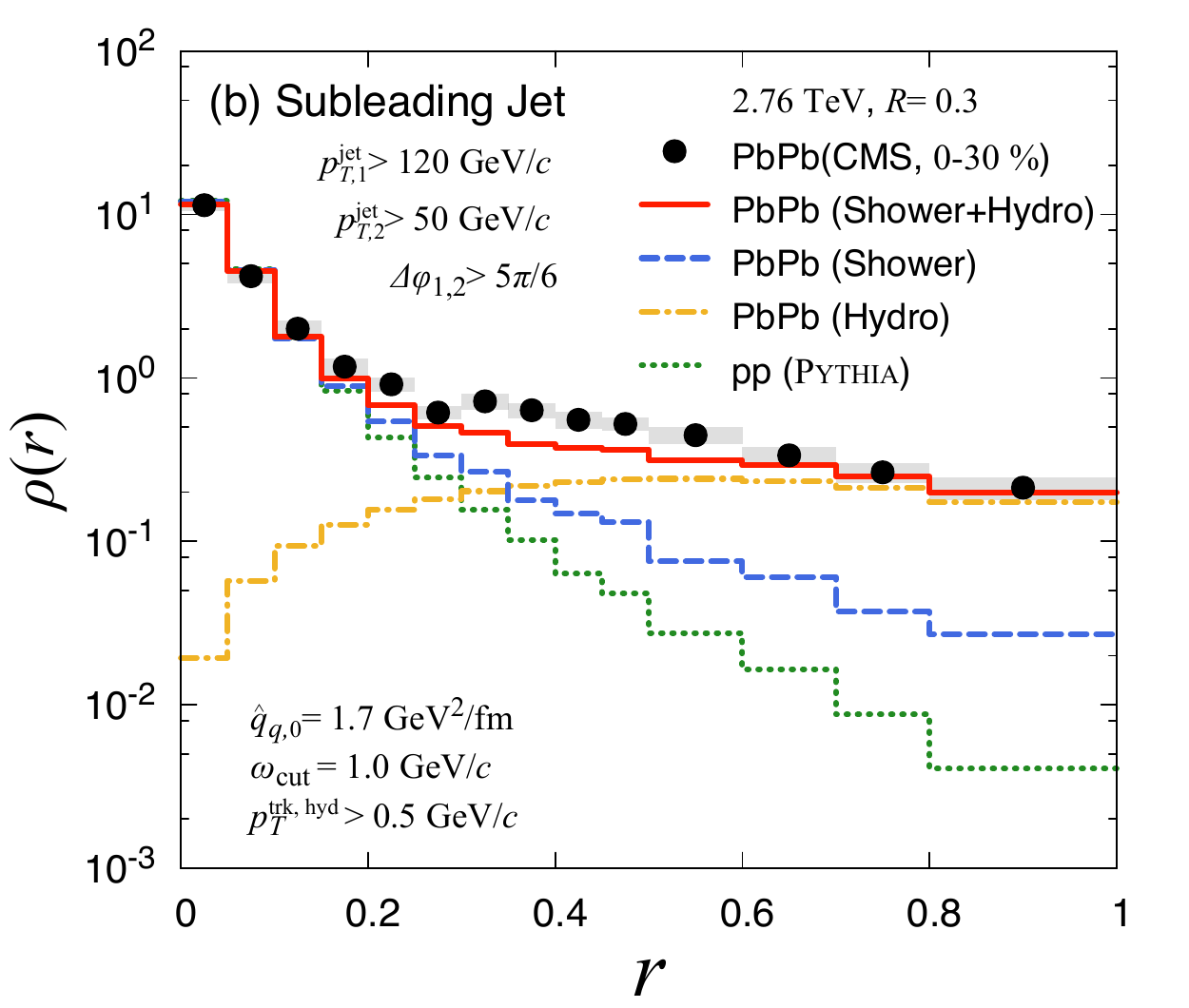}
    \caption{(Color online) Jet shape of (a) the leading and (b) subleading jets in dijet events in 2.76~TeV p+p and central Pb+Pb collisions, compared between contributions from jet shower and jet-induced medium excitation in Pb+Pb collisions. The figures are from Ref.~\cite{Tachibana:2017syd}.}
    \label{fig:Tachibana-dijet_shape}
\end{figure}

Shown in Fig.~\ref{fig:Tachibana-dijet_shape} are jet shapes of both leading and subleading jets in dijet events from a coupled jet-fluid model~\cite{Tachibana:2017syd}, in which elastic and inelastic energy loss of jets are calculated semi-analytically using transport (rate) equations and the lost energy is deposited in the QGP via a source term in a (3+1)-D hydrodynamic model simulation of the bulk medium. One observes that at small $r$, the energy density profile is dominated by energetic partons initiated from jets (labeled as ``shower"). At large $r$, however, soft hadrons produced from the jet-induced hydrodynamic response in the QGP (labeled as ``hydro") dominate the energy density profile. The contributions from jet showers and QGP response to the jet shape cross around $r=0.4$ just outside the jet cone. 

\begin{figure}[tbp]
    \centering
    \includegraphics[width=0.43\textwidth]{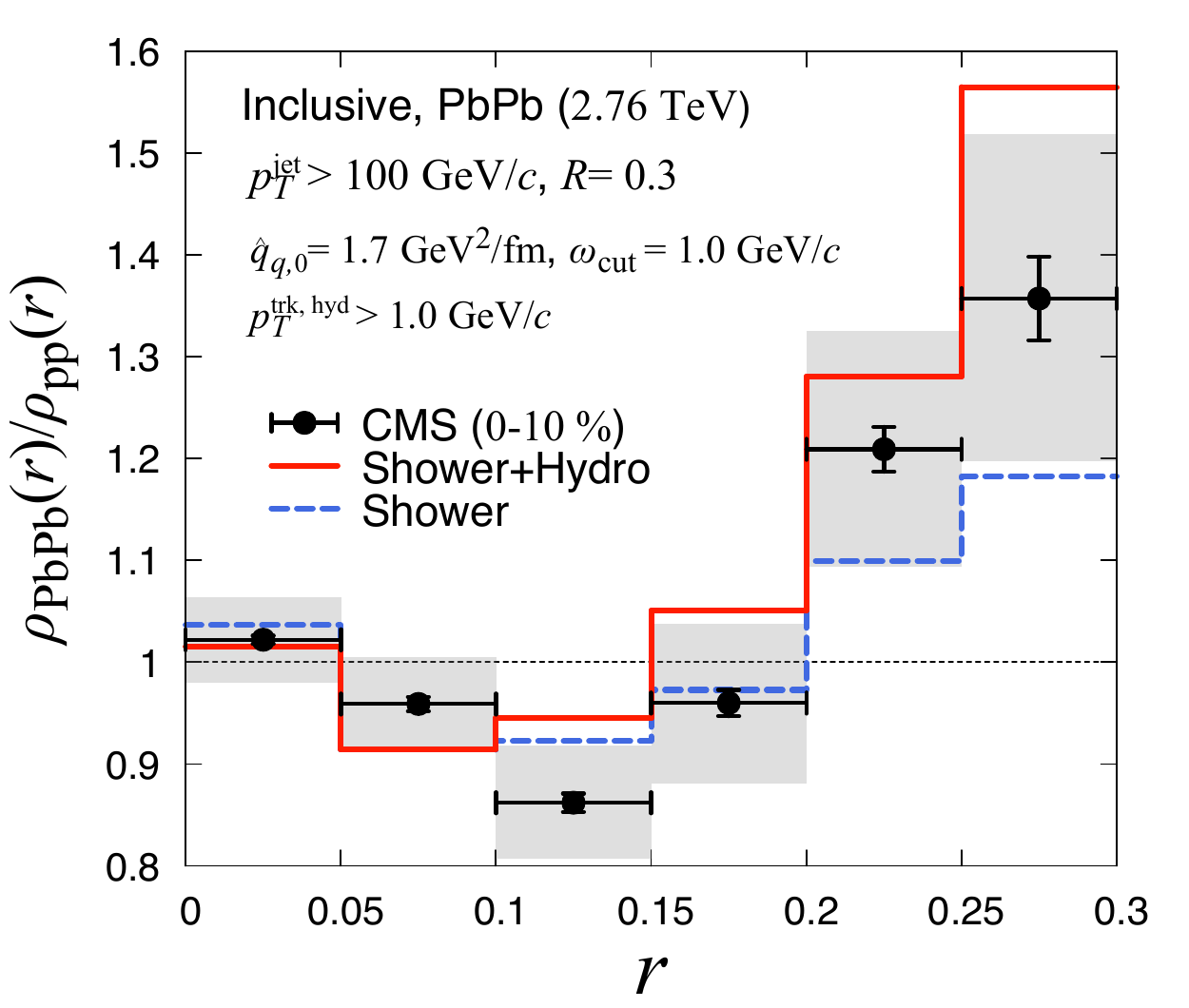}
    \caption{(Color online) The nuclear modification factor of the single inclusive jet shape in central Pb+Pb collisions at 2.76~TeV. The figure is from Ref.~\cite{Tachibana:2017syd}.}
    \label{fig:Tachibana-nm_jet_shape}
\end{figure}

To better illustrate effects of jet-induced medium response on nuclear modification of the jet shape, the ratio between jet profiles in Pb+Pb and p+p collisions is shown in Fig.~\ref{fig:Tachibana-nm_jet_shape}. One observes a significant enhancement of this nuclear modification factor at large $r$ after including contributions from jet-induced medium excitation. Since the jet shape defined in Eq.~(\ref{eq:defJetShape}) is a self-normalized quantity, an enhancement at large $r$ will be accompanied by a suppression at smaller $r$. This happens at intermediate $r$ inside the jet-cone since the jet shape at around $r=0$ is dominated by leading hadrons that actually should be enhanced slightly according to what we see in the previous subsection on jet fragmentation function. Similar effects of medium response on the jet shape have also been found in Refs.~\cite{Casalderrey-Solana:2016jvj,KunnawalkamElayavalli:2017hxo,Tachibana:2018yae,Park:2018acg} with different treatments of jet-induced medium excitation as discussed in Sec.~\ref{sec:medium_response}.

It is worth noting that even without jet-induced medium response, one can still observe an enhancement of the jet shape at large $r$ since jet-medium interactions, both elastic scatterings and medium-induced splittings, broaden the energy distribution towards larger $r$. In fact, as shown in Fig.~\ref{fig:Tachibana-nm_jet_shape}, current experimental data within $r<0.3$ is incapable of precisely constraining the effects of medium response, and several calculations without medium response~\cite{Ramos:2014mba,Lokhtin:2014vda,Chien:2015hda} can provide reasonable descriptions of the data as well. Much stronger effects of medium response are expected at $r>R$ where the energy density profile is dominated by soft hadron production from jet-induced medium excitation. Thus, more sophisticated comparisons between data and theoretical calculations in this region are required in the future. One bottleneck in current model calculations is the lack of a reliable event generator for p+p collisions. As shown in Ref.~\cite{Kumar:2019bvr}, while \textsc{Pythia}-based event generators produce satisfactory $p_\mathrm{T}$ spectra of inclusive hadrons and jets, clear deviations from experimental data are observed in jet substructures, especially in the kinematic regions dominated by soft hadrons. This problem needs to be solved before we may expect more accurate predictions on nuclear modification of the jet shape.

\begin{figure}[tbp]
    \centering
    \includegraphics[width=0.43\textwidth]{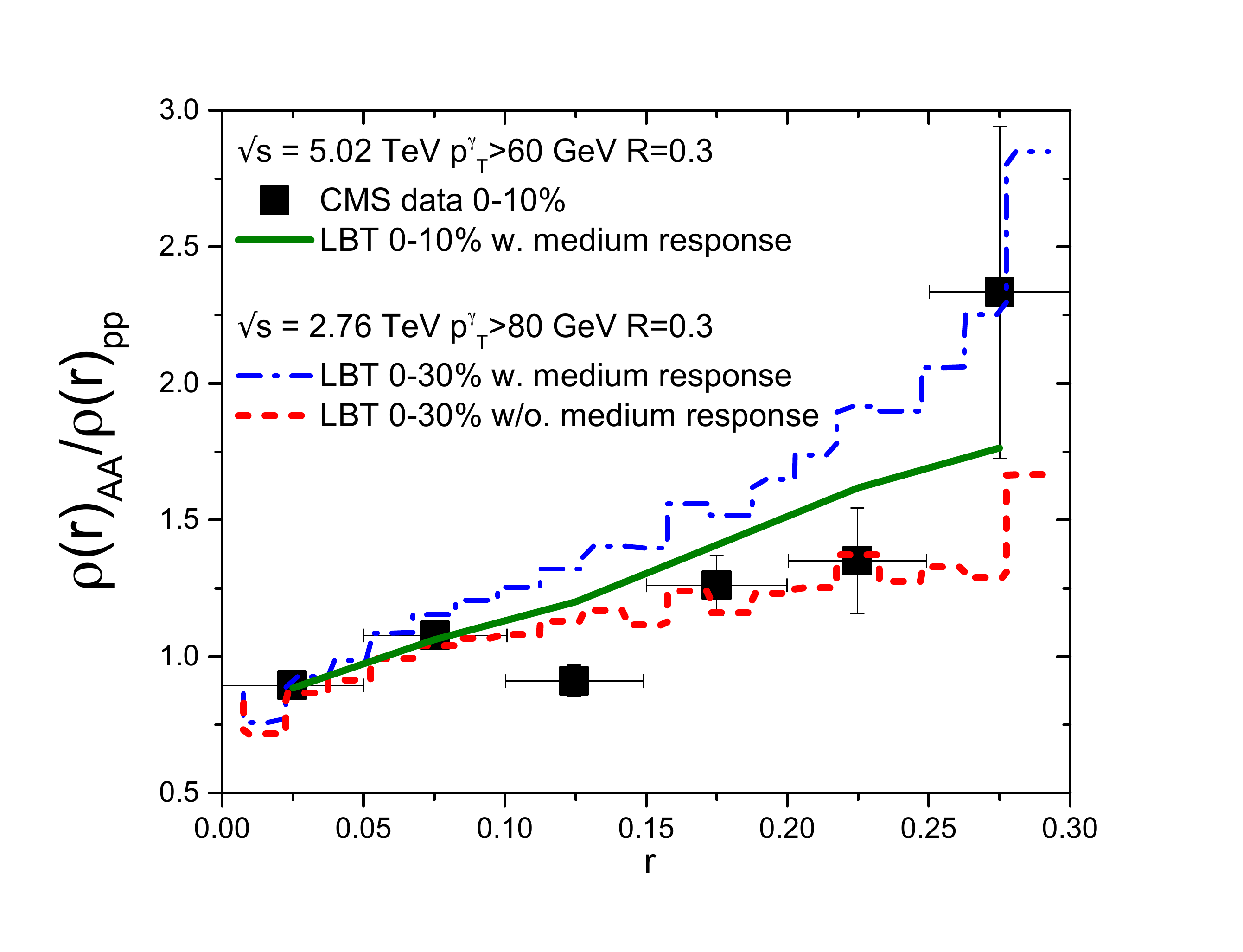}
    \caption{(Color online) The nuclear modification factor of the $\gamma$-triggered jet shape in central Pb+Pb collisions at 2.76~TeV, compared between with and without including medium response; and in central Pb+Pb collisions at 5.02~TeV, compared to the CMS data. The figure is from Ref.~\cite{Luo:2018pto,Luo:private}.}
    \label{fig:Luo_jet_shape}
\end{figure}

The nuclear modification factor of the $\gamma$-triggered jet shape can also be calculated as shown in Fig.~\ref{fig:Luo_jet_shape} from the \textsc{Lbt} model simulations~\cite{Luo:2018pto} in which effects of medium response are modeled through propagation of recoil and ``negative" partons. Similar to the single inclusive jet shape discussed above, one can see that jet-medium interactions transport energy towards the outer layer of the jet cone and result in an enhancement of this nuclear modification factor at large $r$. Comparing the results for 2.76~TeV Pb+Pb collisions with and without medium response, it is clear that the medium response significantly increases the energy density near the edge of the jet cone while it has negligible impact on the hard core ($r<0.05$) of the jet. The \textsc{Lbt} result for 5.02~TeV Pb+Pb collisions, including contributions from medium response, is consistent with the CMS data~\cite{Sirunyan:2018ncy}.

Another interesting observation of the jet shape is that a small enhancement of the jet shape at very small $r$ for single inclusive jets at $p_\mathrm{T}^\mathrm{jet}>100$~GeV~\cite{Chatrchyan:2013kwa} does not exist for $\gamma$-triggered jets at $p_\mathrm{T}^\mathrm{jet}>30$~GeV~\cite{Sirunyan:2018ncy}. This was first expected to result from different broadening of quark and gluon jets: the hard core of a quark jet (which dominates $\gamma$-triggered jets) is less broadened inside the QGP than that of a gluon jet (which contributes most to single inclusive jets). However, this is shown incorrect in recent studies~\cite{Chang:2016gjp,Chang:2019sae}. Although the shape of a quark jet is narrower than that of a gluon jet, the shape of their mixture is not necessarily in between due to its definition [Eq.~(\ref{eq:defJetShape})]: it is self-normalized to each jet event separately. Instead of the jet flavor (quark vs. gluon) effect, different $p_\mathrm{T}$ regimes are found to be the main reason for different jet shapes. For both single inclusive and $\gamma$-triggered jets, while a small enhancement of the jet shape exists at small $r$ for high $p_\mathrm{T}$ jets, it does not for low $p_\mathrm{T}$ jets. This can be easily tested with future measurements of single inclusive jets at lower $p_\mathrm{T}$ or $\gamma$-jets at higher $p_\mathrm{T}$.

\begin{figure}[tbp]
    \centering
    \includegraphics[width=0.43\textwidth]{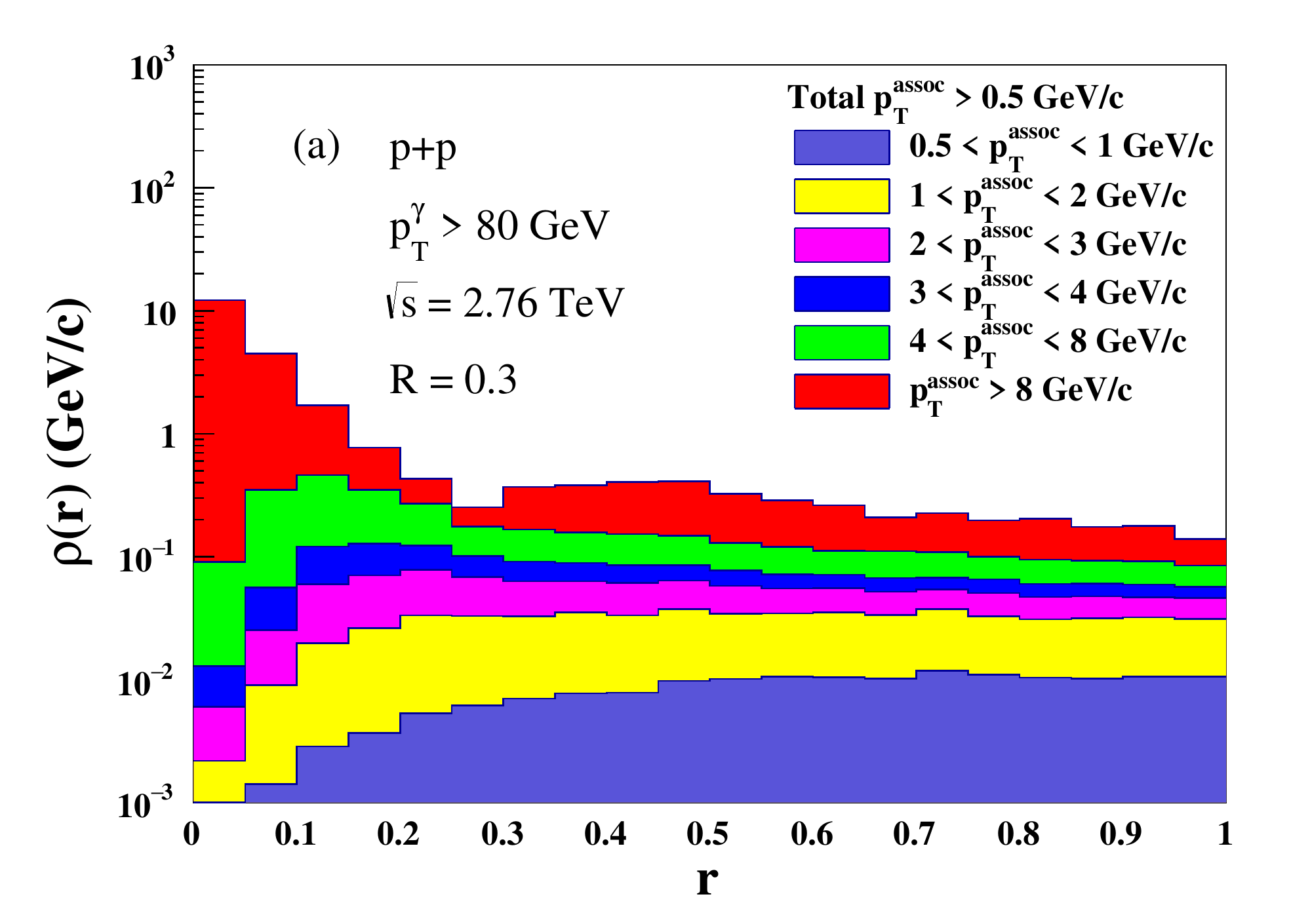}\\
    \vspace{-0.4cm}
    \includegraphics[width=0.43\textwidth]{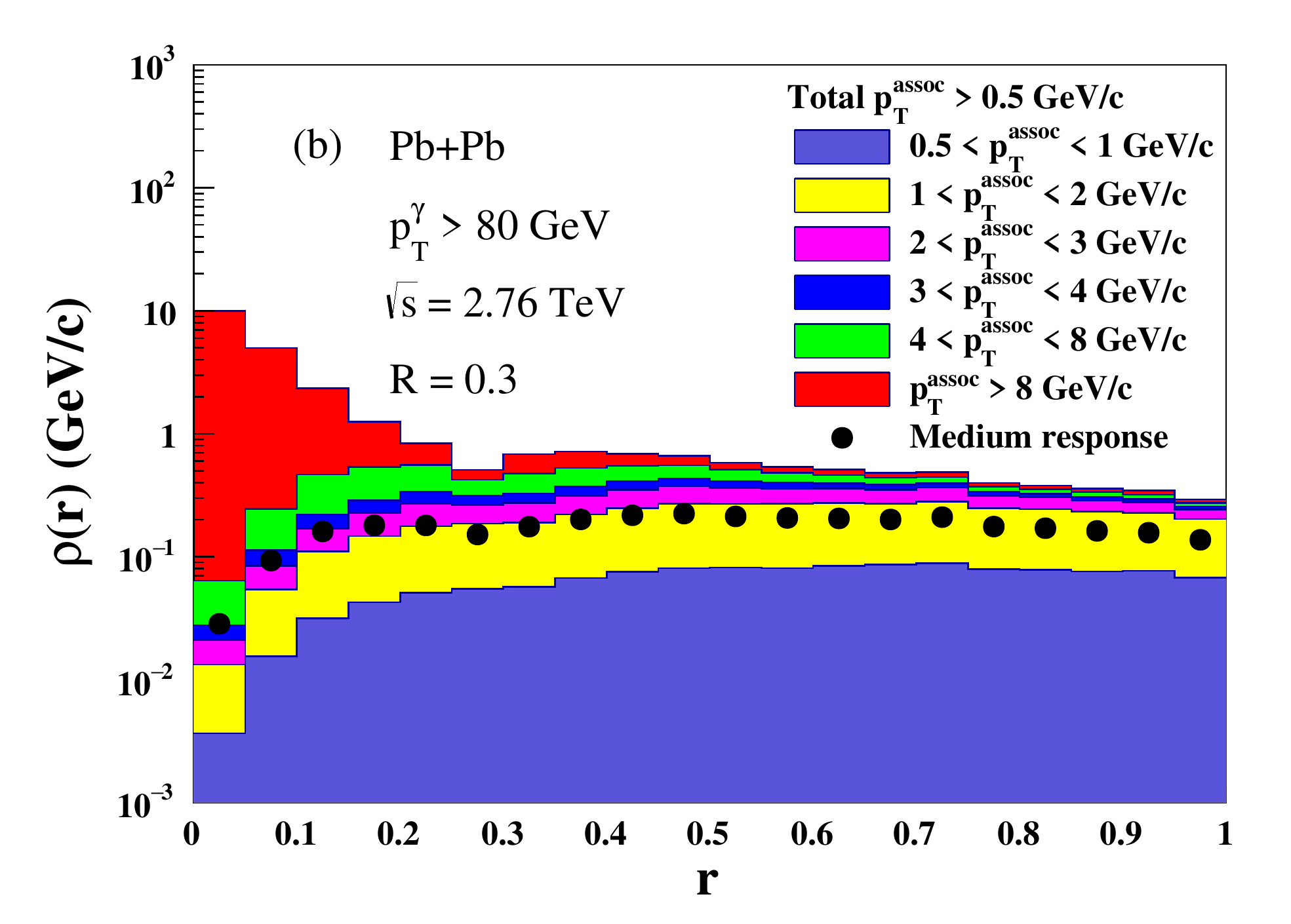}
    \caption{(Color online) Contributions to the jet shape of $\gamma$-jets in (a) p+p and (b) 0-30\% Pb+Pb collisions at 2.76~TeV from partons with different $p_\mathrm{T}$. The solid circles in the lower panel show the total contribution from medium response. The figures are from Ref.~\cite{Luo:2018pto}.}
    \label{fig:Luo-energyFlow}
\end{figure}

For a more detailed investigation of the jet shape, one may separate contributions from particles within differentiated $p_\mathrm{T}$ bins~\cite{Khachatryan:2016tfj}, as illustrated in Fig.~\ref{fig:Luo-energyFlow} from the \textsc{Lbt} model calculation for $\gamma$-jets~\cite{Luo:2018pto}. Comparing between p+p and central Pb+Pb collisions, we observe a suppression of high $p_\mathrm{T}$ particles at large $r$ while a significant enhancement of low $p_\mathrm{T}$ particles over the full $r$ range. This comparison demonstrates how the jet energy is transported from energetic partons to soft ones via scatterings and medium-induced splittings. This is referred to as the ``energy flow" within jets. In the lower panel, the total contribution from jet-induced medium excitation, shown as solid circles, constitutes a significant amount of energy density especially at large $r$. As a result, medium response is crucial for understanding how the jet energy loss is re-distributed when jets propagate through the QGP. A similar observable for studying this energy re-distribution is known as the ``missing $p_\mathrm{T}$" in dijet events~\cite{Khachatryan:2015lha}. One may study the imbalance of the summed particle $p_\mathrm{T}$ between the leading jet hemisphere and the subleading jet hemisphere and learn how this imbalanced (missing) $p_\mathrm{T}$ is recovered with the increase of the cone size around the dijet axis. As suggested in Ref.~\cite{Hulcher:2017cpt}, including jet-induced medium excitation is essential for recovering the lost energy from dijet systems and obtaining balanced transverse momenta between the two hemispheres at large cone size.

\subsection{Jet splitting function}
\label{subsec:jetSplitting}

As discussed in Secs.~\ref{sec:theory} and~\ref{sec:models}, parton energy loss and nuclear modification of jets are closely related to the medium-induced parton splitting functions. All observables that have been discussed do not provide direct constraints on the splitting function. However, with the introduction of the soft drop jet grooming algorithm~\cite{Larkoski:2014wba,Larkoski:2015lea,Dasgupta:2013ihk,Frye:2016aiz}, we are now able to eliminate soft hadrons at wide angles and identify the hard splitting within a groomed jet, providing a direct study on the splitting function~\cite{Sirunyan:2017bsd,Kauder:2017cvz,D.CaffarrifortheALICE:2017xsw}. Contributions from hadronization and underlying events are suppressed with the soft drop, and a more direct comparison between data and perturbative QCD calculations becomes possible.

\begin{figure}[tbp]
    \centering
    \includegraphics[width=0.43\textwidth]{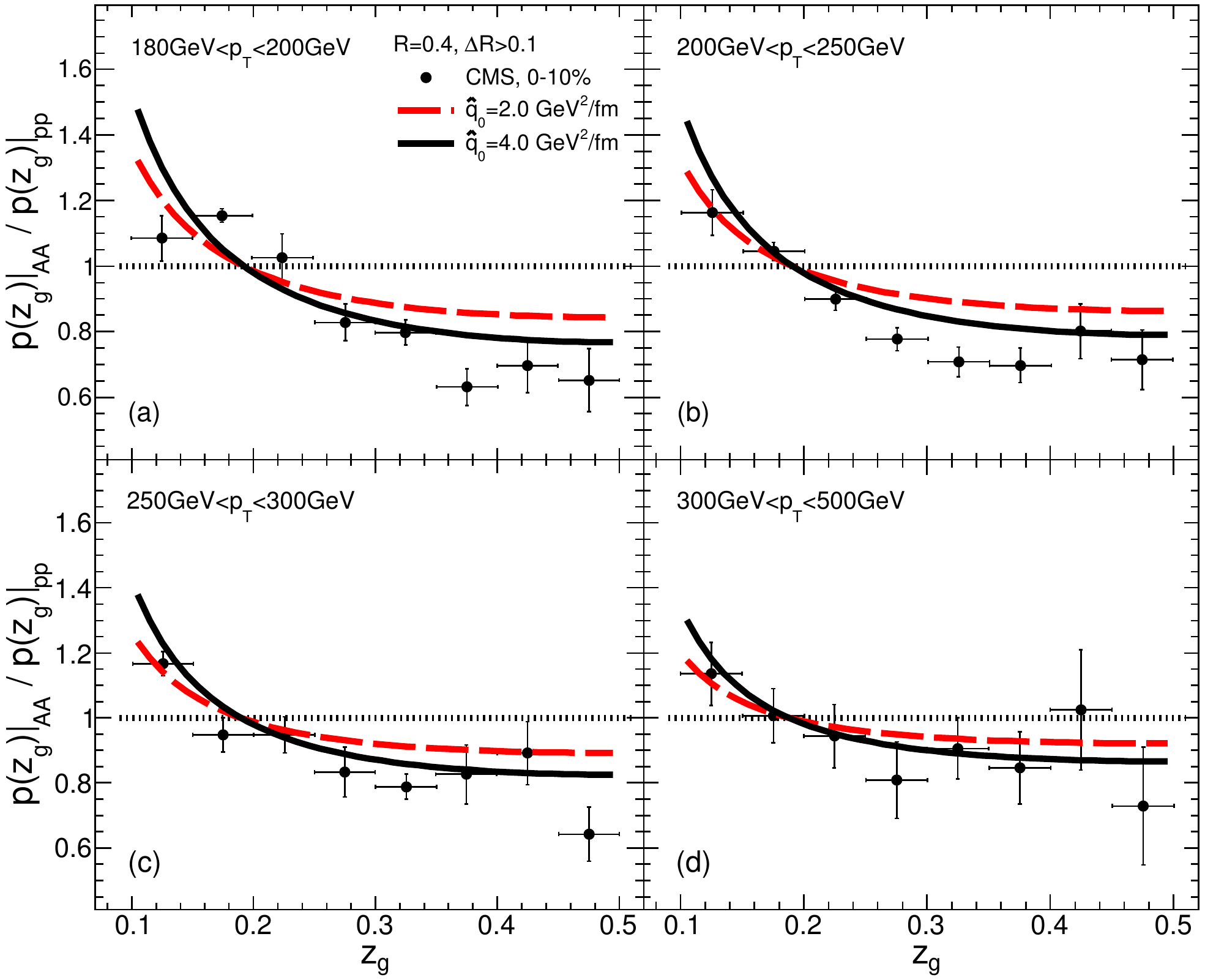}
    \caption{(Color online) Nuclear modification of the groomed jet $z_g$ distribution in 0-10\% Pb+Pb collisions at 5.02~TeV, compared between different jet $p_\mathrm{T}$ intervals. The figure is from Ref.~\cite{Chang:2017gkt}.}
    \label{fig:Chang-Pzg-CMS}
\end{figure}

\begin{figure}[tbp]
    \centering
    \includegraphics[width=0.45\textwidth]{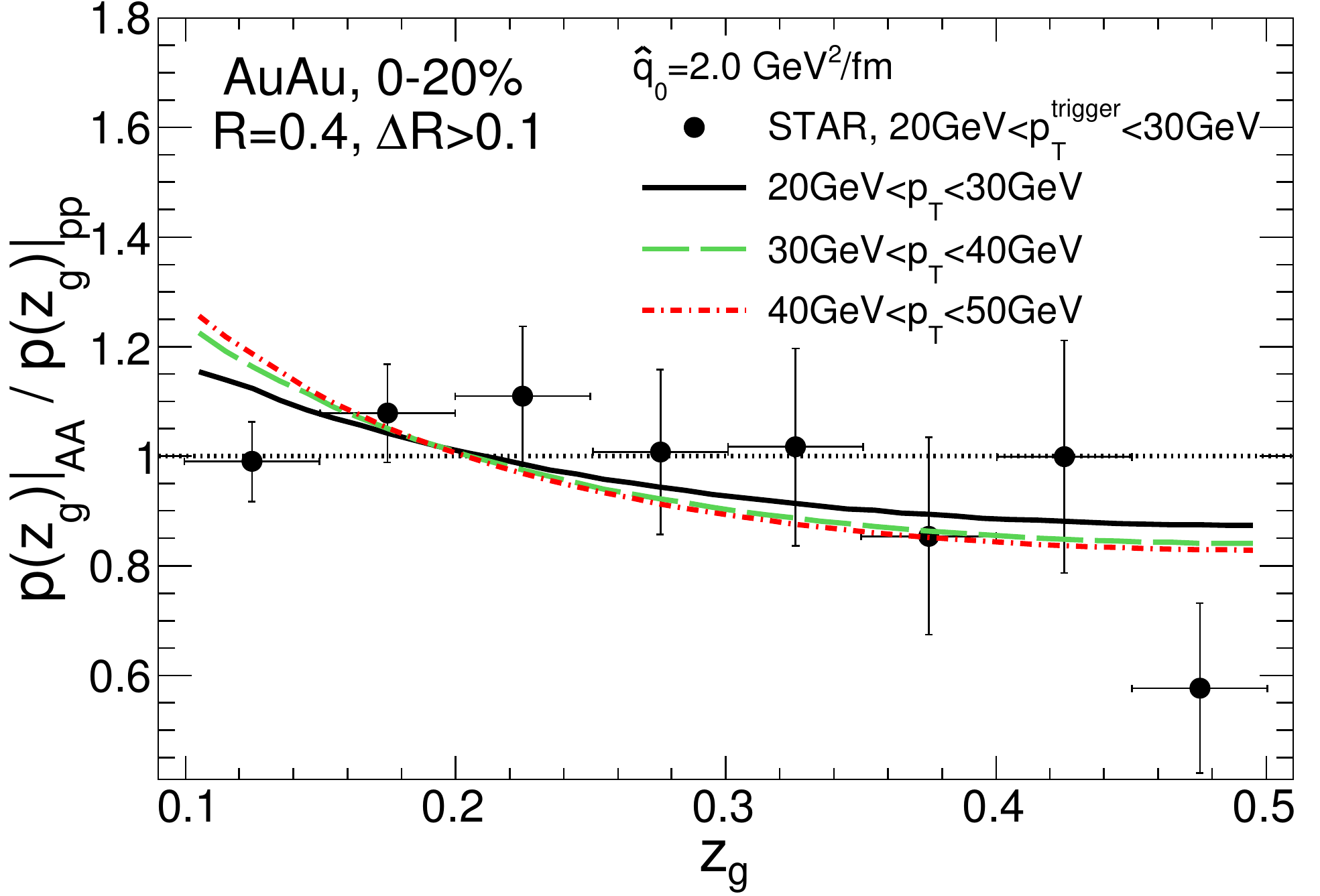}
    \caption{(Color online) Nuclear modification of the groomed jet $z_g$ distribution in 0-20\% Au+Au collisions at 200~GeV, compared between different jet $p_\mathrm{T}$ intervals. The figure is from Ref.~\cite{Chang:2017gkt}.}
    \label{fig:Chang-Pzg-STAR}
\end{figure}

In the soft drop procedure, as adopted by CMS and STAR Collaboration for analyses of heavy-ion collisions, a full jet constructed using radius $R$ via the anti-$k_\mathrm{T}$ algorithm is first re-clustered using the Cambridge-Aachen (C/A) algorithm and then de-clustered in the reverse order by dropping the softer branch until two hard branches are found to satisfy the following condition:
\begin{equation}
\label{eq:defSoftDrop}
\frac{\min({p_{\mathrm{T}1},p_{\mathrm{T}2}})}{p_{\mathrm{T}1}+p_{\mathrm{T}2}} \equiv z_g > z_\mathrm{cut}\left(\frac{\Delta R}{R}\right)^\mathrm{\beta},
\end{equation}
where $p_\mathrm{T1}$ and $p_\mathrm{T2}$ are the transverse momenta of the two subjets at a particular step of declustering, $\Delta R$ is their angular separation, $z_\mathrm{cut}$ is the lower cutoff of the momentum sharing $z_g$. Both CMS~\cite{Sirunyan:2017bsd} and STAR~\cite{Kauder:2017cvz} measurements take $z_\mathrm{cut}=0.1$ and $\beta=0$, the former also requires $\Delta R\ge 0.1$ while the latter does not. With this setup, the self-normalized momentum sharing distribution
\begin{equation}
\label{eq:defPzg}
p(z_g)\equiv\frac{1}{N_\mathrm{evt}}\frac{dN_\mathrm{evt}}{dz_g}
\end{equation}
is used to characterize the parton splitting function, with $N_\mathrm{evt}$ being the number of jet events in which two qualified hard subjets are found.  

\begin{figure}[tbp]
    \centering
    \includegraphics[width=0.43\textwidth]{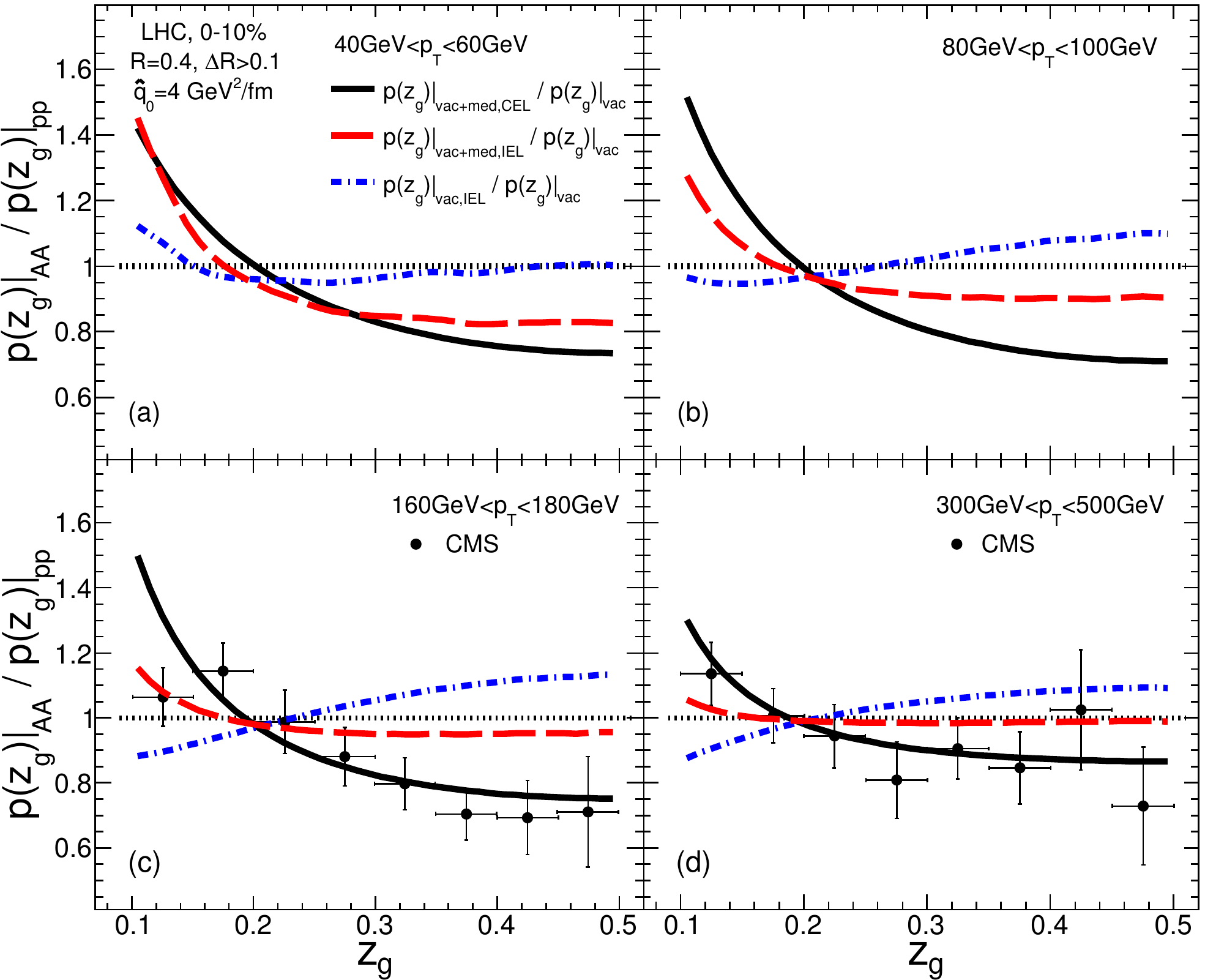}
    \caption{(Color online) Effects of coherent vs. decoherent jet energy loss on the nuclear modification factor of the groomed jet $z_g$ distribution in 0-10\% Pb+Pb collisions at 5.02~TeV. The figure is from Ref.~\cite{Chang:2017gkt}.}
    \label{fig:Chang-Pzg-coherence}
\end{figure}

\begin{figure*}[tbp]
    \centering
    \includegraphics[width=0.65\textwidth]{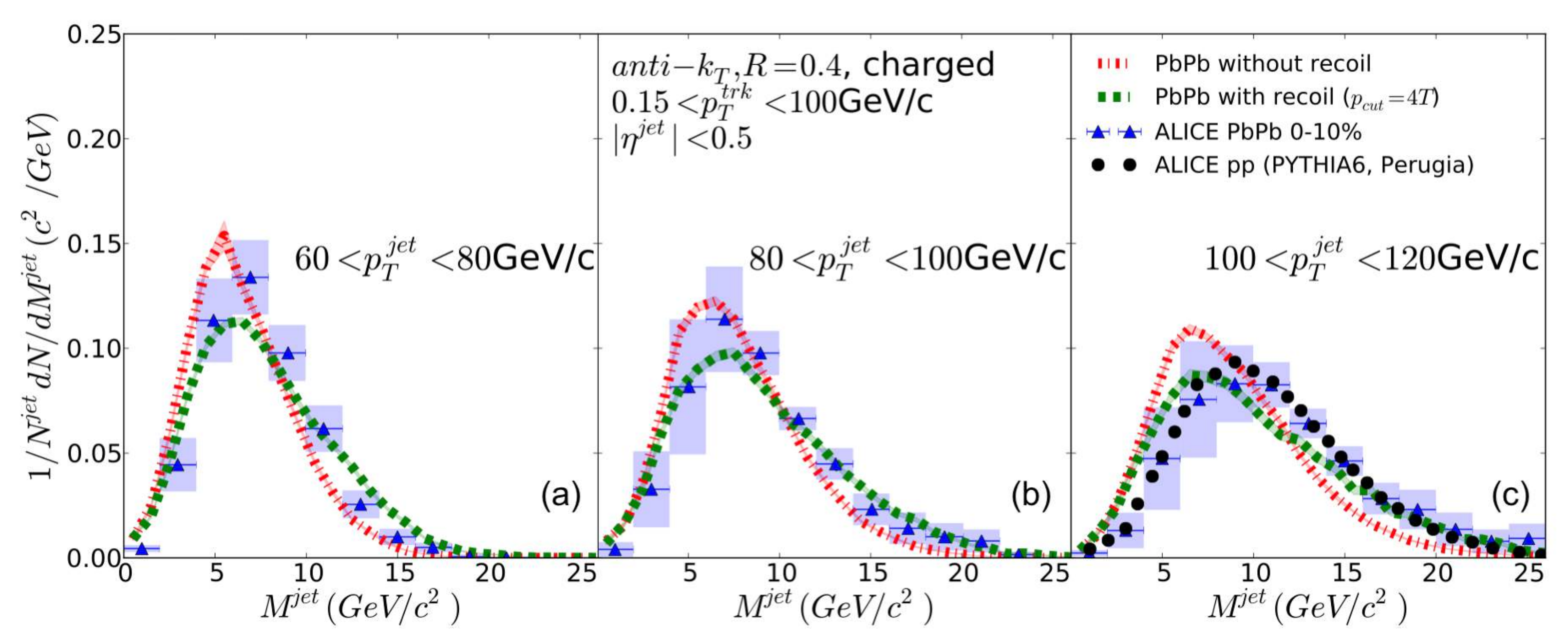}
    \includegraphics[width=0.26\textwidth]{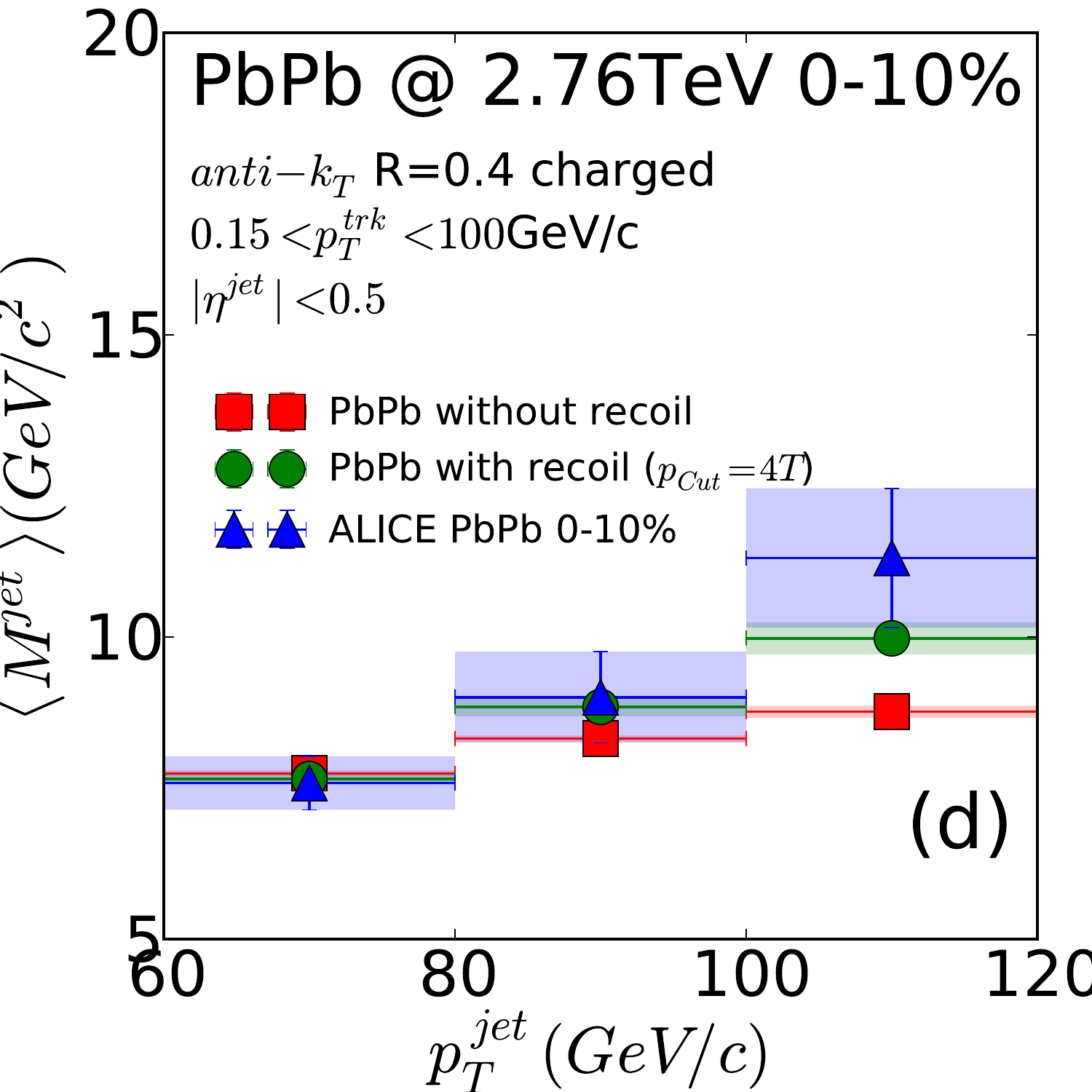}
    \caption{(Color online) (a)-(c) The jet mass distribution in 0-10\% Pb+Pb collisions at 2.76~TeV, compared between with and without including recoil contributions, and between different jet $p_\mathrm{T}$ intervals. (d) The average jet mass as a function of the jet $p_\mathrm{T}$. The figures are from Ref.~\cite{Park:2018acg}.}
    \label{fig:Park-MARTINI-mass}
\end{figure*}

This momentum sharing distribution has been investigated with different theoretical approaches to jet quenching in heavy-ion collisions~\cite{Chien:2016led,Mehtar-Tani:2016aco,KunnawalkamElayavalli:2017hxo,Chang:2017gkt,Milhano:2017nzm,Li:2017wwc,Caucal:2019uvr}. For example, as shown in Figs.~\ref{fig:Chang-Pzg-CMS} and~\ref{fig:Chang-Pzg-STAR}, a simultaneous description of the nuclear modification factor of $p(z_g)$ can be obtained based on the high-twist energy loss formalism~\cite{Chang:2017gkt}. One interesting observation is that while the nuclear modification of the splitting function is strong for large $p_{\rm T}$ jets at the LHC, it increases as the jet $p_\mathrm{T}$ decreases and becomes rather weak again at lower $p_{\rm T}$ at RHIC. Apart from the different $\Delta R$ in the implementations  of soft-drop in CMS and STAR measurements, authors in Ref.~\cite{Chang:2017gkt} find a non-monotonic dependence of this nuclear modification factor on the jet $p_\mathrm{T}$ even when they use the same $\Delta R$ cut. Considering that the extracted splitting function has contributions from both the vacuum and medium-induced splitting [shown by Eq.~(\ref{eq:splitPtot})], one expects vacuum splitting to become more dominant as the jet $p_\mathrm{T}$ increases,  and therefore a weaker medium modification as observed in Fig.~\ref{fig:Chang-Pzg-CMS}. On the other hand, the medium-induced splitting $P_\mathrm{med}(z)$ and the vacuum one $P_\mathrm{vac}(z)$ share a similar $1/z$ dependence  in the low energy limit, the medium modification of the observed splitting function vanishes as  jet $p_\mathrm{T}$ approaches to this low limit. Since $p(z_g)$ is defined as a self-normalized quantity, its nuclear modification effects appear to decrease with jet $p_\mathrm{T}$ at RHIC, as seen in Fig.~\ref{fig:Chang-Pzg-STAR}. The competition between these two effects results in the non-monotonic $p_\mathrm{T}$ dependence as observed in the current data. Note that both effects are seen in the calculations within the high-twist energy loss formalism and could be model dependent. Therefore, this non-monotonic behavior should be further tested with high $p_\mathrm{T}$ jets at RHIC or low $p_\mathrm{T}$ jets at the LHC.

The nuclear modification of $p(z_g)$ has also been proposed as a possible probe of the color (de)coherence effects in jet-medium interactions~\cite{Mehtar-Tani:2016aco,Chang:2017gkt} as illustrated in Fig.~\ref{fig:Chang-Pzg-coherence}, where different assumptions of nuclear modification are compared with the CMS data. In this study, ``vac+med, CEL" denotes medium-modified parton splitting functions with coherent energy loss for the two subjets; ``vac+med, IEL" denotes the medium-modified splitting function with incoherent (or independent) energy loss for two subjets within a jet; and ``vac, IEL" denotes vacuum-like splittings of partons followed by incoherent energy loss of subjets. Figure~\ref{fig:Chang-Pzg-coherence} shows that only results with ``vac+med, CEL" splitting can describe the experimental data.

Within the high-twist formalism~\cite{Chang:2017gkt}, partons lose smaller fractional energy at higher energy (the energy dependence of the energy loss is less than a linear one) after traversing the QGP.  As a result, independent energy loss of the two subjets in a given jet event leads to a larger $z_g$ fraction as compared to a coherent energy loss.  This leads to an enhancement in the nuclear modification factor at large $z_g$ after shifting (smaller $z_g$) and re-self-normalizing $p(z_g)$ with incoherent energy loss. Current experimental data seem to prefer the assumption of coherent energy loss within the high-twist formalism. Note that such a conclusion is specific to an energy loss theory and depends on the extent of the $z_g$ shift caused by incoherent energy loss. According to Ref.~\cite{Caucal:2019uvr}, a reasonable description of data may also be obtained using the BDMPS energy loss formalism with an incoherent energy loss assumption.

Although implementing soft drop jet grooming is expected to suppress contributions from soft hadrons within jets, Refs.~\cite{KunnawalkamElayavalli:2017hxo,Milhano:2017nzm} still show a sizable effect of medium response on the $z_g$ distribution. Therefore, variations on soft drop grooming, $z_\mathrm{cut}$, $\beta$, $R$ and $\Delta R$ in Eq.~(\ref{eq:defSoftDrop}) need to be investigated in more detail to better separate hard and soft contributions.

\subsection{Jet mass}
\label{subsec:jetMass}

In some jet-quenching models, such as \textsc{Q-Pythia} and \textsc{Yajem-rad} as discussed in Sec.~\ref{subsec:med_showers_virtual}, the parton splitting functions are not directly modified by the medium. Instead, an enhanced parton virtuality by parton-medium scatterings is assumed, which induces additional vacuum-like splittings, leading to an effective jet energy loss. The variation of partons' virtuality can be explored using the jet mass~\cite{Acharya:2017goa} defined as
\begin{equation}
\label{eq:defJetMass}
M=\sqrt{E^2-p_\mathrm{T}^2-p_z^2},
\end{equation}
where $E$, $p_\mathrm{T}$ and $p_z$ are the total energy, transverse momentum and longitudinal momentum of a given jet, respectively.

There are several theoretical studies searching for medium  effects on the jet mass distribution~\cite{KunnawalkamElayavalli:2017hxo,Park:2018acg,Chien:2019lfg,Casalderrey-Solana:2019ubu}. Figure~\ref{fig:Park-MARTINI-mass} presents the \textsc{Martini} model calculation of the jet mass distribution in  central Pb+Pb collisions at 2.76~TeV~\cite{Park:2018acg} with and without medium response. Compared to the p+p baseline, jet energy loss without recoil partons from the medium response shifts the mass distribution towards smaller values, while medium recoil partons tend to increase the jet mass. As shown in panel (c), no significant variation of the mass distribution has been found in the experimental data in Pb+Pb collisions from that in p+p. In panel (d) of Fig.~\ref{fig:Park-MARTINI-mass}, one observes that the average jet mass and the contribution from recoil partons both increase with the jet $p_\mathrm{T}$. Similar findings have been seen in Refs.~\cite{KunnawalkamElayavalli:2017hxo,Casalderrey-Solana:2019ubu} as well.

\begin{figure}[tbp]
    \centering
    \includegraphics[width=0.43\textwidth]{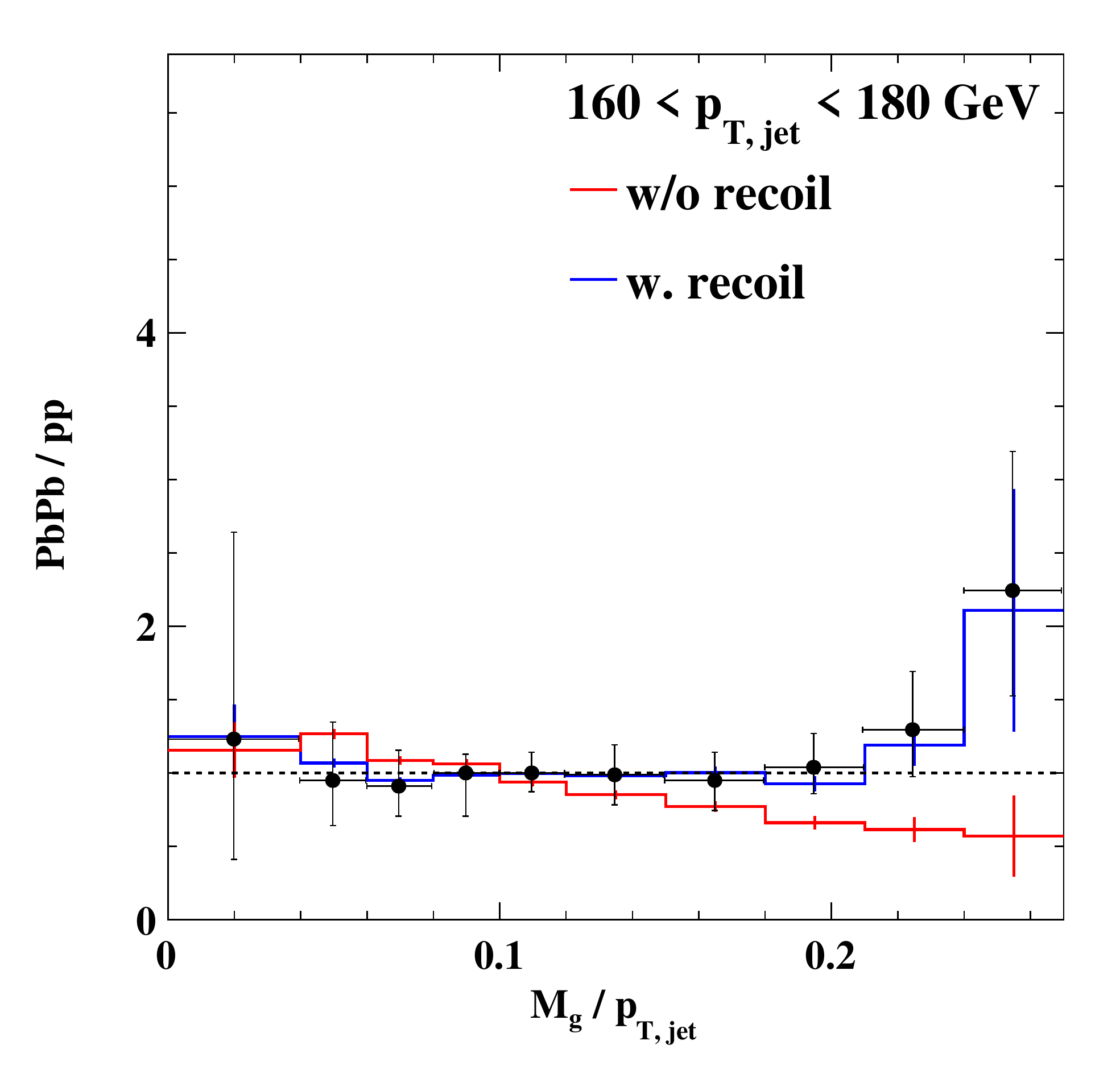}
    \caption{(Color online) The nuclear modification factor of the groomed jet mass distribution in 0-10\% Pb+Pb collisions at 5.02~TeV, from \textsc{Lbt} simulations with and without contributions from recoil partons. The figure is from Ref.~\cite{Luo:2019lgx}.}
    \label{fig:Luo-groomed-jet-mass}
\end{figure}

To suppress contributions from soft hadrons and study the virtuality scale of hard splittings, the mass distributions of groomed jets have been measured recently~\cite{Sirunyan:2018gct}. However, as shown by the \textsc{Lbt} model calculation~\cite{Luo:2019lgx} in Fig.~\ref{fig:Luo-groomed-jet-mass}, the effects of jet-induced medium excitation may still be large in jets with large masses. Without the contribution from recoil partons, a suppression of the mass distribution due to parton energy loss is observed at the large groomed mass. To the contrary, taking into account recoil partons results in an enhancement. Again, one should note that while the importance of incorporating jet-induced medium excitation for understanding jet-medium interactions has been commonly recognized, its quantitative contributions in theoretical calculations depend on the detailed model implementations. One should also keep in mind that its signals emerging from experimental data will also depend on the choices of various kinematic cuts.

\subsection{Heavy flavor jets}
\label{subsec:HFJets}

As discussed in Sec.~\ref{subsec:heavyHadron}, heavy quarks are valuable probes of the mass effects on parton energy loss. However, due to the NLO (or gluon fragmentation) contribution to heavy flavor production, it is hard to observe the mass hierarchy directly in the suppression factors $R_\mathrm{AA}$ of light and heavy flavor hadrons at high $p_\mathrm{T}$. Similarly, after including the gluon fragmentation process, the suppression factors $R_\mathrm{AA}$ of single inclusive jets and heavy-flavor-tagged jets are also found to be similar~\cite{Huang:2013vaa}. Photon-tagged or $B$-meson-tagged $b$-jets are proposed in Ref.~\cite{Huang:2015mva} to increase the fraction of $b$-jets that originate from prompt $b$-quarks relative to single inclusive $b$-jets. These await future experimental investigations.

\begin{figure}[tbp]
    \centering
    \includegraphics[width=0.5\textwidth]{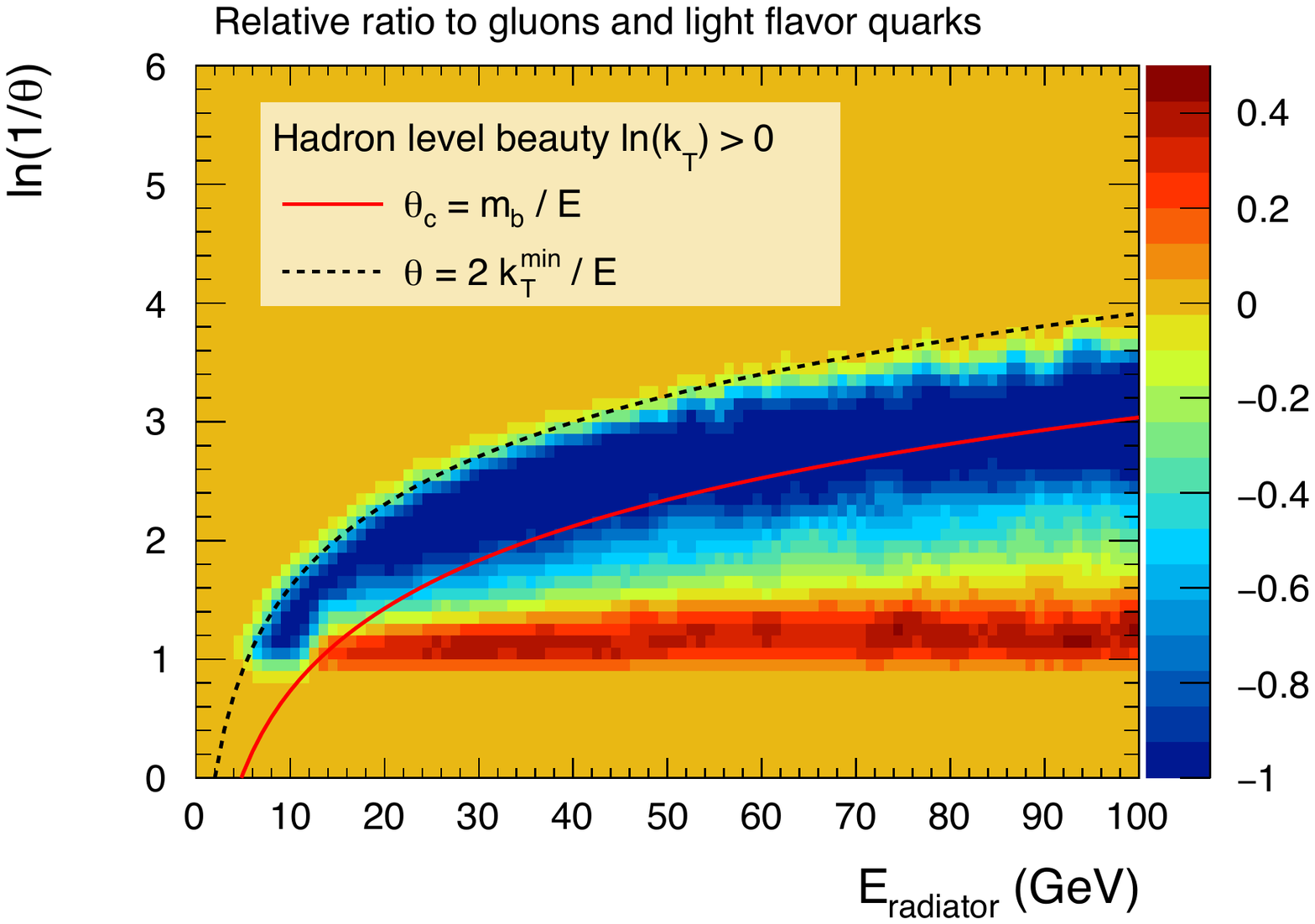}
    \caption{(Color online) The relative difference between $b$-jets and inclusive jets on the correlation between the splitting angle and the radiator energy. The figure is from Ref.~\cite{Cunqueiro:2018jbh}.}
    \label{fig:Cunqueiro-HF-Lund}
\end{figure}

Instead of the integrated spectra of heavy flavor jets, recent studies indicate that one can use their substructures to provide more direct insights into the ``dead cone effect"~\cite{Dokshitzer:2001zm} that suppresses the radiative energy loss of massive quarks. Shown in Fig.~\ref{fig:Cunqueiro-HF-Lund} is the relative difference between $b$-jets and inclusive jets on the correlation between the splitting angle and the radiator energy~\cite{Cunqueiro:2018jbh}. In this work, \textsc{Pythia} is used to generate heavy and light flavor jet events and the soft drop jet grooming is applied. For heavy flavor jets, the branch containing the heavy flavor is always followed when one moves backwards through the jet clustering history. The relative transverse momentum $k_\mathrm{T}$ and angle $\theta$ of the other branch, together with the energy $E_\mathrm{radiator}$ of the parent (radiator) that splits into these two branches, are mapped into the density distribution as shown in Fig.~\ref{fig:Cunqueiro-HF-Lund}. 
With a proper kinematic cut, e.g. $\ln(k_\mathrm{T})>0$, one can clearly observe a suppression of gluon radiation from heavy quarks between the lower boundary given by the kinematic cut $\theta=2k_\mathrm{T}^\mathrm{min}/E_\mathrm{radiator}$ and the dead cone size $\theta_c=m_b/E_\mathrm{radiator}$ defined with the $b$-quark mass $m_b$. This simulation result serves as a promising guidance for analyzing experimental data in order to directly observe the dead cone of heavy quarks. Similar studies should also be extended to both event generators and experimental analyses of A+A collisions for the purpose of understanding the medium modification of the dead cone, which is the origin of the mass hierarchy of parton energy loss inside the QGP.

We can also study the angular distributions of gluon radiation from heavy and light flavor partons  using the jet shape.  A similar quantity -- the angular distribution of $D$ mesons with respect to the jet axis -- has been explored in both experimental measurements~\cite{Sirunyan:2019dow} and theoretical calculations~\cite{Wang:2019xey}. Comparing p+p and Pb+Pb collisions at the LHC, one observes an enhancement of the nuclear modification factor of this angular distribution at large distance to the jet axis. By utilizing jet as a reference for the heavy quark motion, this provides a novel direction to learn how heavy quarks diffuse and lose energy inside the QGP. To further understand how mass and flavor affect parton-medium interactions and obtain better constraints on jet-quenching theories, it is necessary to continue searching for observables of jet substructures with which heavy and light flavor jets can be directly compared.

\section{Summary and outlook}
\label{sec:summary}

During the last two decades of experimental and theoretical studies of high-energy heavy-ion collisions at both RHIC and LHC, jet quenching has provided a powerful tool to explore the properties of the dense QCD matter formed during the violent collisions at unprecedented energies. The extraction of the jet transport coefficient, which is about 2 (1) orders of magnitude larger than that in a cold nuclear (hot hadronic) matter, provided a quantitative evidence for the formation of the QGP at the center of the heavy-ion collisions at RHIC and LHC. We have entered an era of quantitative study of the properties of QGP through a wide variety of probes, including jet quenching and associated phenomena. We have presented in this review recent progresses in theoretical and phenomenological studies of jet quenching and jet-induced medium response in heavy-ion collisions. 

We have reviewed new developments in theoretical calculations of large angle gluon radiation induced by jet-medium interactions and the connection to existing results under different approximations in the high-twist approach. The implementation of the induced gluon radiation under different approximations in jet transport models are also discussed. We have reviewed phenomenological studies within these jet transport models on suppression of single inclusive light and heavy flavor hadrons, single inclusive jets, modification of $\gamma$-hadron, $\gamma/Z^0$-jet correlations and jet substructures due to jet quenching. Special emphases have been given to effects of jet-induced medium response in current experimental measurements of these observables.

Though an unambiguous signal of Mach-cone  caused by the jet-induced medium response is still elusive in current experimental measurements, there are a wide variety of phenomena in the experimental observation that point to the effects of jet-induced medium response during the jet propagation in the QGP medium.
\begin{itemize}
\item The most striking phenomenon caused by the medium response is the enhancement of soft hadrons in the $\gamma$-hadron correlation and fragmentation functions of single inclusive jets and $\gamma$-jets, especially when the factional momentum in the fragmentation function is defined by the momentum of the triggered photon. The onset of the enhancement starts at a momentum fraction that decreases with the photon's momentum, indicating an intrinsic scale related to the medium, not the jet energy.
\item Another striking effect of medium response is the enhancement of the jet profile toward the edge of the jet-cone and at large angles outside the jet-cone. This enhancement is found to be mostly contributed by soft hadrons/partons from jet-induced medium response. Recoil partons from the medium response also enhance the jet mass distribution in the large mass region.
\item The most unique feature of jet-induced medium response is the depletion of soft $\gamma$-hadron correlation in the direction of the trigger photon due to the diffusion wake of jet-induced medium response. Measurement of such depletion, however, requires statistical subtraction of a large background.
\item The indirect effect of jet-induced medium response is the jet energy dependence of the net jet energy loss and the $p_\mathrm{T}$ dependence of the jet suppression factor $R_\mathrm{AA}$. Inclusion of medium response significantly reduces and changes the jet energy dependence of the net jet energy loss. Therefore, without jet-induced medium response, the $p_\mathrm{T}$ dependence of $R_\mathrm{AA}$ will be different. Including medium response also changes the cone-size dependence of the jet $R_\mathrm{AA}$.
\end{itemize}

Some of the above listed phenomena may be explained by some alternative mechanisms such as large angle and de-coherent emission of radiated gluons~\cite{MehtarTani:2011tz,Mehtar-Tani:2014yea,Mehtar-Tani:2016aco}. However, it is important to have a coherent and systematic explanation of all the observed phenomena in jet quenching measurements. Since jet-induced medium response is partly caused by elastic scatterings between jet shower and medium partons, which is a well established mechanism of jet-medium interactions in most theoretical descriptions, any established models for jet transport should include them. Only then one can look for additional mechanisms that can also contribute to these jet quenching phenomena. 

Experimentally, one still needs data with higher precision since some unique features of medium response need to be confirmed with better background subtractions. The tension between ATLAS and CMS experimental data on the cone-size dependence of jet suppression still needs to be resolved, in particular for smaller cone size and lower jet $p_\mathrm{T}$. We also need to explore more novel measurements that can identify medium response, especially the long-sought-after Mach-cone which can help provide more direct constraints on the EoS of the QGP matter produced in relativistic heavy-ion collisions.


\section*{Acknowledgments}

We are grateful to Weiyao Ke, Guang-You Qin, Yasuki Tachibana and Daniel Pablos for helpful discussions. We thank Wei Chen, Yayun He, Tan Luo and Yuanyuan Zhang for their collaboration on many results presented in this review. This work is supported by the Director, Office of Energy Research, Office of High Energy and Nuclear Physics, Division of Nuclear Physics, of the U.S. Department of Energy under grant Nos. DE-SC0013460 and DE-AC02-05CH11231, the National Science Foundation (NSF) under grant Nos. ACI-1550300 and ACI-1550228 within the framework of the JETSCAPE Collaboration, and the National Natural Science Foundation of China (NSFC) under grant Nos. 11935007, 11221504 and 11890714.

\section*{References}
\bibliographystyle{iopart-num}
\bibliography{SCrefs}

\end{document}